\documentclass[pra,superscriptaddress,preprintnumbers, preprint, showpacs,amsmath,amssymb]{revtex4}
\usepackage{morefloats}
\usepackage{psfrag}
\usepackage{graphicx}
\usepackage{subfigure}

\def\be{\begin{equation}}       \def\ee{\end{equation}}
\def\bea{\begin{eqnarray}}      \def\eea{\end{eqnarray}}

\begin{document}

\title{Tuning entanglement and ergodicity in two-dimensional spin systems using impurities and anisotropy}

\author{Gehad Sadiek \footnote{Corresponding author: gehad@ksu.edu.sa}}
\affiliation{Department of Physics, King Saud University, Riyadh 11451, Saudi Arabia and\\
Department of Physics, Ain Shams University, Cairo 11566, Egypt}
\author{Qing Xu}
\affiliation{Department of Chemistry and Birck Nanotechnology center,
Purdue University, West Lafayette, Indiana 47907, USA}
\author{Sabre Kais}
\affiliation{Department of Chemistry and Birck Nanotechnology center,
Purdue University, West Lafayette, Indiana 47907, USA}

\begin{abstract}
We consider the entanglement in a two dimensional $XY$ model in an external magnetic field $h$. The model consists of a set of 7 localized spin-$\frac{1}{2}$ particles in a two dimensional triangular lattice coupled through nearest neighbor exchange interaction $J$. We examine the effect of single and double impurities in the system as well as the degree of anisotropy on the nearest neighbor entanglement and ergodicity of the system. We have found that the entanglement of the system at the different degrees of anisotropy mimics that of the one dimensional spin systems at the extremely small and large values of the parameter $\lambda = h/J$. The entanglement of the Ising and partially anisotropic system show phase transition in the vicinity of $\lambda = 2$ while the entanglement of the isotropic system suddenly vanishes there. Also we investigate the dynamic response of the system containing single and double impurities to an external exponential magnetic field at different degrees of anisotropy. We have demonstrated that the ergodicity of the system can be controlled by varying the strength and location of the impurities as well as the degree of anisotropy of the coupling.

\end{abstract}

\pacs{03.67.Mn, 03.65.Ud, 75.10.Jm}

\maketitle

\section{Introduction}

Quantum entanglement is a corner stone in the structure of quantum theory with no classical analog \cite{Peres1993}. Entanglement is a nonlocal correlation between two (or more) quantum systems such that the description of their states has to be done with reference to each other even if they are spatially well separated. Particular fields where entanglement is considered as a crucial resource are quantum teleportation, cryptography and quantum computation \cite{Nielsen2000,Boumeester2000}, where it provides the physical basis for manipulating the linear superposition of the quantum states used to implement the different computational algorithms. On the other hand, many questions regarding the behavior of the complex quantum systems significantly rely on a deep understanding and a good quantification of the entanglement \cite{Sondhi1997,Osborne2002,HuangZ2004,ZhangJF2009,Sadiek2010,Alkurtass2011}. Particularly, entanglement is considered as the physical resource responsible for the long range correlations taking place in many-body systems during quantum phase transitions. There has been great interest in studying the different sources of errors in quantum computing and their effect on quantum gate operations \cite{Jones2003, Cummins2003}. Different approaches have been proposed for protecting quantum systems during the computational implementation of algorithms such as quantum error correction \cite{Shor1995} and  decoherence-free subspace \cite{Bacon2000,DiVincenzo2000}. Nevertheless, realizing a practical protection against the different types of induced decoherence is still a hard task. Therefore, studying the effect of naturally existing sources of errors such as impurities and lack of isotropy in coupling between the quantum systems implementing the quantum computing algorithms is a must. Furthermore, considerable efforts should be devoted to utilizing such sources to tune the entanglement rather than eliminating them. The effect of impurities and anisotropy of coupling between neighbor spins in a one dimensional spin system has been investigated \cite{Osenda2003}. It was demonstrated that the entanglement can be tuned in a class of one-dimensional systems by varying the anisotropy of the coupling parameter as well as by introducing impurities into the spin system. For a physical quantity to be eligible for an equilibrium statistical mechanical description it has to be ergodic, which means that its time average coincides with its ensemble average. To test ergodicity for a physical quantity one has to compare the time evolution of its physical state to the corresponding equilibrium state. There has been an intensive efforts to investigate ergodicity in one-dimensional spin chains where it was demonstrated that the entanglement, magnetization, spin-spin correlation functions are non-ergodic in Ising and XY spin chains for finite number of spins as well as at the thermodynamic limit \cite{Barouch1970,Sen(De)2004,HuangZ2006,Sadiek2010}.

Studying quantum entanglement in two-dimensional systems face more obstacles in comparison to the one dimensional case, particularly the rapid increase in the dimension of the Hilbert spaces which lead to much larger scale calculations relying mainly on the numerical methods. The existence of exact solutions has contributed enormously to the understanding of the entanglement for 1D systems \cite{Lieb1961,Sachdev2001,HuangZ2006,Sadiek2010}. In a previous work, the entanglement in a 19-site two-dimensional transverse Ising model at zero temperature \cite{XuQ2010} was studied. The spin-$1/2$ particles are coupled through an exchange interaction $J$ and subject to an external time-independent magnetic field $h$. It was demonstrated that for such a class of systems the entanglement can be tuned by varying the parameter $\lambda=h/J$ and also by introducing impurities into the system, which showed a quantum phase transition at a  critical value of the parameter $\lambda$ in the vicinity of $2$. Recently, we have investigated the time evolution of entanglement in a two dimensional triangular transverse Ising system with seven spins in an external magnetic field \cite{XuQ2011}. Different time dependent forms of the magnetic field were applied. The system have demonstrated different responses based on the type of applied field, where for a smoothly changing magnetic field the system entanglement follows the profile of the field very closely.  

In this paper, we consider the entanglement in a two-dimensional $XY $triangular spin system, where the nearest neighbor spins are coupled through an exchange interaction $J$ and subject to an external magnetic field $h$. We consider the system at different degrees of anisotropy to test its effect on the system entanglement and dynamics.
The number of spins in the system is 7 with a number of impurities existing. We consider two different cases of impurities, the first case is a single impurity existing either at the border of the system or at the center with the coupling strength between the impurity spin and its neighbors different from that between the rest of the spins. The second case is double impurities, existing both at the border or one at the border and one at the center. We consider the coupling between the two impurities as $J'$, which is different from the coupling $J''$ between each one of them and its neighbors, while the interaction among the other spins is $J$.

We show that the entanglement profile of the system at different degrees of anisotropy has great resemblance to that of the one dimensional spin systems as the parameter $\lambda \rightarrow 0$ and $\infty$. On the other hand, both the Ising and the partially anisotropic systems show phase transition behavior in the vicinity of $\lambda=2$ but the isotropic system show sharp step variations in the same region before suddenly vanishing. Examining the effect of an external exponential magnetic field on the time evolution of the entanglement showed that the ergodicity of the system can be tuned by varying the strength and location of the impurities and the degree of anisotropy in the system.

This paper is organized as follows. In the next section we present our model and quantification of entanglement. In sec. III we consider the case of a single impurity. In sec IV we study the system with a double impurity. We conclude in sec V.
\section{Model and quantification of entanglement}
We consider a set of 7 localized spin-$\frac{1}{2}$ particles in a two dimensional triangular lattice coupled through exchange 
interaction $J$ and subject to an external time-dependent magnetic field of strength $h(t)$. All the particles are identical 
except one (or two) of them which are considered impurities. The Hamiltonian for such a system is given by 
\begin{equation}
\label{Sch_equ}
H=-\frac{(1+\gamma)}{2}\sum_{<i,j>}J_{i,j}\sigma_{i}^x\sigma_{j}^x -\frac{(1-\gamma)}{2}\sum_{<i,j>}J_{i,j}\sigma_{i}^y\sigma_{j}^y  - h(t) \sum_{i} \sigma_{i}^z,
\end{equation}
\begin{figure}[htbp]
\begin{minipage}[c]{\textwidth}
 \centering
   \includegraphics[width=8 cm]{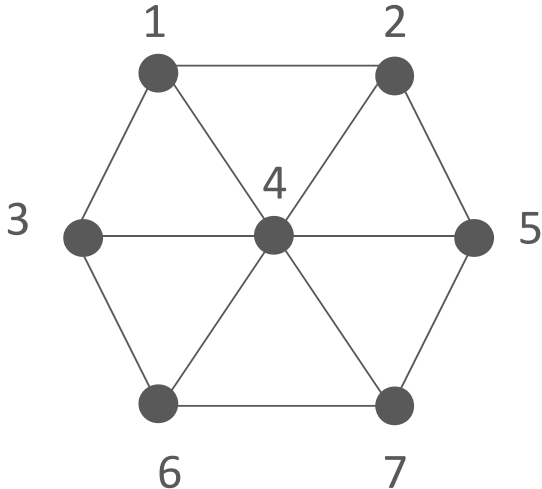}
  \caption{{\protect\footnotesize The two dimensional triangular spin lattice in presence of an external transverse magnetic field.}}
 \label{Model}
 \end{minipage}
 \end{figure}
where $<i,j>$ is a pair of nearest-neighbors sites on the lattice, $J_{i,j}=J$ for all sites except the sites nearest to an impurity site. For a single impurity, the coupling between the impurity and its neighbors $J_{i,j}=J'=(\alpha+1)J$, where $\alpha$ measures the strength of the impurity. For double impurities $J_{i,j}=J'=(\alpha_1+1)J$ is the coupling between the two impurities and $J_{i,j}=J''=(\alpha_2+1)J$ is the coupling between any one of the two impurities and its neighbors while the coupling is just $J$ between the rest of the spins.

For this model it is convenient to set $J=1$. For a system of 7 spins, its Hilbert space is huge with $2^7$ dimensions, yet it is exactly diagonalizable using the standard computational techniques. Exactly solving Schrodinger equation of the Hamiltonian (\ref{Sch_equ}), yielding the system energy eigenvalues ${E_i}$ and eigenfunctions ${\psi_i}$. The density matrix of the system is defined by
\begin{equation}
\label{dens_matrix}
\rho = |\psi_0 \rangle \langle \psi_0 | \; ,
\end{equation}
where $|\psi_0\rangle$ is the ground state energy of the entire spin system. 
We confine our interest to the entanglement between two spins, at any sites $i$ and $j$ \cite{Osterloh2002}. All the information needed in this case, at any moment $t$, is contained in the reduced density matrix $\rho_{i, j}(t)$ which can be obtained from the entire system density matrix by integrating out all the spins states except $i$ and $j$. We adopt the entanglement of formation, as a well known  measure of entanglement where Wootters \cite{Wooters1998} has shown that, for a pair of binary qubits, the concurrence $C$, which goes from $0$ to $1$, can be taken
as a measure of entanglement. The concurrence between two sites $i$ and
$j$ is defined as
\begin{equation}
\label{concurrence}
C(\rho)=max\{0,\epsilon_1-\epsilon_2-\epsilon_3-\epsilon_4\},
\end{equation}
where the $\epsilon_i$'s are the eigenvalues of the Hermitian matrix
$R\equiv\sqrt{\sqrt{\rho}\tilde{\rho}\sqrt{\rho}}$ with
$\tilde{\rho}=(\sigma^y \otimes
\sigma^y)\rho^*(\sigma^y\otimes\sigma^y)$ and $\sigma^y$ is the
Pauli matrix of the spin in y direction.
For a pair of qubits the entanglement can be written as,
\begin{equation}
\label{entanglement}
E(\rho)=\epsilon(C(\rho)),
\end{equation}
where $\epsilon$ is a function of the ``concurrence'' $C$
\begin{equation}
\epsilon(C)=h\left(\frac{1-\sqrt{1-C^2}}{2}\right),
\end{equation}
where $h$ is the binary entropy function
\begin{equation}
h(x)=-x\log_{2}x-(1-x)log_{2}(1-x).
\end{equation}

In this case, the entanglement of formation is given in terms of another entanglement measure, the concurrence $C$. The dynamics of entanglement is evaluated using the same techniques applied in our previous work \cite{XuQ2011}. Specifically, we apply the step-by-step time-evolution projection technique, which was proved to give the same exact result as the matrix transformation technique, where both techniques were introduced in \cite{XuQ2011}, but 20 times faster. In this technique we assume that our system is initially, at $t_0$, in the ground state at zero temperature $|\phi\rangle$ with energy, say, $\varepsilon$ in an external magnetic field with strength $a$. The magnetic field is turned to a new value $b$ and the system Hamiltonian becomes $H$ with $N$ eigenpairs $E_i$ and $|\psi_{i}\rangle$. The original state $|\phi\rangle$ can be expanded in the basis $\{|\psi_{i}\rangle\}$:
\be
|\phi\rangle=c_{1}|\psi_{1}\rangle+c_{2}|\psi_{2}\rangle+...+c_{N}|\psi_{N}\rangle,
\ee
where
\be
c_i=\langle\psi_i|\phi\rangle.
\ee
When $H$ is independent of time between $t$ and $t_0$ then we can write
\be
U(t,\, t_{0})\,|\psi_{i,t_0}\rangle=e^{-iH(t>t_0)(t-t_{0})/\hbar}|\psi_{i,t_0}\rangle=e^{-iE_{i}(t-t_{0})/\hbar}|\psi_{i,t_0}\rangle,
\ee
where $U(t,\, t_{0})$ is the time evolution operator. The ground state will evolve with time as
\bea\label{projection_sum}
|\phi(t)\rangle &=& c_{1}|\psi_{1}\rangle e^{-iE_{1}(t-t_0)}+c_{2}|\psi_{2}\rangle e^{-iE_{2}(t-t_0)}+...+c_{N}|\psi_{N}\rangle e^{-iE_{N}(t-t_0)}\nonumber\\
&=& \sum_{i=1}^{N}c_{i}|\psi_{i}\rangle e^{-iE_{i}(t-t_0)}.
\eea
and the pure state density matrix becomes
\be
\rho(t)=|\phi(t)\rangle\langle\phi(t)|.
\ee
Simply any complicated function can be treated as a collection of step functions. When the state evolves to the next step just repeat the procedure to get the next step results. Of course the lack of smoothness in the magnetic field function imposes a challenging obstacle in the calculations but this can be overcome by choosing a proper small enough time step. Because the size of our 7-site system is still manageable, in our actual calculations, we included all the $2^7=128$ states in every step, without any truncation of the higher energy eigenstates. This ensures us no approximation in this step. But the method itself is aiming at larger size system, like 19 sites XY model. By then, due to the computation limit, cutting off higher energy eigenstates might be a necessary action.
\section{Single impurity}
\subsection{Static system with border impurity}
\begin{figure}[htbp]
\begin{minipage}[c]{\textwidth}
 \centering
   \subfigure{\includegraphics[width=8 cm]{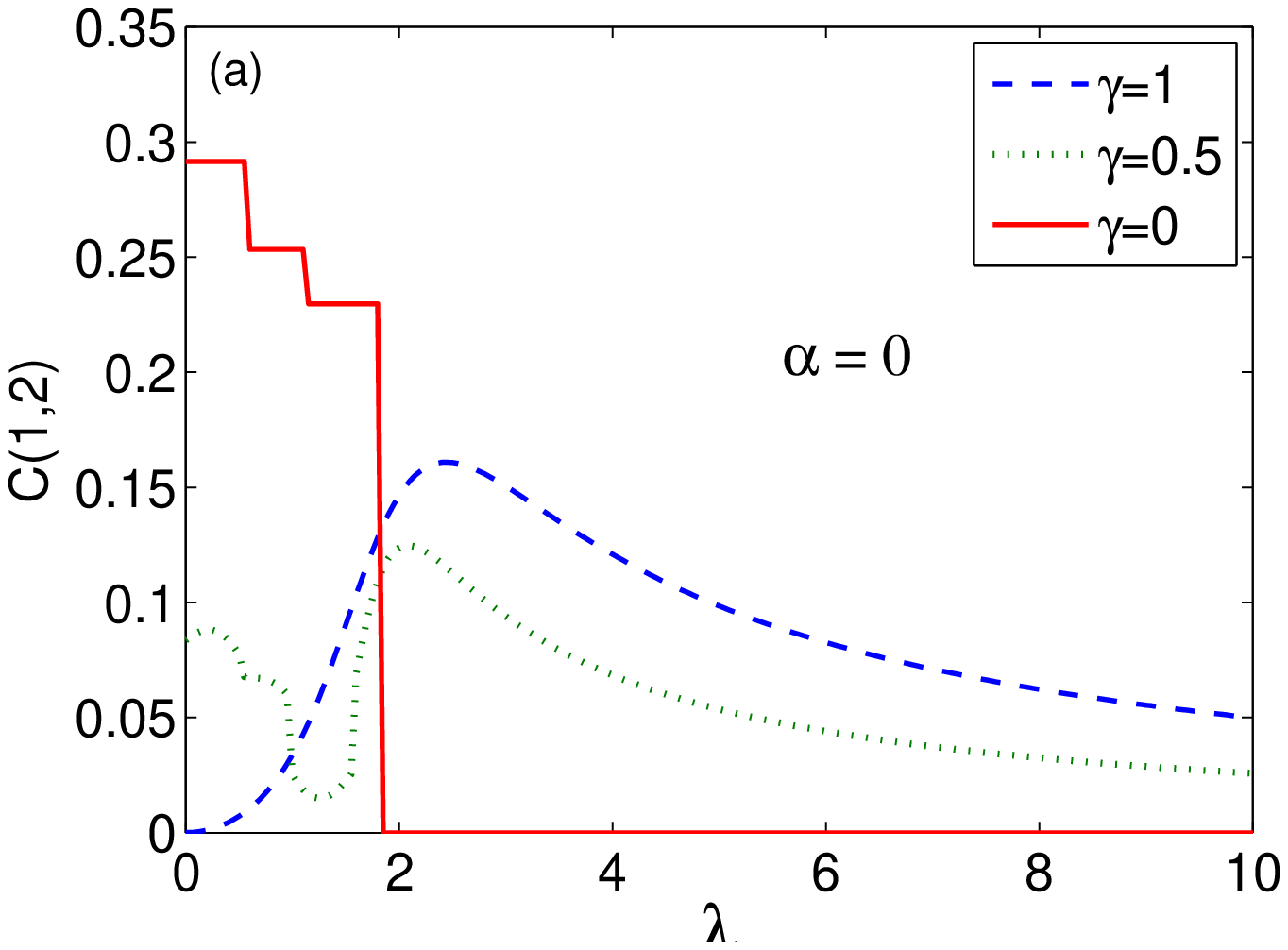}}\quad
   \subfigure{\includegraphics[width=8 cm]{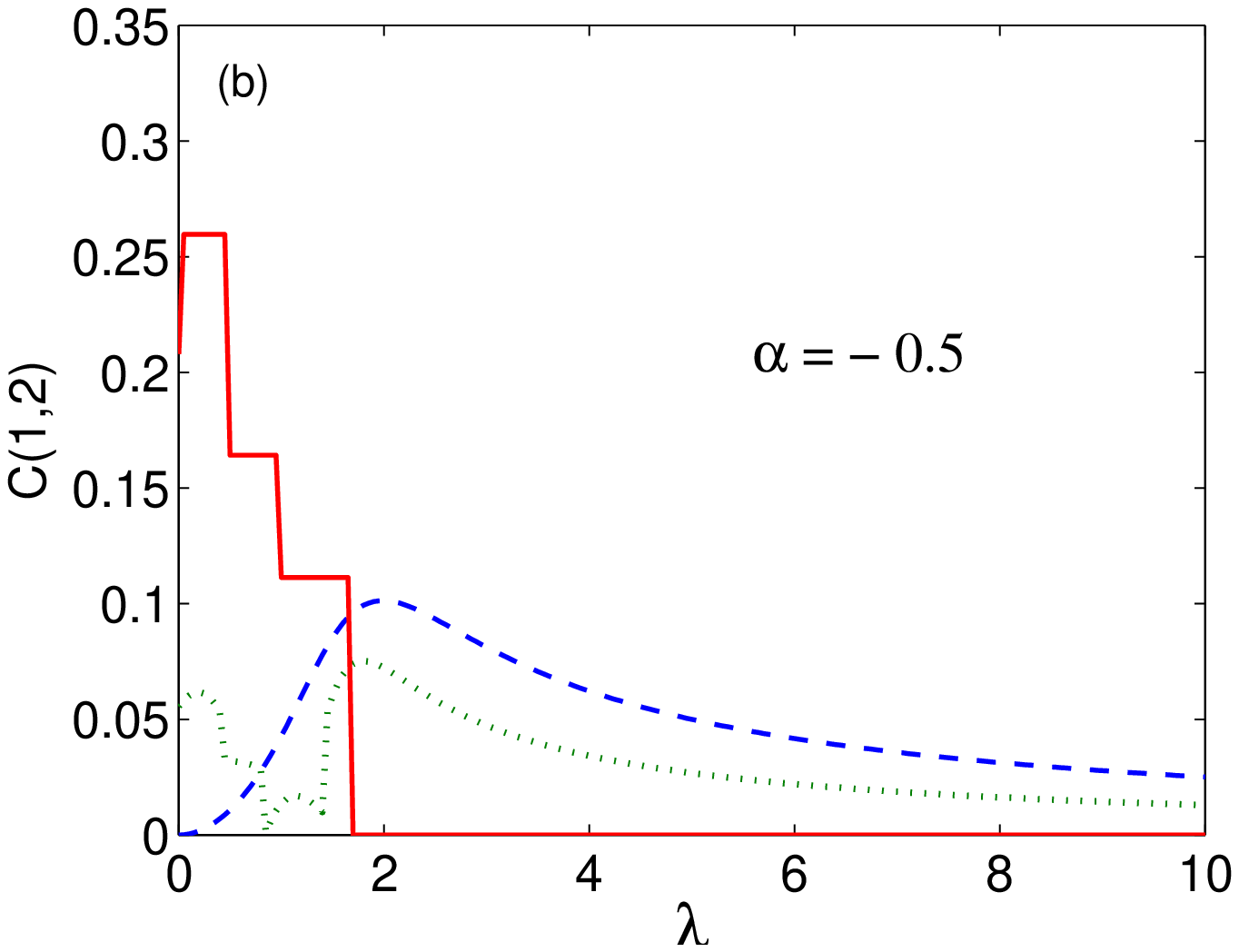}}\\
   \subfigure{\includegraphics[width=8 cm]{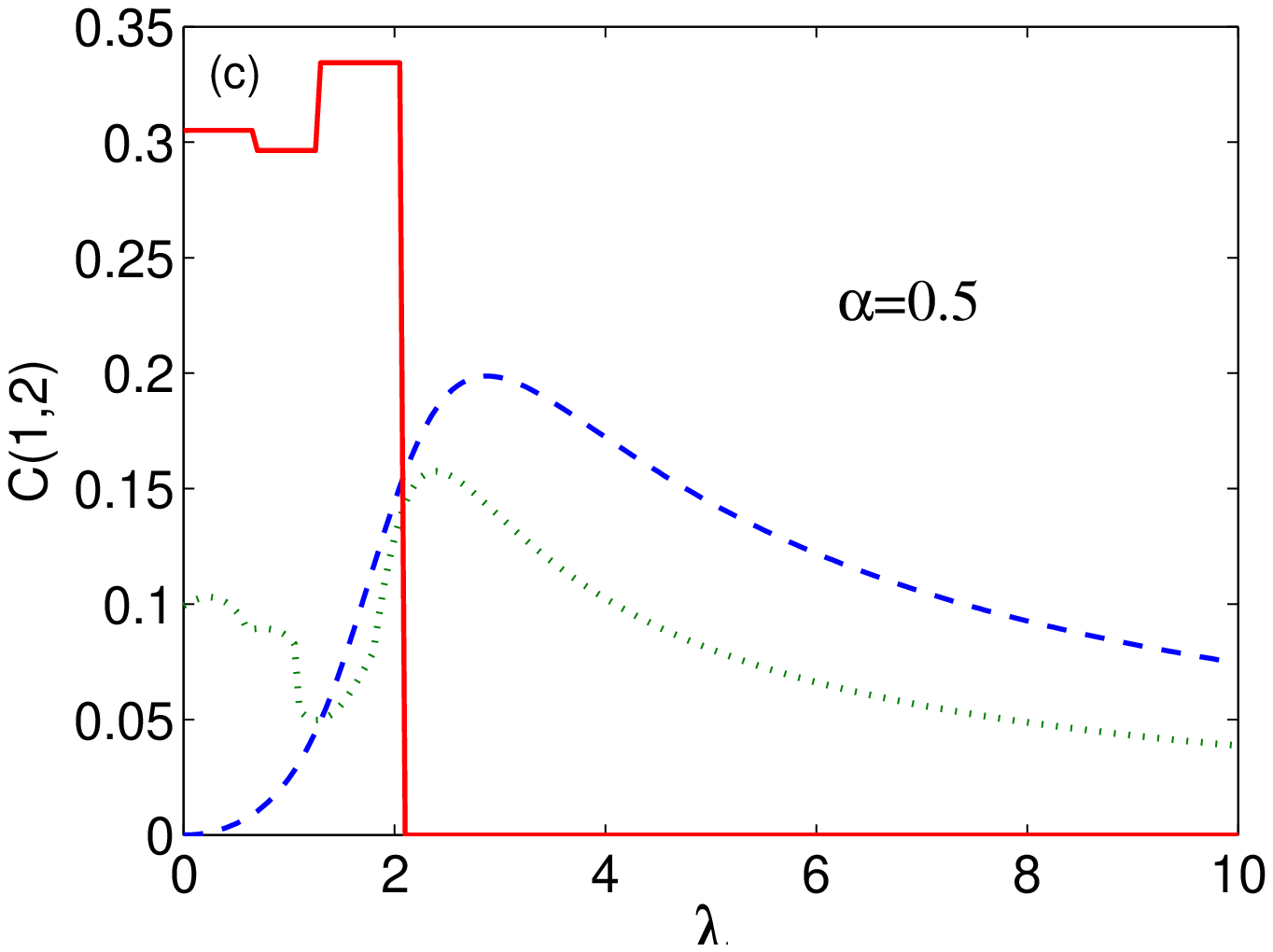}}\quad
   \subfigure{\includegraphics[width=8 cm]{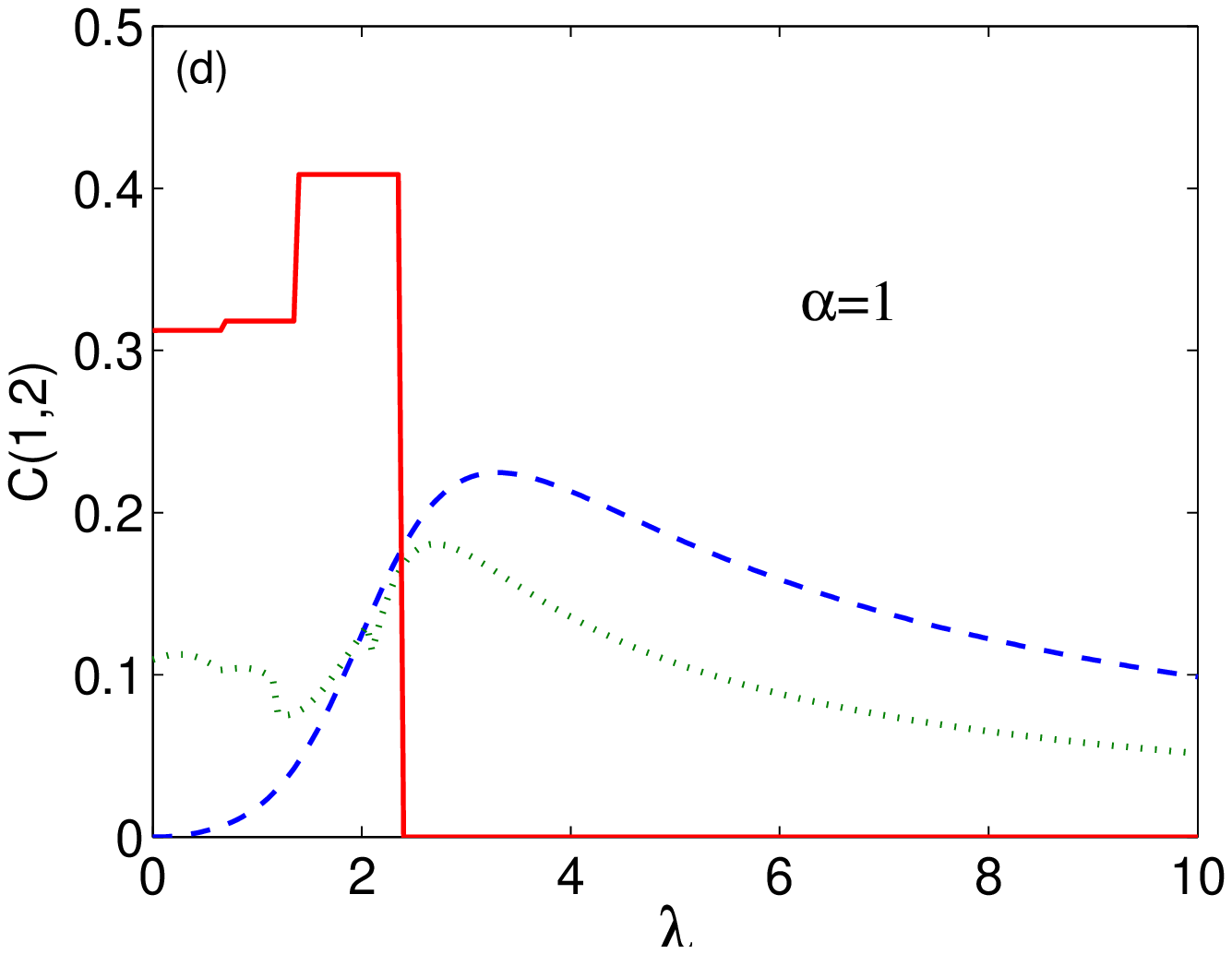}}   
   \caption{{\protect\footnotesize (Color online) The concurrence $C(1,2)$ versus the parameter $\lambda$ with a single impurity at the border site 1 with different impurity coupling strengths $\alpha = -0.5, 0, 0.5, 1$ for different degrees of anisotropy $\gamma = 1, 0.5, 0$ as shown in the subfigures. The legend for all subfigures is as shown in subfigure (a).}}
 \label{B_C12}
 \end{minipage}
\end{figure}

We define a dimensionless coupling parameter $\lambda=h/J$ and we set $J=1$ throughout this paper for convenience. We start by considering the effect of a single impurity located at the border site 1. The concurrence between the impurity site 1 and site 2, $C(1,2)$, versus the parameter $\lambda$ for the three different models, Ising ($\gamma=1$), partially anisotropic ($\gamma=0.5$) and isotropic XY ($\gamma=0$) at different impurity strengths ($\alpha= -0.5, 0, 0.5, 1$) is in fig.~\ref{B_C12}. 
Firstly, the impurity paramter $\alpha$ is set to zero. For the corresponding Ising model, the concurrence $C(1,2)$, in fig.~\ref{B_C12}(a), demonstrates the usual phase transition behavior where it starts at zero value and increases gradually as $\lambda$ increases reaching a maximum at $\lambda \approx 2$ then decays as $\lambda$ increases further. As the degree of anisotropy decreases the behavior of the entanglement changes, where it starts with a finite value at $\lambda=0$ and then shows a step profile for the small values of $\lambda$. For the partially anisotropic case, the step profile is smooth and the entanglement mimics the Ising case as $\lambda$ increases but with smaller magnitude. The entanglement of isotropic XY system shows a sharp step behavior then suddenly vanishes before reaching $\lambda=2$. Interestingly, the entanglement behavior of the two-dimensional spin system at the different degrees of anisotropy mimics the behavior of the one-dimensional spin system at the same degrees of anisotropy at the extreme values of the parameter $\lambda$. The ground state of the one-dimensional Ising model is characterized by a quantum phase transition that takes place at the critical value $h/J = 1$ \cite{Osborne2002, Sadiek2010} which corresponds to a maximum entanglement in the system. The order parameter is the magnetization $\langle \sigma^x \rangle$ which is different from zero for $J \geq h$ and zero otherwise. The ground state is paramagnetic when $J/h \rightarrow 0$ where the spins get aligned in the magnetic field direction, the $z$-direction. It is ferromagnetic when $J/h\rightarrow\infty$ where the spins are aligned in the $x$-direction. Both cases cause zero entanglement. Comparing the entanglement behavior in the two-dimensional Ising spin system with the one-dimensional system, one can see a great resemblance except that the critical value becomes $h/J \approx 2$ in the two dimensional case as shown in fig.~\ref{B_C12}. On the other hand, for the partially anisotropic and isotropic $XY$ systems, the entanglements of the two-dimensional and one-dimensional system agrees at the extreme values of $\lambda$ where it vanishes for $h >> J$ and reaches a finite value for $h << J$. The former case corresponds to an alignment of the spins in the $z$-direction, paramagnetic state, while the latter case corresponds to alignment in the $x$ and $y$-directions which a ferromagnetic state.

The effect of a weak impurity ($J'<J$), $\alpha=-0.5$, is shown in fig.~\ref{B_C12}(b) where the entanglement behavior is the same as before except that the entanglement magnitude is reduced compared with the pure case. On the other hand,  considering the effect of a strong impurity ($J'> J$), where $\alpha=0.5$ and 1, as shown in fig.~\ref{B_C12}(c) and fig.~\ref{B_C12}(d) respectively, one can see that the entanglement profile for $\gamma =1$ and 0.5 have the same overall behavior as in the pure and weak impurity cases except that the entanglement magnitude becomes higher as the impurity gets stronger and the peaks shift toward higher $\lambda$ values. Nevertheless, the isotropic $XY$ system behaves differently from the previous cases where it starts to increase first in a step profile before suddenly dropping to zero again, which will be explained latter.
\begin{figure}[htbp]
\begin{minipage}[c]{\textwidth}
 \centering
   \subfigure{\includegraphics[width=8 cm]{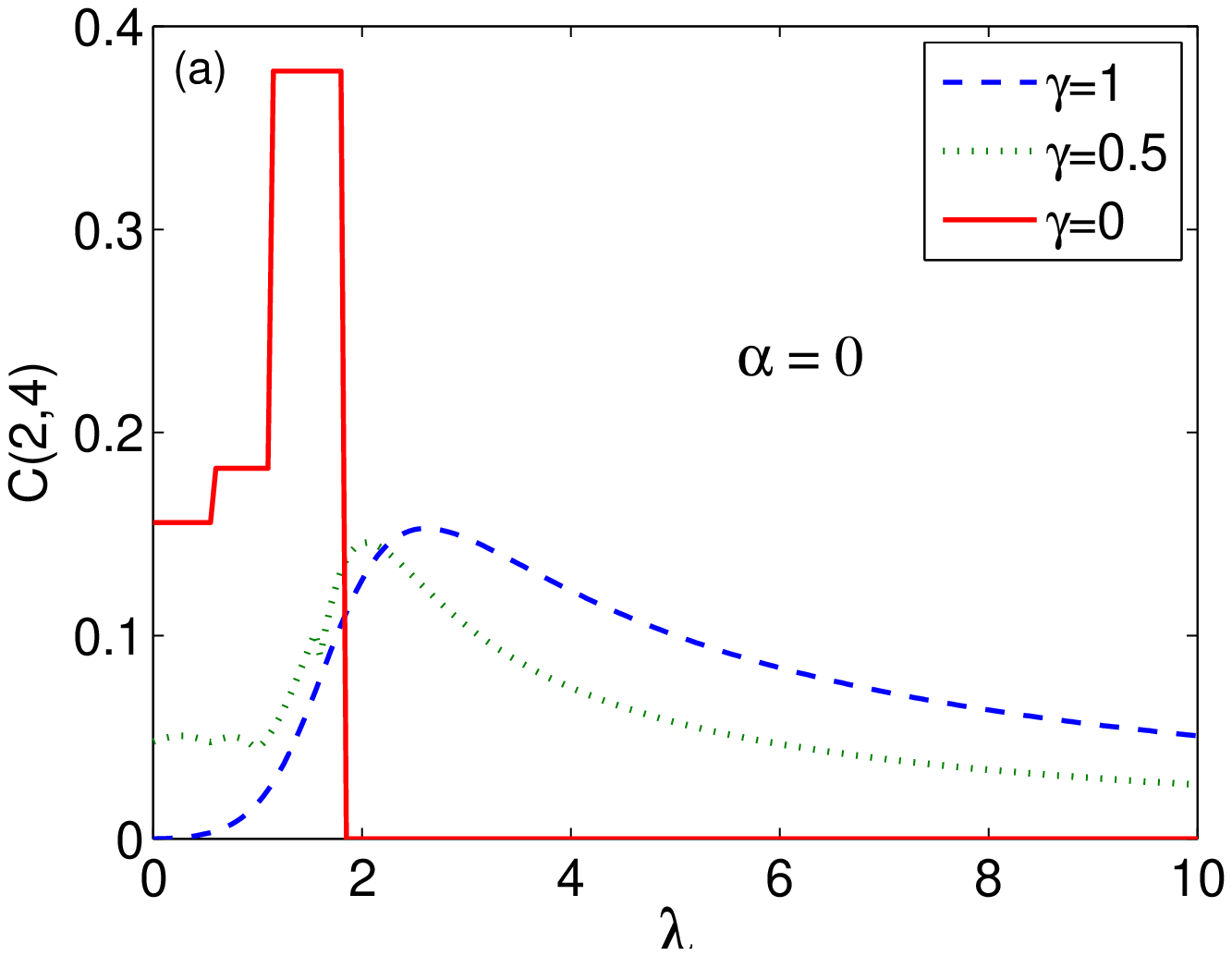}}\quad
   \subfigure{\includegraphics[width=8 cm]{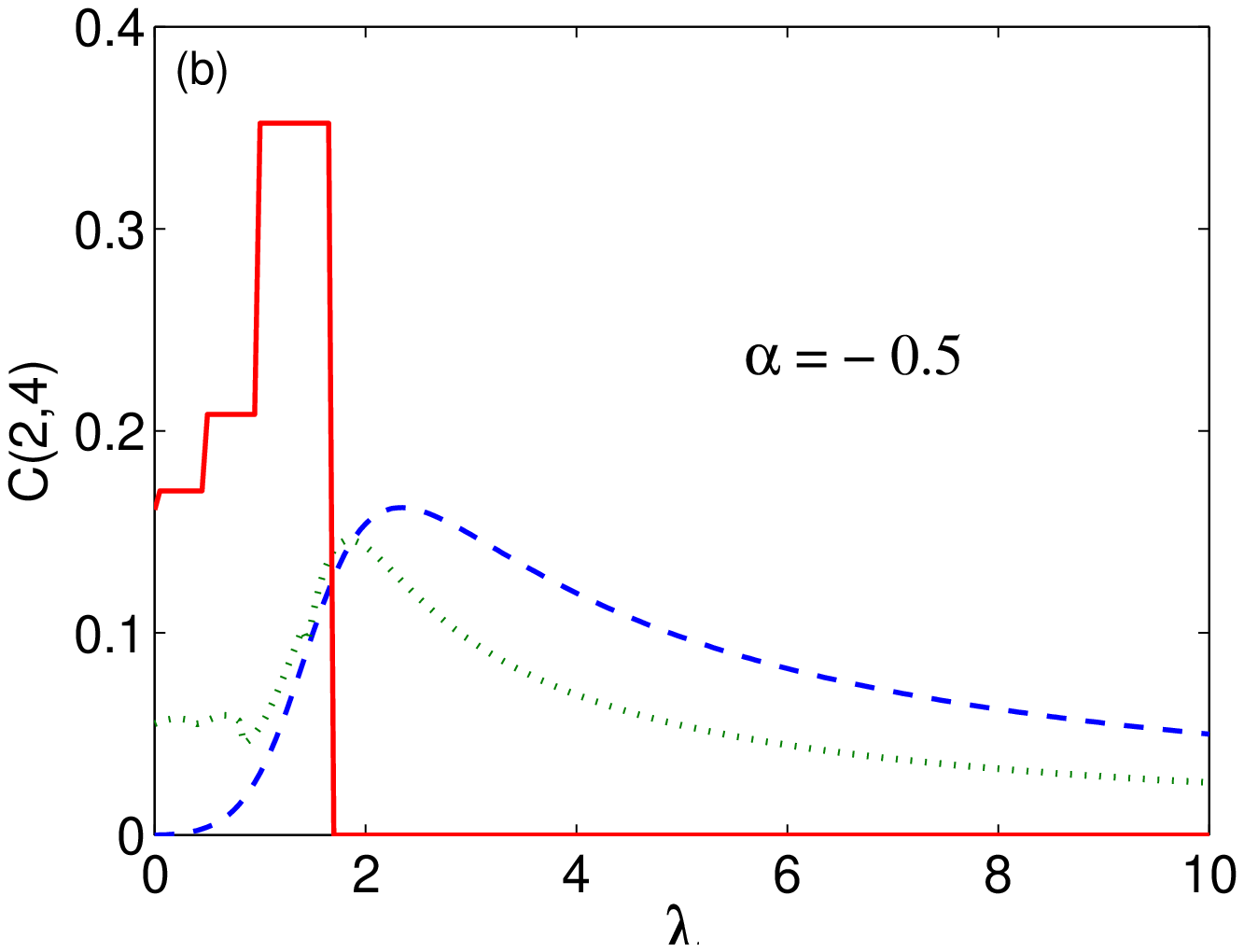}}\\
   \subfigure{\includegraphics[width=8 cm]{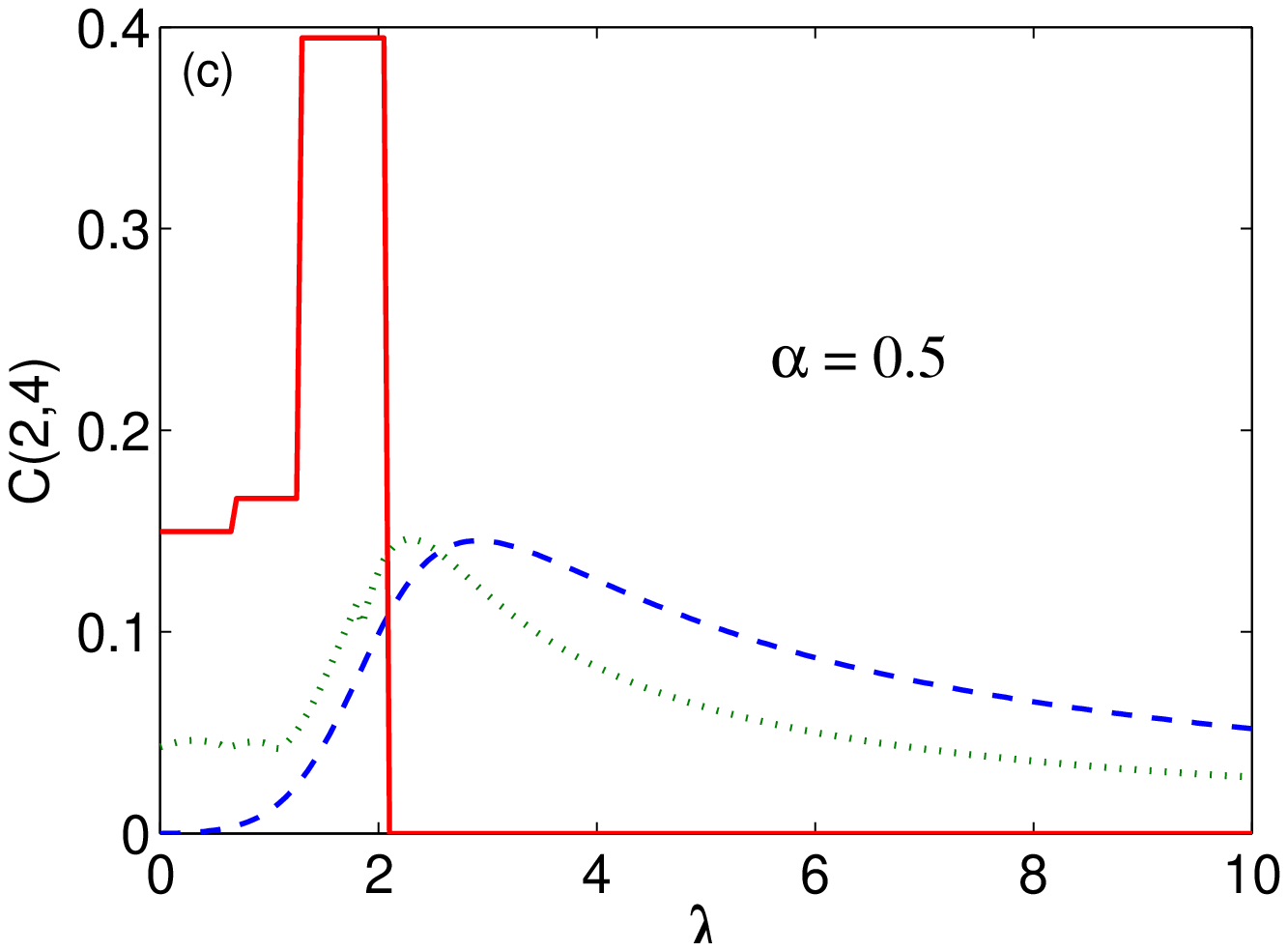}}\quad
   \subfigure{\includegraphics[width=8 cm]{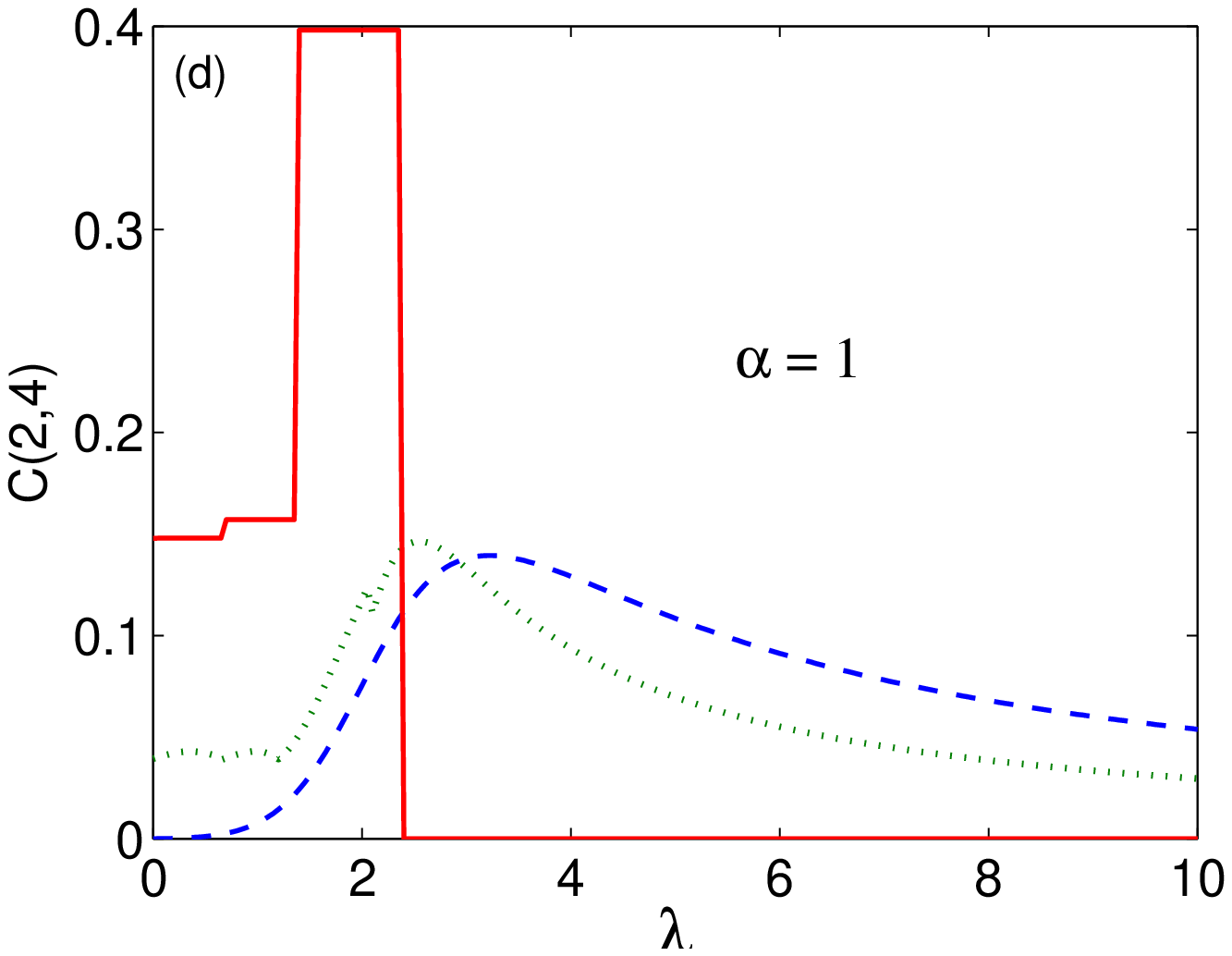}}
   \caption{{\protect\footnotesize (Color online) The concurrence $C(2,4)$ versus the parameter $\lambda$ with a single impurity at the border site 1 with different impurity coupling strengths $\alpha = -0.5, 0, 0.5, 1$ for different degrees of anisotropy $\gamma = 1, 0.5, 0$ as shown in the subfigures. The legend for all subfigures is as shown in subfigure (a).}}
 \label{B_C24}
 \end{minipage}
\end{figure}
To study the entanglement between two sites, none of them is impurity, we consider $C(2,4)$ which is depicted in fig.~\ref{B_C24}. There are two main differences between the behavior of $C(2,4)$ and $C(1,2)$. Firstly, the magnitude of the entanglement envelope is higher for $C(2,4)$ for $\gamma=0.5$ and $0$ (but not $\gamma=1$) when $\alpha=0$, while $C(2,4)$ is greater than $C(1,2)$ for all $\gamma$ values for the weak impurity case, $\alpha=0$. This is an interesting result as internal sites entanglement should be smaller in value than the edge sites. Secondly, the entanglement of the isotropic XY case increases in a multi-step profile for all values of $\alpha$ before suddenly dropping to zero. 
\subsection{Static system with center impurity}
\begin{figure}[htbp]
\begin{minipage}[c]{\textwidth}
\centering
   \subfigure{\includegraphics[width=8 cm]{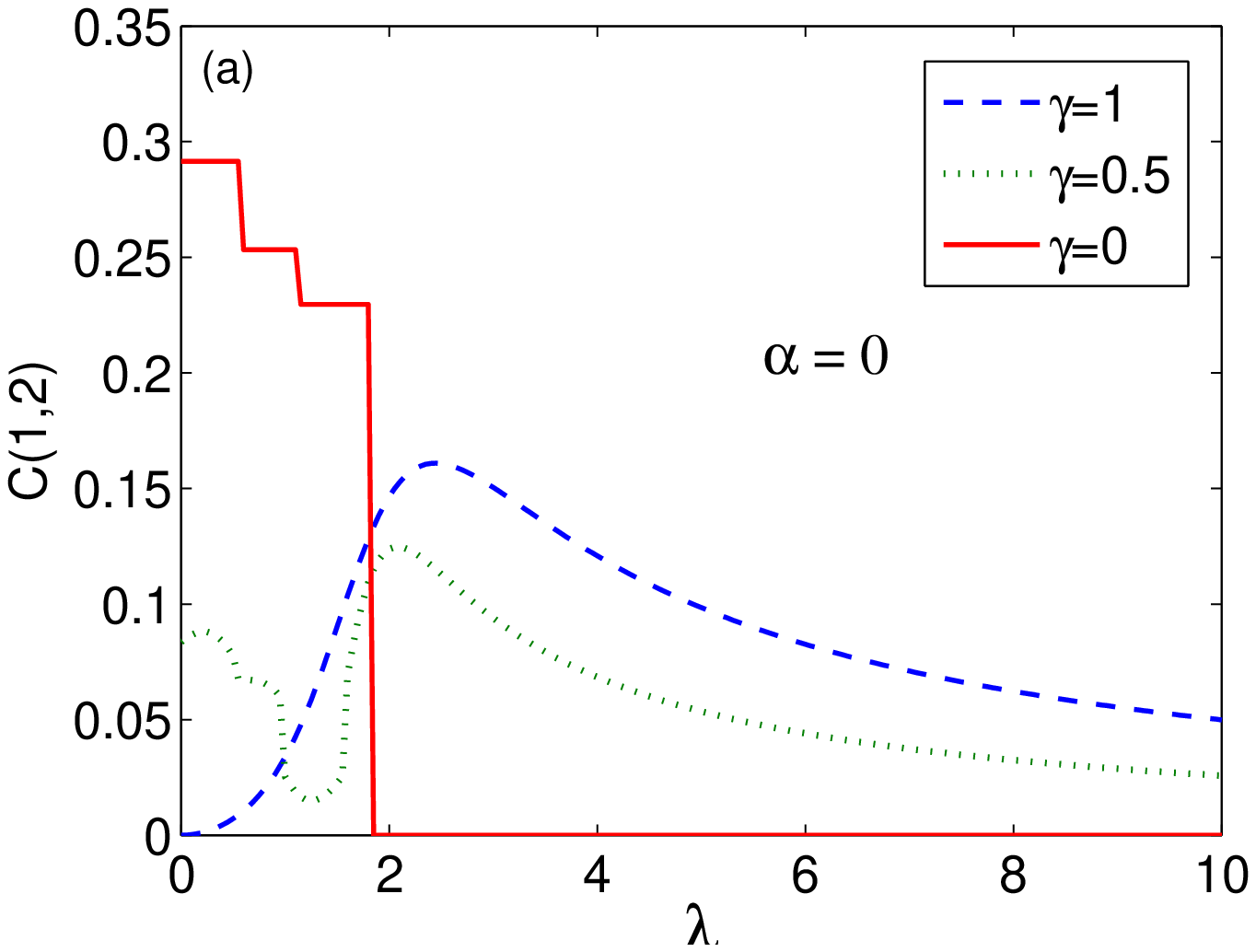}}\quad
   \subfigure{\includegraphics[width=8 cm]{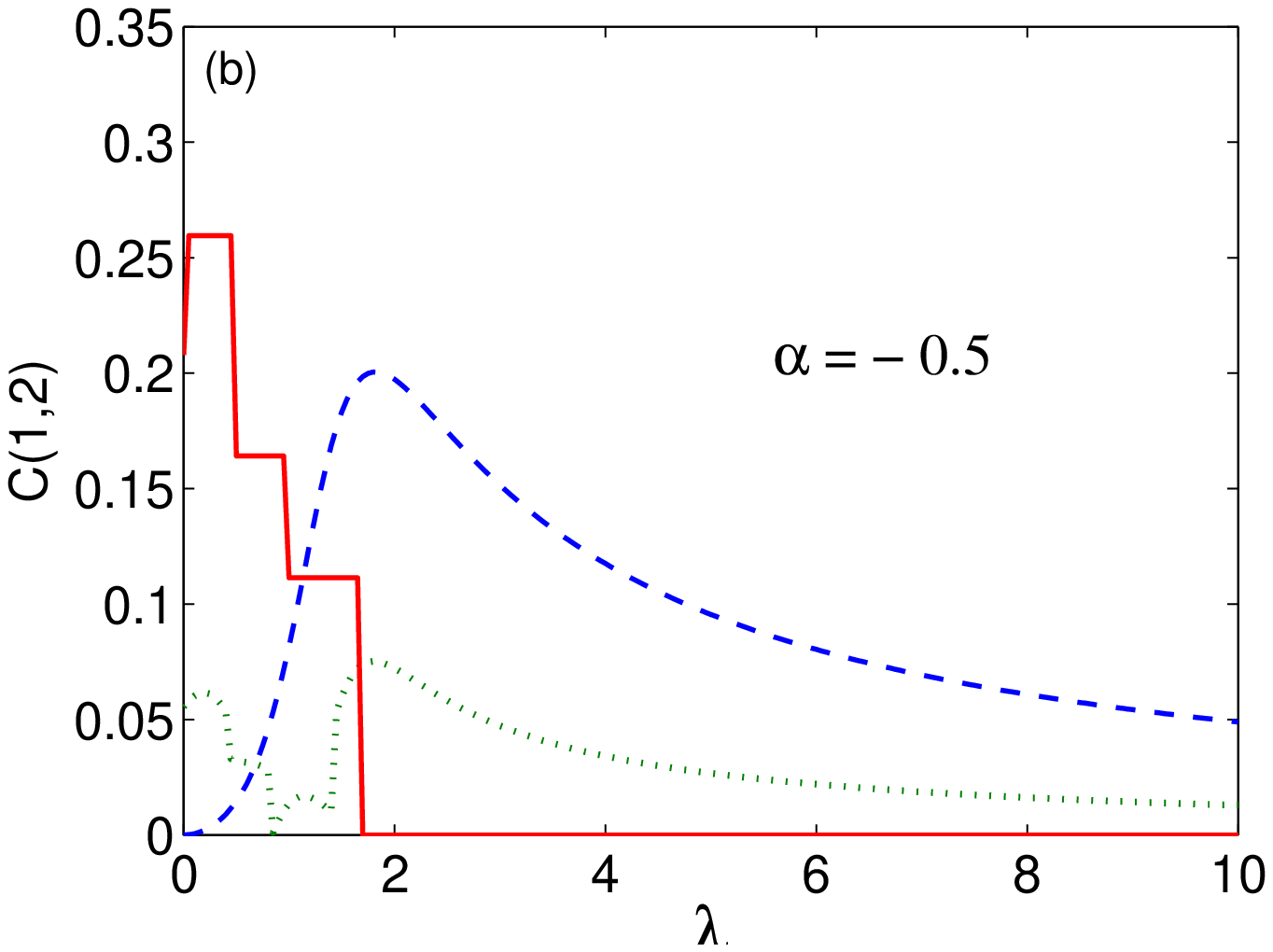}}\\
   \subfigure{\includegraphics[width=8 cm]{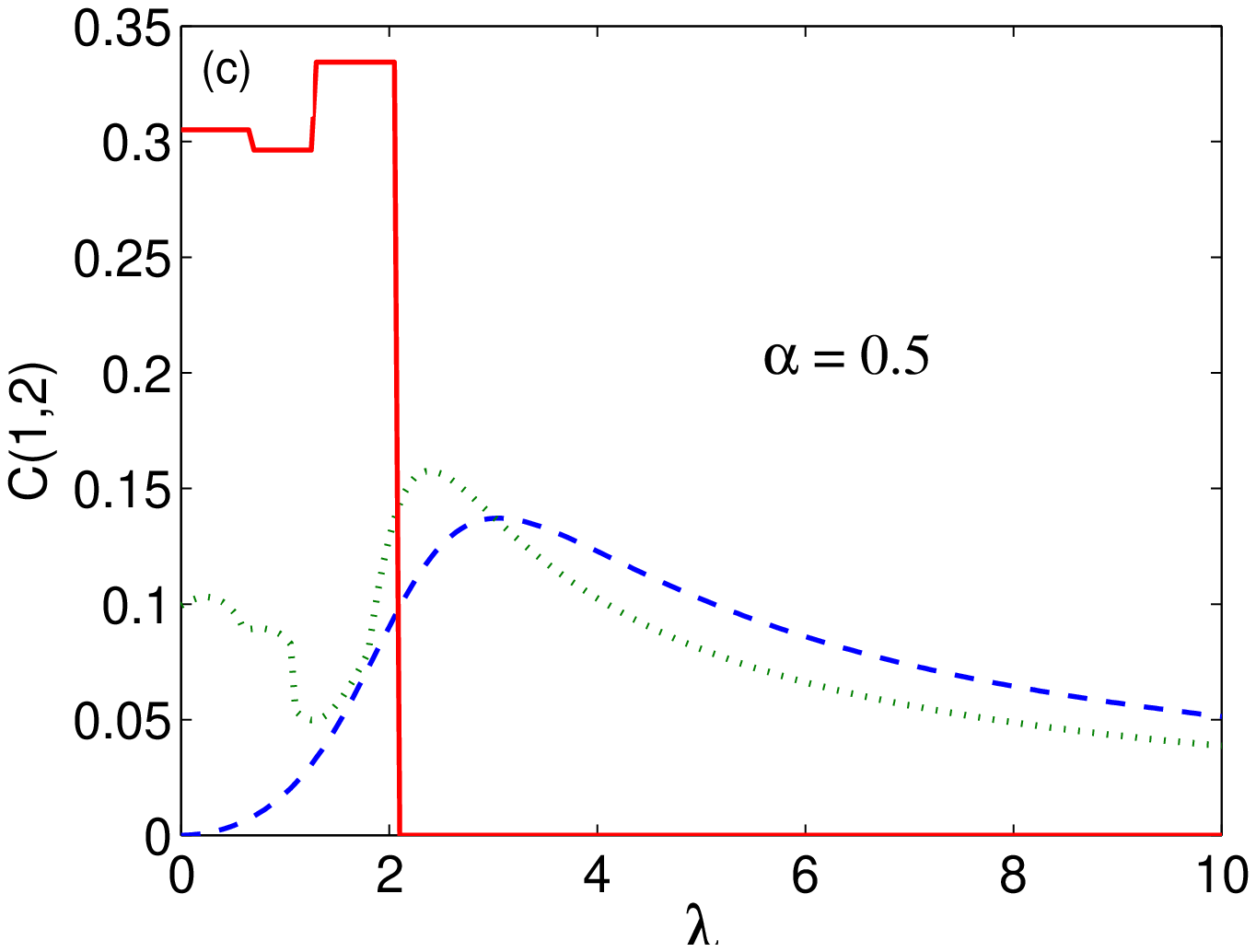}}\quad
   \subfigure{\includegraphics[width=8 cm]{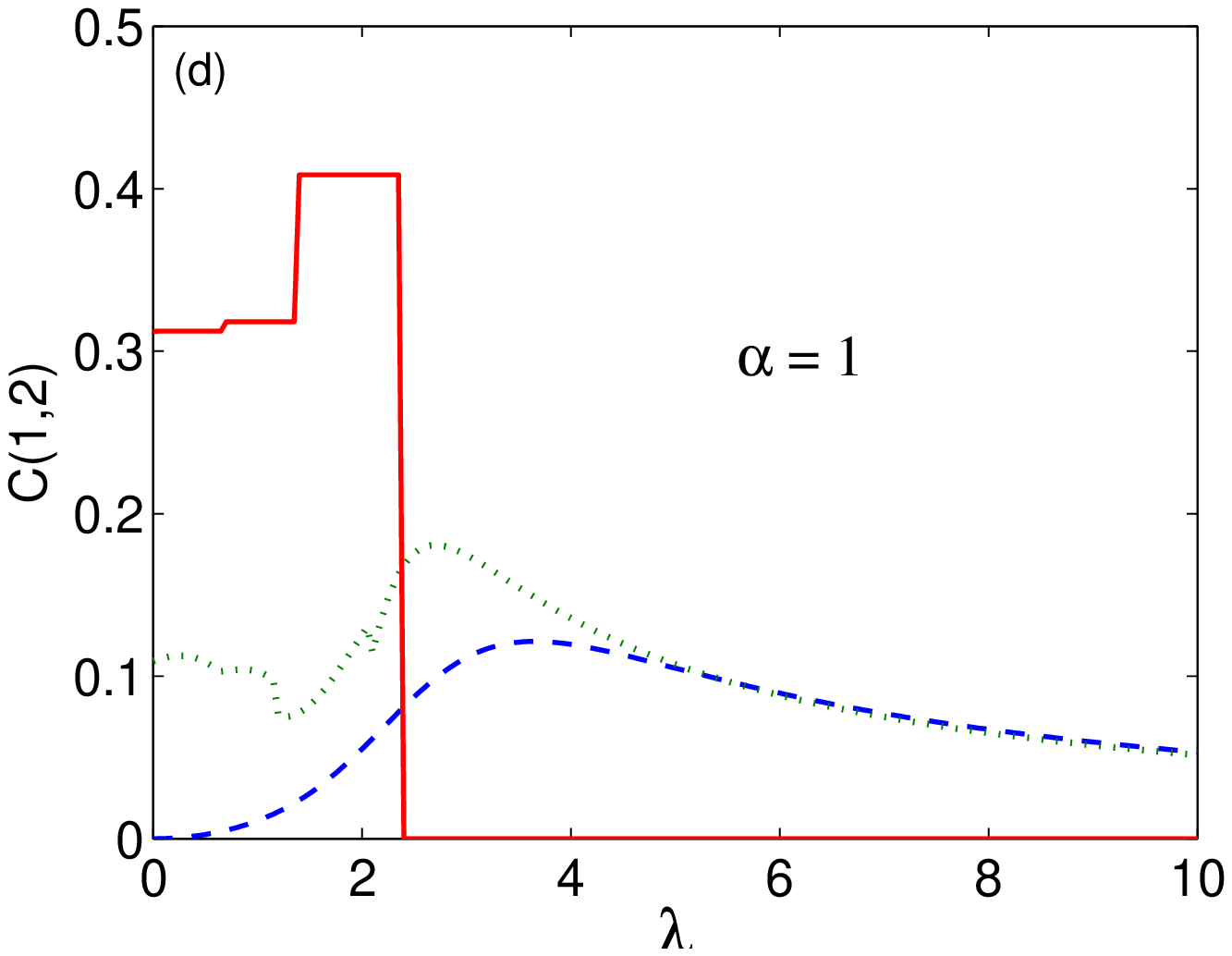}}
    \caption{{\protect\footnotesize (Color online) The concurrence $C(1,2)$ versus the parameter $\lambda$ with a single impurity at the central site 4 with different impurity coupling strengths $\alpha = -0.5, 0, 0.5, 1$ for different degrees of anisotropy $\gamma = 1, 0.5, 0$ as shown in the subfigures. The legend for all subfigures is as shown in subfigure (a).}}
 \label{C_C12}
 \end{minipage}
\end{figure}

\begin{figure}[htbp]
\begin{minipage}[c]{\textwidth}
\centering
   \subfigure{\includegraphics[width=8 cm]{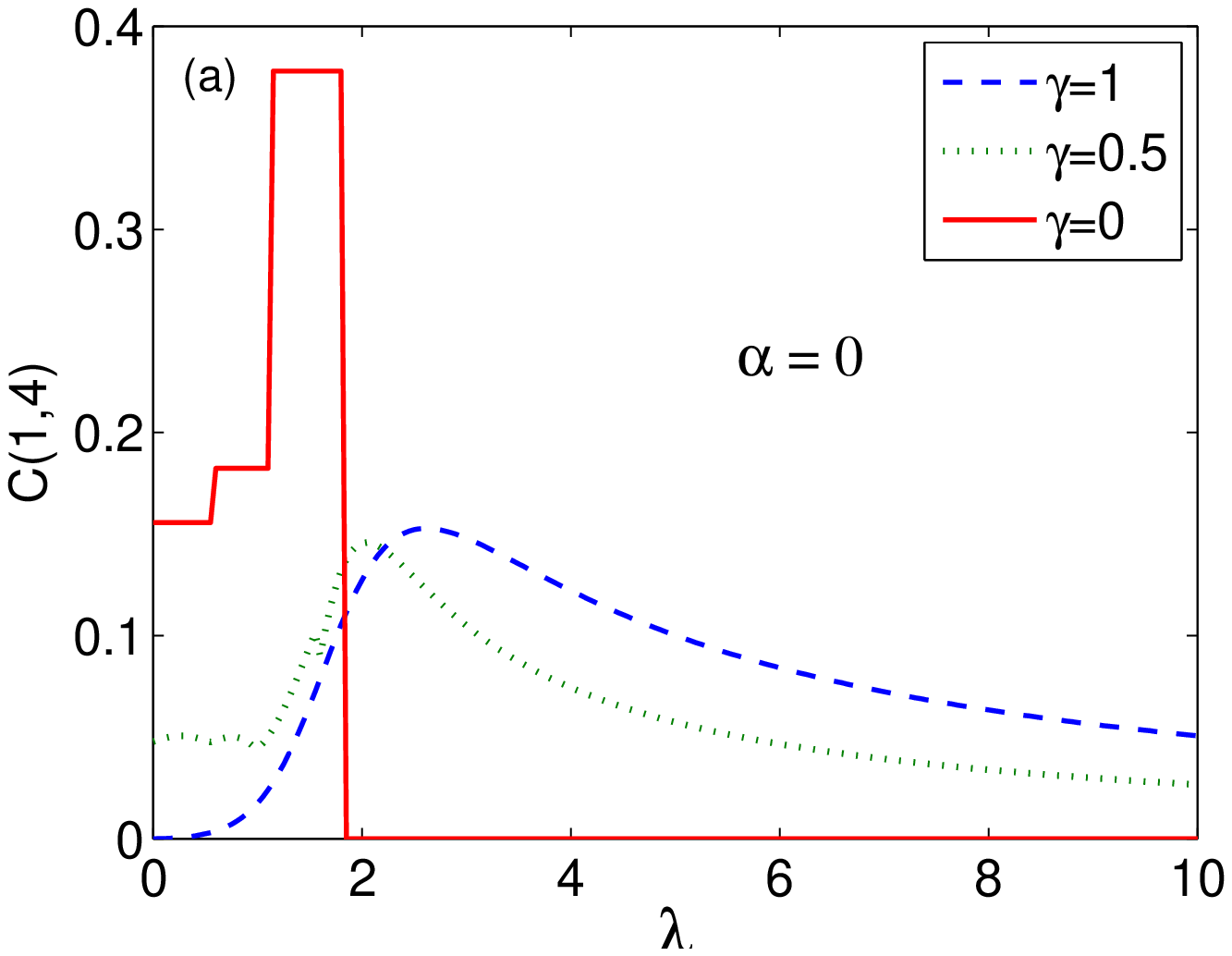}}\quad
   \subfigure{\includegraphics[width=8 cm]{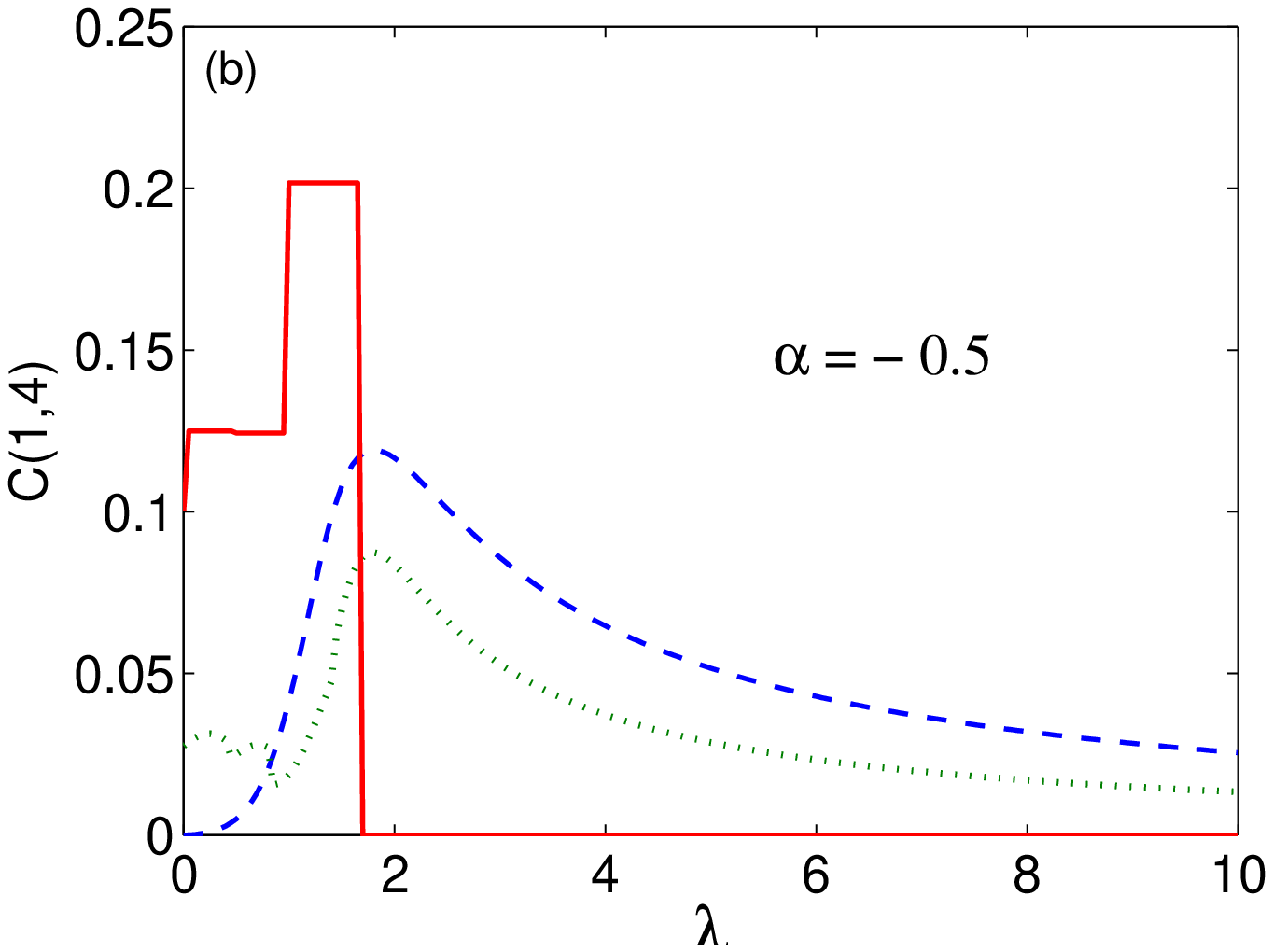}}\\
   \subfigure{\includegraphics[width=8 cm]{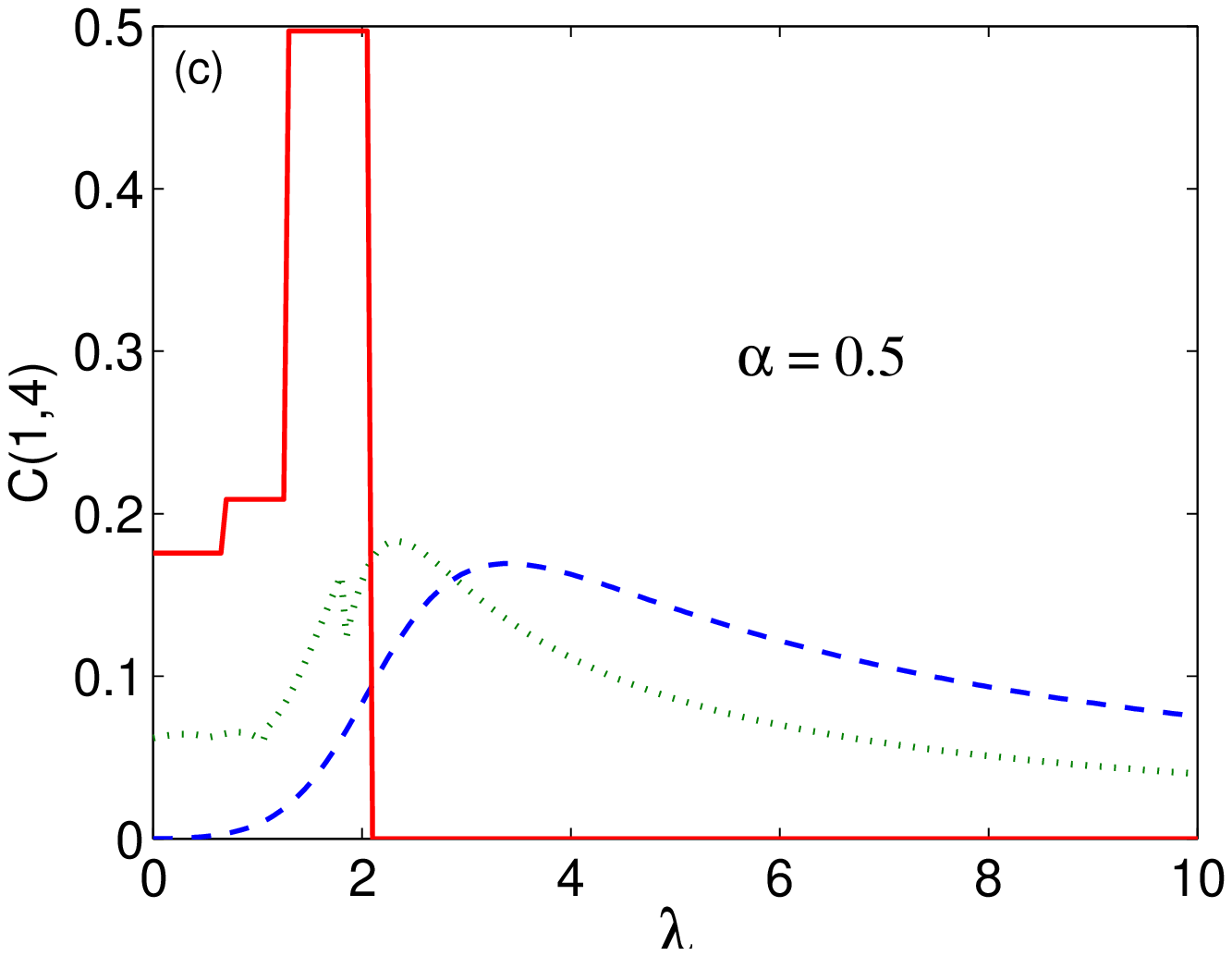}}\quad
   \subfigure{\includegraphics[width=8 cm]{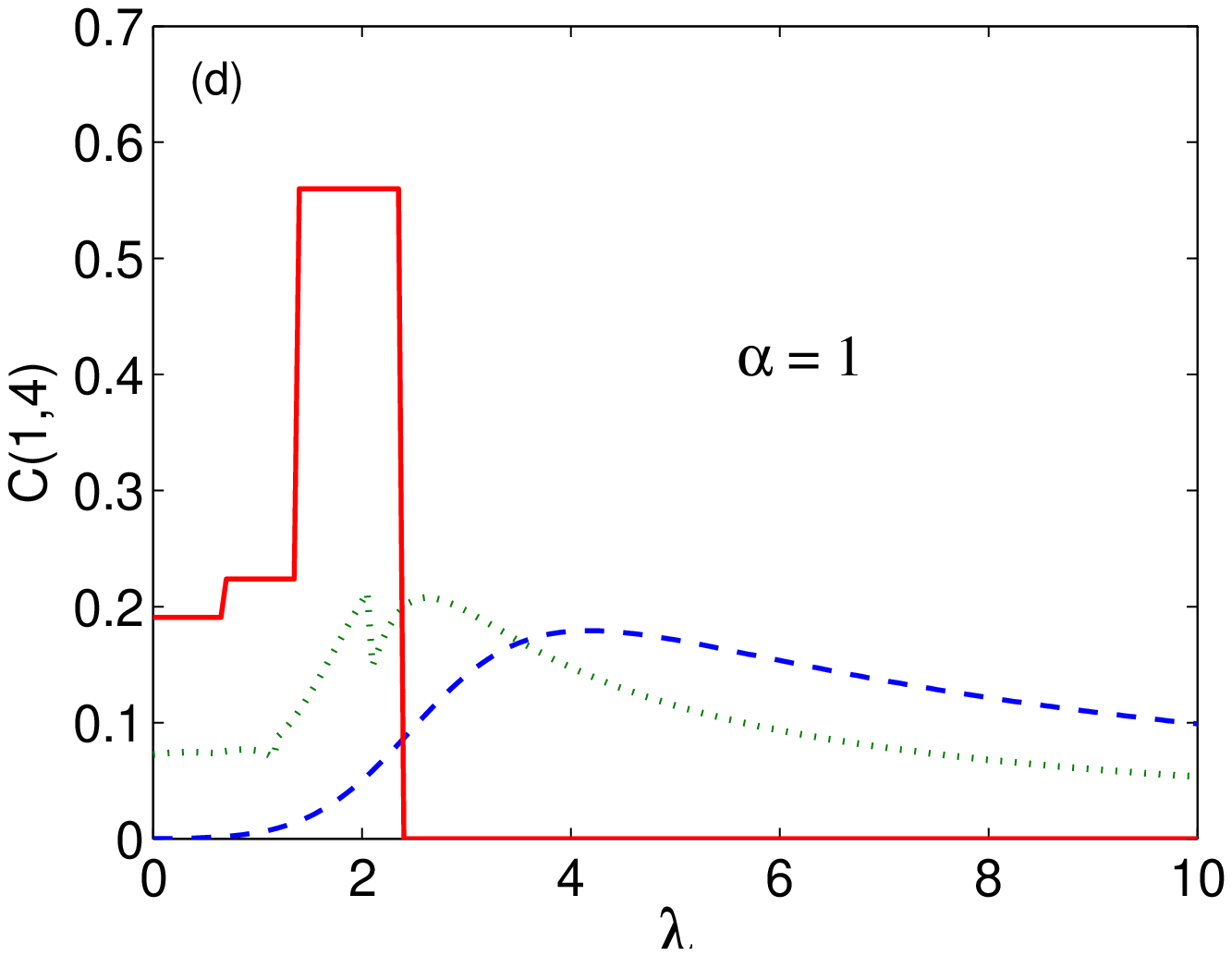}}
      \caption{{\protect\footnotesize (Color online) The concurrence $C(1,4)$ versus the parameter $\lambda$ with a single impurity at the border site 4 with different impurity coupling strengths $\alpha = -0.5, 0, 0.5, 1$ for different degrees of anisotropy $\gamma = 1, 0.5, 0$ as shown in the subfigures. The legend for all subfigures is as shown in subfigure (a).}}
 \label{C_C14}
 \end{minipage}
\end{figure}

To explore the effect of the impurity location we investigate the case of a single impurity spin located at site 4, instead of site 2, where we plot the concurrences $C(1,2)$ and $C(1,4)$ in figs.~\ref{C_C12} and~\ref{C_C14} respectively. 
Interestingly, while changing the impurity location has almost no effect on the behavior of the entanglement $C(1,2)$ of the partially anisotropic and isotropic XY systems, it has a great impact on that of the Ising system where the peak value of the entanglement increases significantly in the weak impurity case and decreases as the impurity gets stronger as shown in fig.~\ref{C_C12}. Now considering the entanglement between the central impurity site 4 and the edge site 1, and comparing with the results in fig.~\ref{B_C24} of the entanglement between the edge site 2 and central site 4, one can see that the entanglement $C(1,4)$ profile for all degrees of anisotropy is very close to the $C(2,4)$. Nevertheless the entanglement $C(1,4)$ magnitude is lower for weak impurity case and higher for the strong impurity which means that the central impurity made a significant change to the entanglement magnitude.
\subsection{Effect of system energy gap on entanglement}
\begin{figure}[htbp]
\begin{minipage}[c]{\textwidth}
\centering
   \subfigure{\includegraphics[width=8 cm]{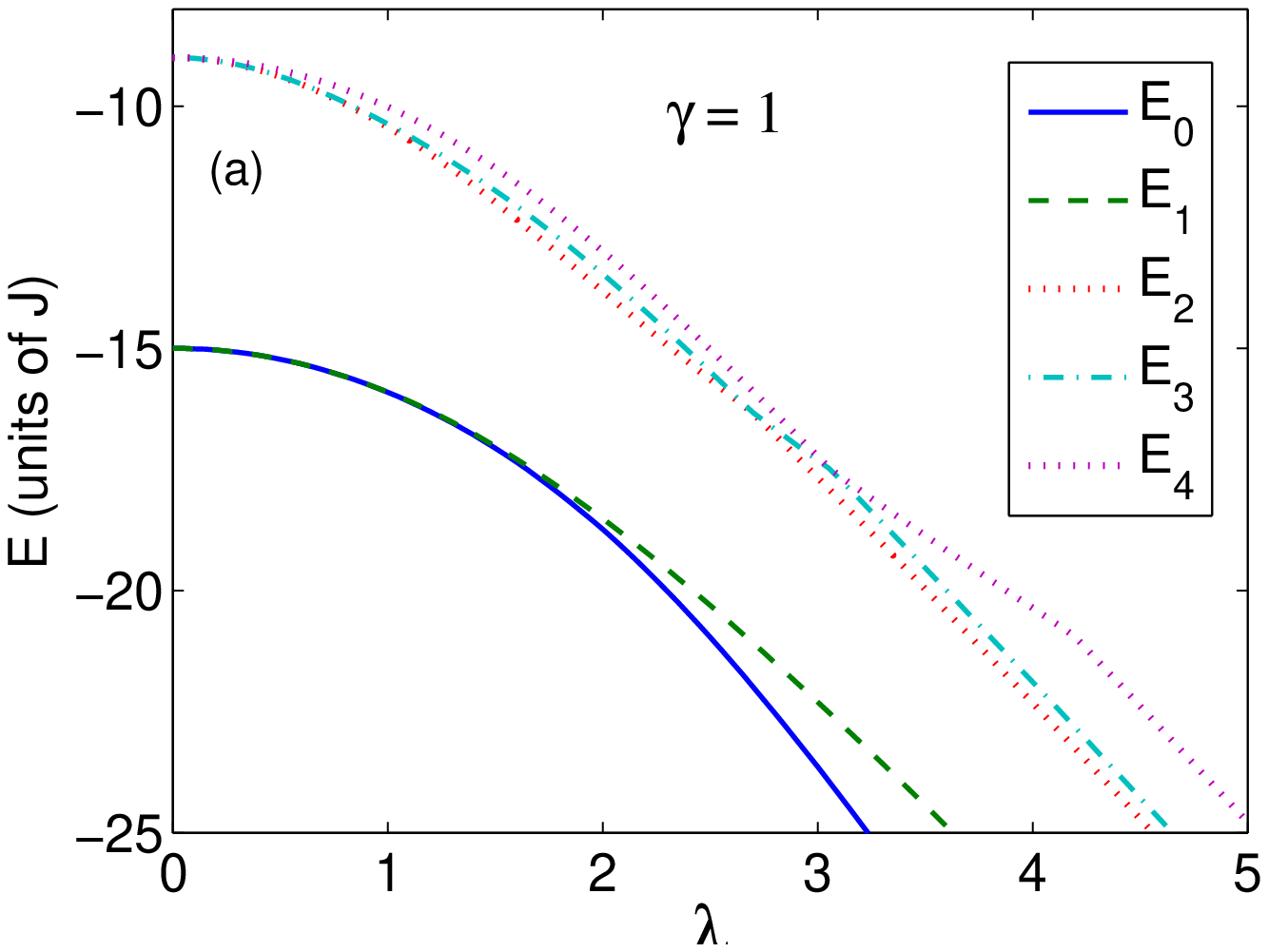}}\quad
   \subfigure{\includegraphics[width=8 cm]{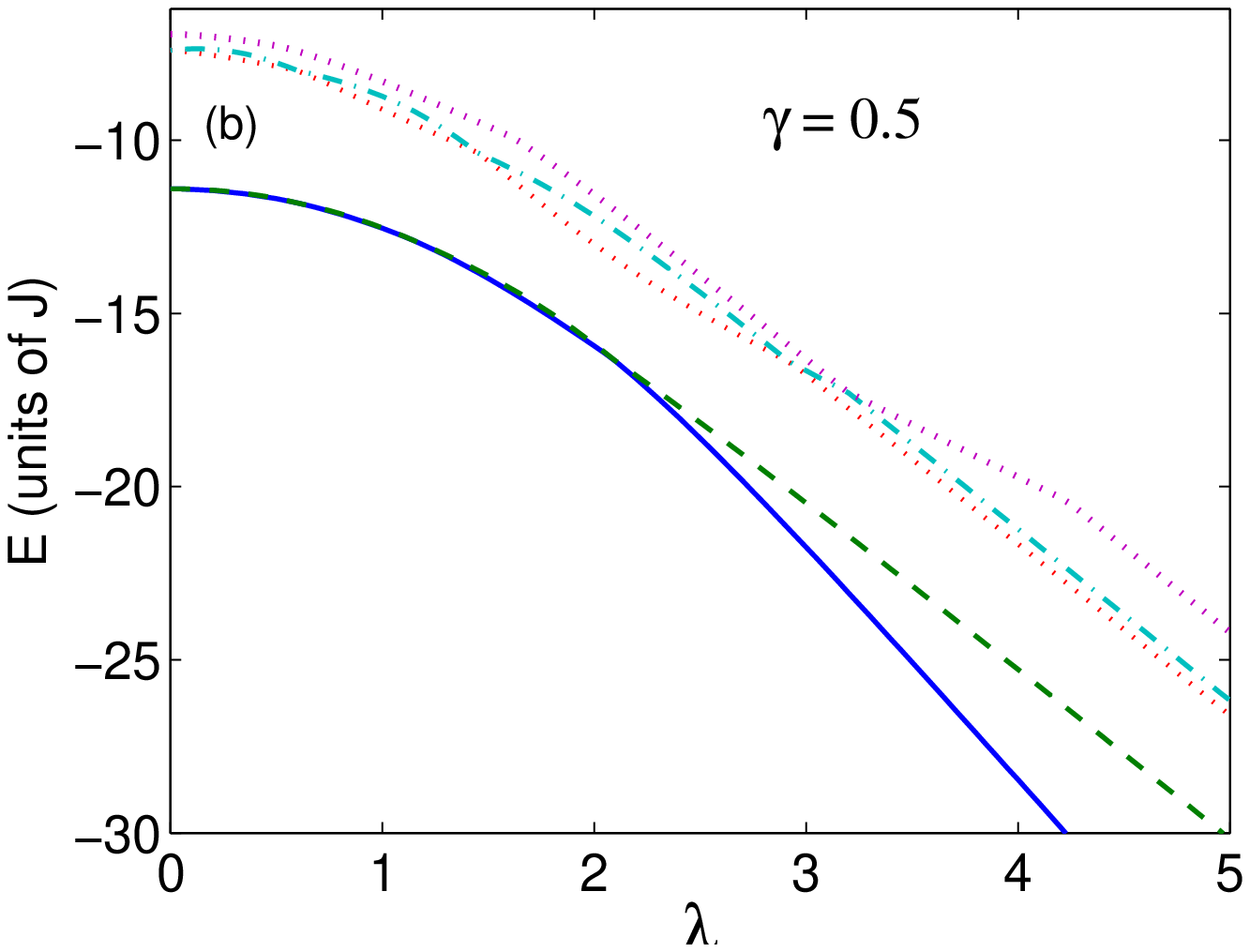}}\\
   \subfigure{\includegraphics[width=8 cm]{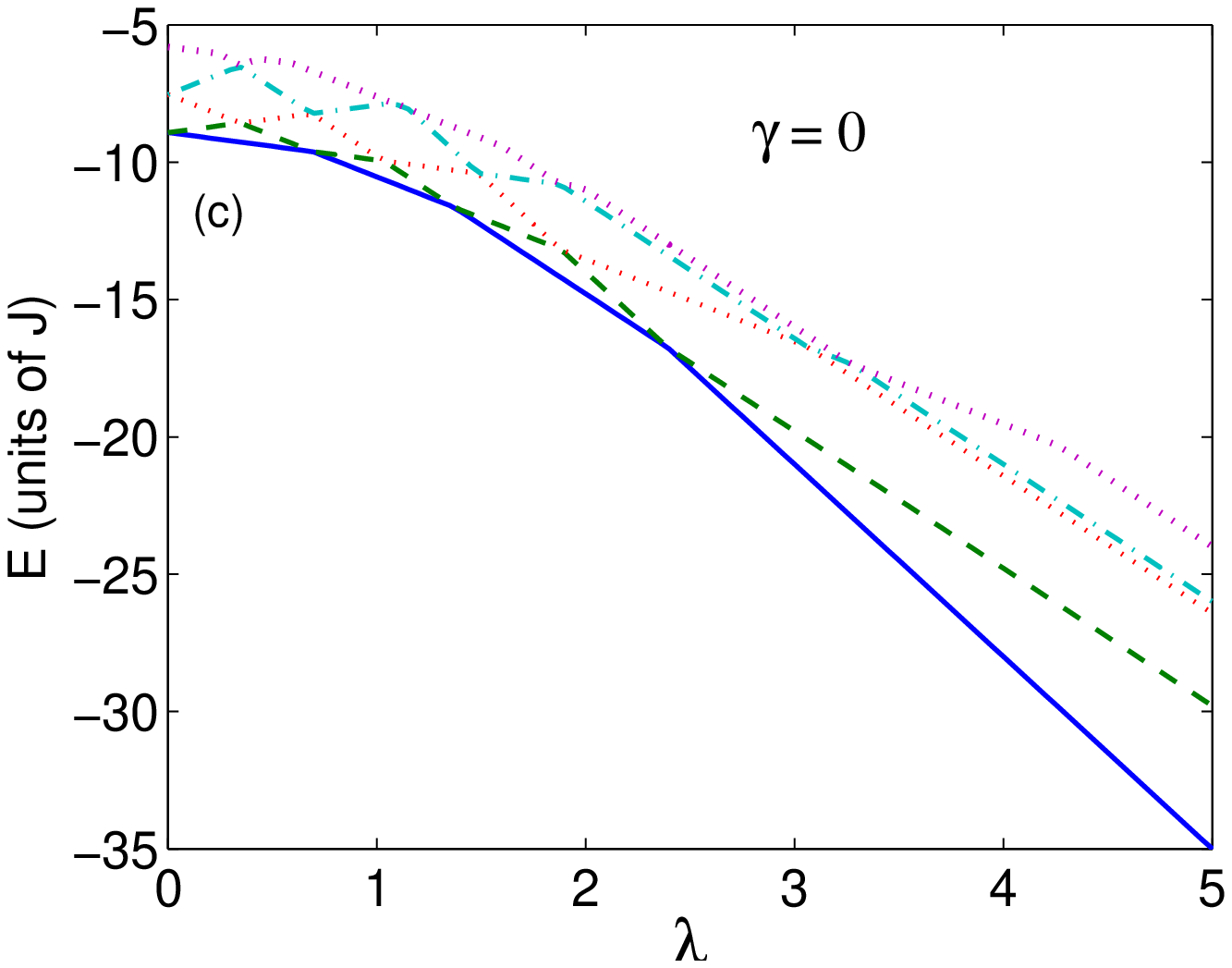}}
   \caption{{\protect\footnotesize (Color online) The energy spectrum versus the parameter $\lambda$ with a single impurity at the border site 1 with impurity coupling strength $\alpha = 1$ for different degrees of anisotropy $\gamma = 1, 0.5, 0$ as shown in the subfigures. The legend for all subfigures is as shown in subfigure (a).}}
 \label{B_E}
 \end{minipage}
\end{figure}
\begin{figure}[htbp]
\begin{minipage}[c]{\textwidth}
\centering
   \subfigure{\includegraphics[width=8 cm]{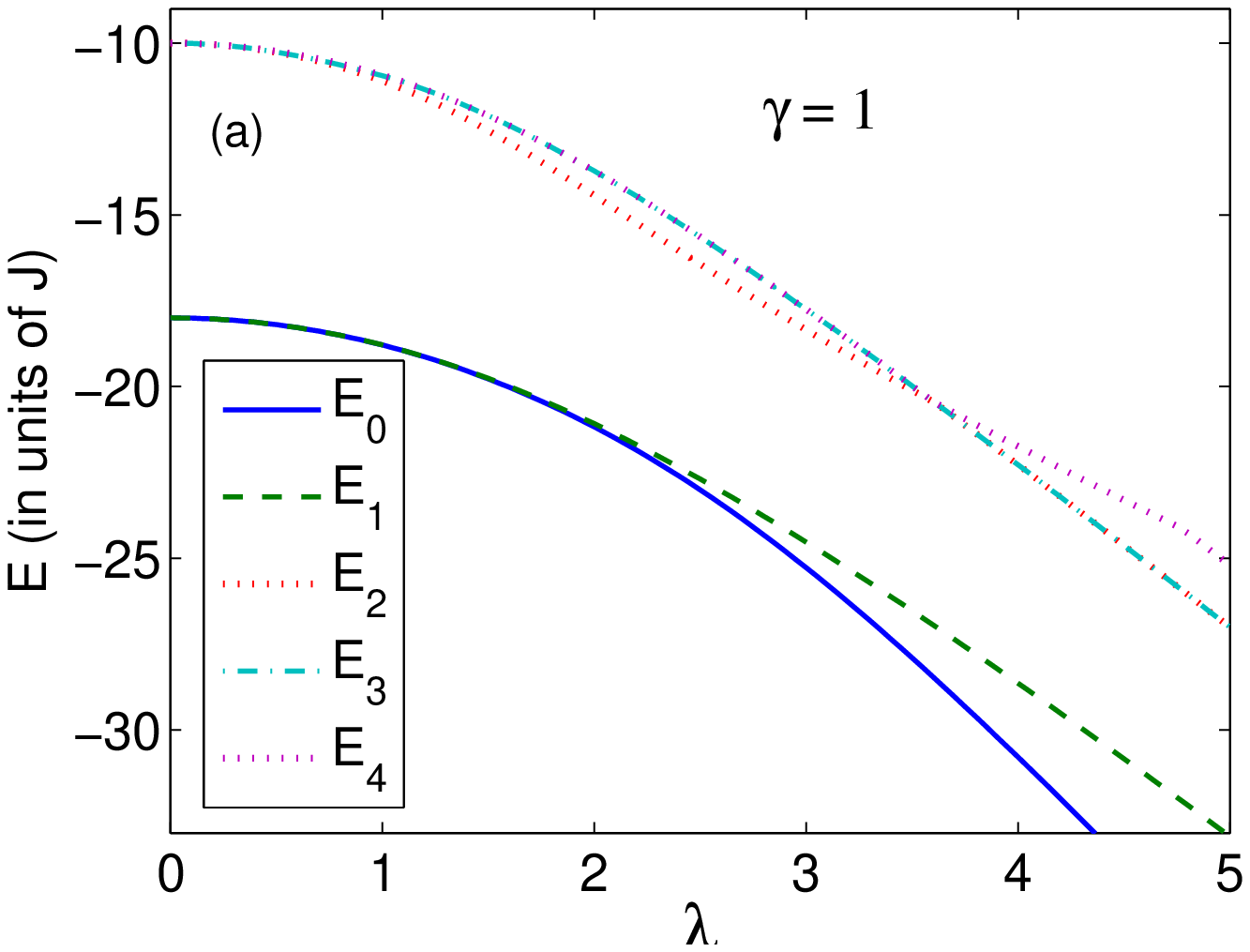}}\quad
   \subfigure{\includegraphics[width=8 cm]{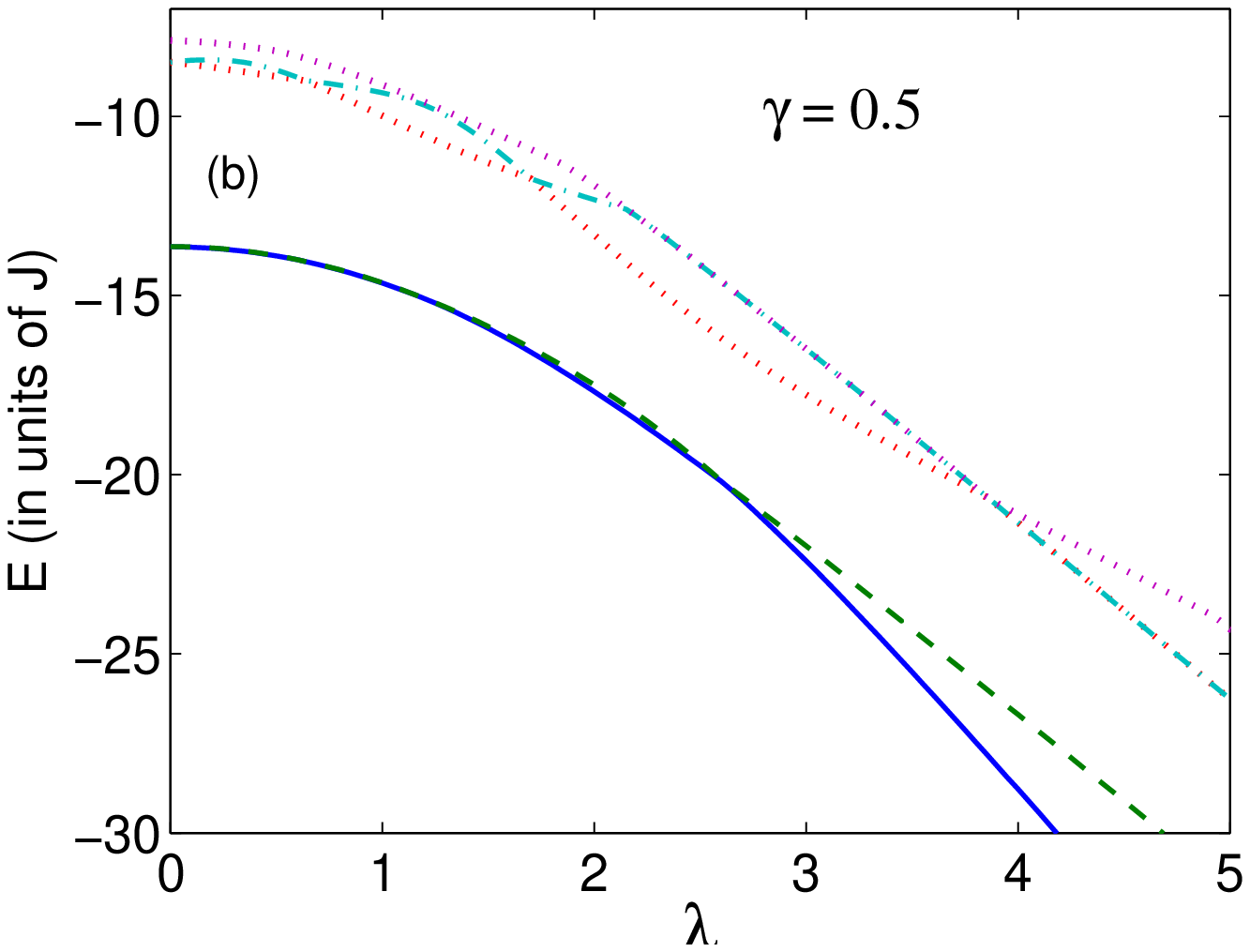}}\\
   \subfigure{\includegraphics[width=8 cm]{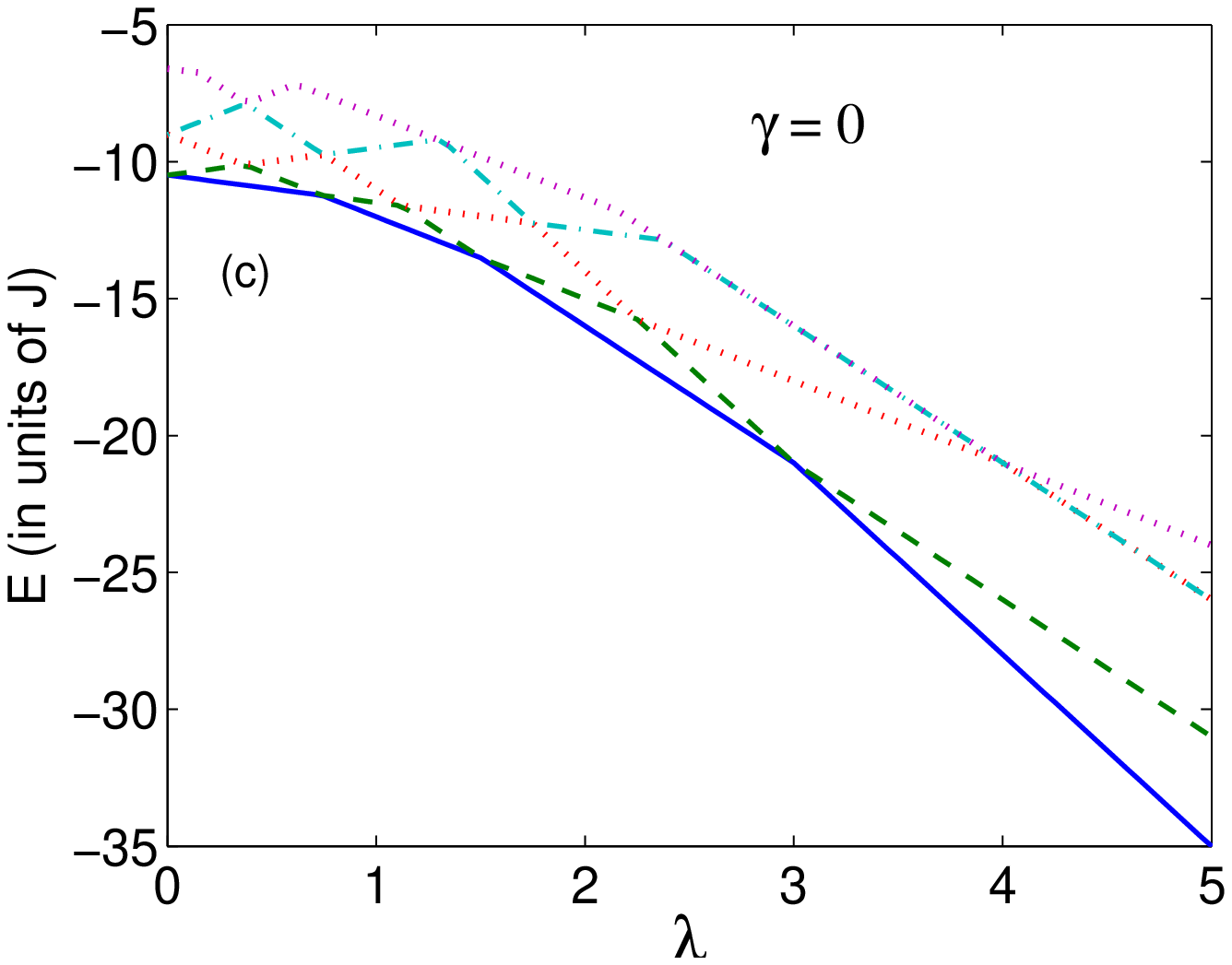}}
   \caption{{\protect\footnotesize (Color online) The energy spectrum versus the parameter $\lambda$ with a single impurity at the central site 4 with impurity coupling strength $\alpha = 1$ for different degrees of anisotropy $\gamma = 1, 0.5, 0$ as shown in the subfigures. The legend for all subfigures is as shown in subfigure (a).}}
 \label{C_E}
 \end{minipage}
\end{figure}

To explain the distinct behavior of the entanglement corresponding to the different degrees of anisotropy $\gamma$ we depict the lowest few energy eigenvalues of the system at the different $\gamma$ values for the two cases of border and central impurities in fig.~\ref{B_E} and fig.~\ref{C_E} respectively. As can be noticed in fig.~\ref{B_E}(a), the energies of the ground state and the first excited state of the Ising system coincide at the beginning at the small values of $\lambda$ until a specific value where they deviate from each other. This is corresponding to the transition from the degenerated ground state to non-degenerated one, from paramagnetic to ferromagnetic order, by breaking the $Z_2$ symmetry, which explains the phase transition curve observed in the Ising case. 
The energy spectrum of the partially anisotropic $XY$ system is a little bit different at the small values of $\lambda$ where the ground state and the first excited state coincide at the beginning but then deviates slightly from each other before recombining again and at last separate from each other completely, this behavior is repeated quite few times depending on the impurity strength, as illustrated in fig.~\ref{B_E}(b). This energy spectrum behavior explains the roughness in the ascending part of the entanglement curves of the partially anisotropic $XY$ system corresponding to subsequent transitions between the ground state and the first excited state taking place before reaching the maximum entanglement point. In fig.~\ref{B_E}(c), the energy spectrum of the isotropic $XY$ system is explored where clearly the deviations and recombination between the ground and first excited state energies become sharper and more frequent compared with the partially anisotropic system. This distinct behavior of the energy spectrum corresponding to $\gamma = 0$ is the reason for the sharp step behavior of the entanglement as was shown in figs.~\ref{B_C12} and ~\ref{B_C24}. 
\begin{figure}
\subfigure{\includegraphics[width=0.48\textwidth,height=0.3\textheight]{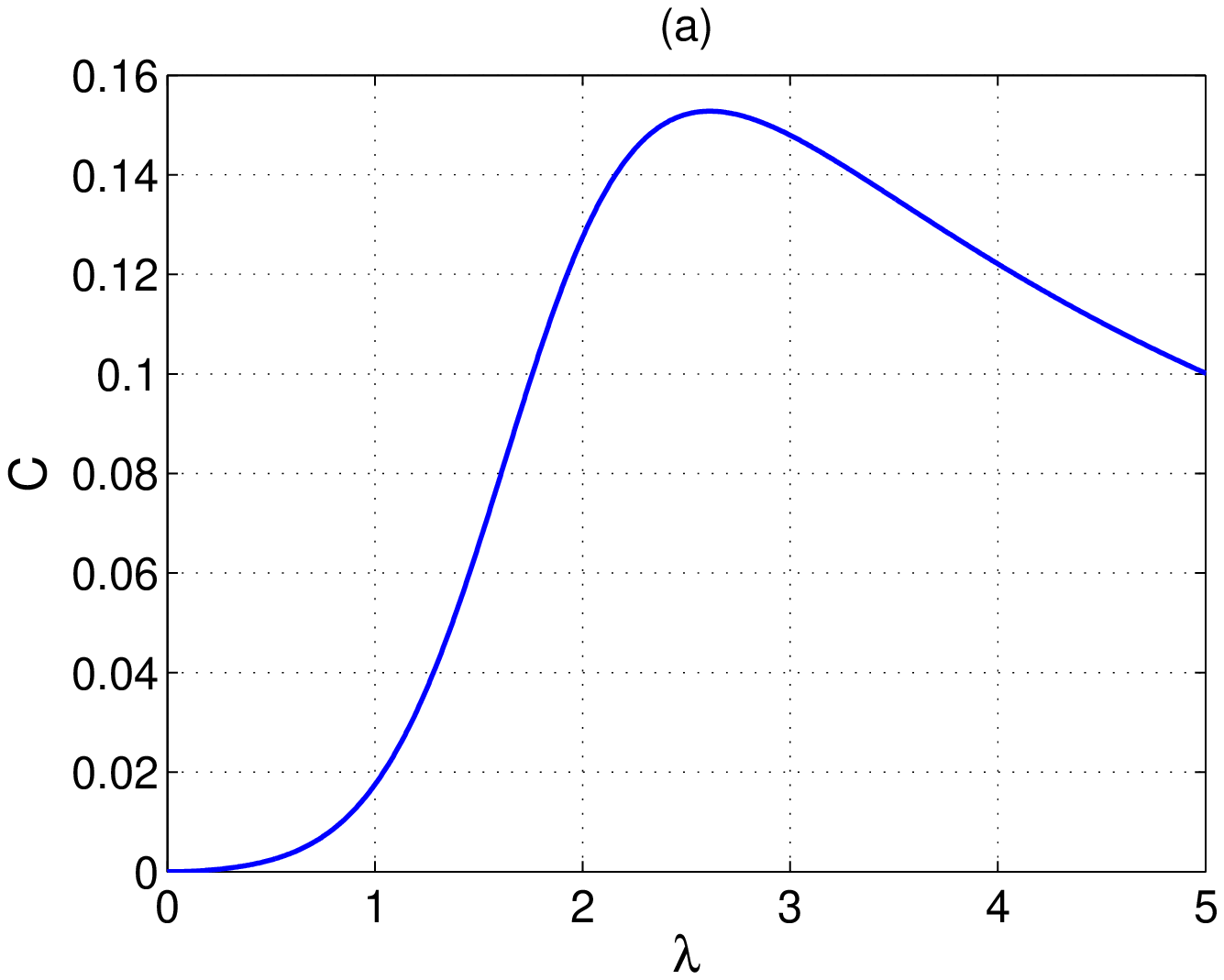}}\quad
\subfigure{\includegraphics[width=0.48\textwidth,height=0.3\textheight]{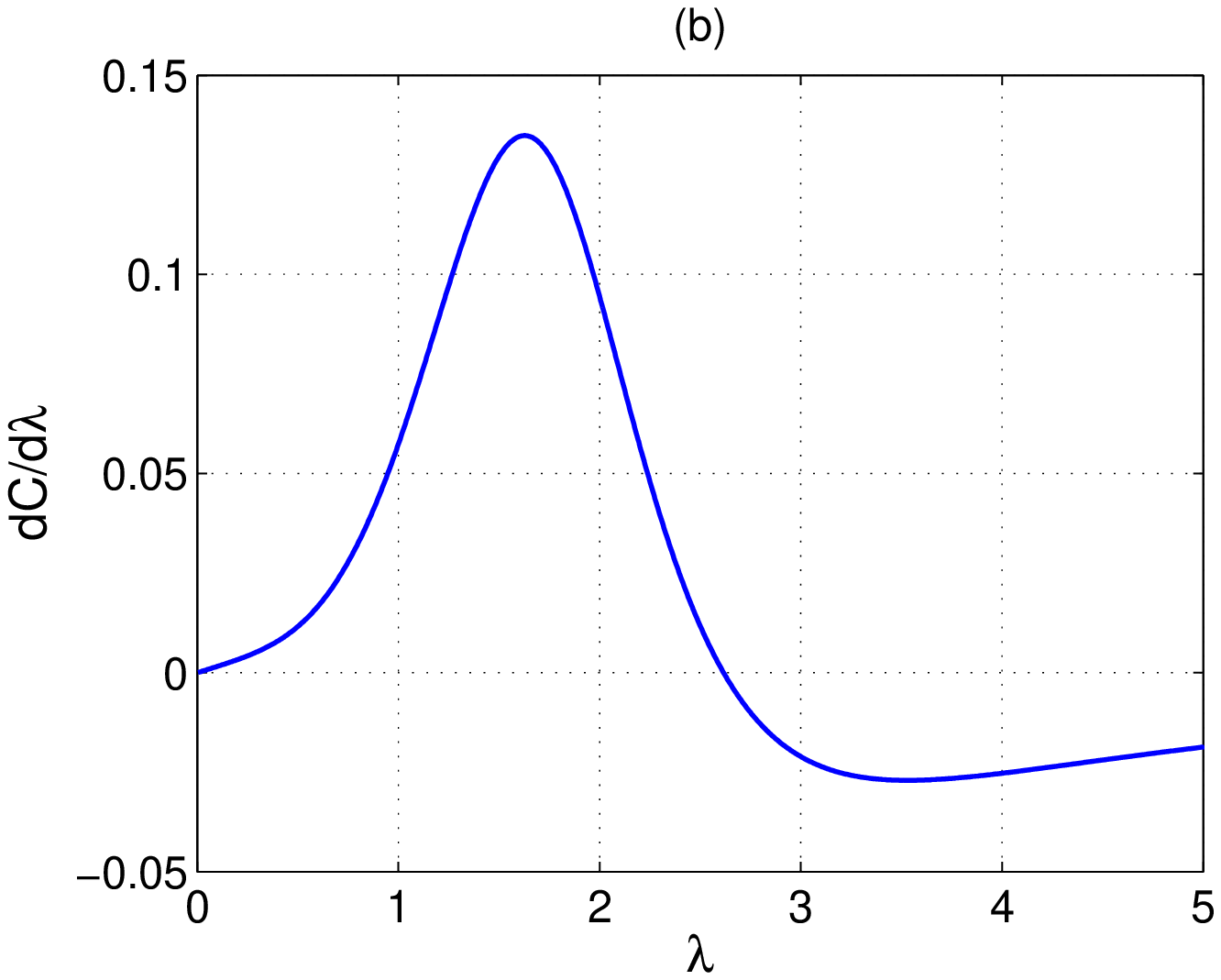}}
\subfigure{\includegraphics[width=0.48\textwidth,height=0.3\textheight]{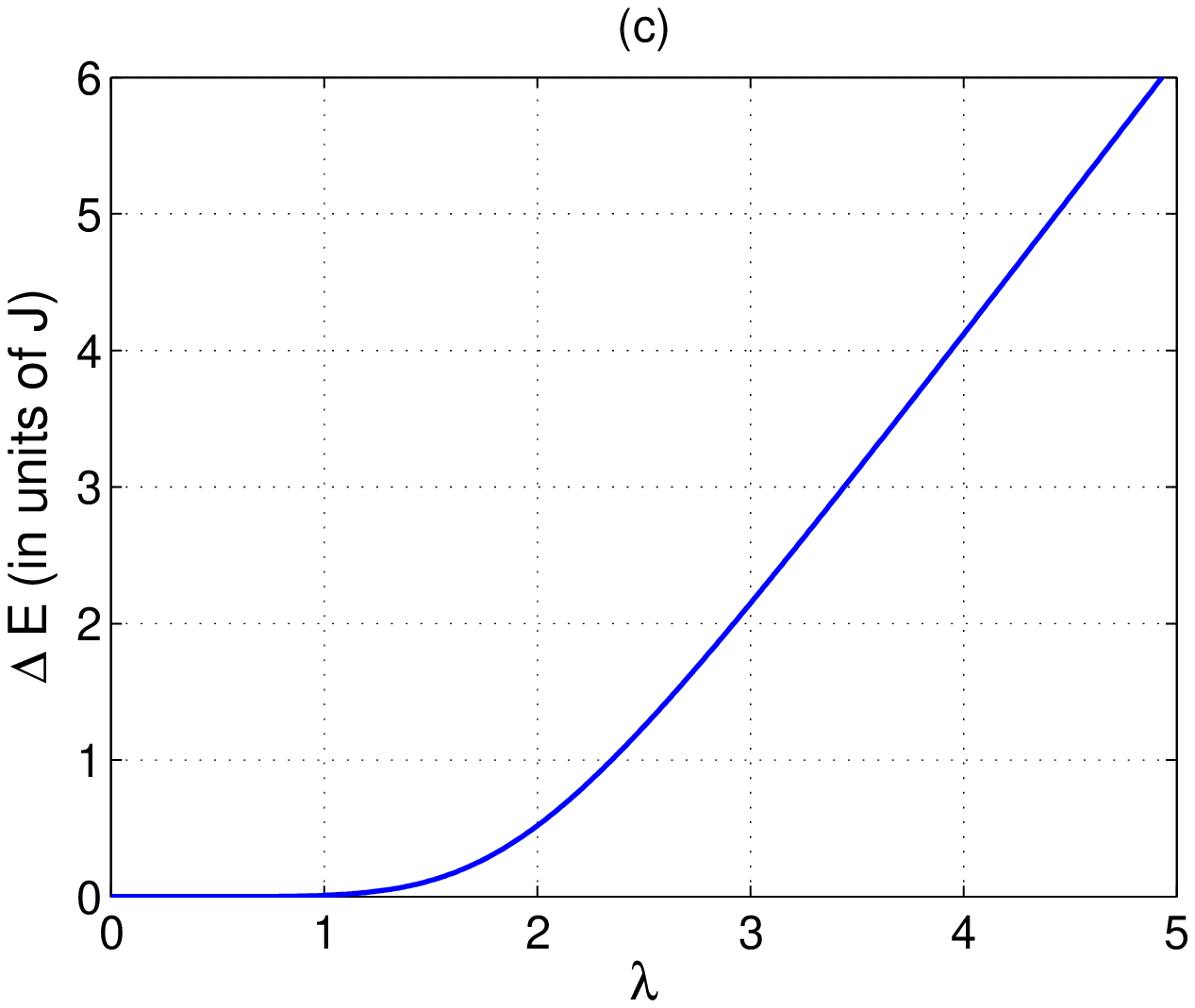}}\quad
\subfigure{\includegraphics[width=0.48\textwidth,height=0.3\textheight]{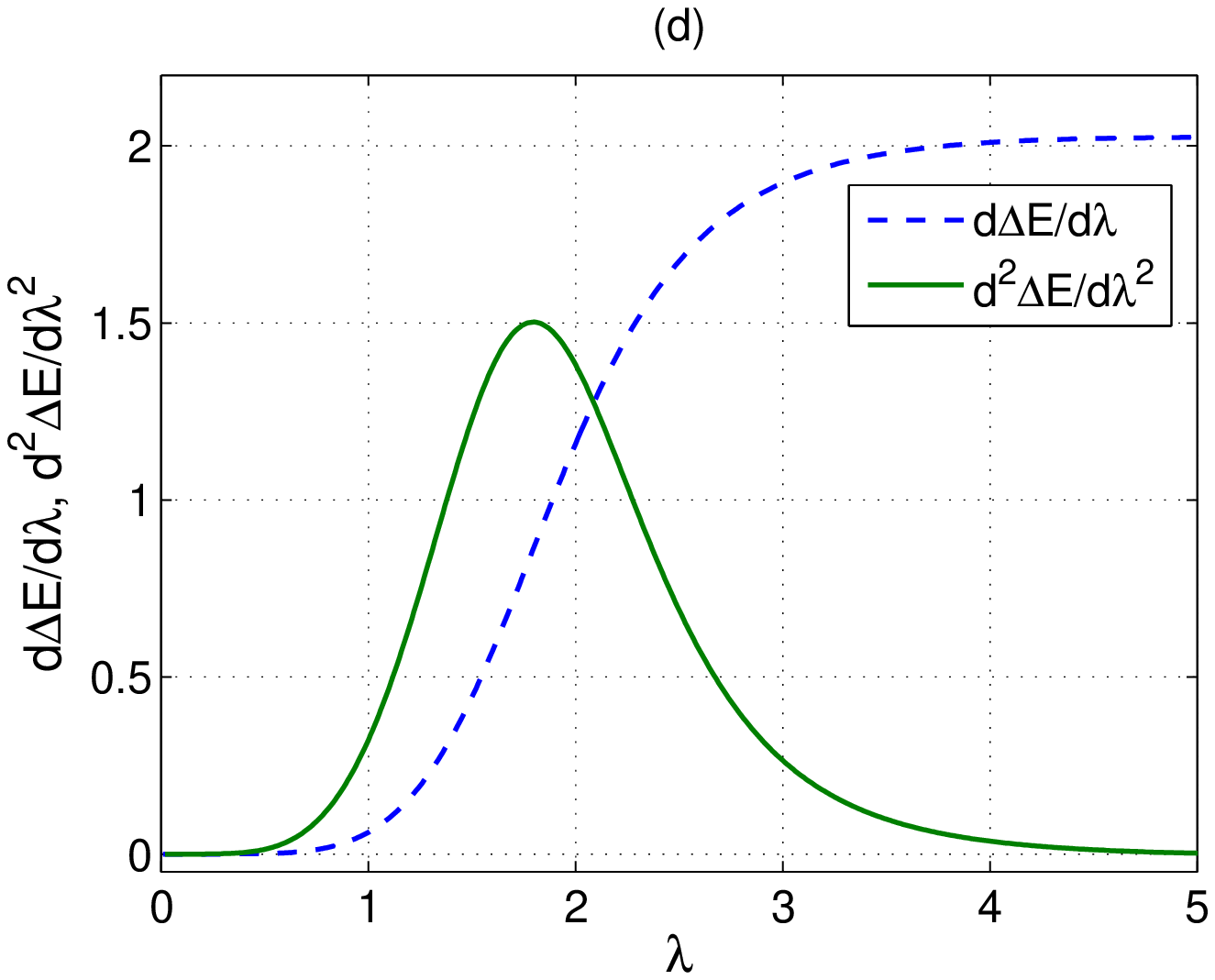}}
\caption{{\protect\footnotesize (Color online) (a) The concurrence $C_{14}$ versus $\lambda$; (b) the first derivative of the concurrence $C_{14}$ with respect to $\lambda$ versus $\lambda$; (c) the energy gap between the ground state and first excited state versus $\lambda$; (d) the first derivative (in units of $J$) and second derivative (in units of $J^2$) of the energy gap with respect $\lambda$ versus $\lambda$ for the pure Ising system ($\gamma=1$ and $\alpha=0$).}}
\label{QQPT1}
\end{figure}
\begin{figure}
\subfigure{\includegraphics[width=0.48\textwidth,height=0.3\textheight]{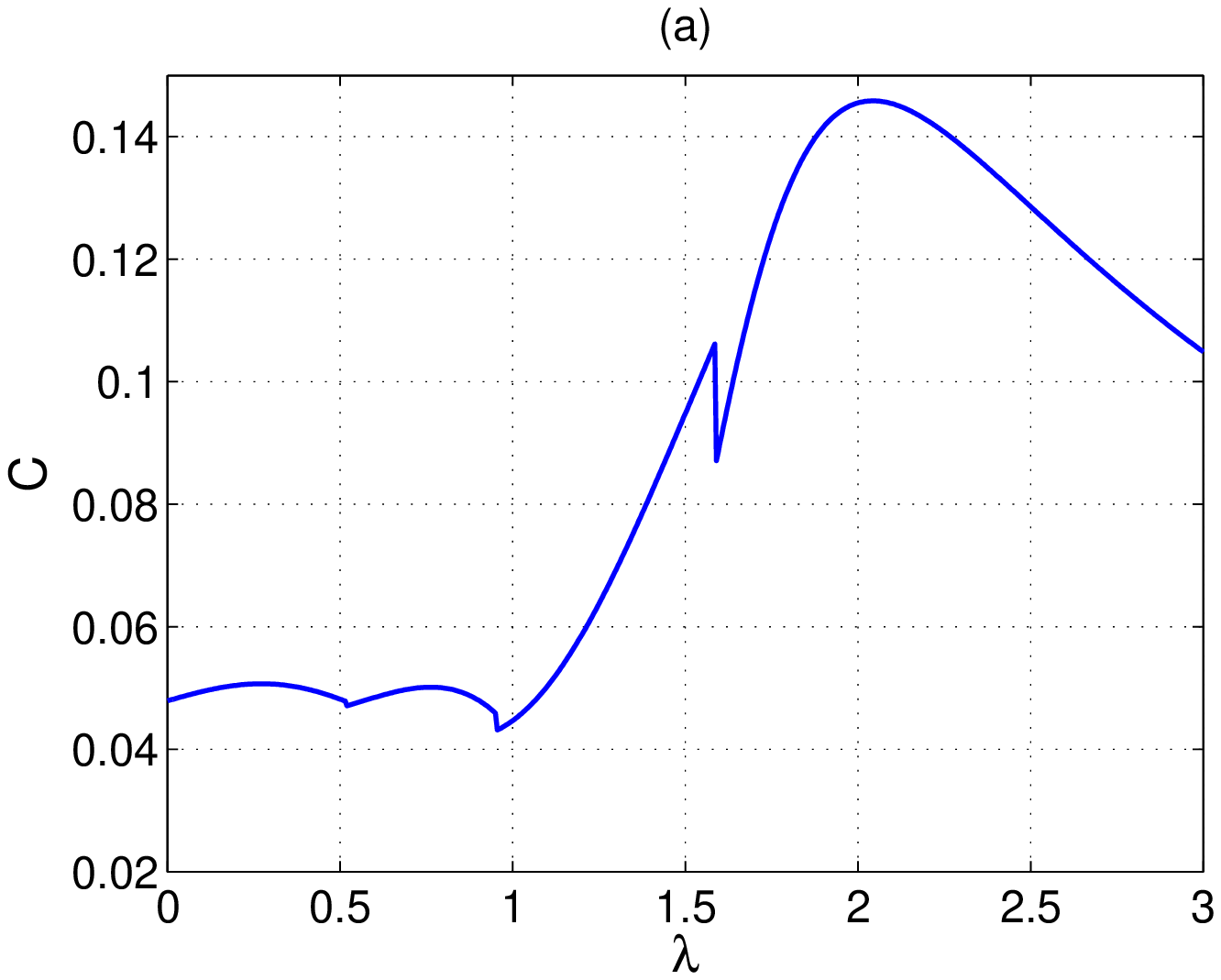}}\quad
\subfigure{\includegraphics[width=0.48\textwidth,height=0.3\textheight]{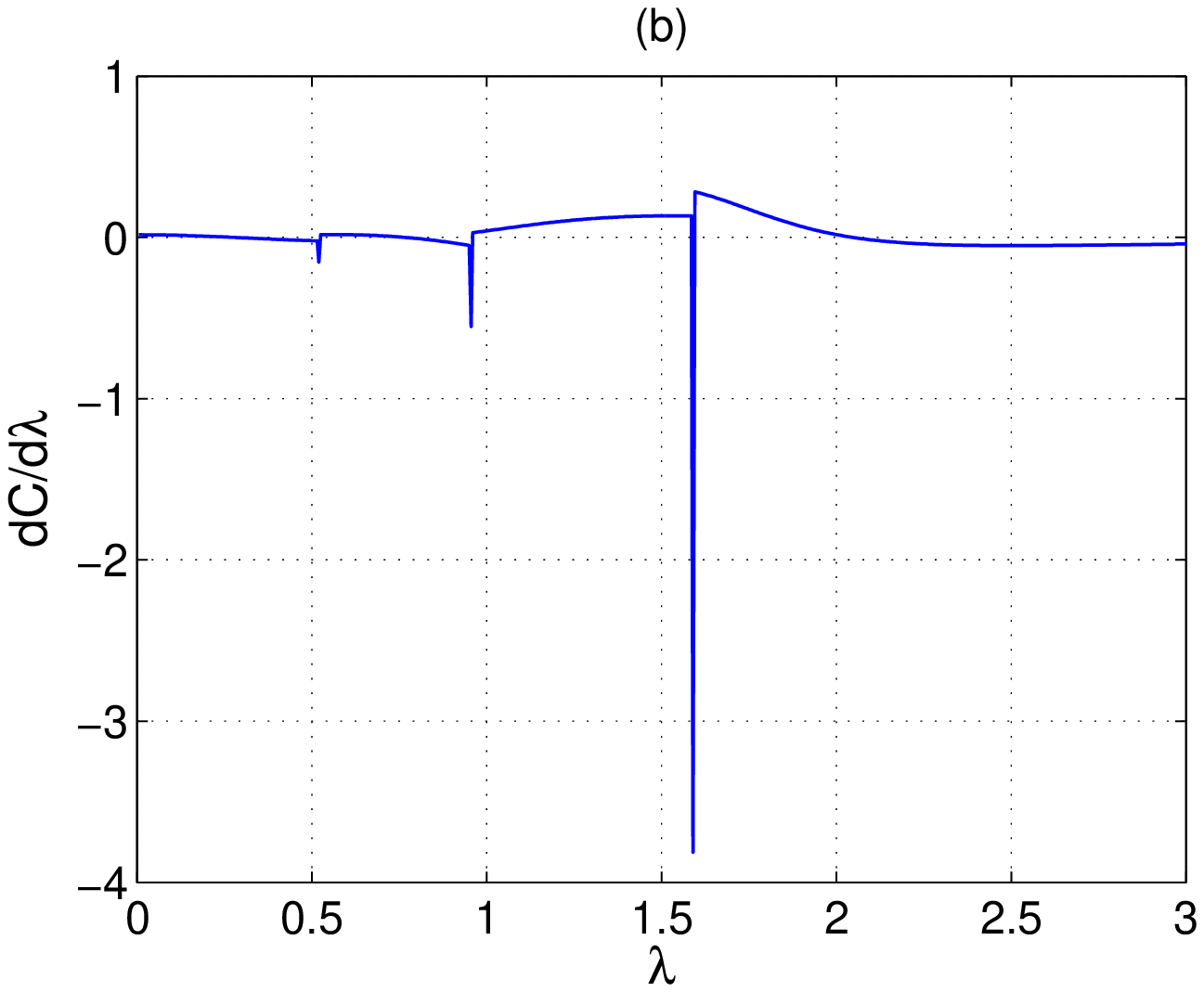}}
\subfigure{\includegraphics[width=0.48\textwidth,height=0.3\textheight]{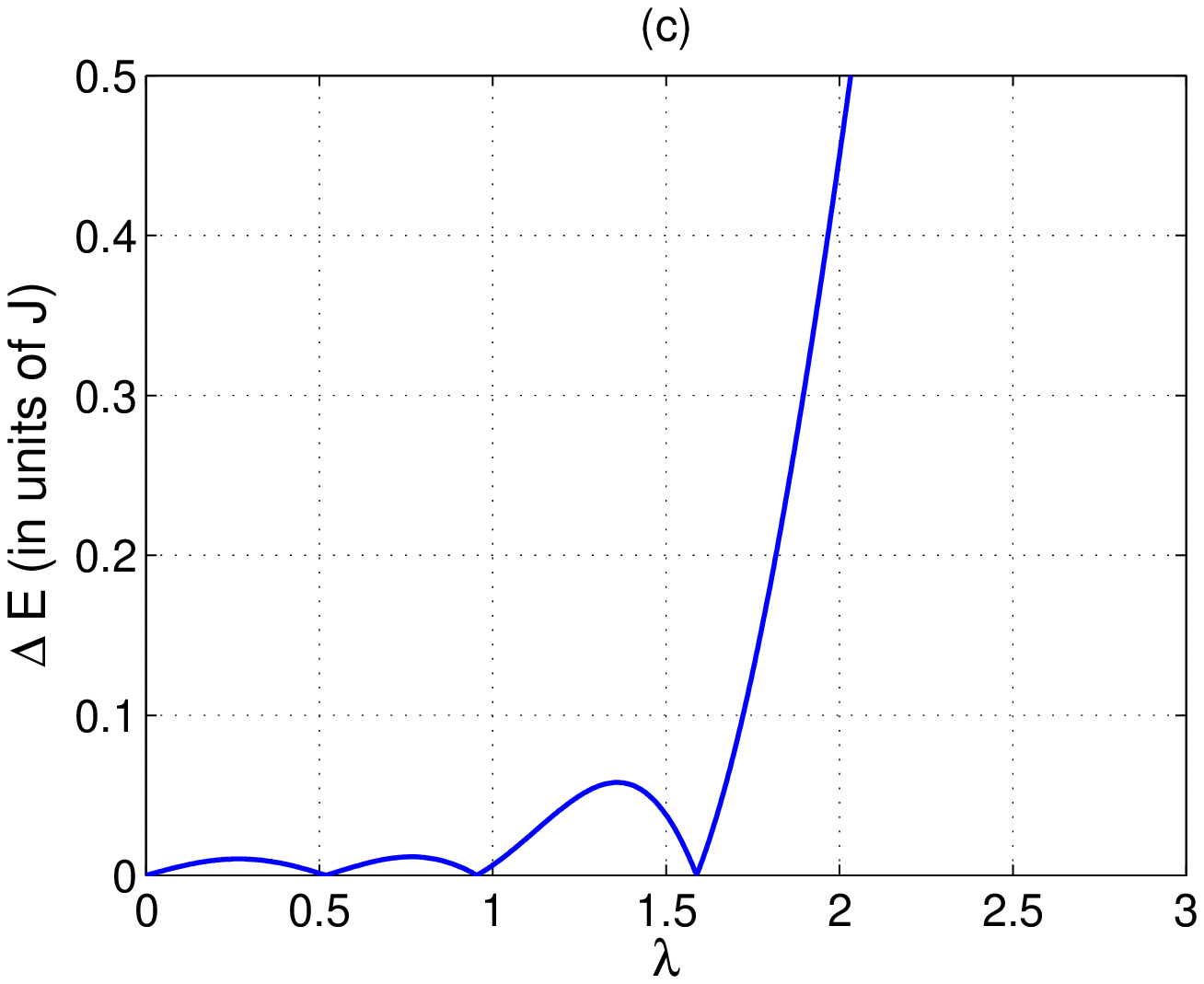}}\quad
\subfigure{\includegraphics[width=0.48\textwidth,height=0.3\textheight]{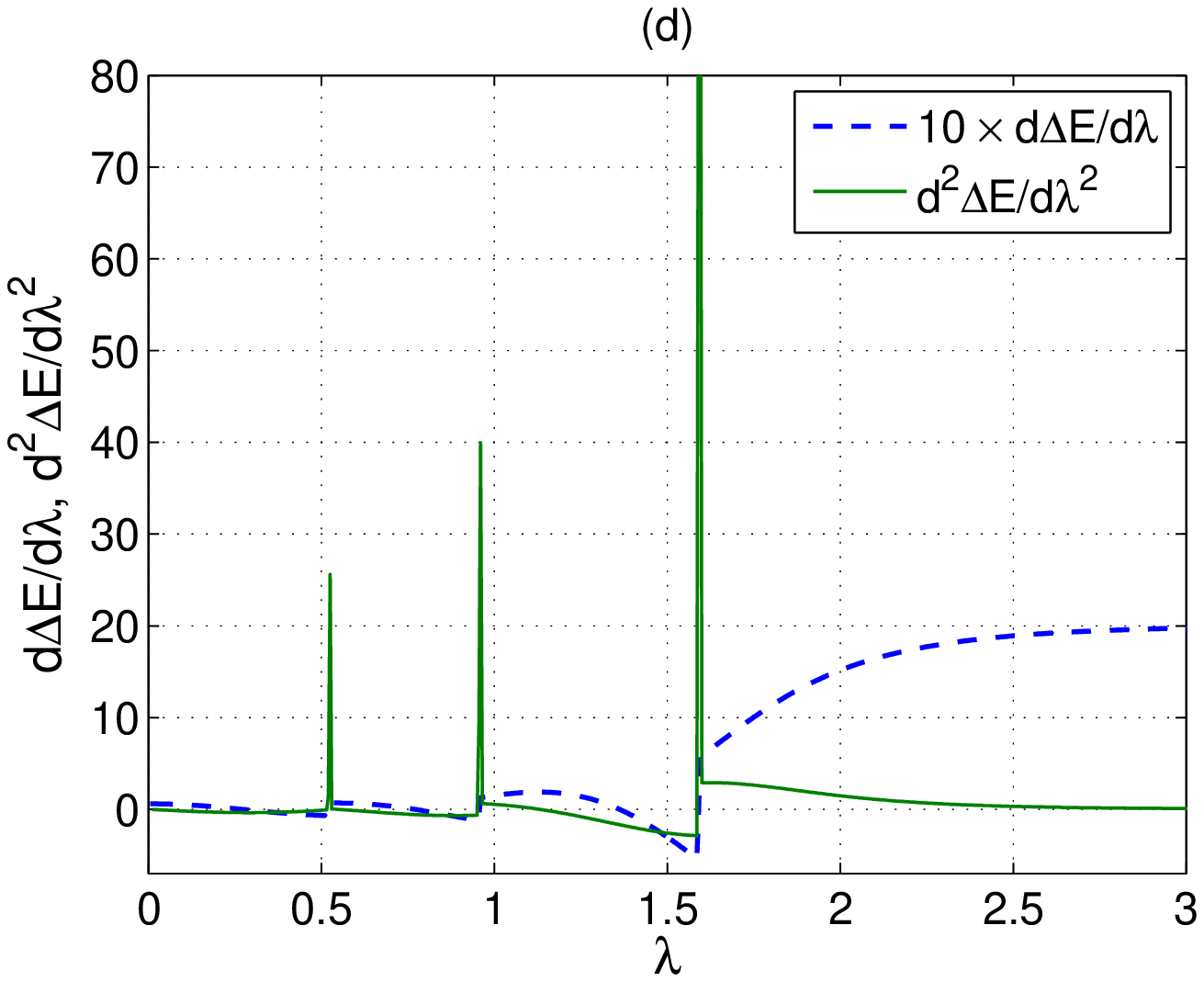}}
\caption{{\protect\footnotesize (Color online) (a) The concurrence $C_{14}$ versus $\lambda$; (b) the first derivative of the concurrence $C_{14}$ with respect to $\lambda$ versus $\lambda$; (c) the energy gap between the ground state and first excited state versus $\lambda$; (d) the first derivative (in units of $J$) and second derivative (in units of $J^2$) of the energy gap with respect $\lambda$ versus $\lambda$ for the pure partially anisotropic system ($\gamma=0.5$ and $\alpha=0$). Notice that the first derivative of energy gap is enlarged 10 times its actual scale for clearness.}}
\label{QQPT2}
\end{figure}
\begin{figure}
\subfigure{\includegraphics[width=0.48\textwidth,height=0.3\textheight]{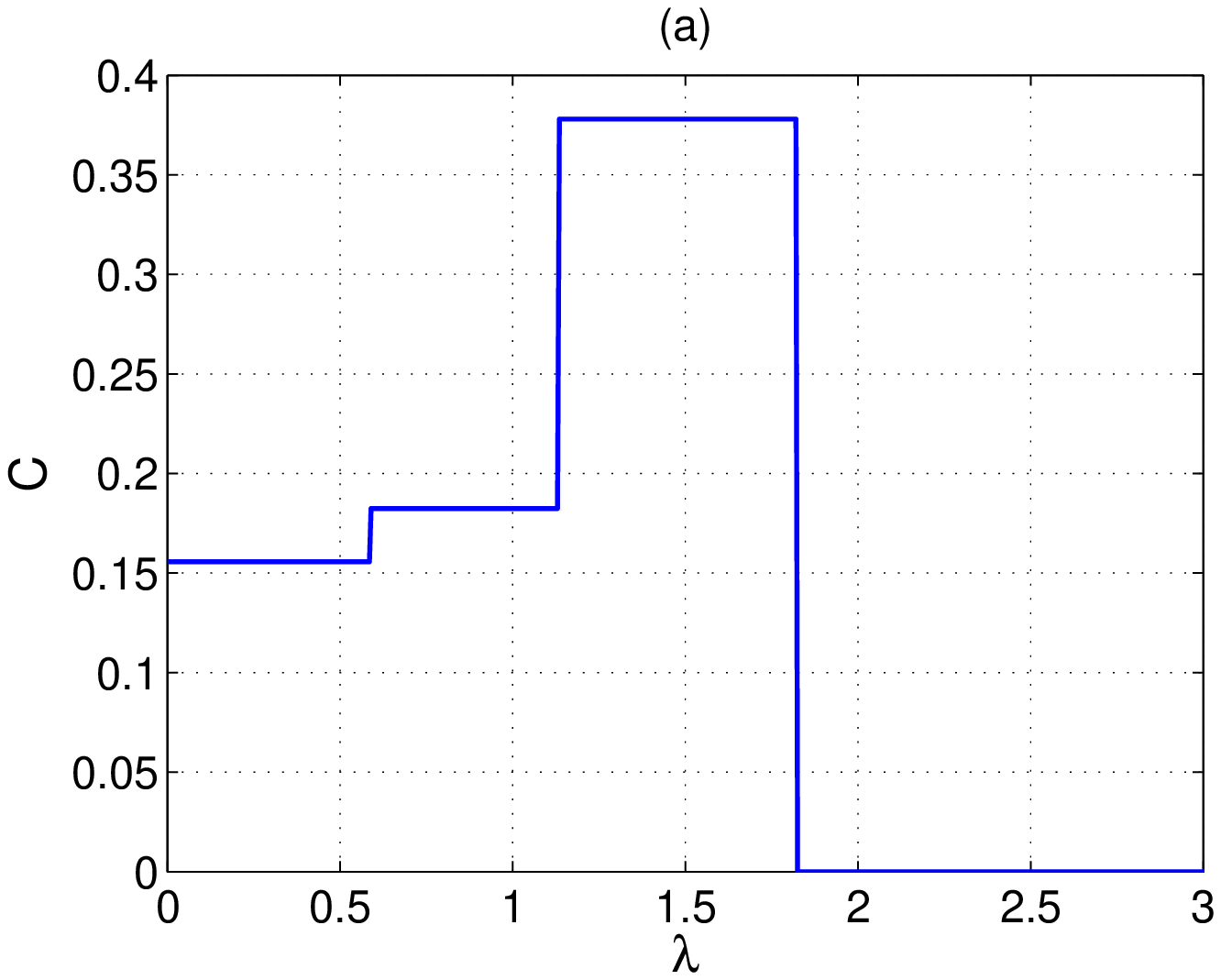}}\quad
\subfigure{\includegraphics[width=0.48\textwidth,height=0.3\textheight]{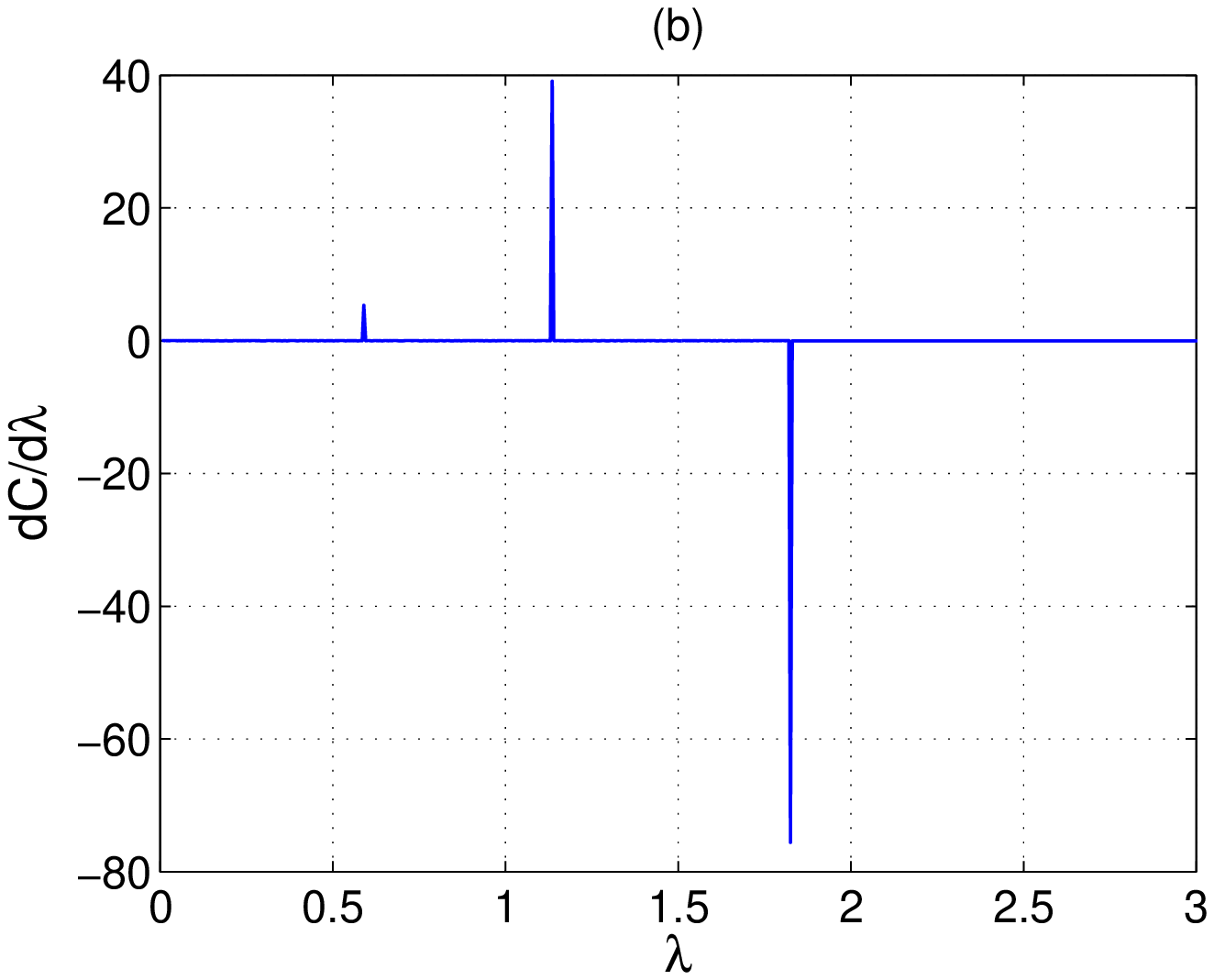}}
\subfigure{\includegraphics[width=0.48\textwidth,height=0.3\textheight]{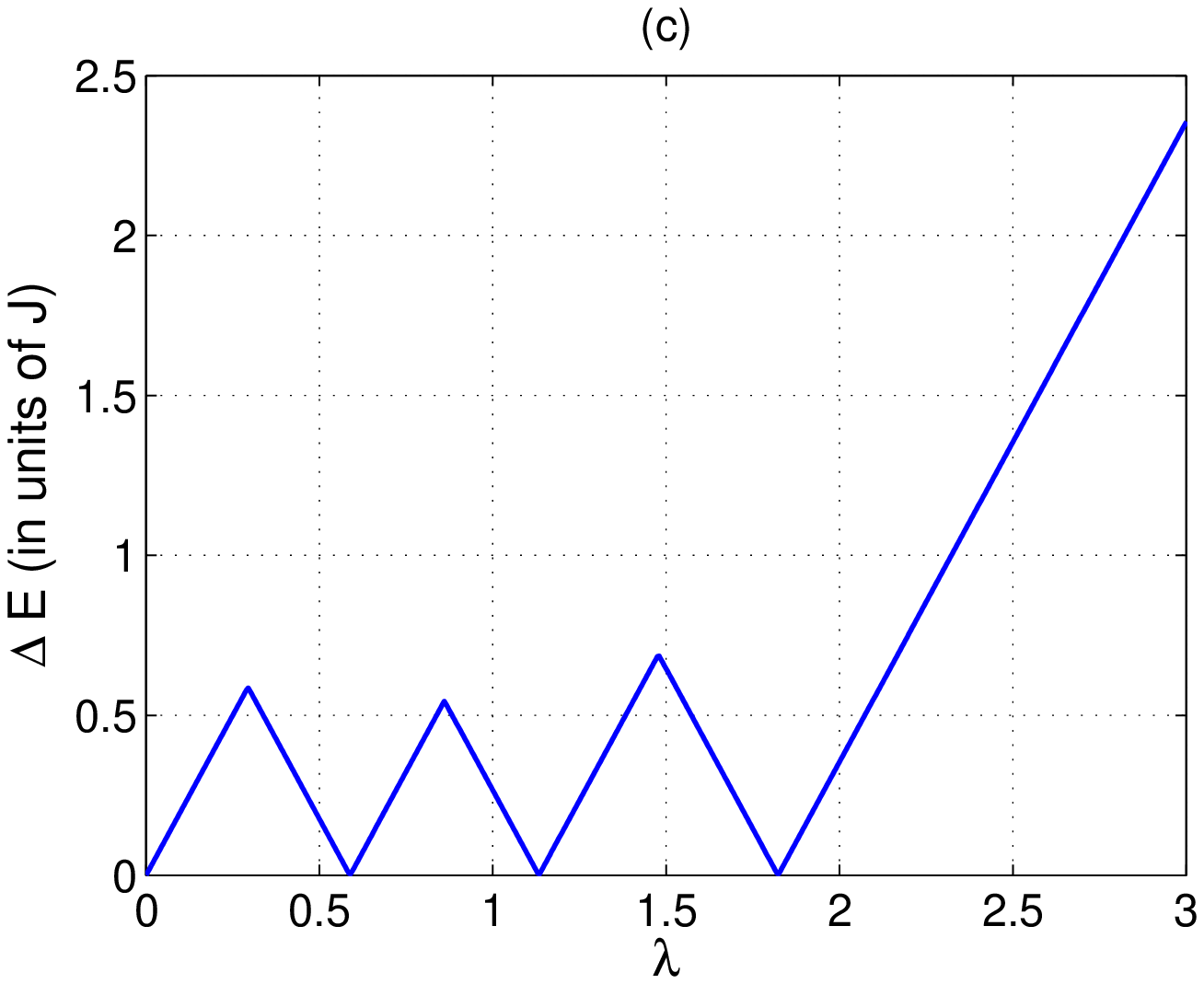}}\quad
\subfigure{\includegraphics[width=0.48\textwidth,height=0.3\textheight]{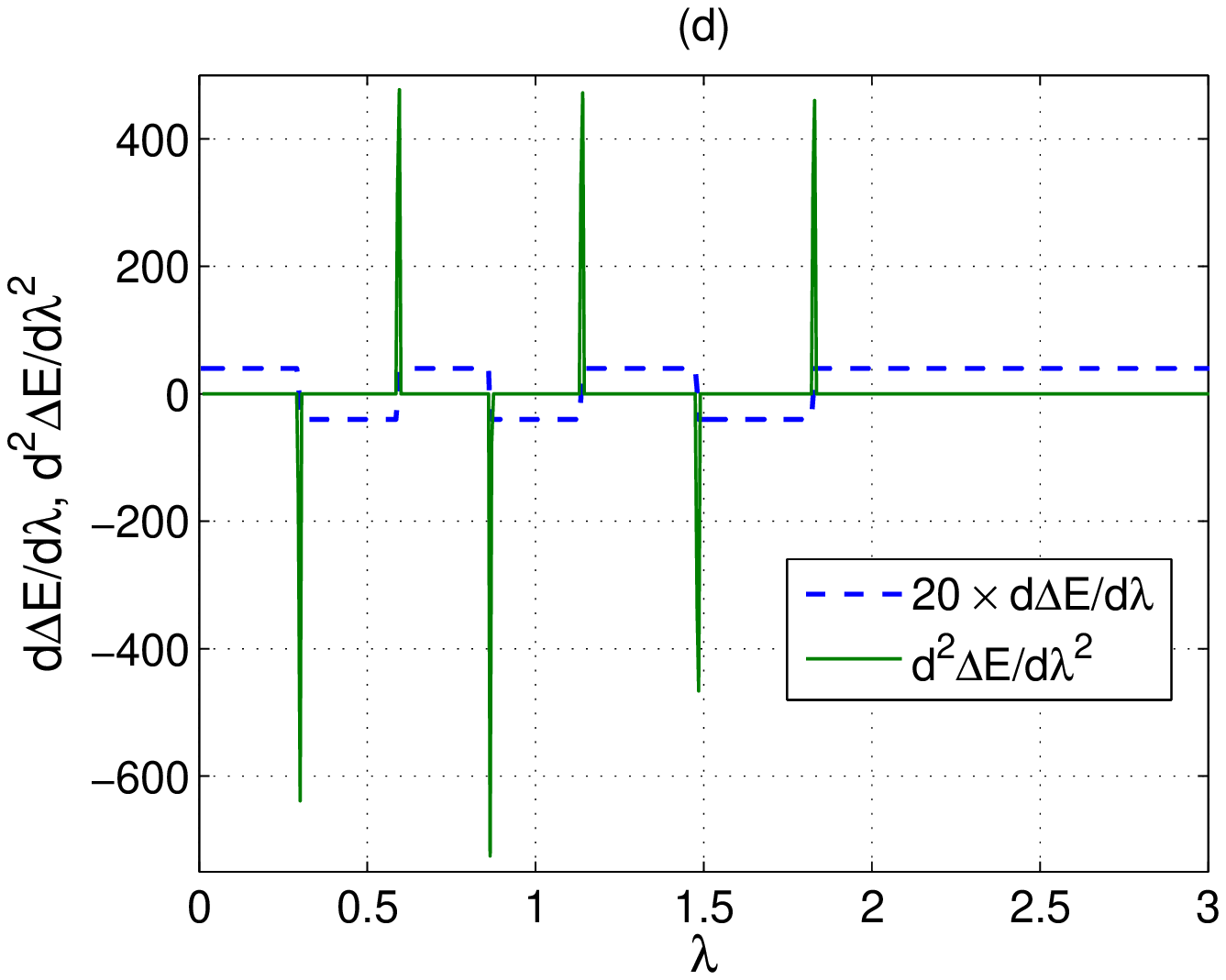}}
\caption{{\protect\footnotesize (Color online) (a) The concurrence $C_{14}$ versus $\lambda$; (b) the first derivative of the concurrence $C_{14}$ with respect to $\lambda$ versus $\lambda$; (c) the energy gap between the ground state and first excited state versus $\lambda$; (d) the first derivative (in units of $J$) and second derivative (in units of $J^2$) of the energy gap with respect $\lambda$ versus $\lambda$ for the pure isotropic system ($\gamma=0$ and $\alpha=0$). Notice that the first derivative of energy gap is enlarged 20 times its actual scale for clearness.}}
\label{QQPT3}
\end{figure}
Critical quantum behavior in a many body system happens either when an actual crossing takes place between the excited state and the ground state or a limiting avoided level-crossing between them exists, i.e. an energy gab between the two states that vanishes in the infinite system size limit at the critical point \cite{Sachdev2001}. When a many body system crosses a critical point, significant changes in both its wave function and ground state energy takes place, which are manifested in the behavior of the entanglement function. The entanglement in one dimensional infinite spin systems, Ising and $XY$, was shown to demonstrate scaling behavior in the vicinity of critical points \cite{Osterloh2002}. The change in the entanglement across the critical point was quantified by considering the derivative of the concurrence with respect to the parameter $\lambda$. This derivative was explored versus $\lambda$ for different system sizes and although it didn't show divergence for finite system sizes, it showed clear anomalies which developed to a singularity at the thermodynamic limit. The ground state of the Heisenberg spin model is known to have a double degeneracy for an odd number of spins which is never achieved unless the thermodynamic limit is reached \cite{Sachdev2001}. Particularly, the Ising 1D spin chain in an external transverse magnetic field has doubly degenerate ground state in a ferromagnetic phase that is gapped from the excitation spectrum by $2 J (1-h/J)$, which is removed at the critical point and the system becomes in a paramagnetic phase. Now let us first consider our two-dimensional finite size Ising spin system. The concurrence $C_{14}$ and its first derivative are depicted versus $\lambda$ in figs.~\ref{QQPT1}(a) and (b) respectively. As one can see, the derivative of the concurrence shows strong tendency of being singular at $\lambda_c = 1.64$. The characteristics of the energy gap between the ground state and the first excited state as a function of $\lambda$ are explored in fig.~\ref{QQPT1}(c). The system shows strict double degeneracy, zero energy gap, only at $\lambda =0$  i.e. at zero magnetic field, but once the magnetic field is on the degeneracy is lifted and an extremely small energy gap develops, which increase very slowly for small magnetic field values but increases abruptly at certain $\lambda$ value. It is important to emphasis here that at $\lambda=0$, regardless of which one of the double ground states is selected for evaluating the entanglement, the same value is obtained. The critical point of a phase transition should be characterized by a singularity in the ground state energy, and an abrupt change in the energy gap of the system as a function of the system parameter as it crosses the critical point. To better understand the behavior of the energy gap across the prospective critical point and identify it, we plot the first and second derivatives of the energy gap as a function of $\lambda$ in fig.~\ref{QQPT1}(d). Interestingly, the first derivative $d\Delta E / d \lambda$ which represents the rate of change of the energy gap as a function of $\lambda$ starts with a zero value at $\lambda=0$ and then increase very slowly before it shows a great rate of change and finally reaches a saturation value. This behavior is best represented by the second derivative $d^2 \Delta E / d \lambda^2$, which shows strong tendency of being singular at $\lambda_c=1.8$, which indicates the highest rate of change the energy gap as a function of $\lambda$. The reason for the small discrepancy between the two values of the $\lambda_c$ extracted from the $dC / d \lambda$ plot and the one of $d^2 \Delta E / d \lambda^2$ is that the concurrence $C_{14}$ is only between two sites and does not represent the whole system in contrary to the energy gap. One can conclude that the rate of change of the energy gap as a function of the system parameter, $\lambda$ in our case, should be maximum across the critical point. Turning to the case of the partially anisotropic spin system, $\gamma=0.5$, presented in fig.~\ref{QQPT2}, one can notice from fig.~\ref{QQPT2}(a) that the concurrence shows few sharp changes, which is reflected in the energy gap plot as an equal number of minima as shown in fig.~\ref{QQPT2}(b). Nevertheless, again there is only one strict double degeneracy at $\lambda=0$ while the other three energy gap minima are non-zero and in the order of $10^{-5}$. It is interesting to notice that the anomalies in both $dC / d \lambda$ and $d^2 \Delta E / d \lambda^2$ are much stronger and sharper compared with the Ising case as shown in figs.~\ref{QQPT2}(c) and (d). Finally the isotropic system which is depicted in fig.~\ref{QQPT3}, shows even sharper energy gap changes as a result of the sharp changes in the concurrence and the anomalies in the derivatives $dC / d \lambda$ and $d^2 \Delta E / d \lambda^2$ are even much stronger than the previous two cases.
\subsection{System dynamics with impurity}
\begin{figure}[htbp]
\begin{minipage}[c]{\textwidth}
\centering
   \subfigure{\includegraphics[width=8 cm]{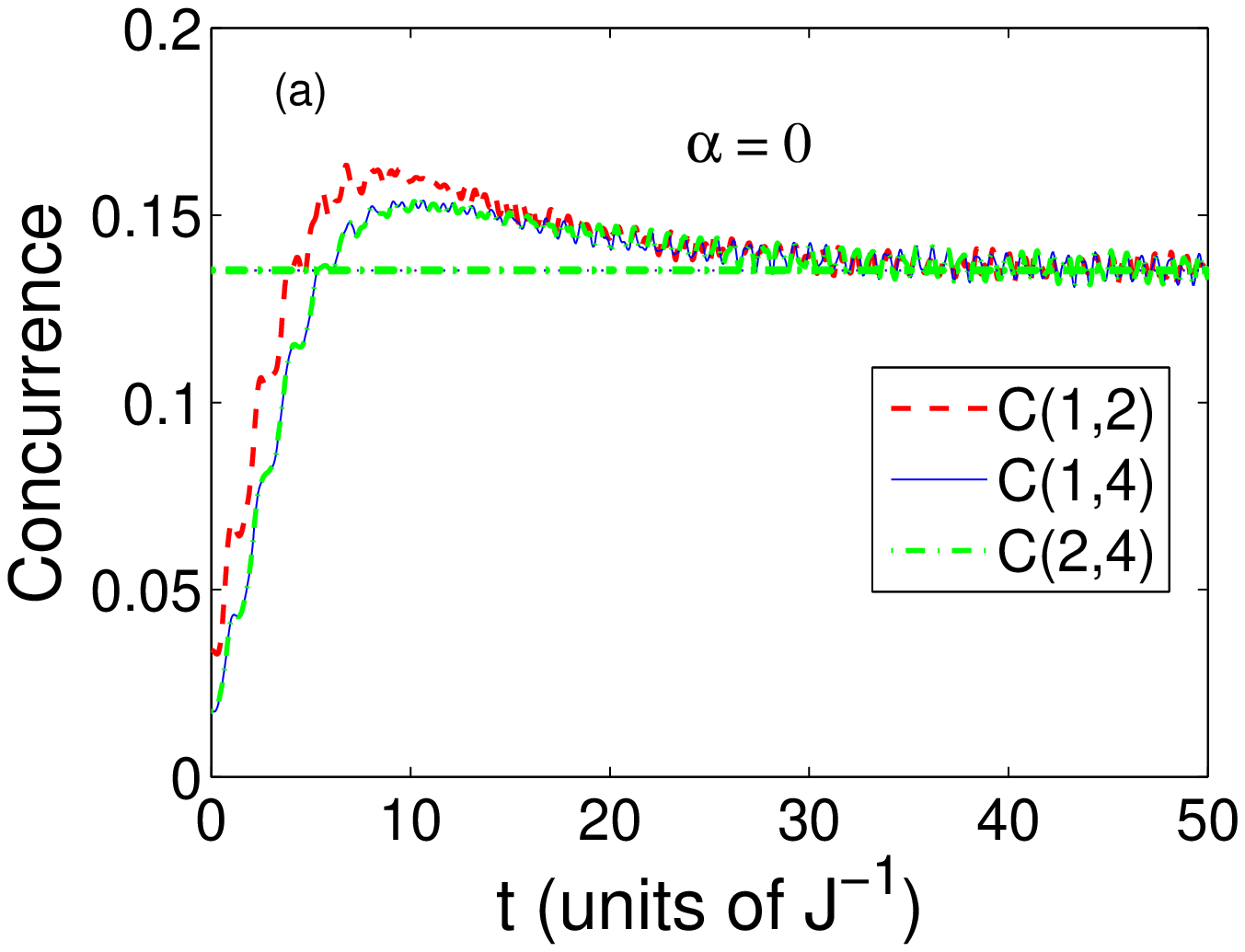}}\quad
   \subfigure{\includegraphics[width=8 cm]{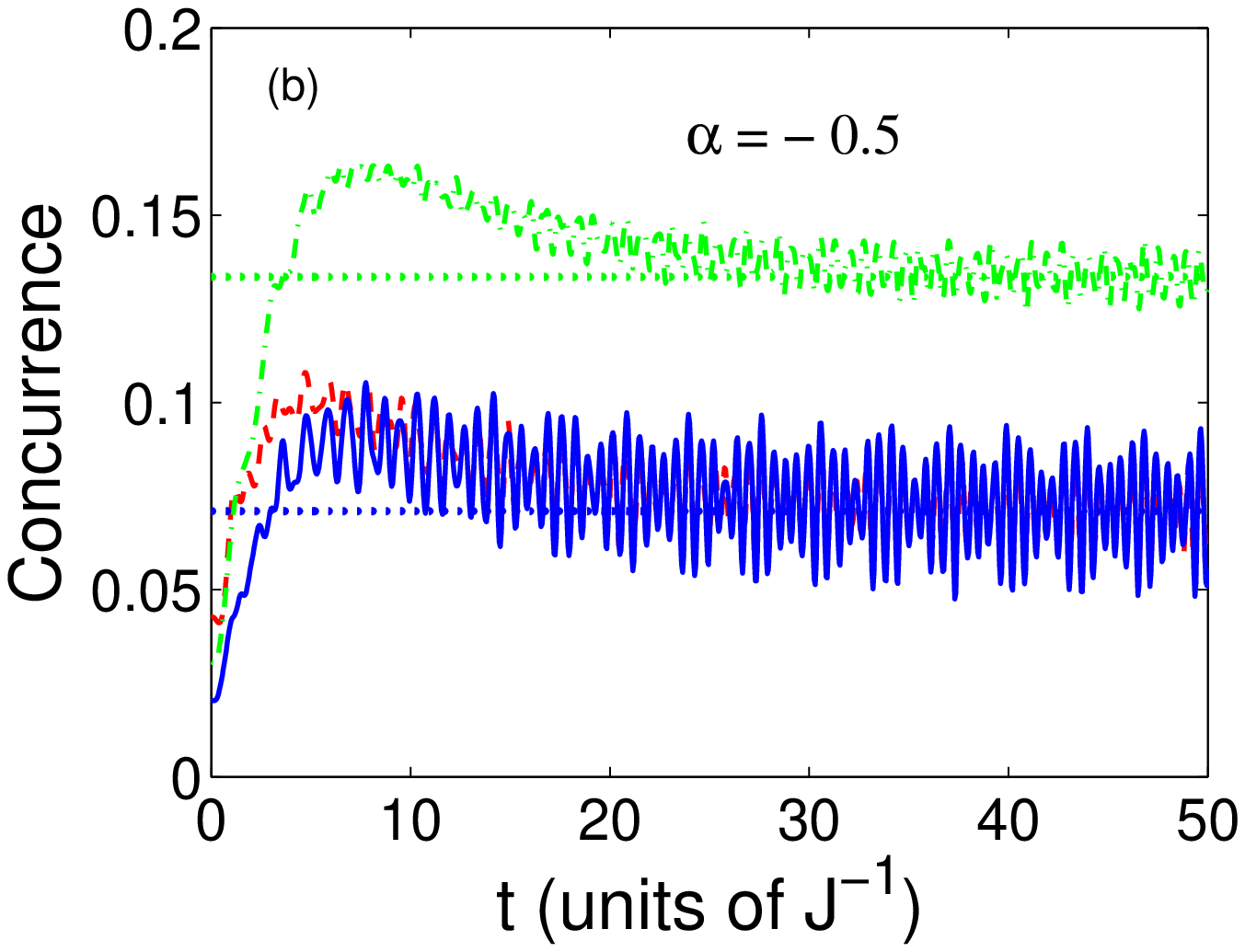}}\\
   \subfigure{\includegraphics[width=8 cm]{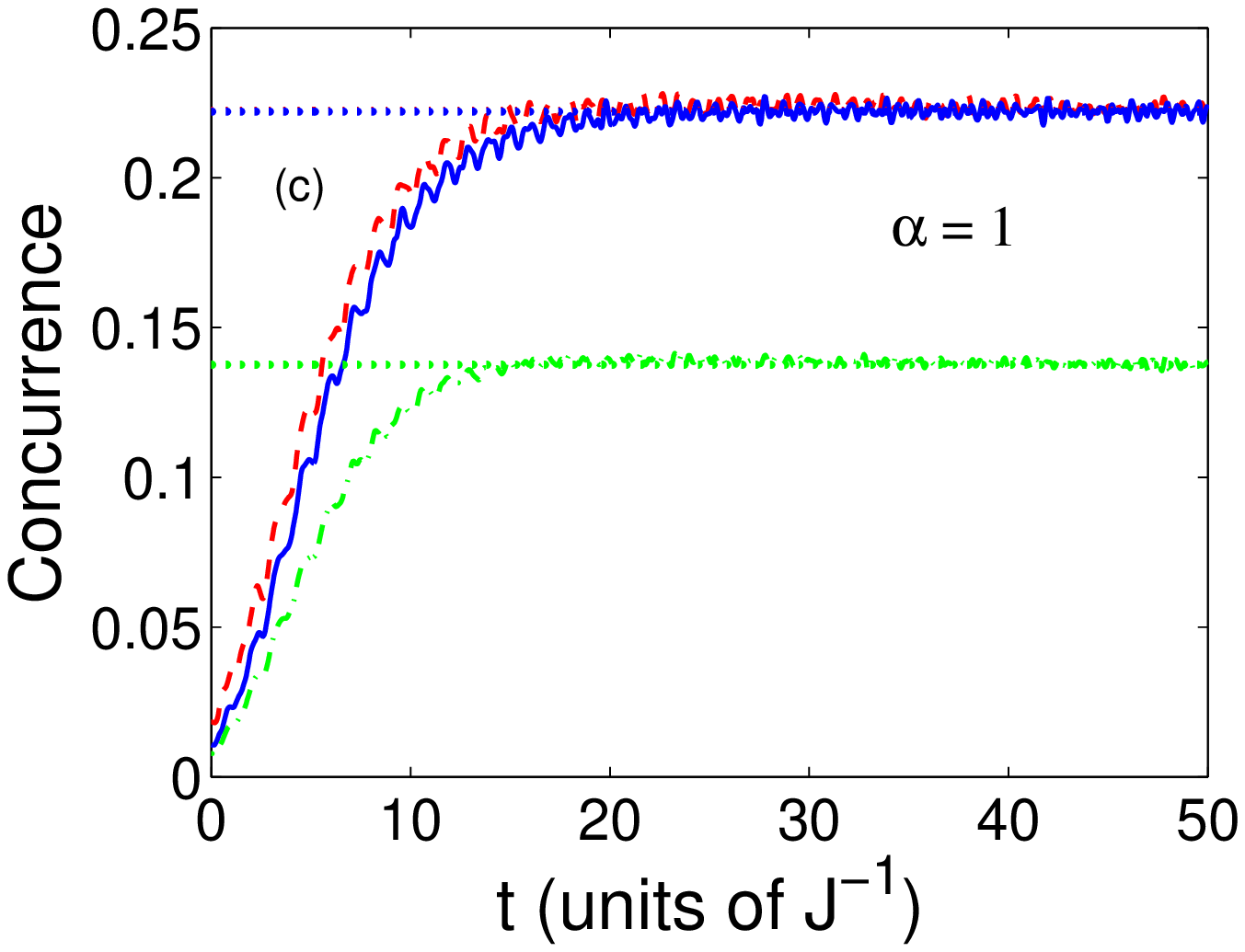}}\quad
   \subfigure{\includegraphics[width=8 cm]{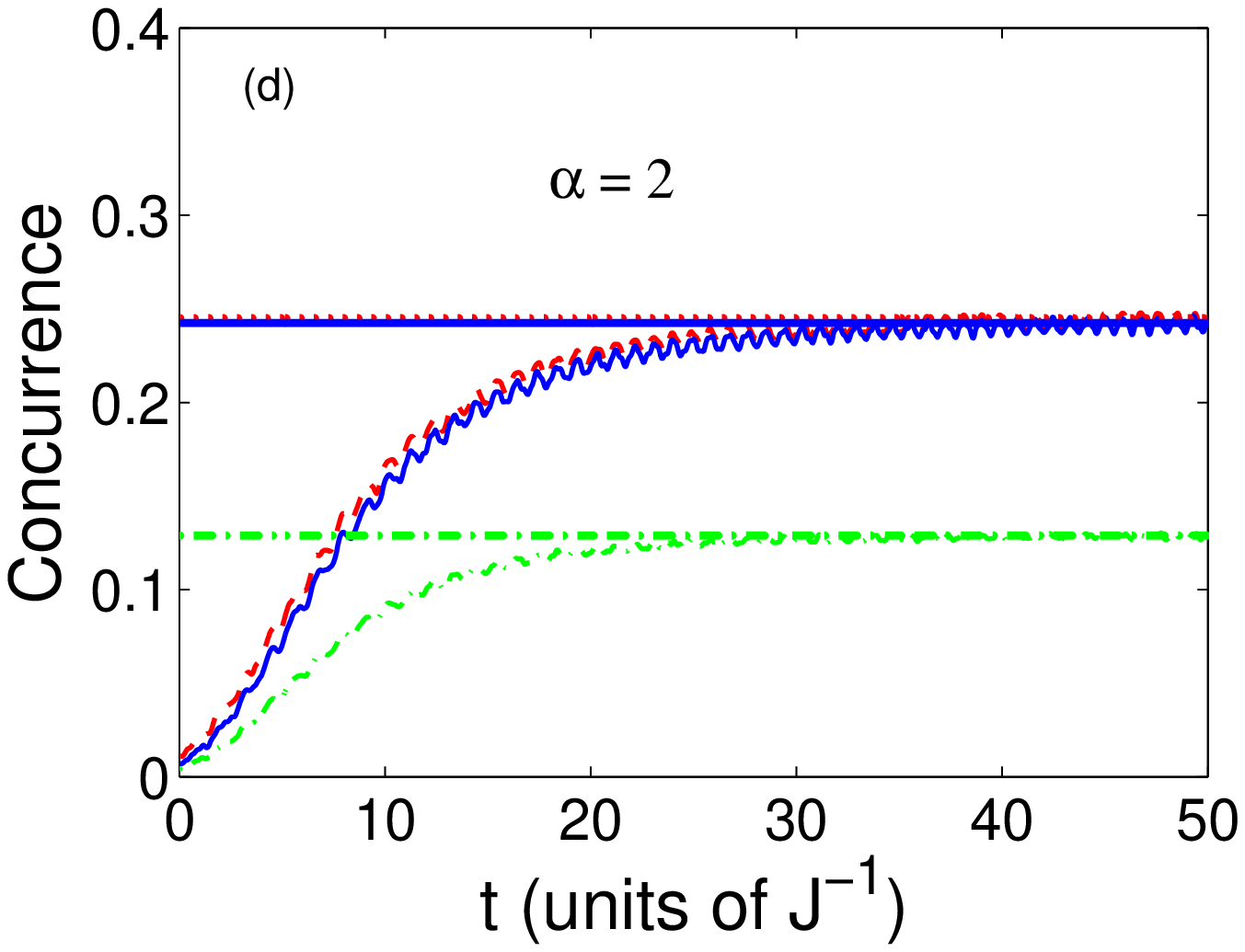}}
   \caption{{\protect\footnotesize (Color online) Dynamics of the concurrences $C(1,2), C(1,4), C(2,4)$ with a single impurity at the border site 1 with different impurity coupling strengths $\alpha = -0.5, 0, 1, 2$ for the two dimensional Ising lattice ($\gamma = 1$) under the effect of an exponential magnetic field with parameters values a=1, b=3.5 and $\omega=0.1$. The straight lines represent the equilibrium concurrences corresponding to constant magnetic field $h=3.5$. The legend for all subfigures is as shown in subfigure (a).}}
 \label{B_Dyn_G_1}
 \end{minipage}
\end{figure}
Now we turn to the dynamics of the two dimensional spin system under the effect of a single impurity and different degrees of anisotropy. We investigate the dynamical reaction of the system to an applied time-dependent magnetic field with exponential form $h(t)= b + (a-b) e^{-w \; t}$ for $t > 0$ and $h(t)=a$ for $t \leq 0$.

We start by considering the Ising system, $\gamma=1$ with a single impurity at the border site 1, which is explored in fig.~\ref{B_Dyn_G_1}, where we set $a=1$, $b=3.5$ and $\omega=0.1$. For the pure case, $\alpha=0$ shown in fig.~\ref{B_Dyn_G_1}(a), the results confirms the ergodic behavior of the system that was demonstrated in our previous work \cite{XuQ2011}, where the asymptotic value of the entanglement coincide with the equilibrium state value at $h(t)=b$. As can be noticed from figs.~\ref{B_Dyn_G_1}(b), \ref{B_Dyn_G_1}(c) and \ref{B_Dyn_G_1}(d) neither the weak nor strong impurities have effect on the ergodicity of the Ising system. Nevertheless, there is a clear effect on the asymptotic value of entanglements $C(1,2)$ and $C(1,4)$ but not on $C(2,4)$ which relates two regular sites. The weak impurity, $\alpha=-0.5$ reduces the asymptotic value of $C(1,2)$ and $C(1,4)$ while the strong impurities, $\alpha= 1, 2$ raise it compared to the pure case.
\begin{figure}[htbp]
\begin{minipage}[c]{\textwidth}
\centering
   \subfigure{\includegraphics[width=8 cm]{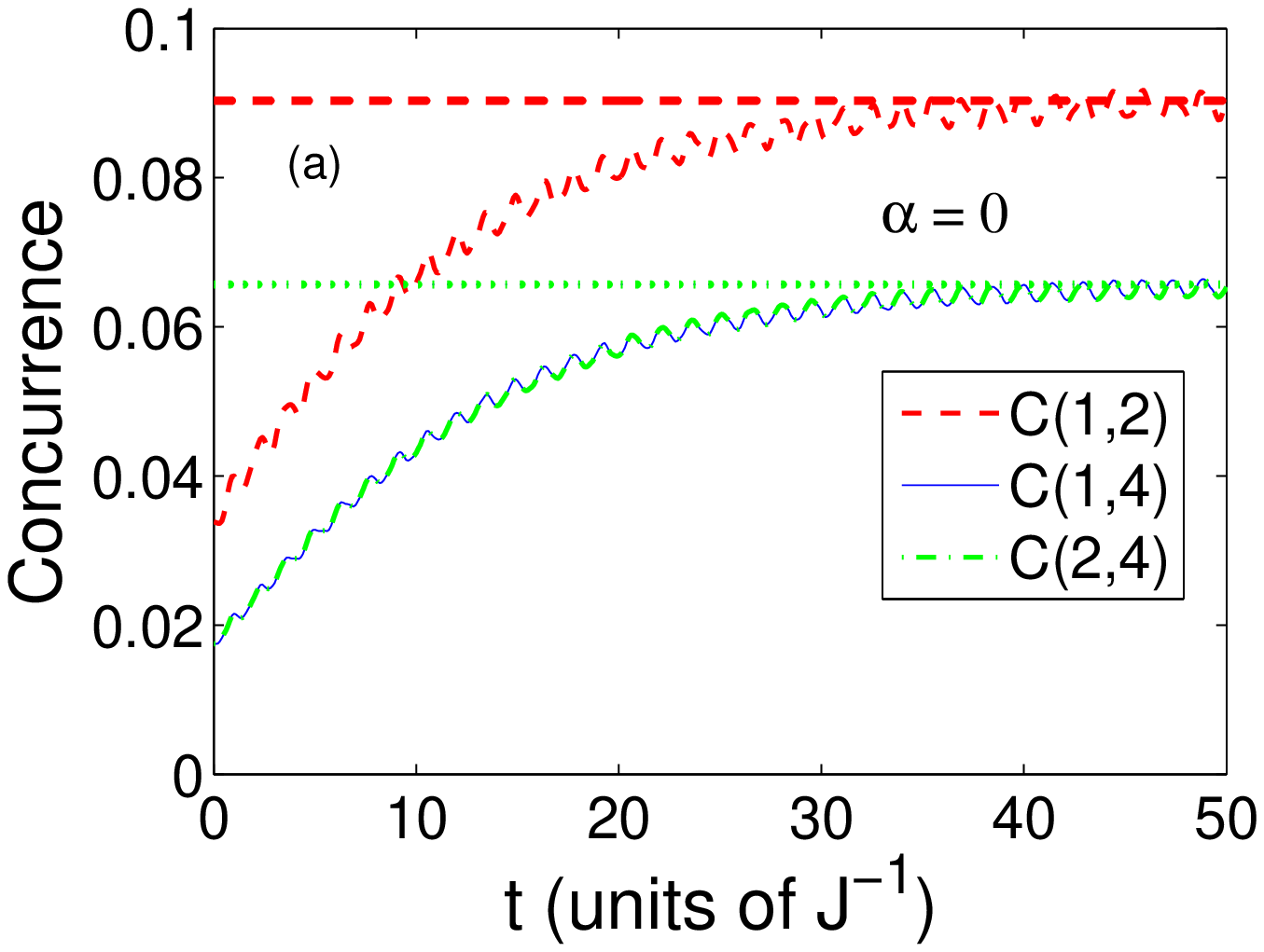}}\quad
   \subfigure{\includegraphics[width=8 cm]{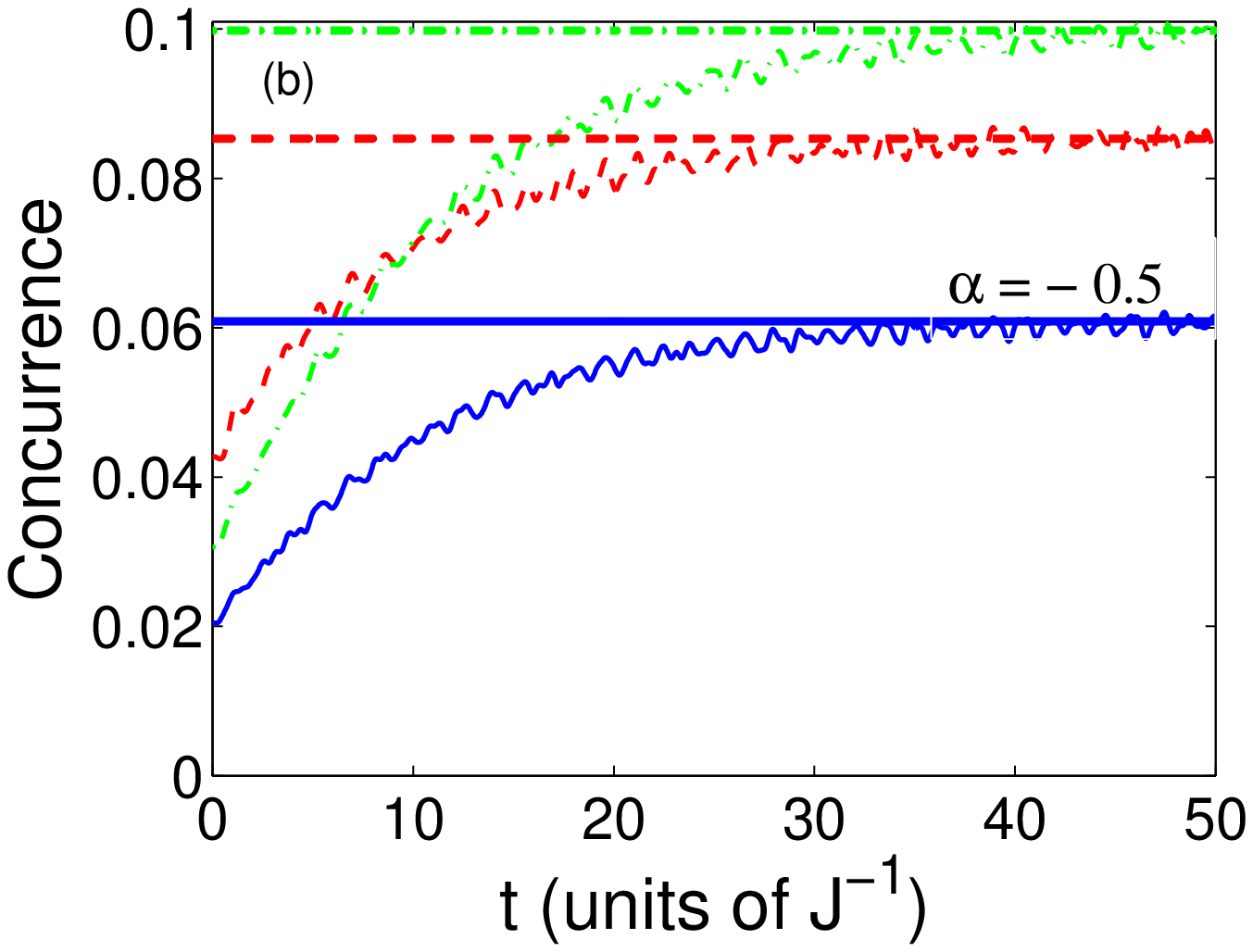}}\\
   \subfigure{\includegraphics[width=8 cm]{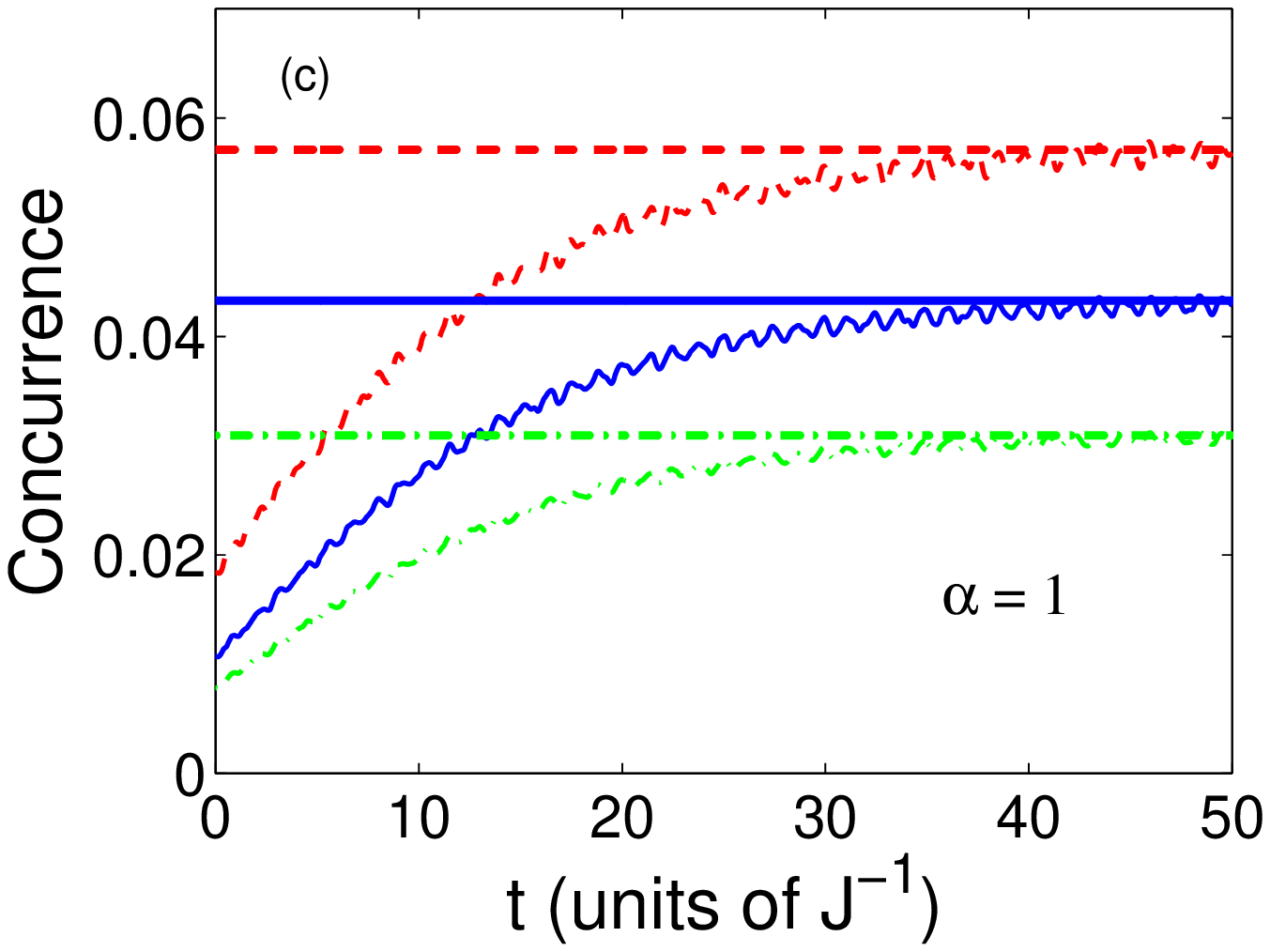}}\quad
   \subfigure{\includegraphics[width=8 cm]{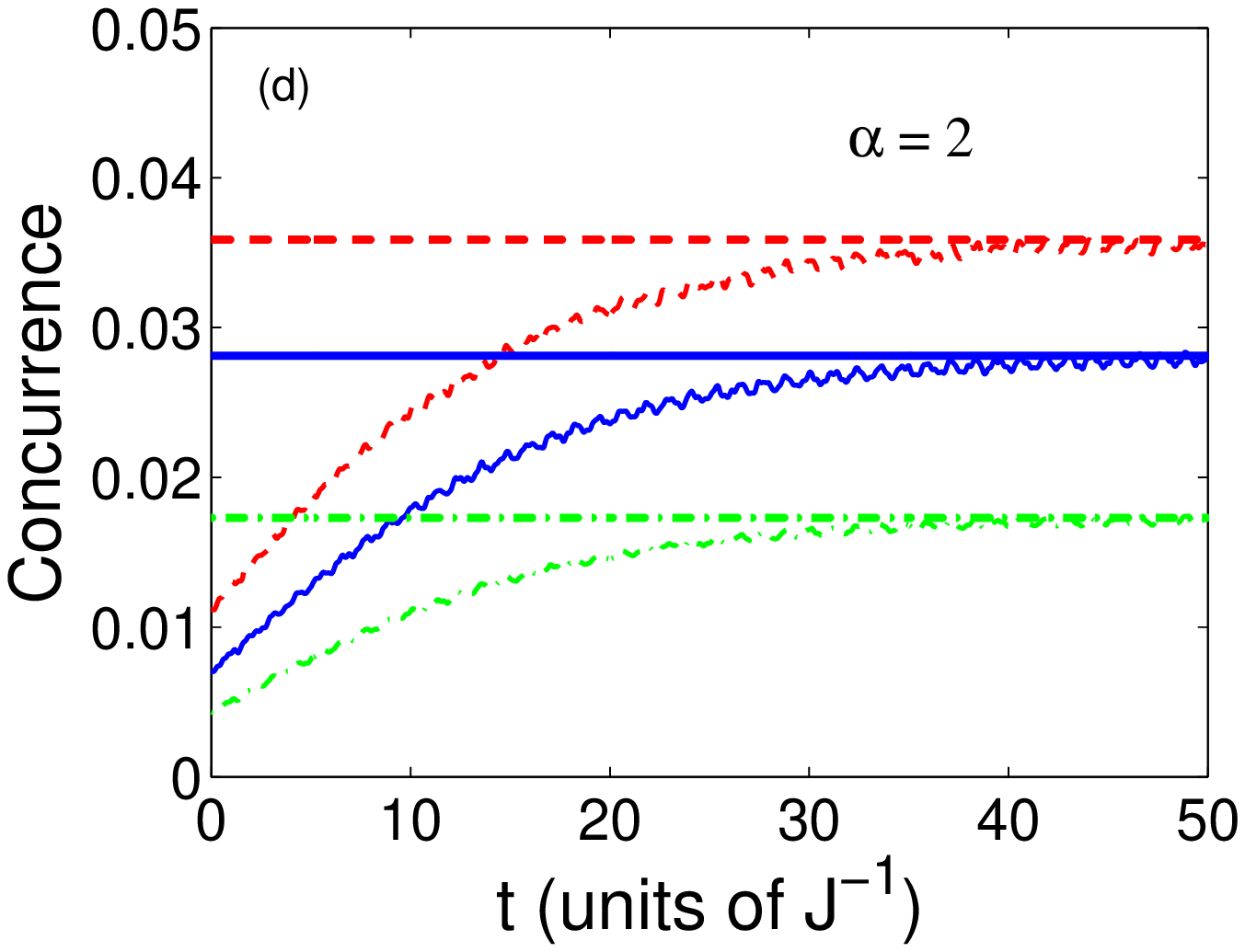}}
   \caption{{\protect\footnotesize (Color online) Dynamics of the concurrences $C(1,2), C(1,4), C(2,4)$ with a single impurity at the border site 1 with different impurity coupling strengths $\alpha = -0.5, 0, 1, 2$ for the two dimensional Ising lattice ($\gamma = 1$) under the effect of an exponential magnetic field with parameters values a=1, b=1.5 and $\omega=0.1$. The straight lines represent the equilibrium concurrences corresponding to constant magnetic field $h=1.5$. The legend for all subfigures is as shown in subfigure (a).}}
 \label{B_Dyn_G_1_15}
 \end{minipage}
\end{figure}
In fig.~\ref{B_Dyn_G_1_15}, we consider the same system but under the effect of a weaker exponential magnetic field with set of parameters $a =1$, $b=1.5$ and $\omega=0.1$. As can be noticed, the entanglement of the Ising system is still showing an ergodic behavior at all impurity strengths. This means that the Ising system with a single border impurity is always ergodic under the effect of different impurity strengths and different magnetic field parameters. Interestingly, the different impurity strengths have different effects on the asymptotic value of the entanglements compared to the previous case under the effect of the new magnetic field. The weak impurity, as shown in fig.~\ref{B_Dyn_G_1_15}(b), raises the asymptotic value of $C(2,4)$ and splits those of $C(1,2)$ and $C(1,4)$ from each other. On the other hand, the strong impurity effects are depicted in fig.~\ref{B_Dyn_G_1_15}(c) and fig.~\ref{B_Dyn_G_1_15}(d) which show that the asymptotic values of all concurrences are reduced significantly as the impurity strength increases, in contrary to the previous case. This emphasis the important role that the magnetic filed parameters play beside the impurity strength in controlling the entanglement behavior.
\begin{figure}[htbp]
\begin{minipage}[c]{\textwidth}
\centering
   \subfigure{\includegraphics[width=8 cm]{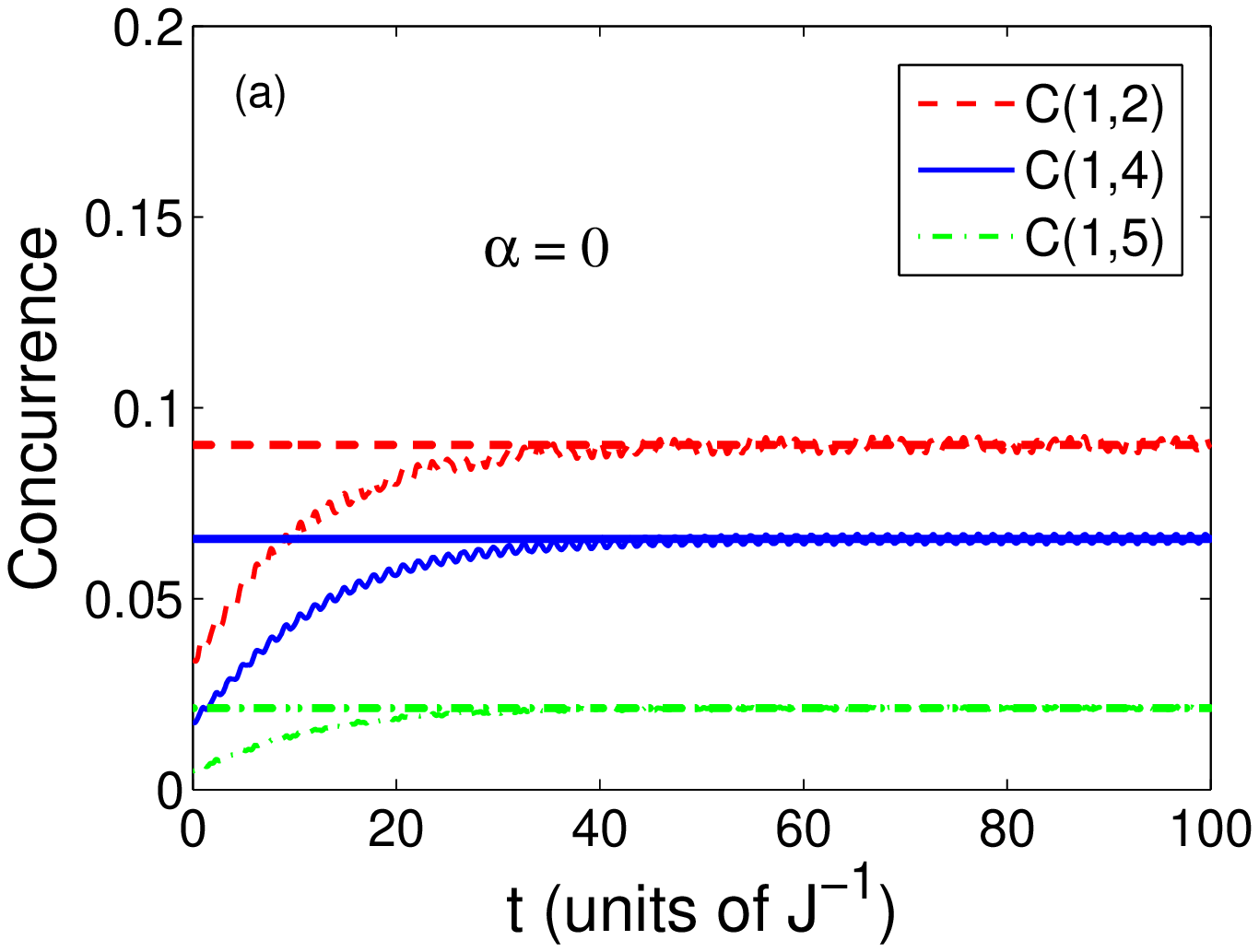}}\quad
   \subfigure{\includegraphics[width=8 cm]{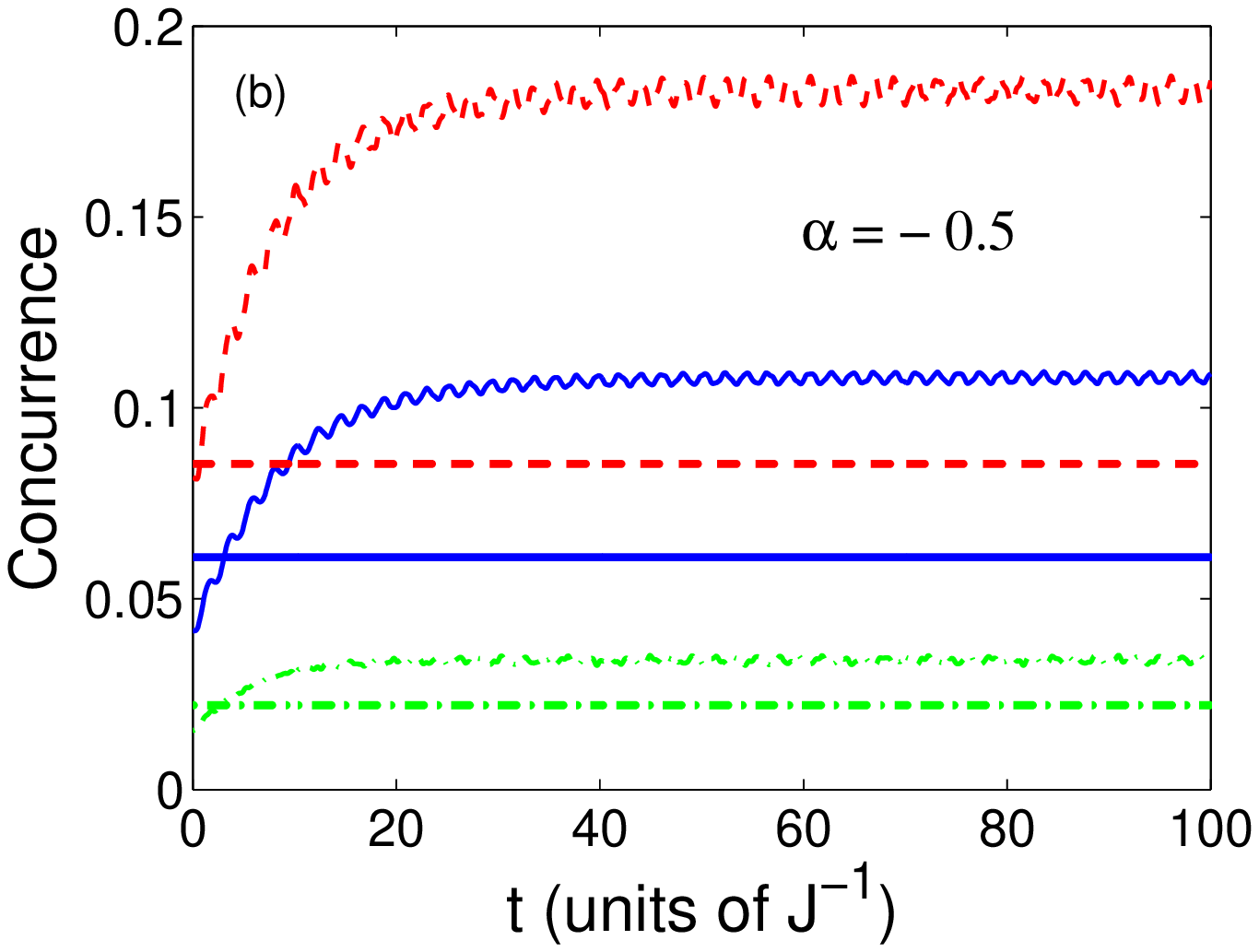}}\\
   \subfigure{\includegraphics[width=8 cm]{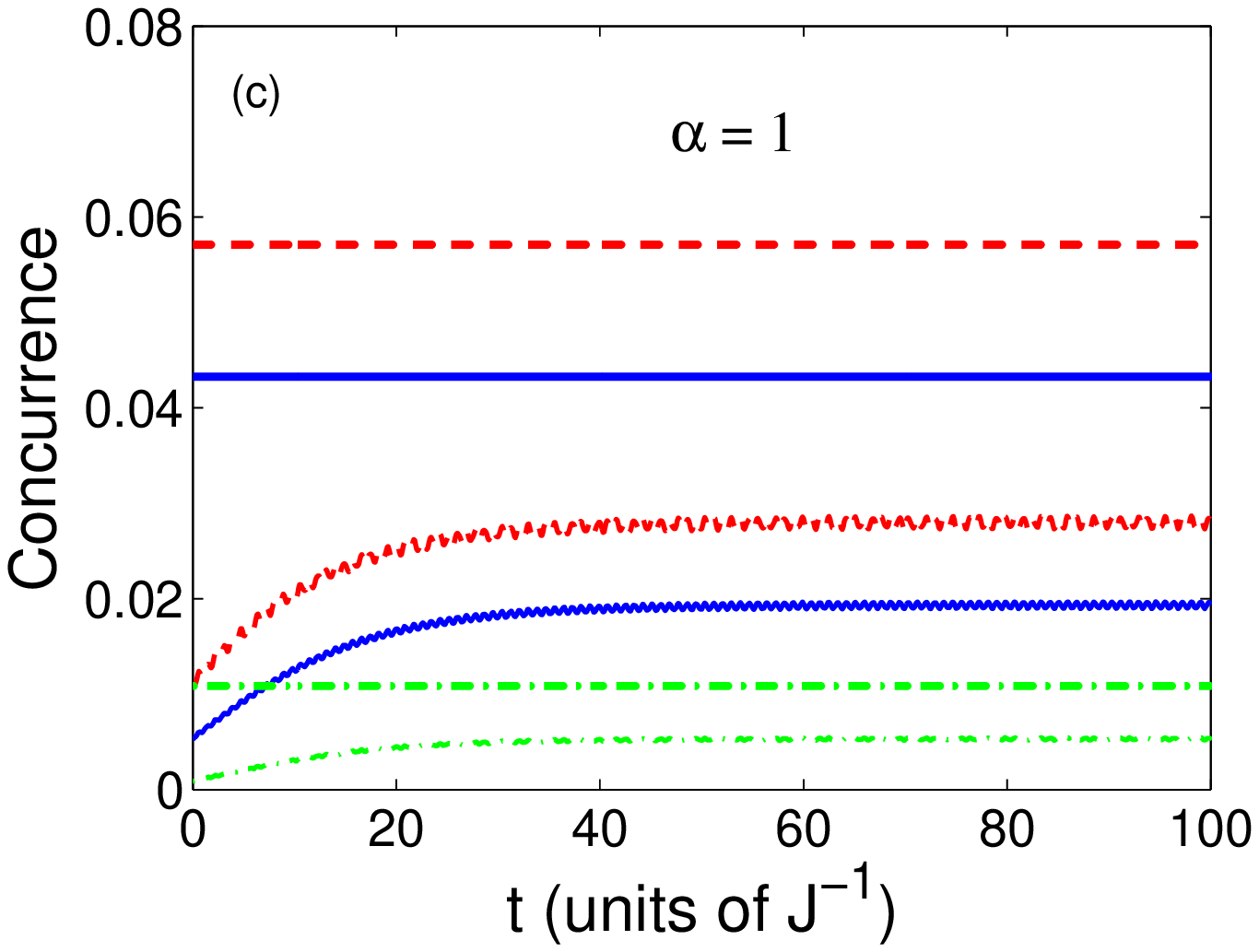}}\quad
   \subfigure{\includegraphics[width=8 cm]{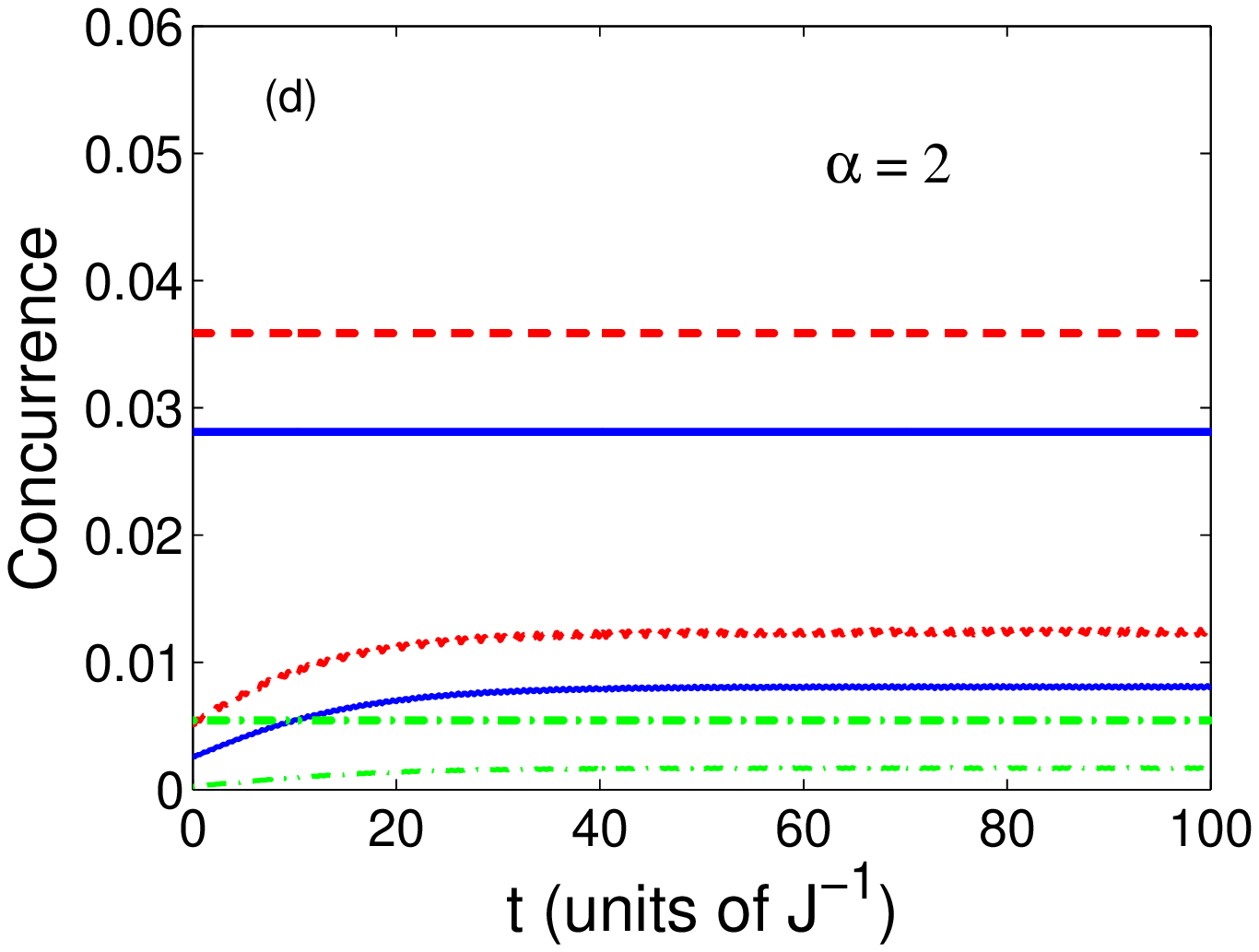}}
   \caption{{\protect\footnotesize (Color online) Dynamics of the concurrences $C(1,2), C(1,4), C(1,5)$ with a single impurity at the central site 4 with different impurity coupling strengths $\alpha = -0.5, 0, 1, 2$ for the two dimensional Ising lattice ($\gamma = 1$) under the effect of an exponential magnetic field with parameters values a=1, b=1.5 and $\omega=0.1$. The straight lines represent the equilibrium concurrences corresponding to constant magnetic field $h=1.5$. The legend for all subfigures is as shown in subfigure (a).}}
 \label{C_Dyn_G_1}
 \end{minipage}
\end{figure}
It is of great interest to examine the effect of the impurity location, which we investigate in fig.~\ref{C_Dyn_G_1} where a single impurity is located at the central site 4 instead of the border site 1 with exponential magnetic field parameters $a =1$, $b=1.5$ and $\omega=0.1$. Very interestingly, the entanglement behavior changes significantly as a result of changing the impurity location. Though the pure Ising system is still ergodic as shown in fig.~\ref{C_Dyn_G_1}(a), the system with weak and strong impurity becomes non-ergodic which is illustrated in figs.~\ref{C_Dyn_G_1}(b), ~\ref{C_Dyn_G_1}(c) and ~\ref{C_Dyn_G_1}(d) respectively. Again the weak impurity raises the asymptotic values wheres the strong impurities reduces them significantly.
\begin{figure}[htbp]
\begin{minipage}[c]{\textwidth}
\centering
   \subfigure{\includegraphics[width=8 cm]{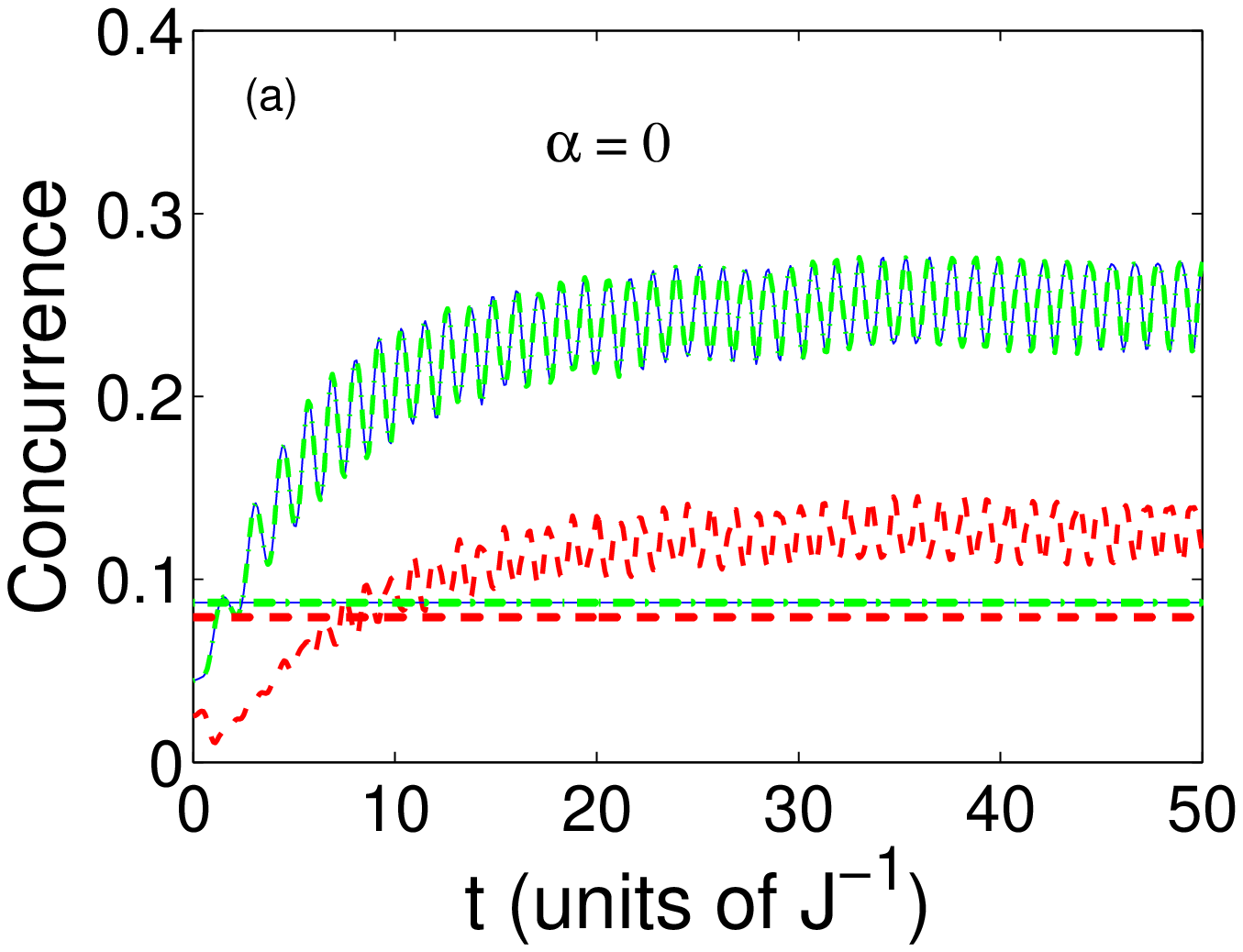}}\quad
   \subfigure{\includegraphics[width=8 cm]{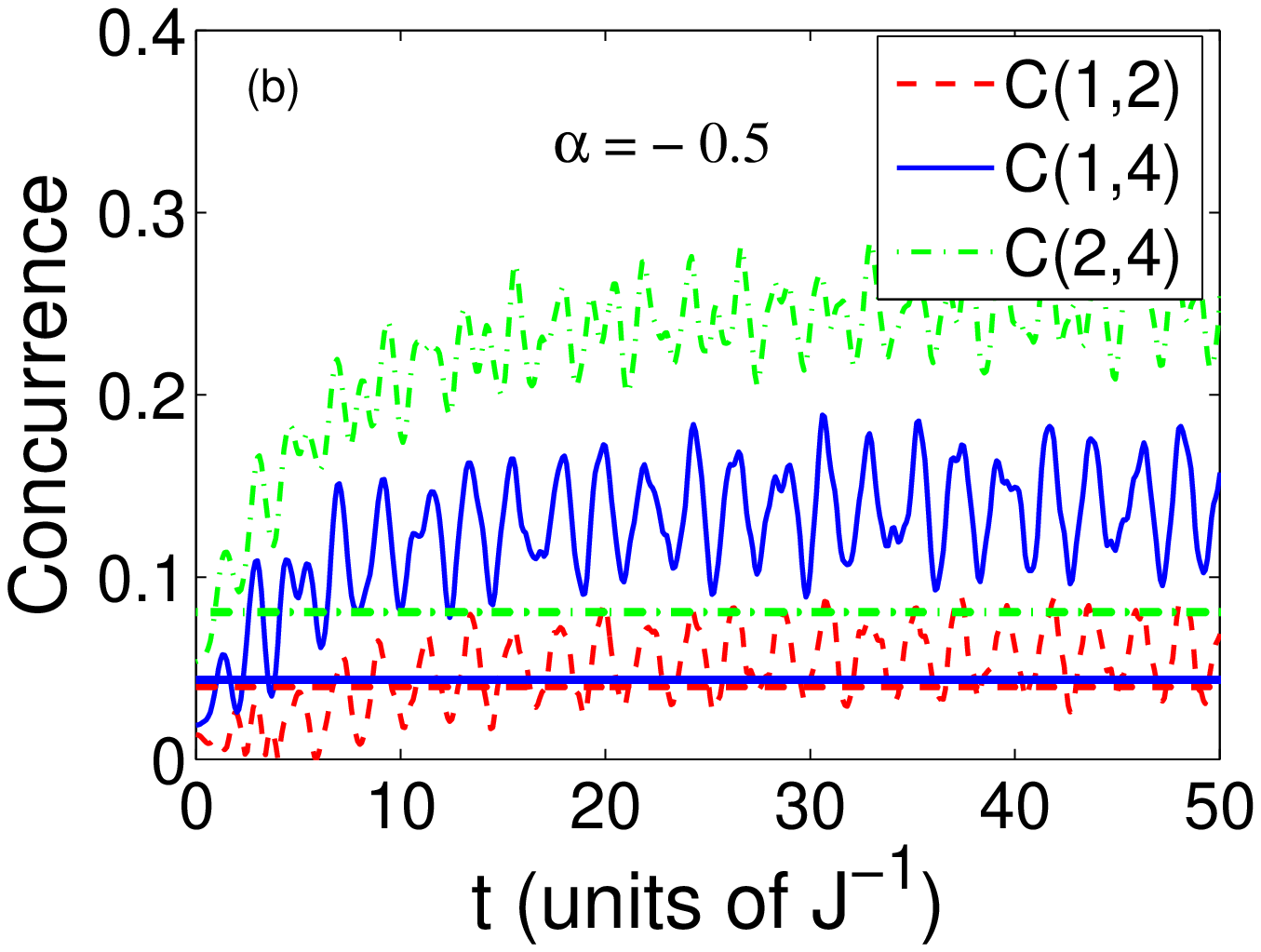}}\\
   \subfigure{\includegraphics[width=8 cm]{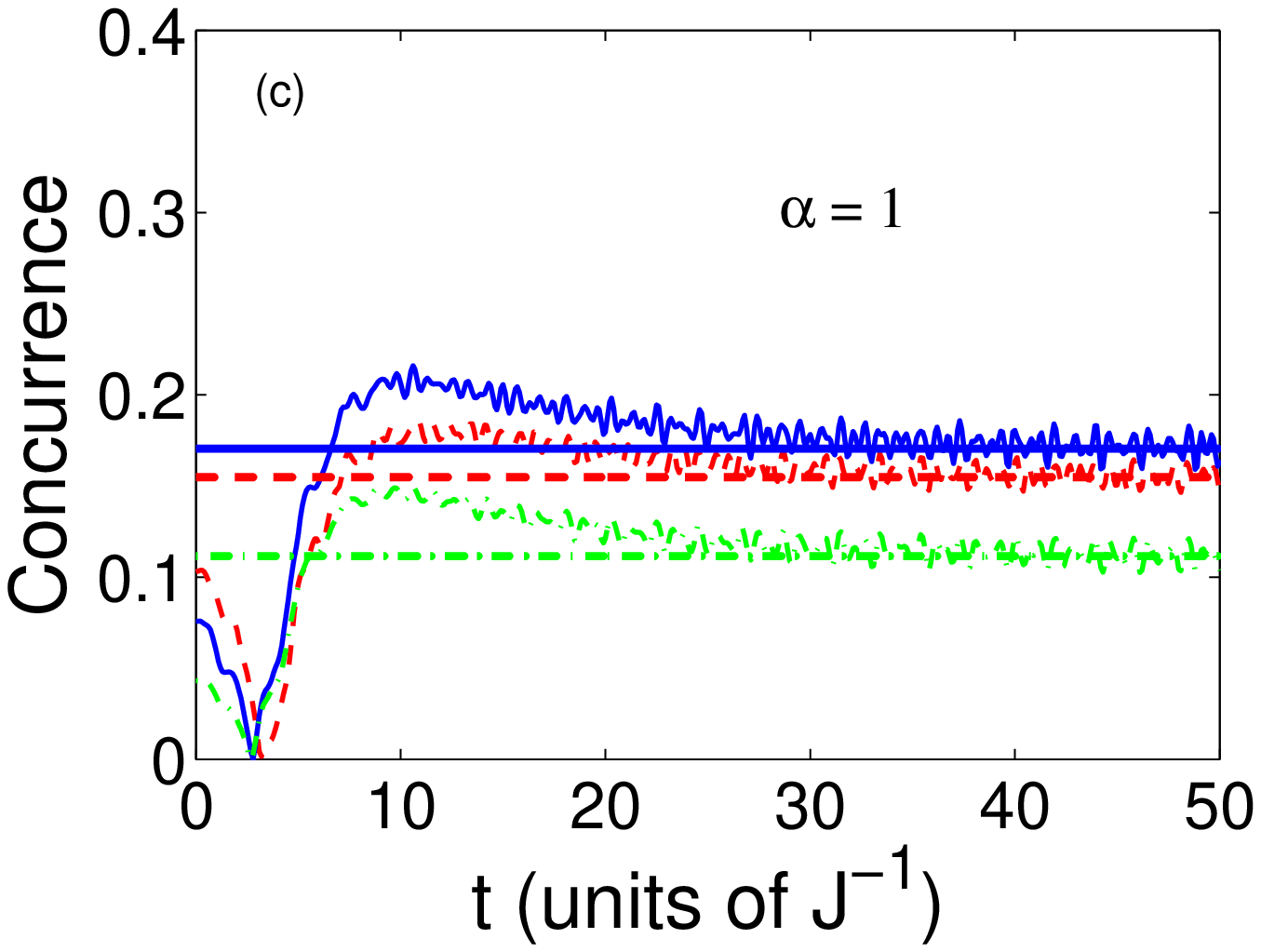}}\quad
   \subfigure{\includegraphics[width=8 cm]{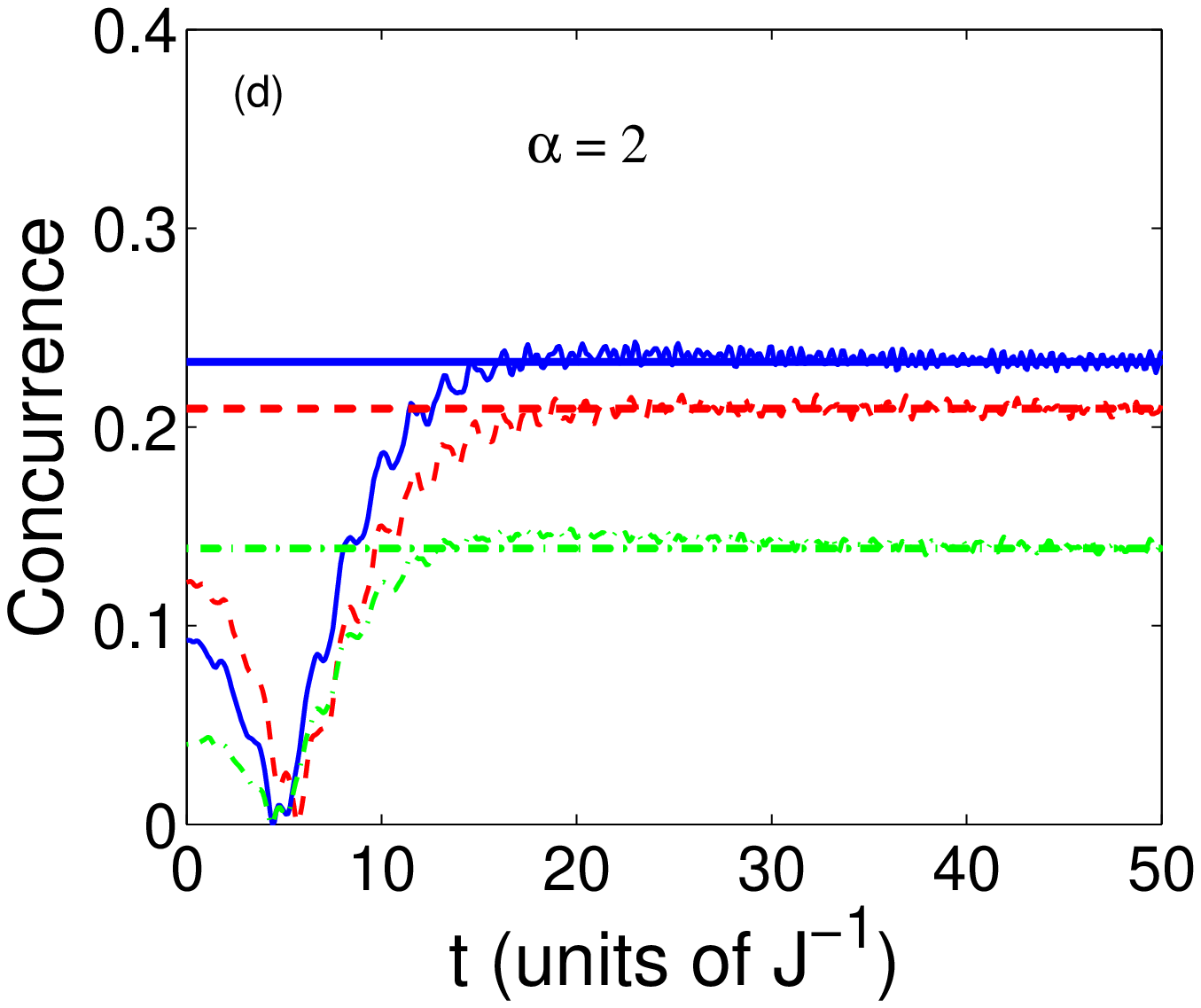}}
   \caption{{\protect\footnotesize (Color online) Dynamics of the concurrences $C(1,2), C(1,4), C(2,4)$ with a single impurity at the border site 1 with different impurity coupling strengths $\alpha = -0.5, 0, 1, 2$ for the two dimensional partially anisotropic lattice ($\gamma = 0.5$) under the effect of an exponential magnetic field with parameters values a=1, b=3.5 and $\omega=0.1$. The straight lines represent the equilibrium concurrences corresponding to constant magnetic field $h=3.5$. The legend for all subfigures is as shown in subfigure (b).}}
 \label{B_Dyn_G_05}
 \end{minipage}
\end{figure}
The dynamics of the partially anisotropic XY system under the effect exponential magnetic field with parameters $a =1$, $b=3.5$ and $\omega = 0.1$, is explored in fig.~\ref{B_Dyn_G_05}. It is remarkable to see that while for both the pure and weak impurity cases, $\alpha=0$ and $-0.5$ , the system is nonergodic as shown in figs.~\ref{B_Dyn_G_05}(a) and ~\ref{B_Dyn_G_05}(b), and it is ergodic in the strong impurity cases $\alpha= 1$ and 2 as illustrated in figs.~\ref{B_Dyn_G_05}(c) and ~\ref{B_Dyn_G_05}(d).
\begin{figure}[htbp]
\begin{minipage}[c]{\textwidth}
\centering
   \subfigure{\includegraphics[width=8 cm]{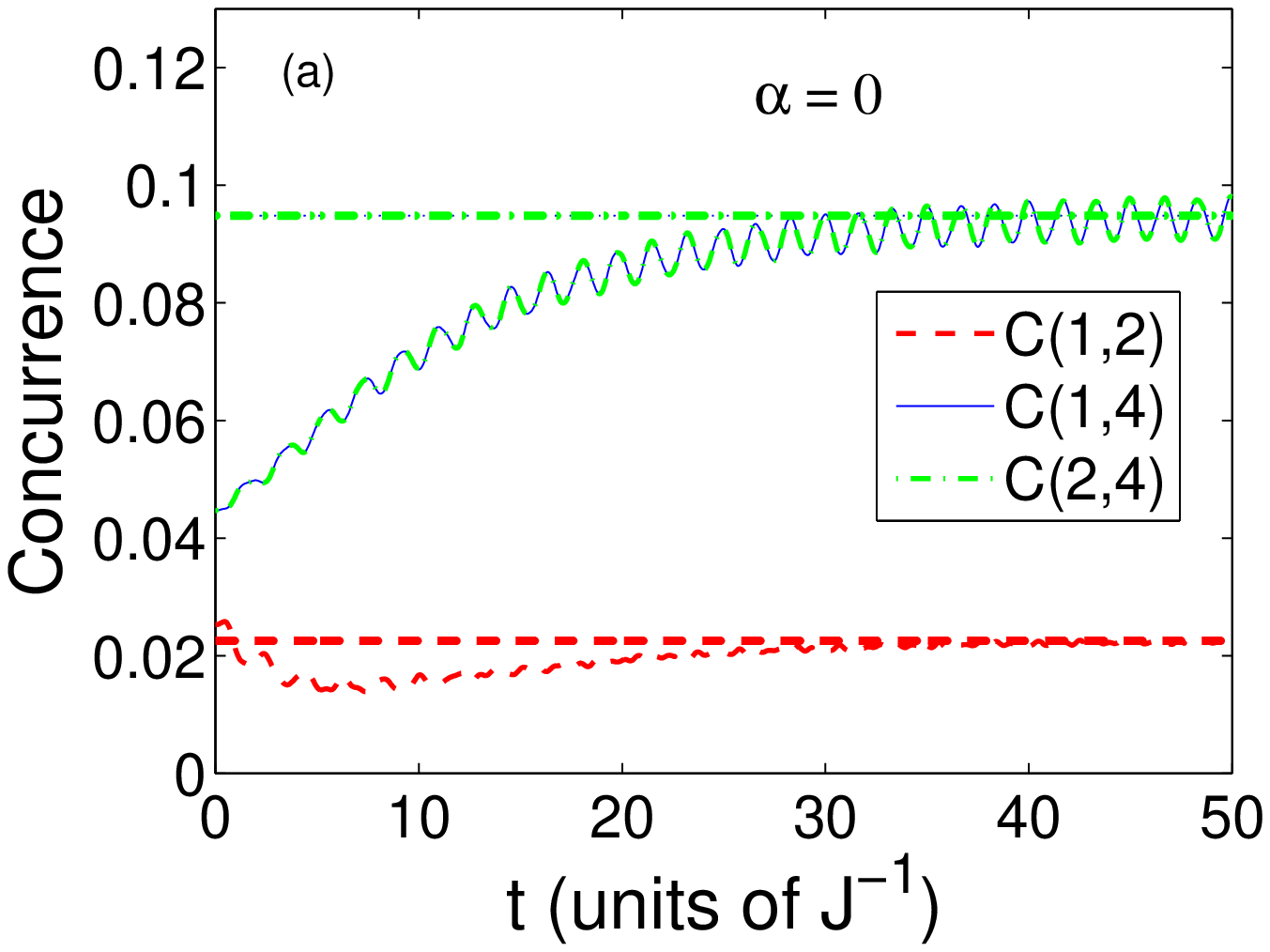}}\quad
   \subfigure{\includegraphics[width=8 cm]{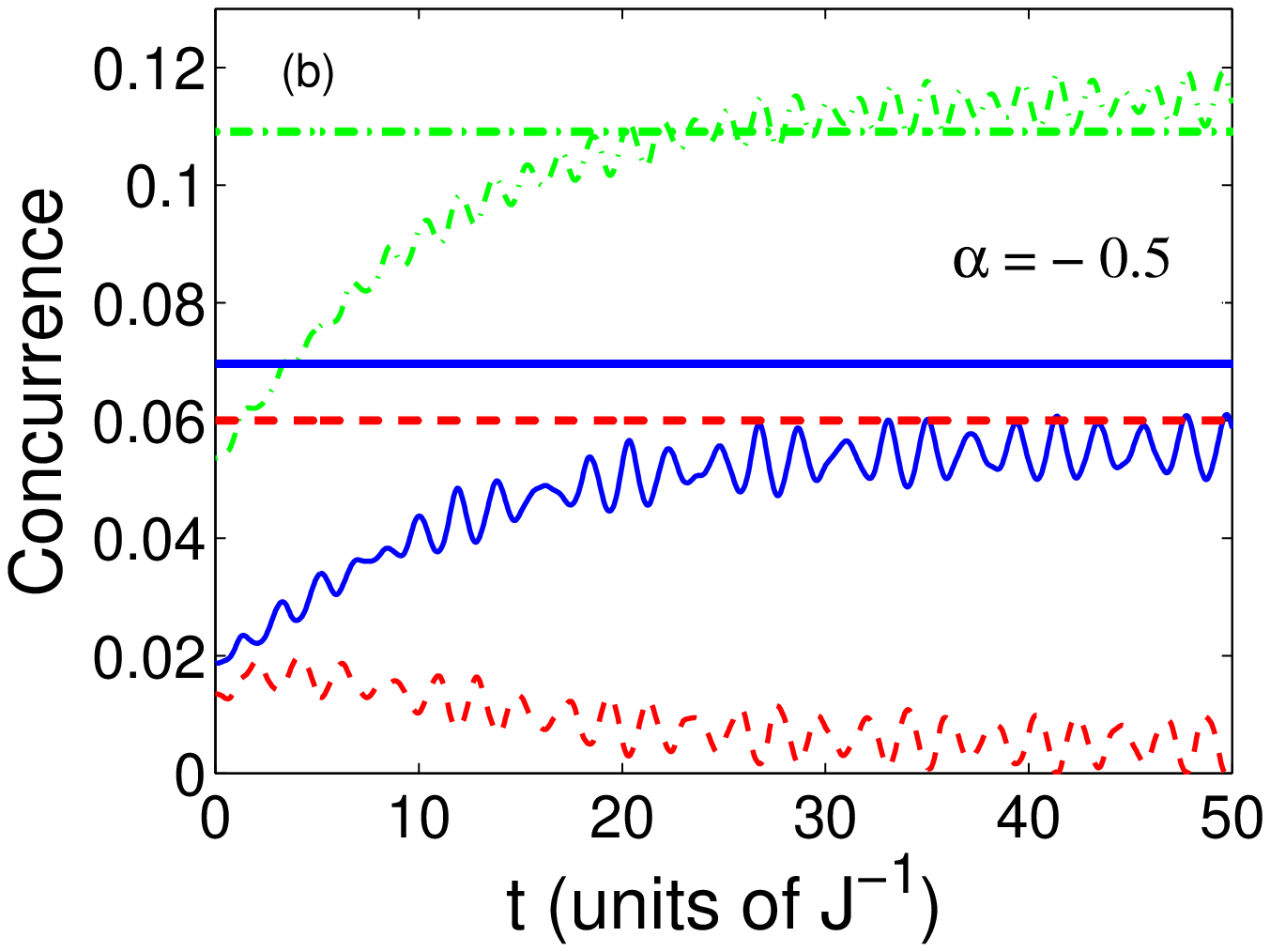}}\\
   \subfigure{\includegraphics[width=8 cm]{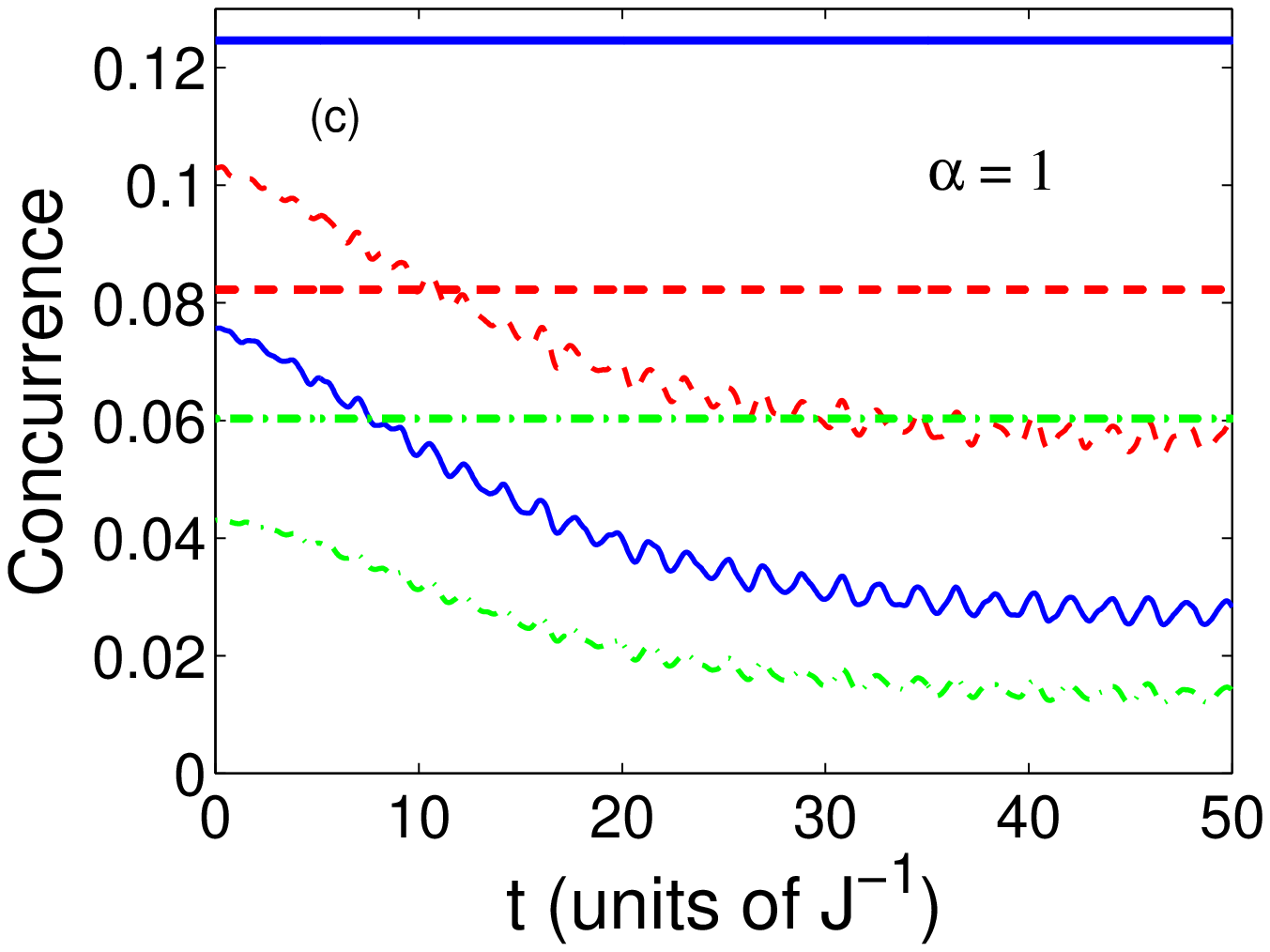}}\quad
   \subfigure{\includegraphics[width=8 cm]{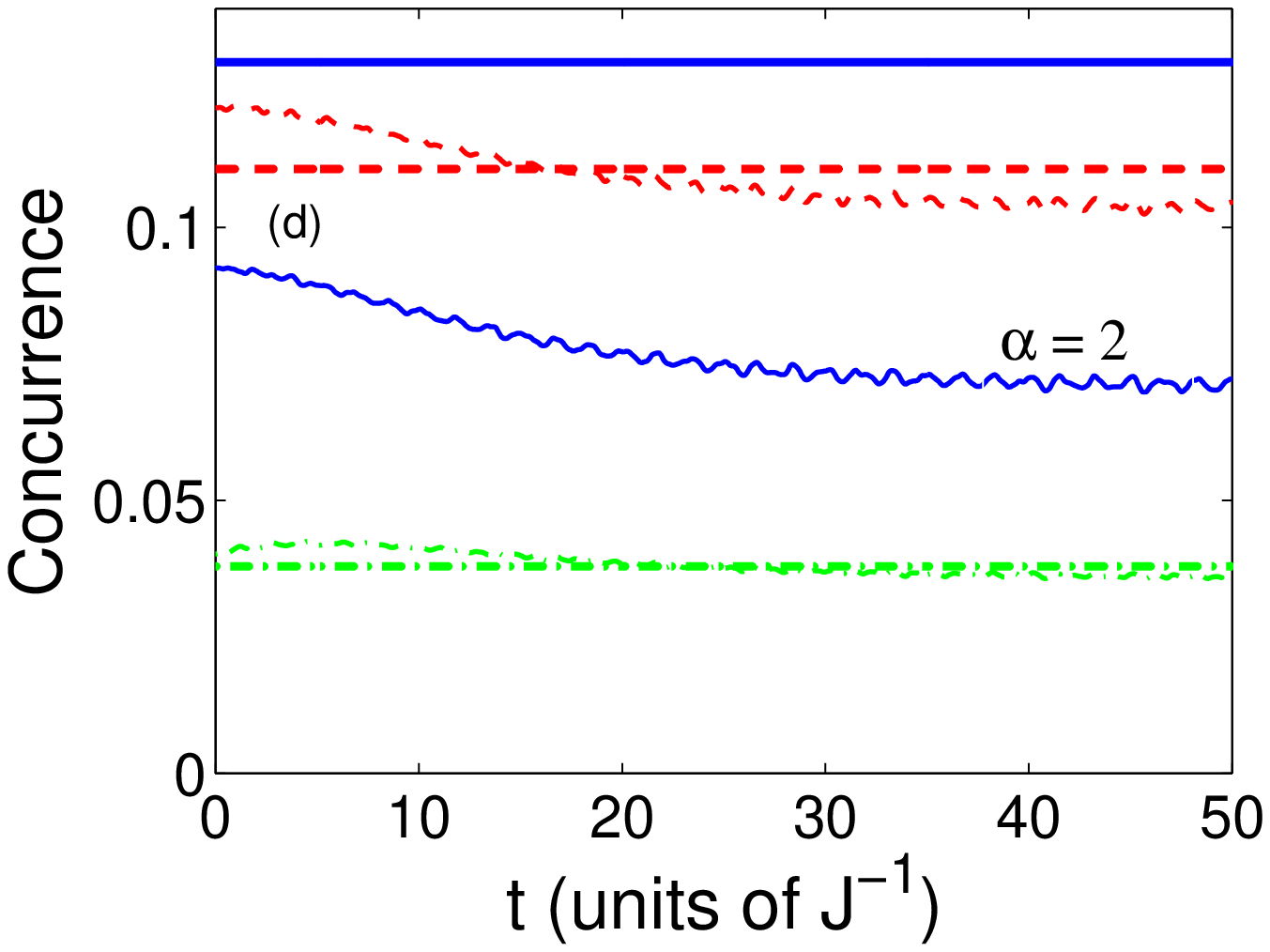}}       \caption{{\protect\footnotesize (Color online) Dynamics of the concurrences $C(1,2), C(1,4), C(2,4)$ with a single impurity at the border site 1 with different impurity coupling strengths $\alpha = -0.5, 0, 1, 2$ for the two dimensional partially anisotropic lattice ($\gamma = 0.5$) under the effect of an exponential magnetic field with parameters values a=1, b=1.5 and $\omega=0.1$. The straight lines represent the equilibrium concurrences corresponding to constant magnetic field $h=1.5$. The legend for all subfigures is as shown in subfigure (a).}}
 \label{B_Dyn_G_05_15}
 \end{minipage}
\end{figure}
\begin{figure}[htbp]
\begin{minipage}[c]{\textwidth}
\centering
   \subfigure{\includegraphics[width=8 cm]{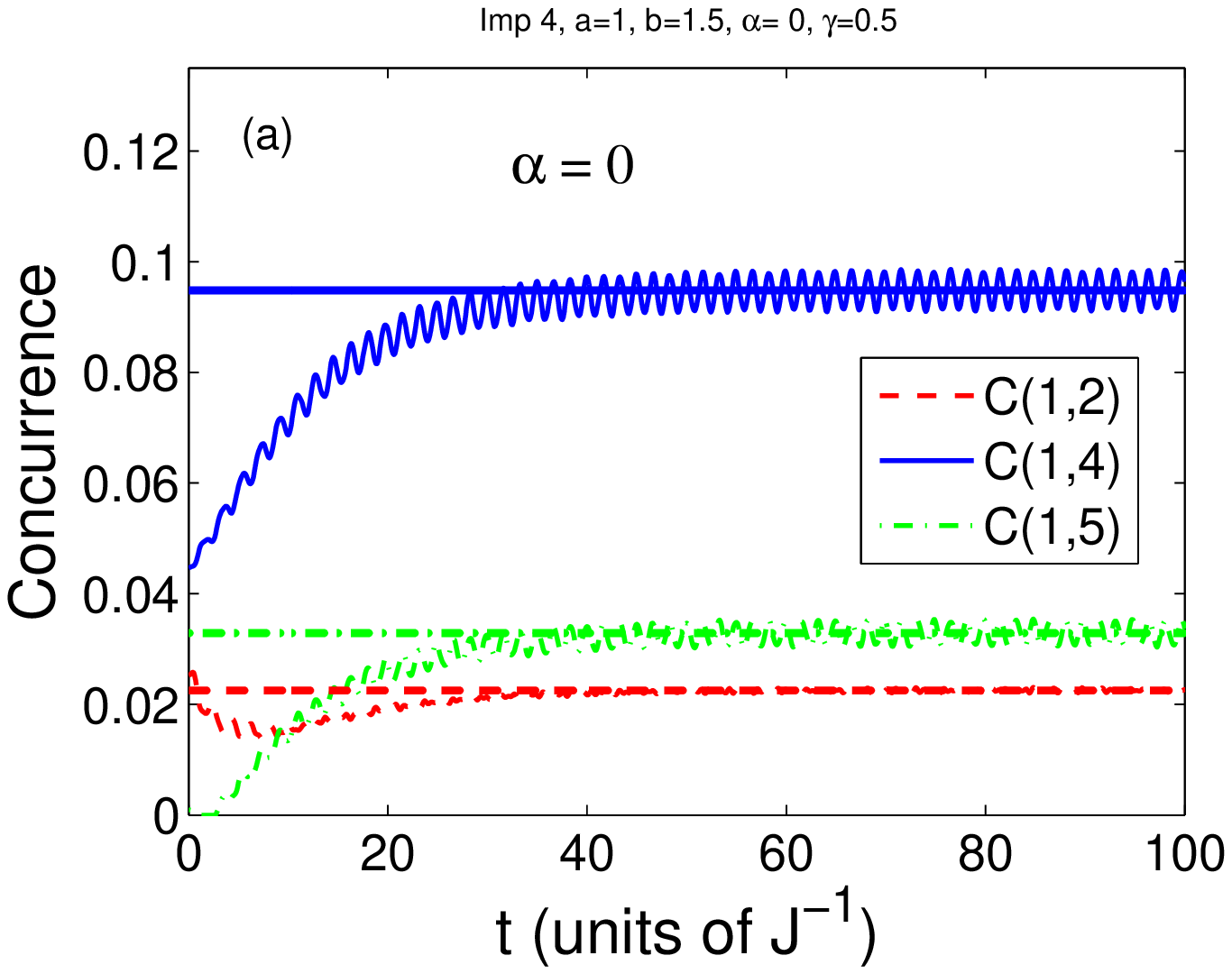}}\quad
   \subfigure{\includegraphics[width=8 cm]{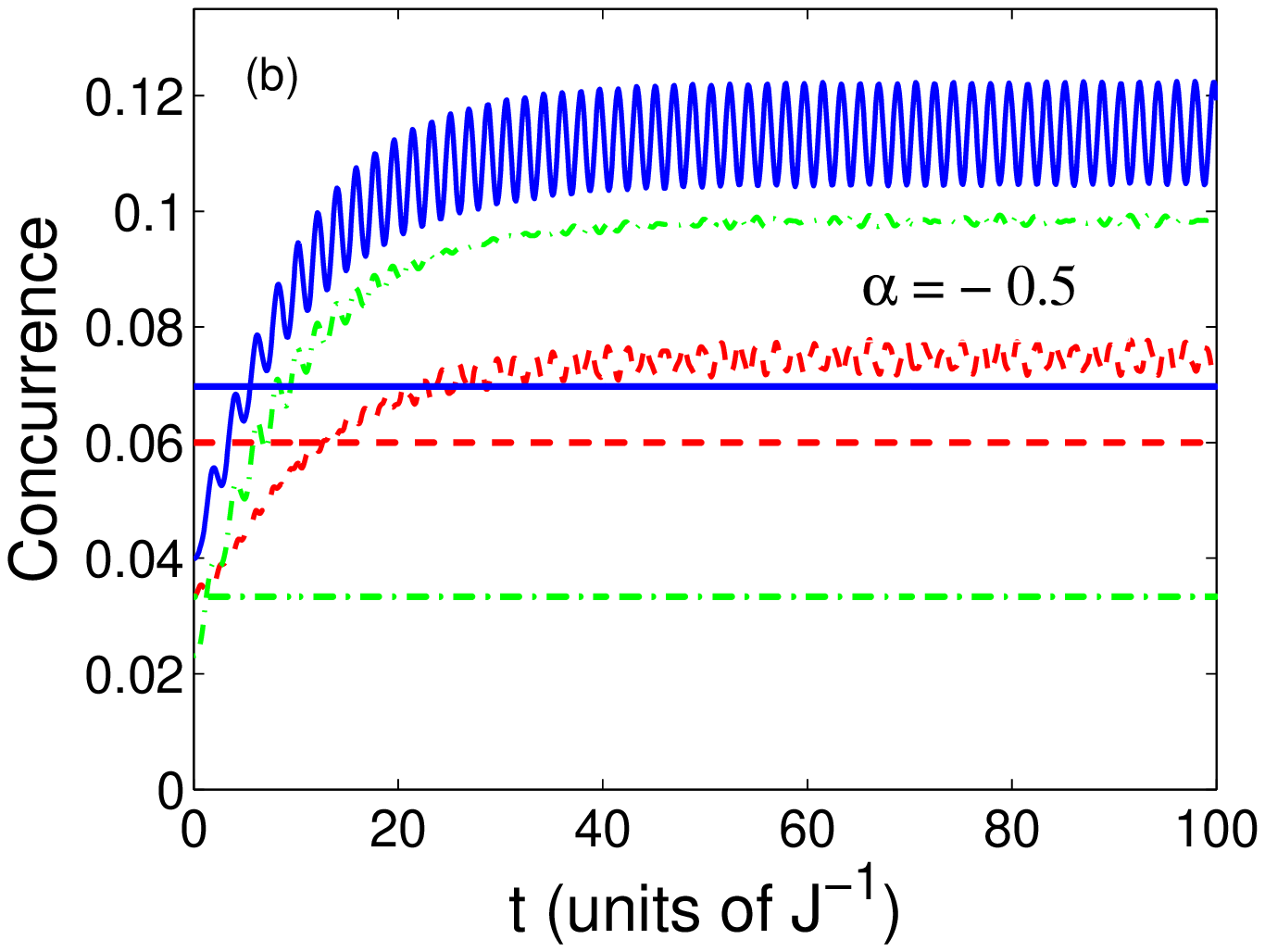}}\\
   \subfigure{\includegraphics[width=8 cm]{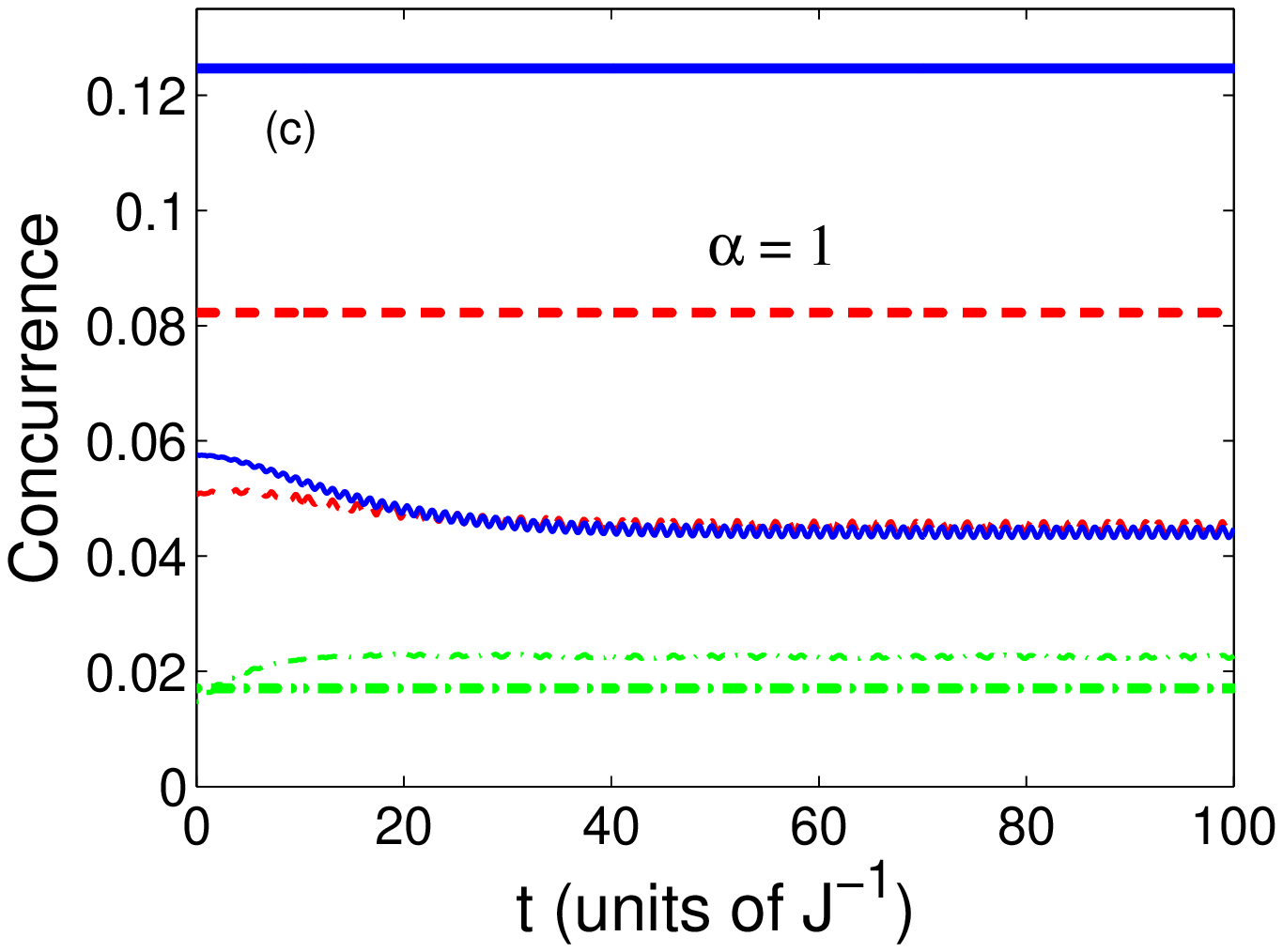}}\quad
   \subfigure{\includegraphics[width=8 cm]{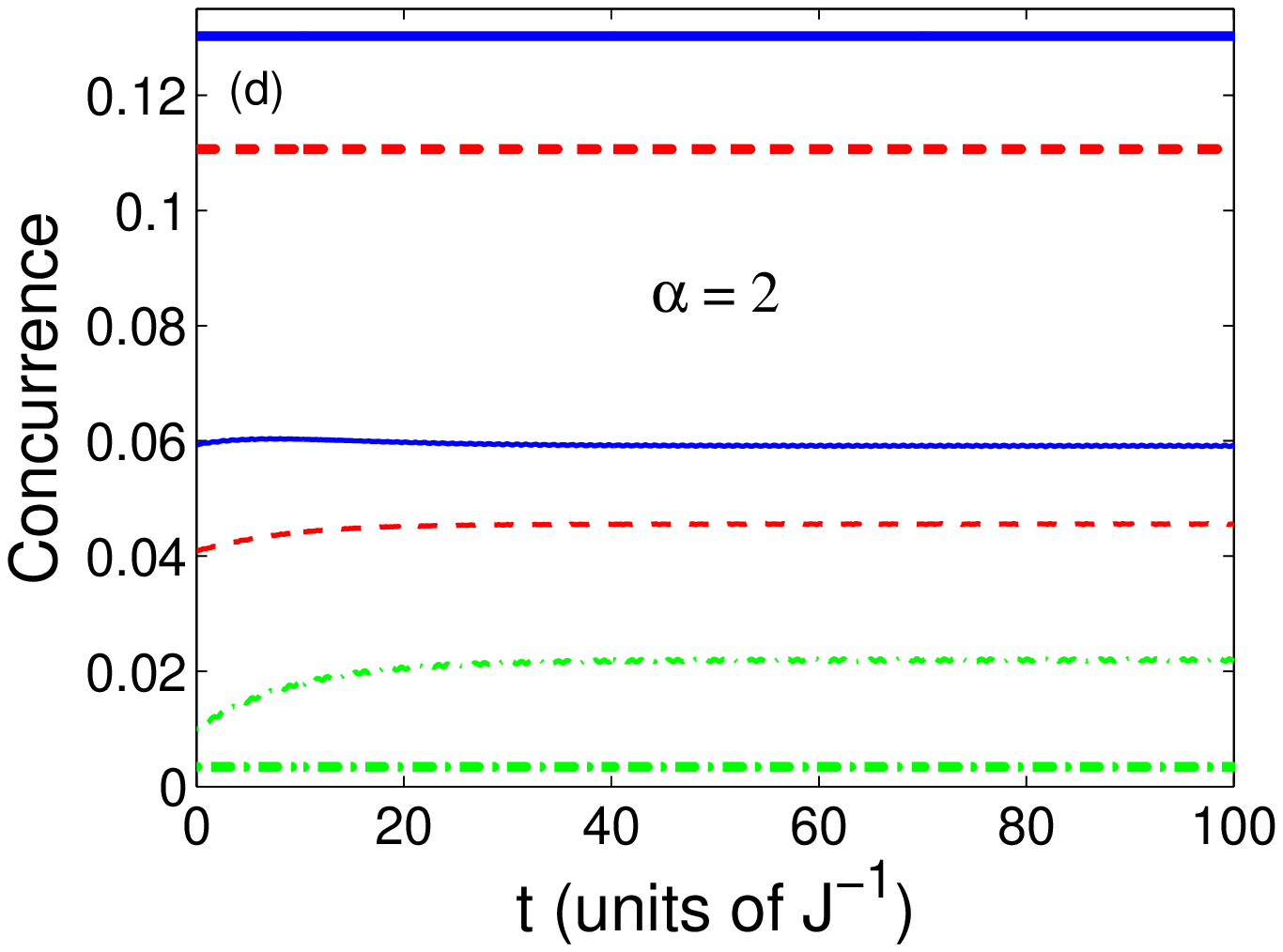}}
   \caption{{\protect\footnotesize (Color online) Dynamics of the concurrences $C(1,2), C(1,4), C(2,4)$ with a single impurity at the central site 4 with different impurity coupling strengths $\alpha = -0.5, 0, 1, 2$ for the two dimensional partially anisotropic lattice ($\gamma = 0.5$) under the effect of an exponential magnetic field with parameters values $a=1, b=1.5$ and $\omega=0.1$. The straight lines represent the equilibrium concurrences corresponding to constant magnetic field $h=1.5$. The legend for all subfigures is as shown in subfigure (a).}}
 \label{C_Dyn_G_05}
 \end{minipage}
\end{figure}
Clearly the asymptotic values of the entanglement are higher in the pure and weak impurity cases compared with the strong impurities. Changing the magnetic field parameter value $b$ to $1.5$, one can observe the great impact in fig.~\ref{B_Dyn_G_05_15}, where only the pure system becomes ergodic while the system with any impurity strength is nonergodic. This means that the magnetic field parameters control ergodicity as well. In fig.~\ref{C_Dyn_G_05}, we study the same system with a single impurity at the central site 4 with magnetic field parameters $a =1$, $b=1.5$ and $\omega = 0.1$. As can be seen from the different subfigures, the system is ergodic only in the pure case and the asymptotic values are higher for the pure and weak impurity systems compared with the strong impurities.
\begin{figure}[htbp]
\begin{minipage}[c]{\textwidth}
\centering
   \subfigure{\includegraphics[width=8 cm]{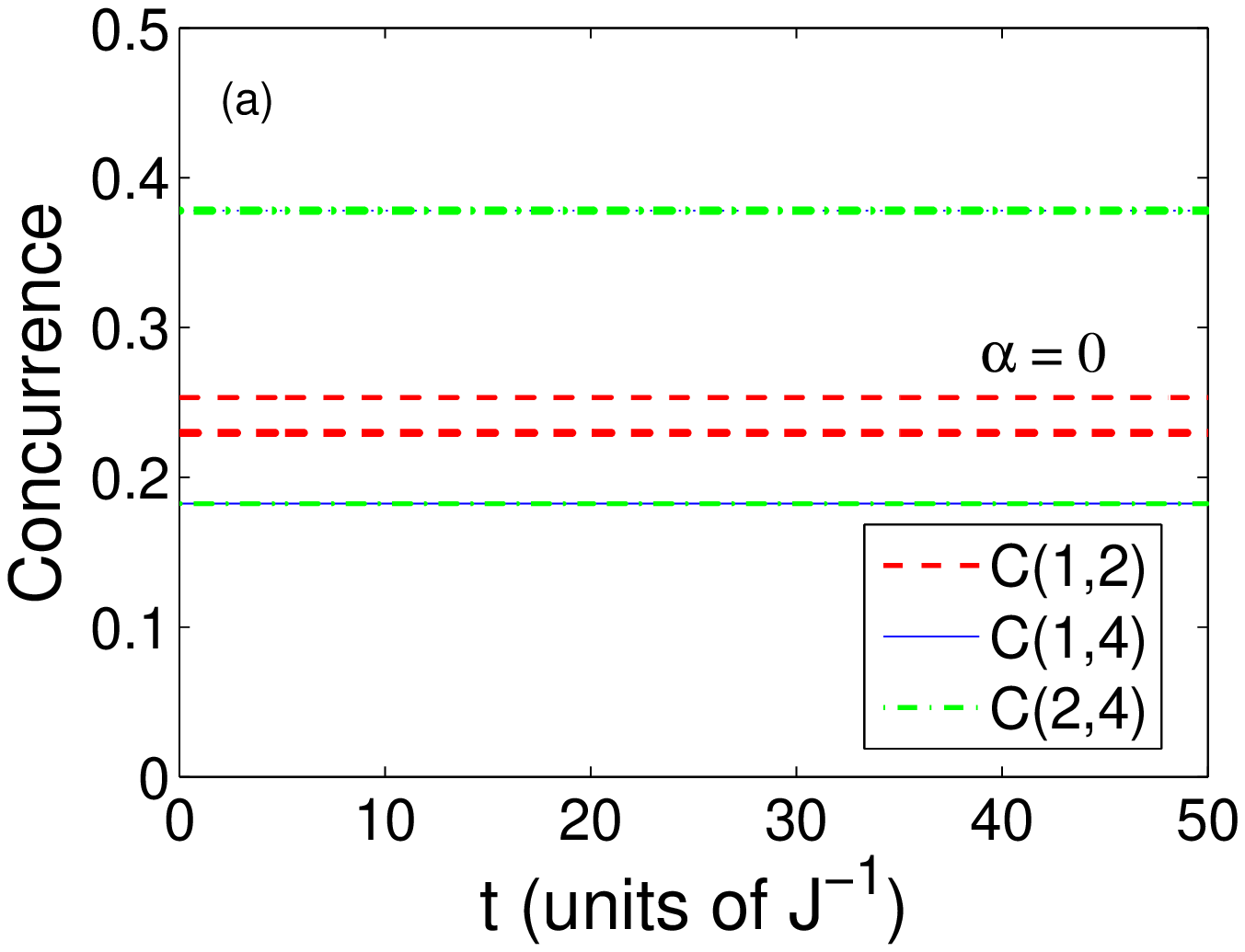}}\quad
   \subfigure{\includegraphics[width=8 cm]{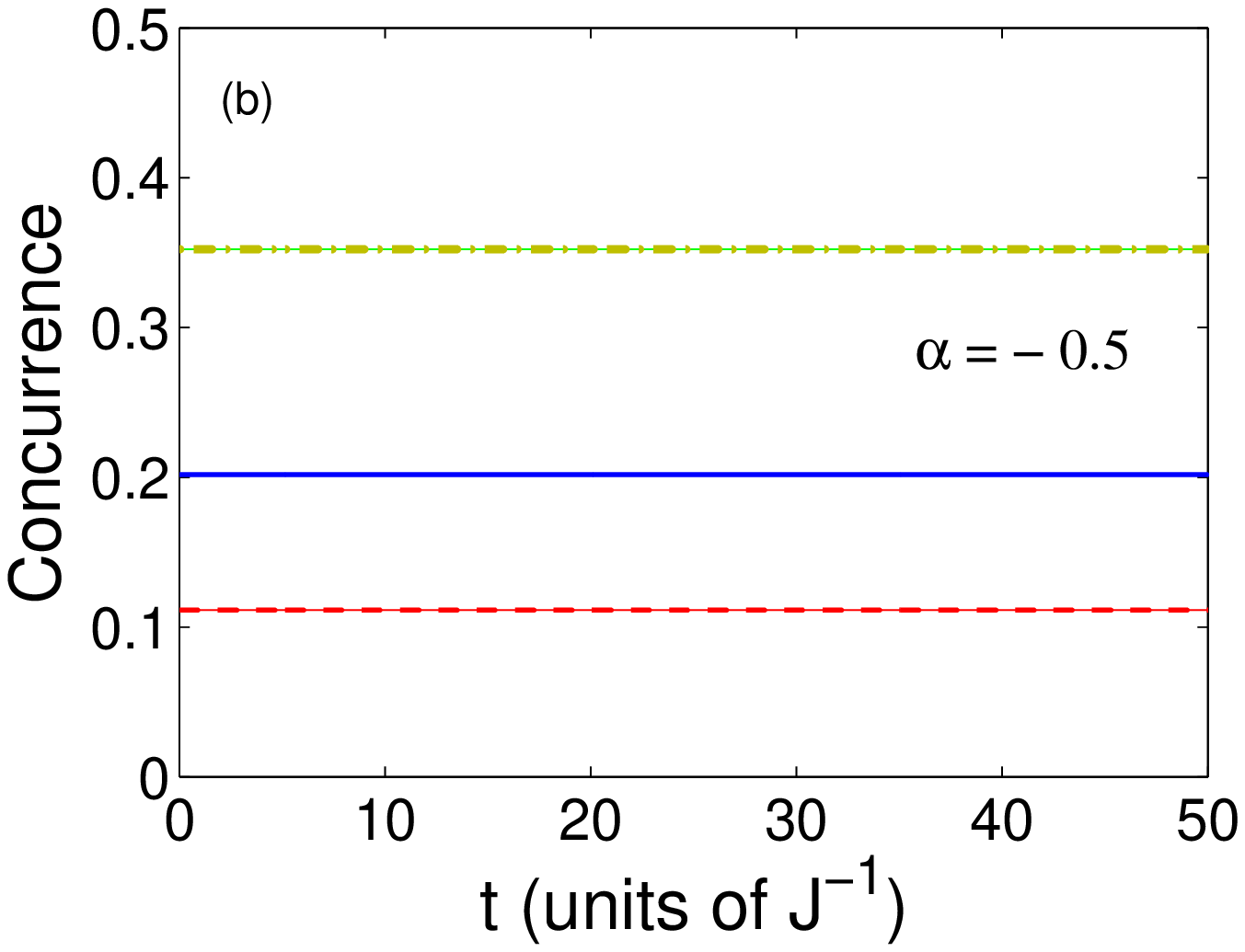}}\\
   \subfigure{\includegraphics[width=8 cm]{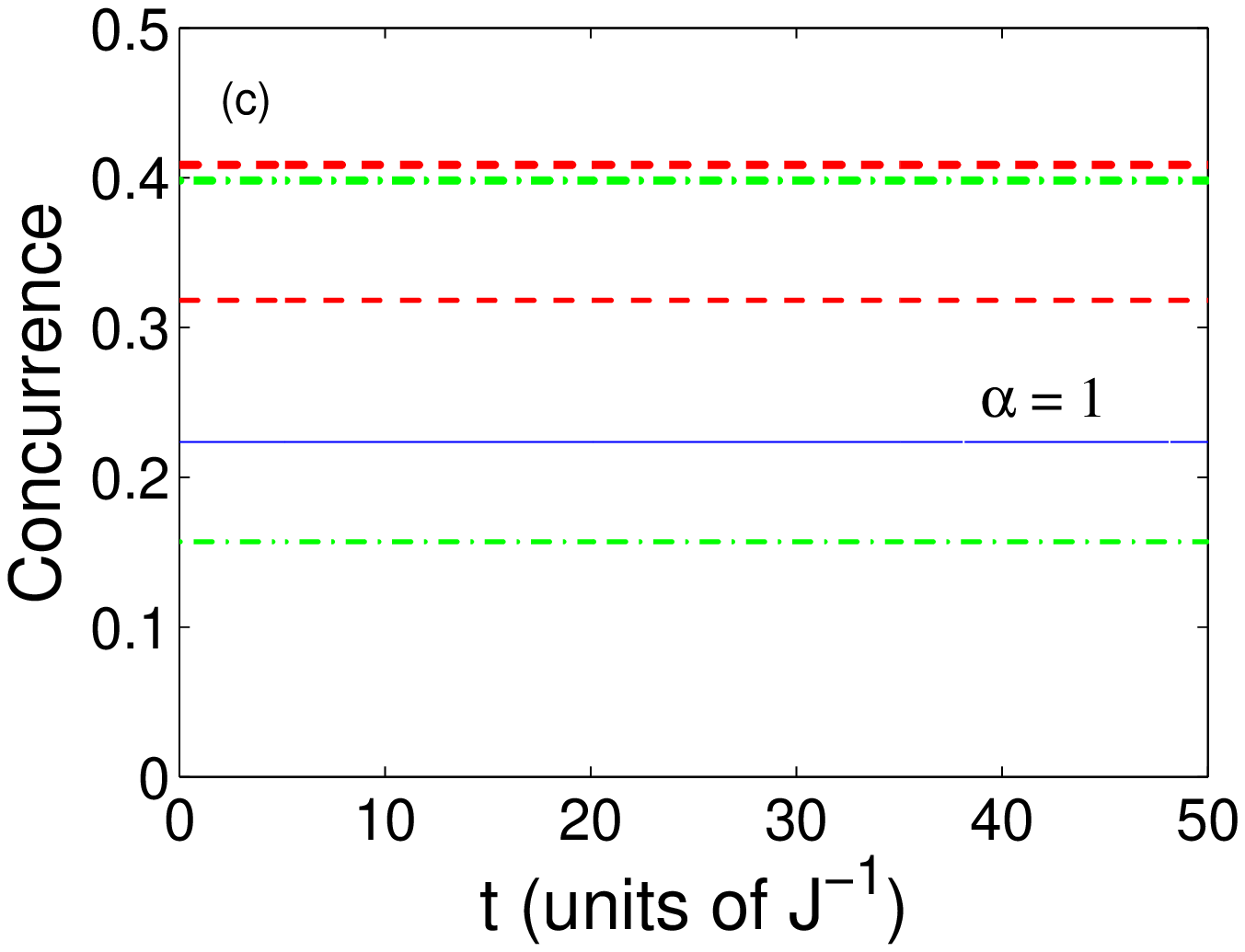}}\quad
   \subfigure{\includegraphics[width=8 cm]{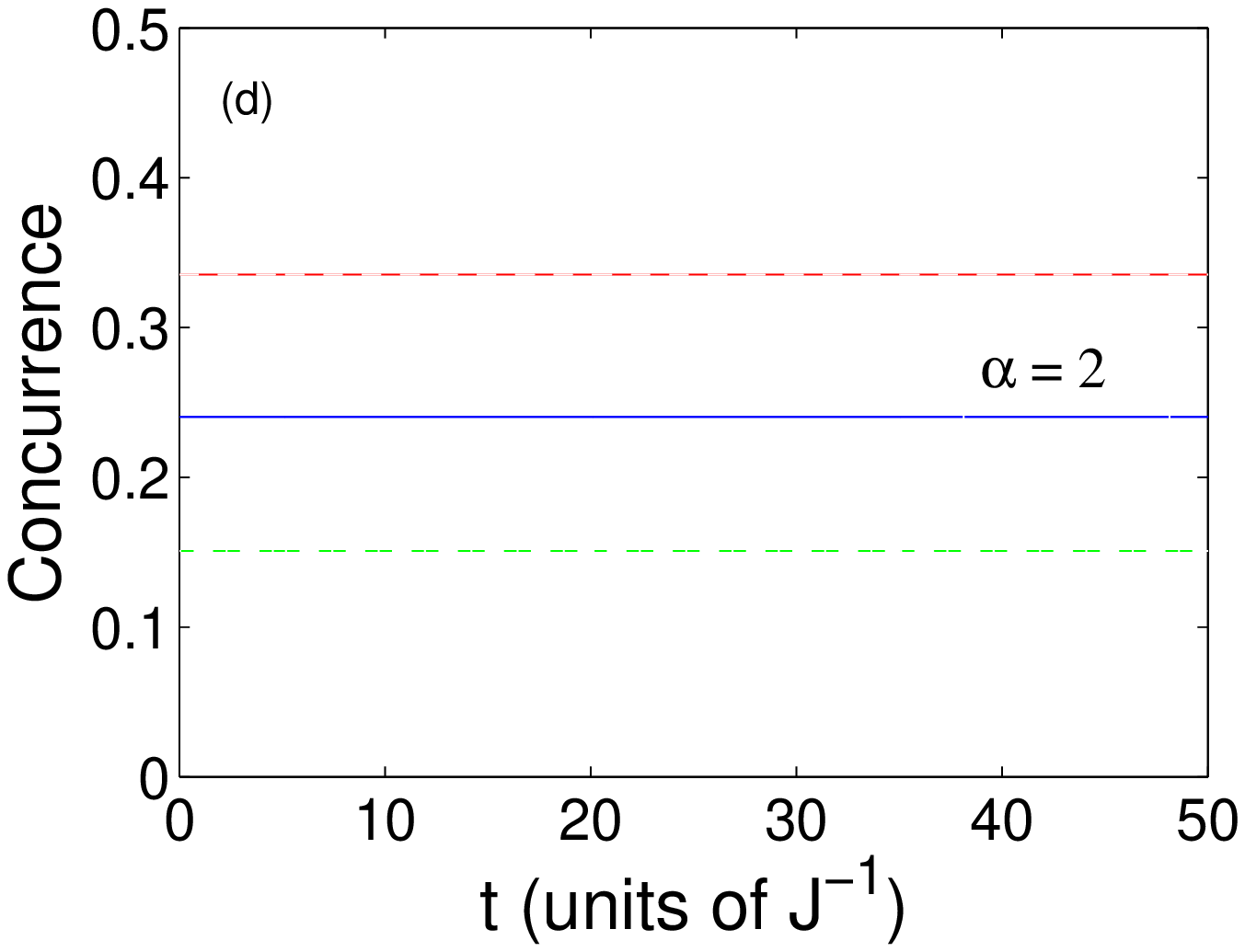}}   \caption{{\protect\footnotesize (Color online) Dynamics of the concurrences $C(1,2), C(1,4), C(2,4)$ with a single impurity at the border site 1 with different impurity coupling strengths $\alpha = -0.5, 0, 1, 2$ for the two dimensional XY lattice ($\gamma = 0$) under the effect of an exponential magnetic field with parameters values $a=1, b=1.5$ and $\omega=0.1$. The straight (thicker) lines represent the equilibrium concurrences corresponding to constant magnetic field $h=1.5$. The legend for all subfigures is as shown in subfigure (a).}}
 \label{B_Dyn_G_0_15}
 \end{minipage}
\end{figure}
\begin{figure}[htbp]
\begin{minipage}[c]{\textwidth}
\centering
   \subfigure{\includegraphics[width=8 cm]{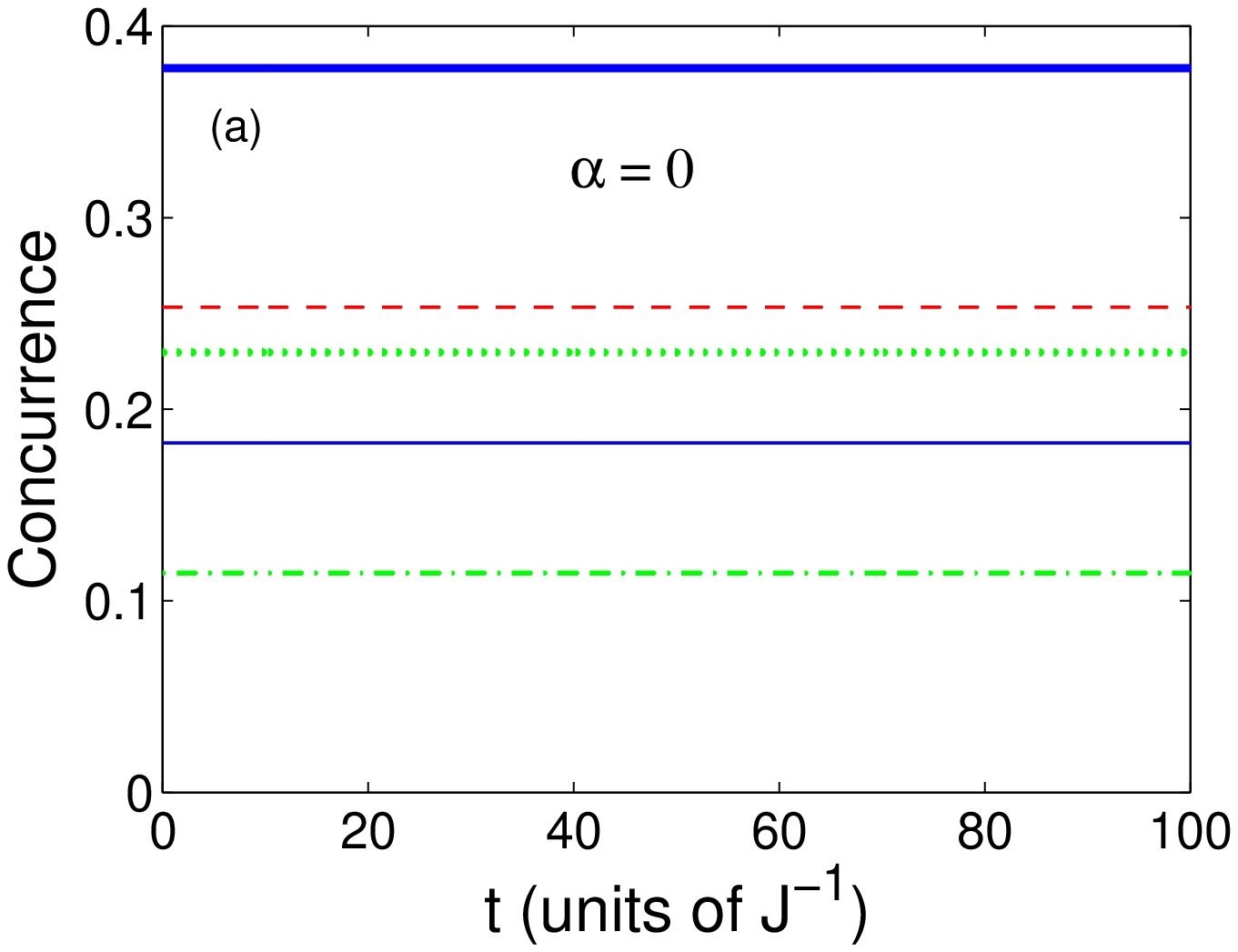}}\quad
   \subfigure{\includegraphics[width=8 cm]{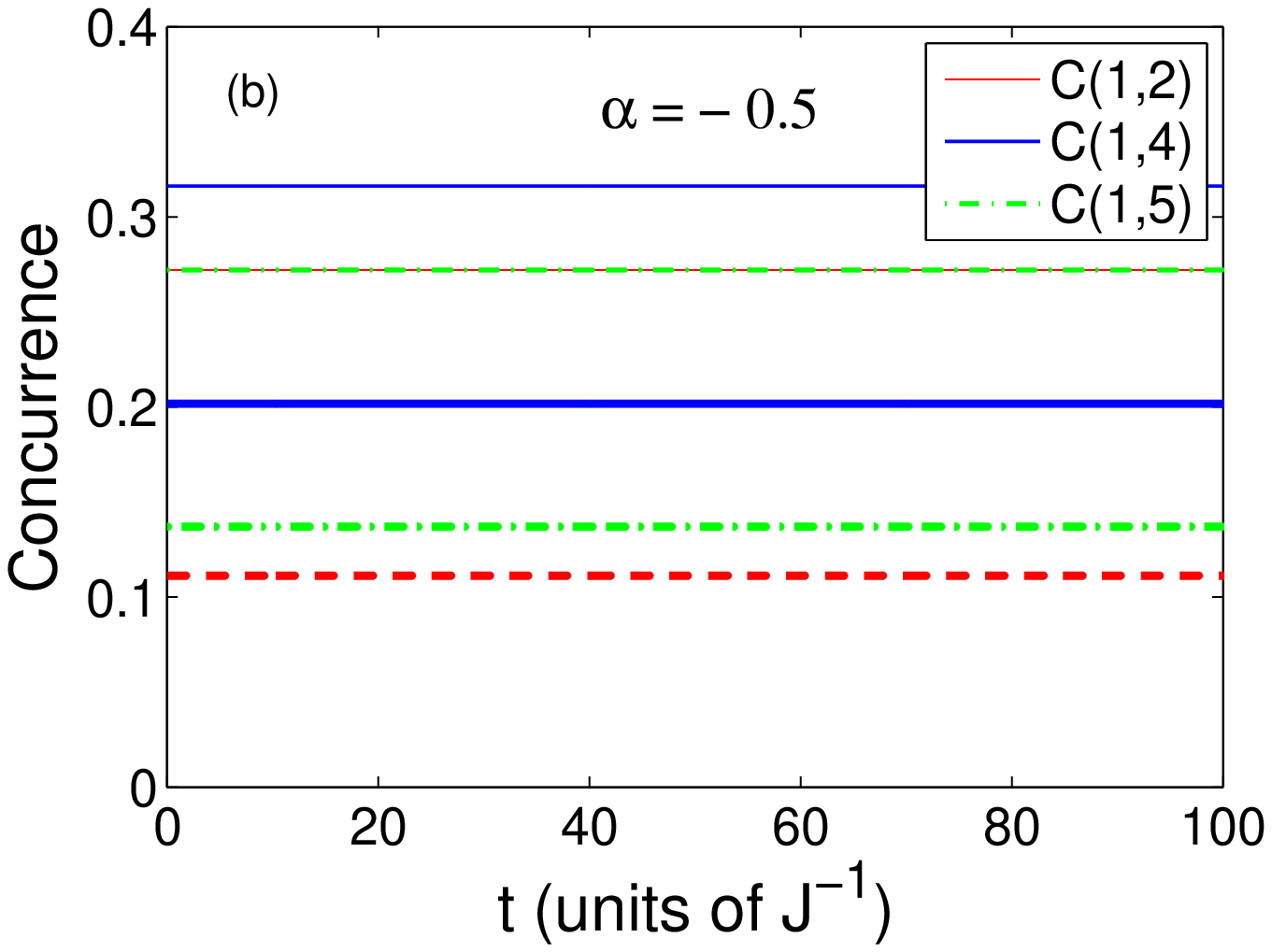}}\\
   \subfigure{\includegraphics[width=8 cm]{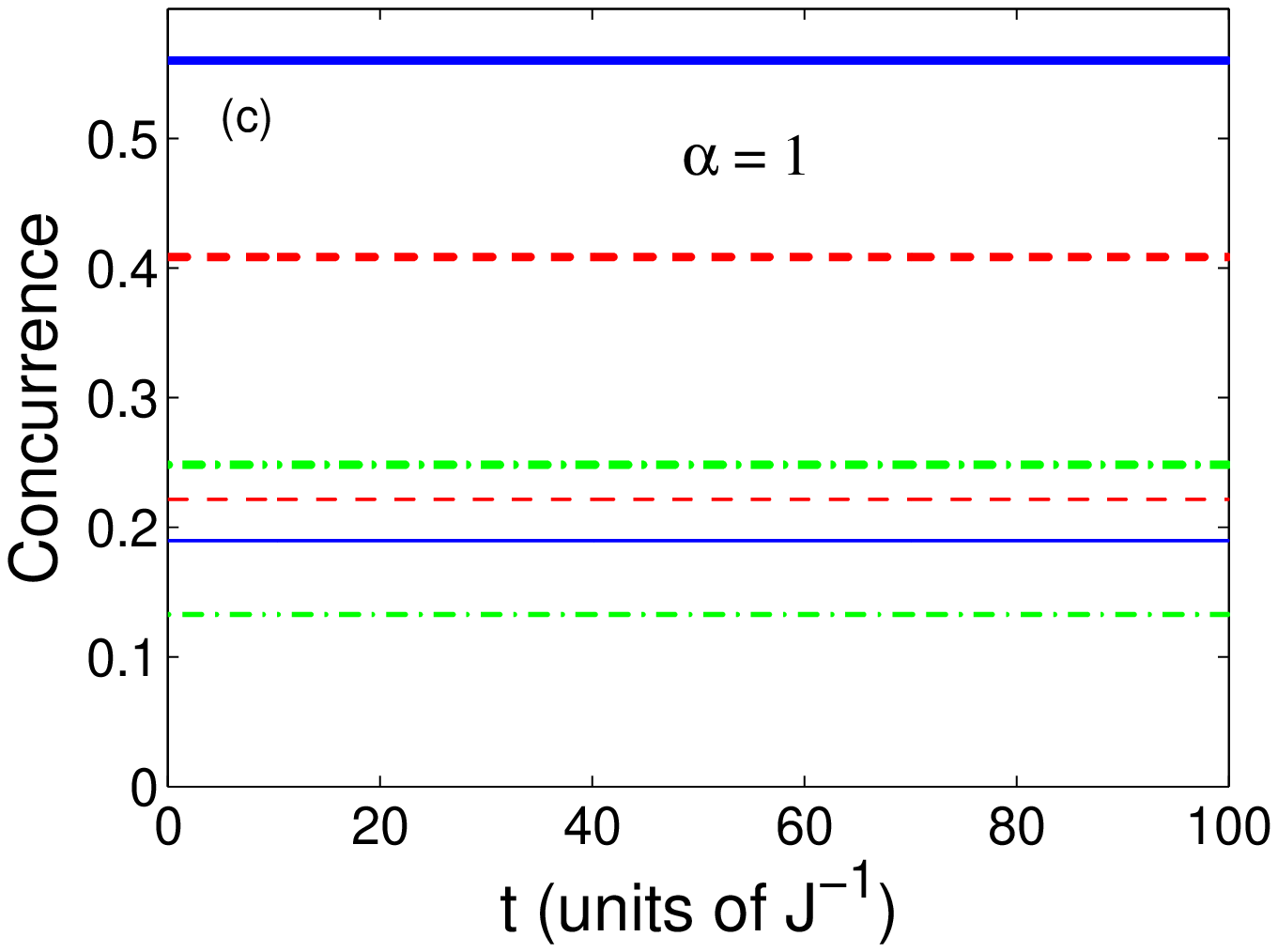}}\quad
   \subfigure{\includegraphics[width=8 cm]{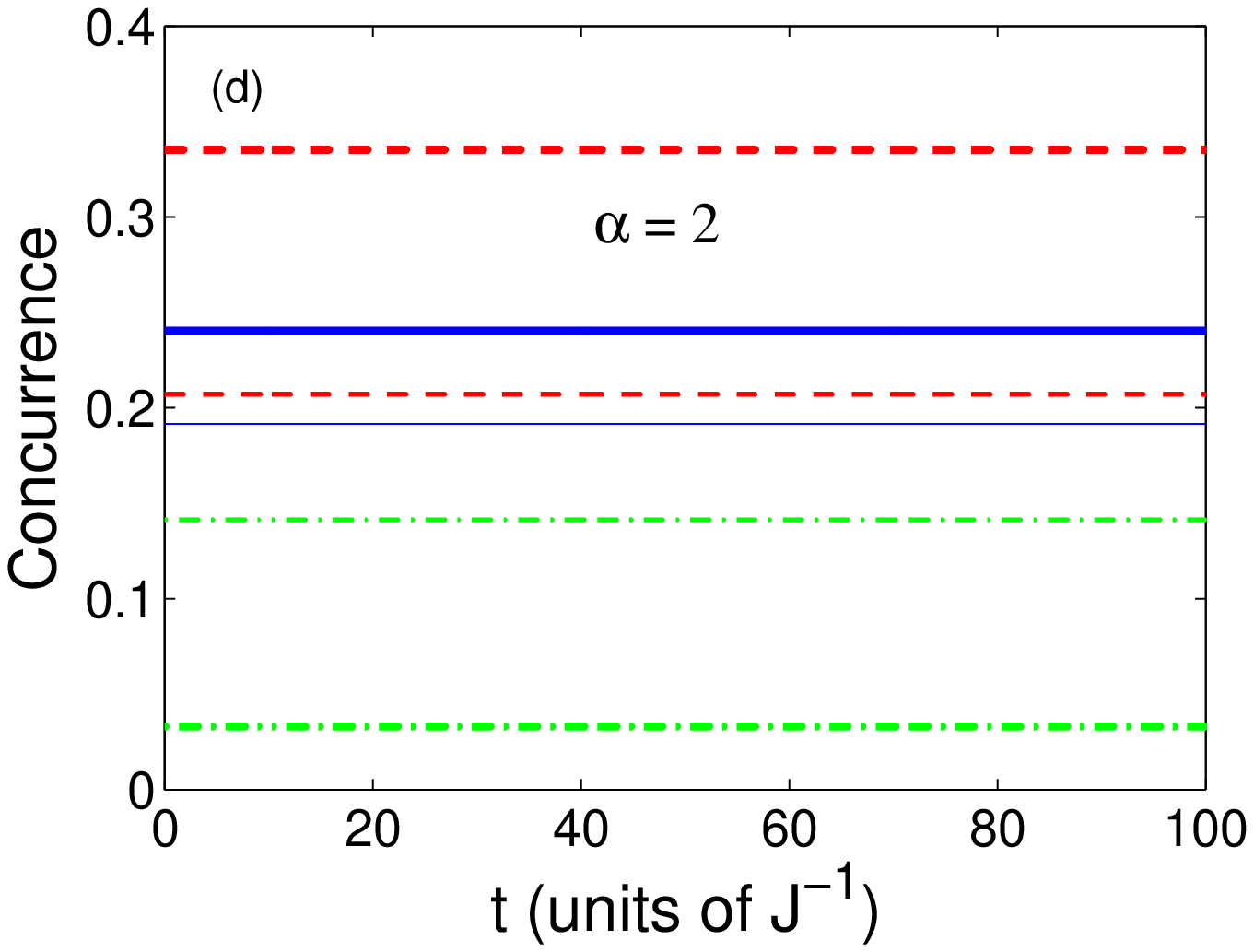}}
   \caption{{\protect\footnotesize (Color online) Dynamics of the concurrences $C(1,2), C(1,4), C(2,4)$ with a single impurity at the central site 4 with different impurity coupling strengths $\alpha = -0.5, 0, 1, 2$ for the two dimensional XY lattice ($\gamma = 0$) under the effect of an exponential magnetic field with parameters values $a=1, b=1.5$ and $\omega=0.1$. The straight (thicker) lines represent the equilibrium concurrences corresponding to constant magnetic field $h=1.5$. The legend for all subfigures is as shown in subfigure (b).}}
 \label{C_Dyn_G_0}
 \end{minipage}
\end{figure}
The complete isotropic $XY$ system with a single border impurity at site 1 under the effect of an exponential magnetic field with parameter values $a =1$, $b=1.5$ and $\omega = 0.1$, is investigated in fig.~\ref{B_Dyn_G_0_15}. The trivial effect of the magnetic field, similar to the one dimensional case results \cite{Sadiek2010}, is clear where the entanglement assumes a constant value for all pair of spins. This trivial effect is the result of the fact that for $\gamma=0$ the exchange coupling terms in the Hamiltonian commute with the magnetic field term. Nevertheless one still can see an effect of the impurity on the ergodicity of the system where for $\alpha=0$ and 1, the system is nonergodic wile for $\alpha=-0.5$ and 2 it is ergodic as shown in fig.~\ref{B_Dyn_G_0_15}. In fact, testing a wide range of $\alpha$ values indicates that for the values approximately in the range $-0.4 \geq \alpha \leq 1.9$ the system is nonergodic, otherwise it is ergodic i.e. for small absolute values of the impurity. Examining the same system under the effect of the same magnetic field but with a single central impurity, for wide range of $\alpha$, demonstrates that the system becomes nonergodic at all values of $\alpha$ which is illustrated in fig.~\ref{C_Dyn_G_0}. 
\section{Double impurities}
\subsection{Static system with impurities}
\begin{figure}[htbp]
\begin{minipage}[c]{\textwidth}
\centering
   \subfigure{\includegraphics[width=8 cm]{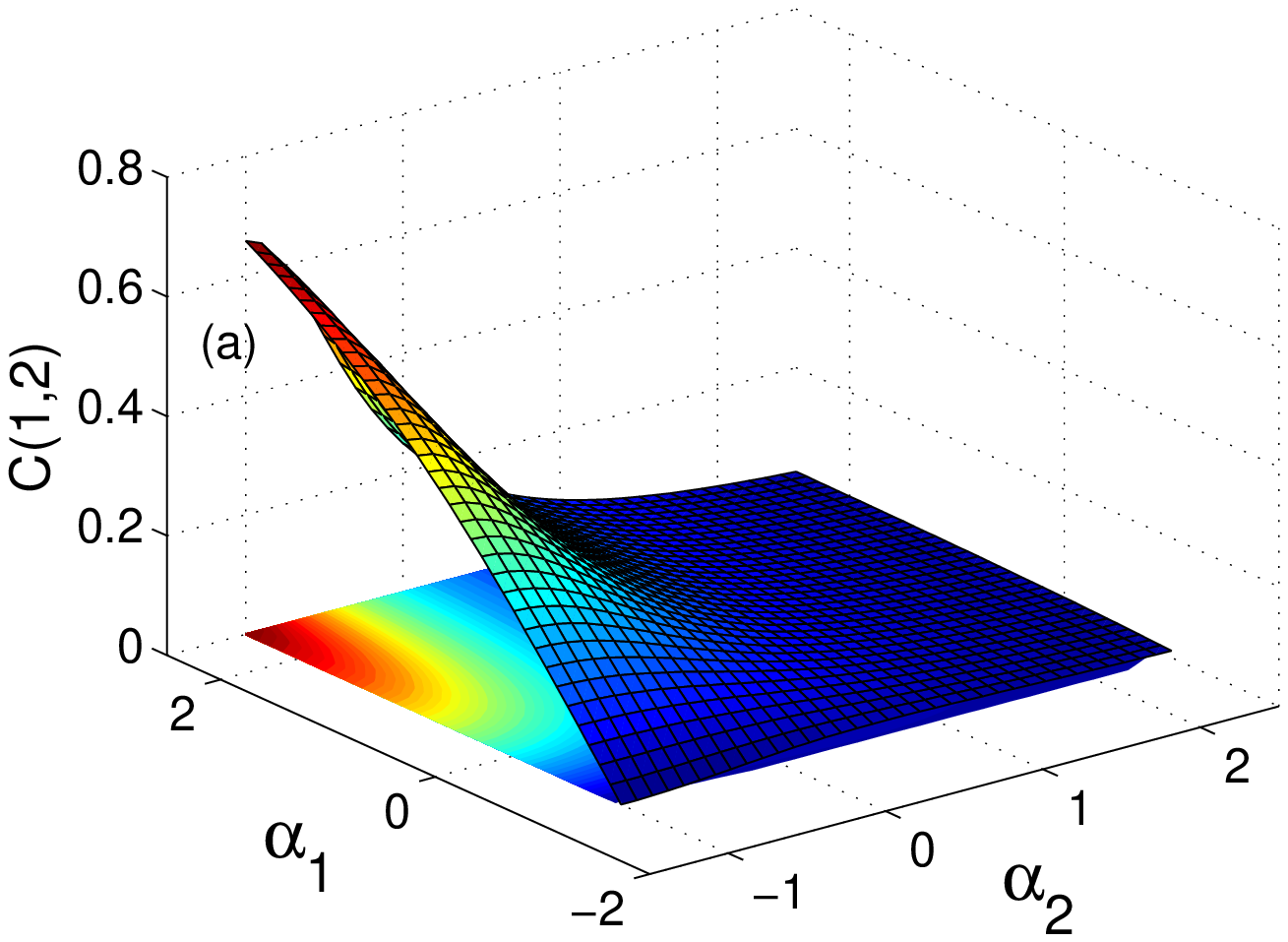}}\quad
   \subfigure{\includegraphics[width=8 cm]{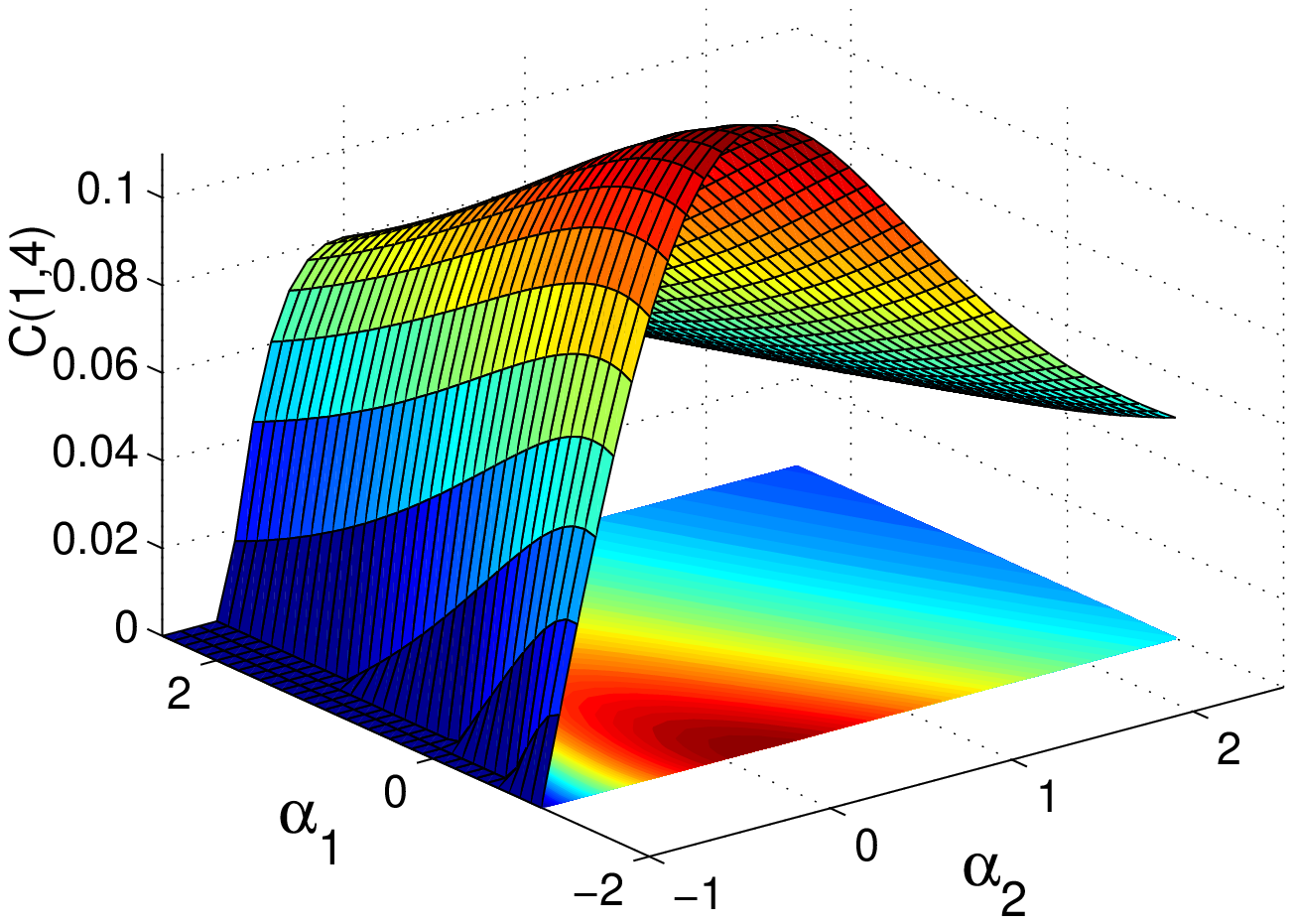}}\\
   \subfigure{\includegraphics[width=8 cm]{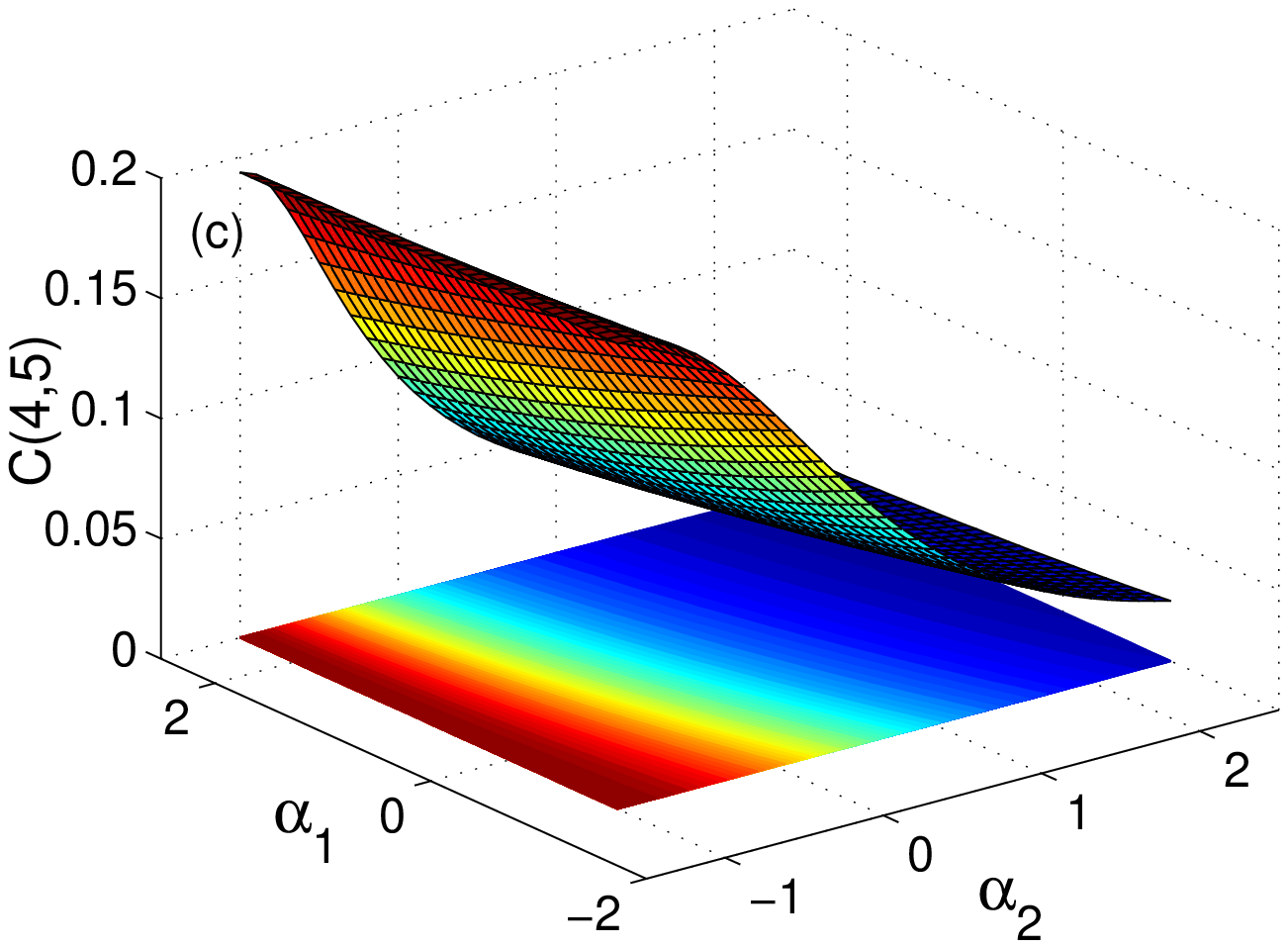}}\quad
   \subfigure{\includegraphics[width=8 cm]{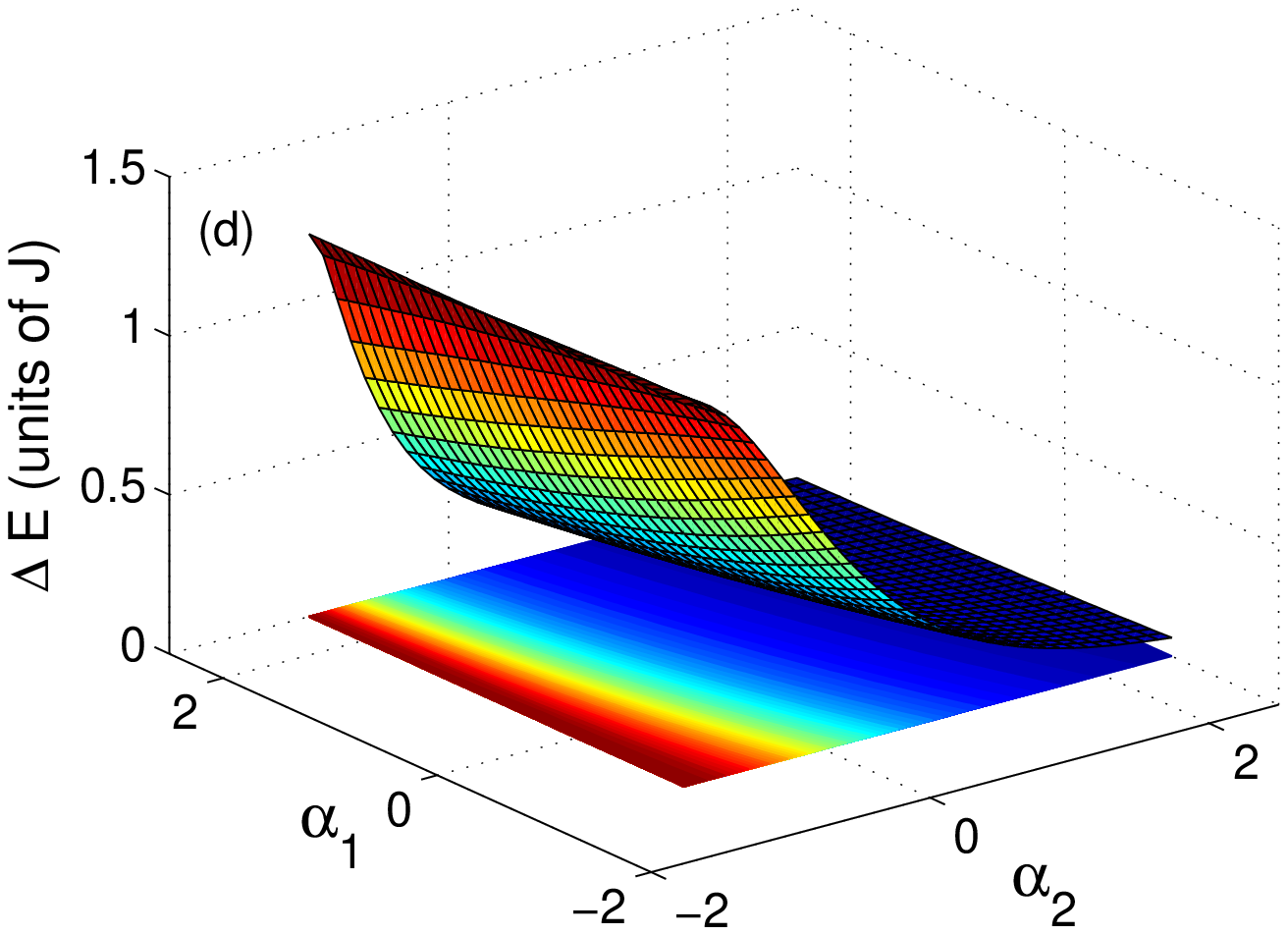}}
   \caption{{\protect\footnotesize (Color online) The concurrence $C(1,2)$, $C(1,4)$, $C(4,5)$ versus the impurity coupling strengths $\alpha_1$ and $\alpha_2$ with double impurities at sites 1 and 2 for the two dimensional Ising lattice ($\gamma = 1$) in an external magnetic field h=2.}}
 \label{Imp12_G1_h2}
 \end{minipage}
\end{figure}
\begin{figure}[htbp]
\begin{minipage}[c]{\textwidth}
\centering
   \subfigure{\includegraphics[width=8 cm]{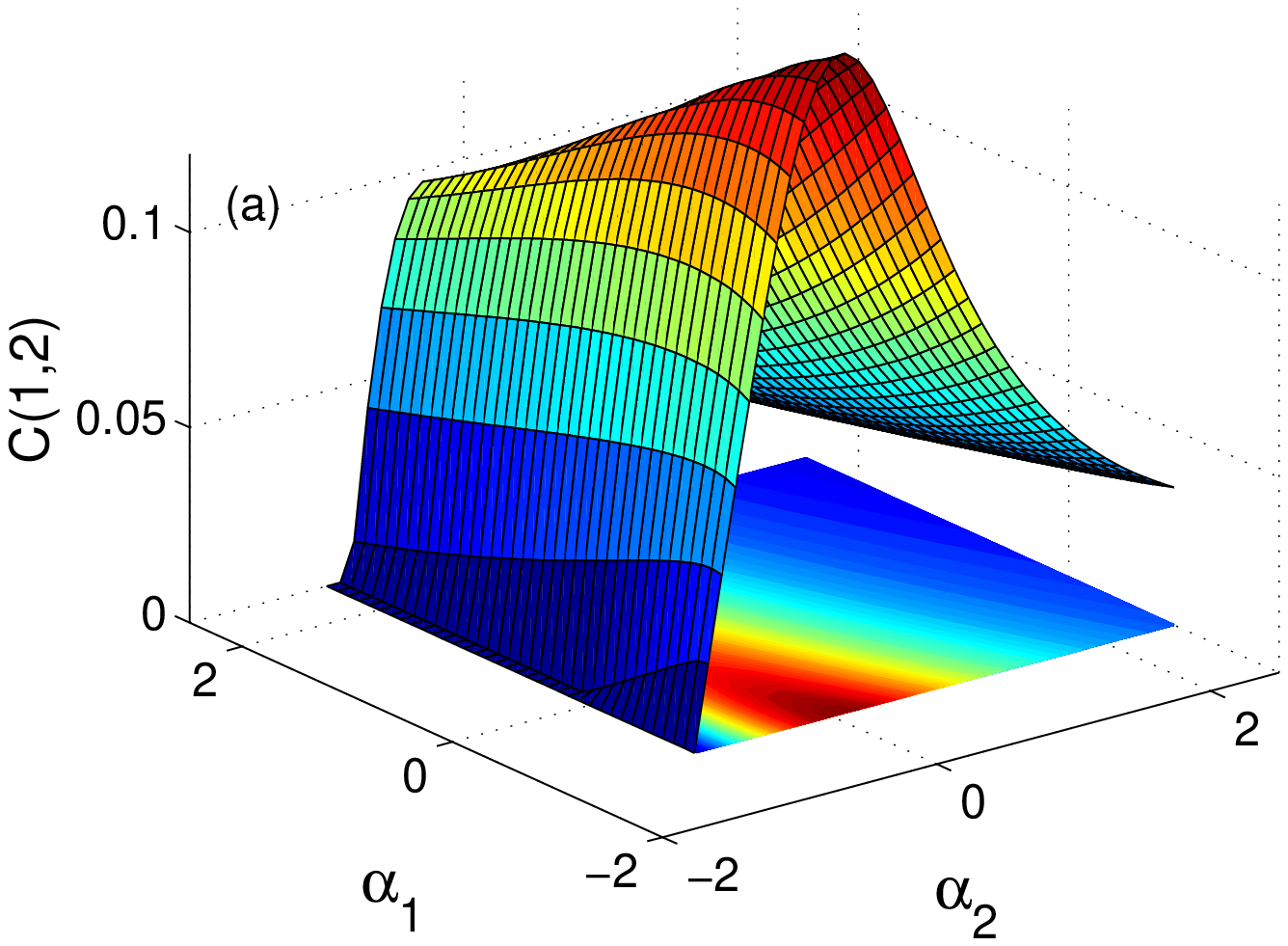}}\quad
   \subfigure{\includegraphics[width=8 cm]{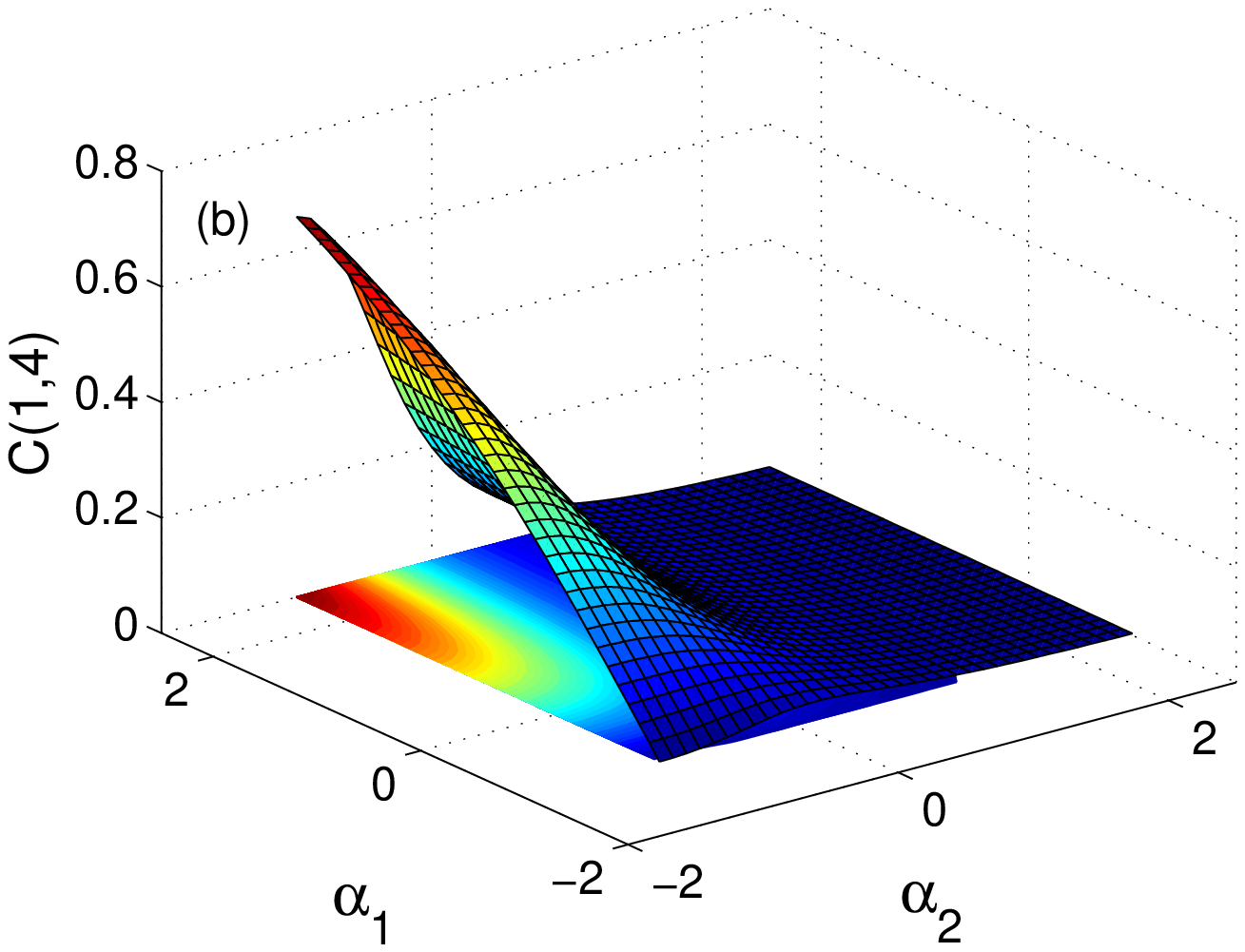}}\\
   \subfigure{\includegraphics[width=8 cm]{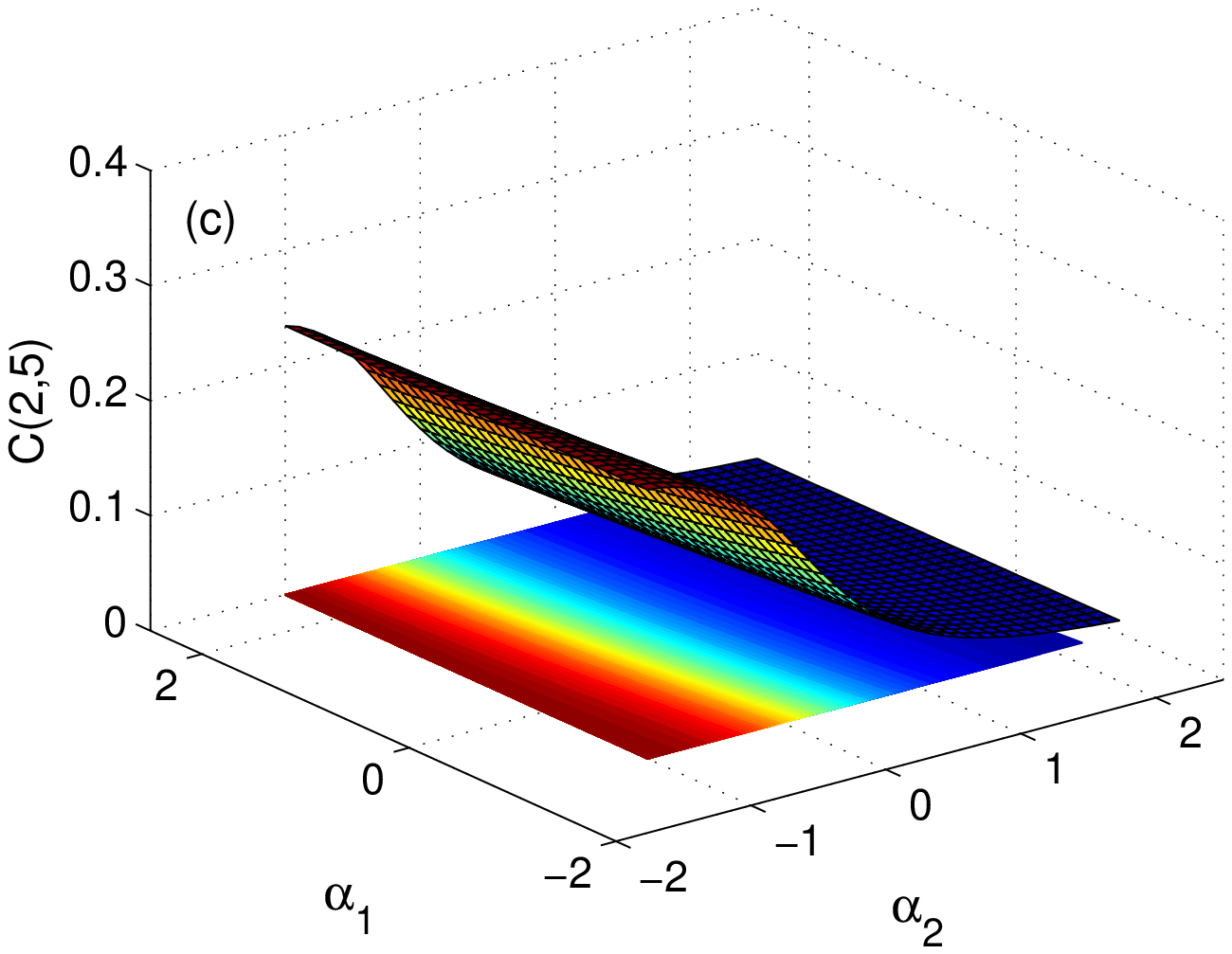}}\quad
   \subfigure{\includegraphics[width=8 cm]{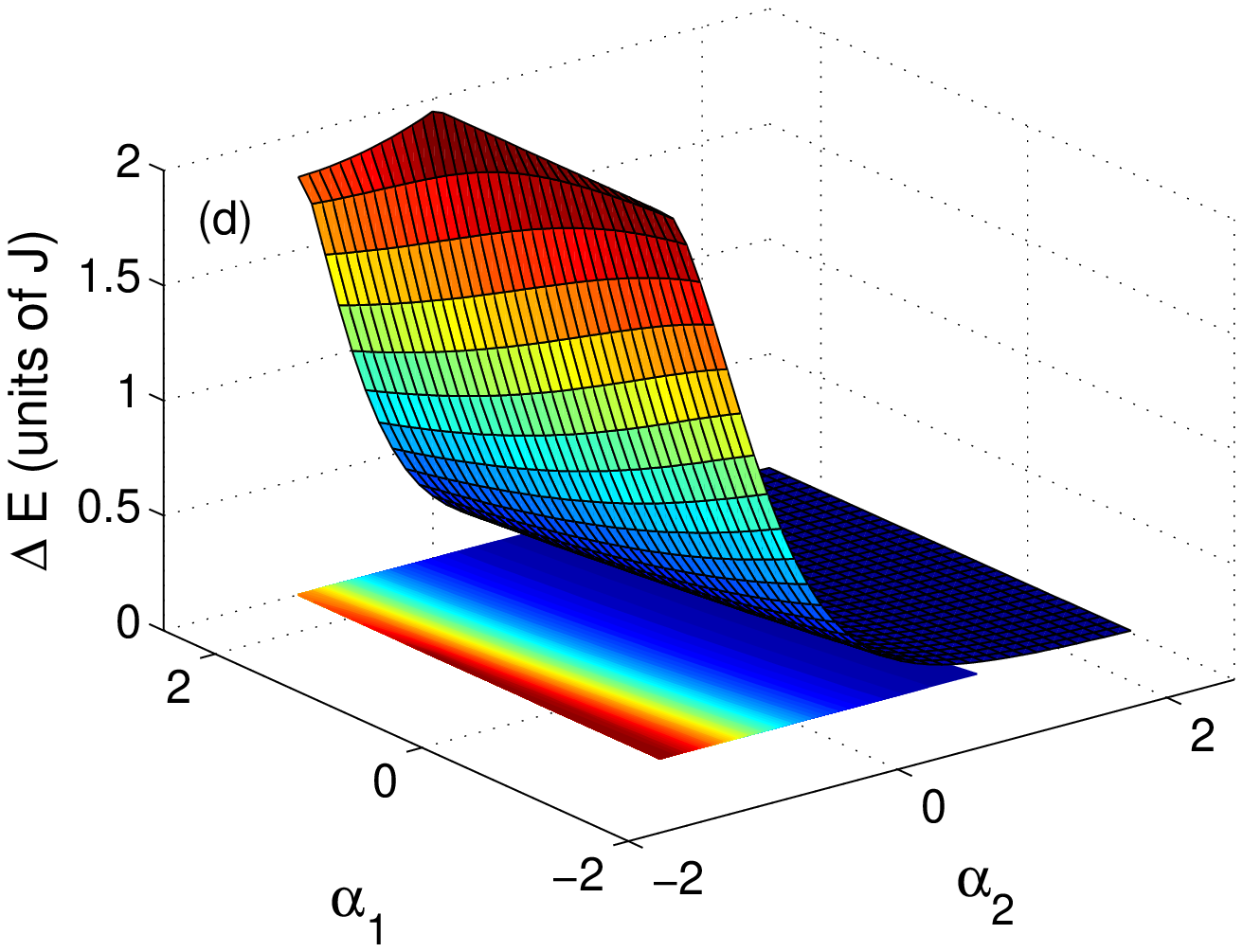}}   \caption{{\protect\footnotesize (Color online) The concurrence $C(1,2)$, $C(1,4)$, $C(2,5)$, and the energy gap $\Delta E$ versus the impurity coupling strengths $\alpha_1$ and $\alpha_2$ with double impurities at sites 1 and 4 for the two dimensional Ising lattice ($\gamma = 1$) in an external magnetic field h=2.}}
 \label{Imp14_G1_h2}
 \end{minipage}
\end{figure}
In this section we study the effect of double impurity, where we start with two located at the border sites 1 and 2. We set the coupling strength between the two impurities as $J' = (1+\alpha_1) J$, between any one of the impurities and its regular nearest neighbors as $J'' = (1+\alpha_2) J$ and between the rest of the nearest neighbor sites on the lattice as $J$. The effect of the impurities strength on the concurrence between different pairs of sites for the Ising lattice is shown in fig.~\ref{Imp12_G1_h2}. In fig.~\ref{Imp12_G1_h2}(a) we consider the entanglement between the two impurity sites 1 and 2 under a constant external magnetic field $h=2$. The concurrence $C(1,2)$ takes a large value when the impurity strengths $\alpha_1$, controlling the coupling between the impurity sites, is large and when $\alpha_2$, controlling coupling between impurities and their nearest neighbors, is weak. As $\alpha_1$ decreases and $\alpha_2$ increases, $C(1,2)$ decreases monotonically until it vanishes. As one can conclude, $\alpha_1$ is more effective than $\alpha_2$ in controlling the entanglement in this case. On the other hand, the entanglement between the impurity site 1 and the regular central site 4 is illustrated in fig.~\ref{Imp12_G1_h2}(b) which behaves completely different from C(1,2). 
\begin{figure}[htbp]
\begin{minipage}[c]{\textwidth}
\centering
   \subfigure{\includegraphics[width=8 cm]{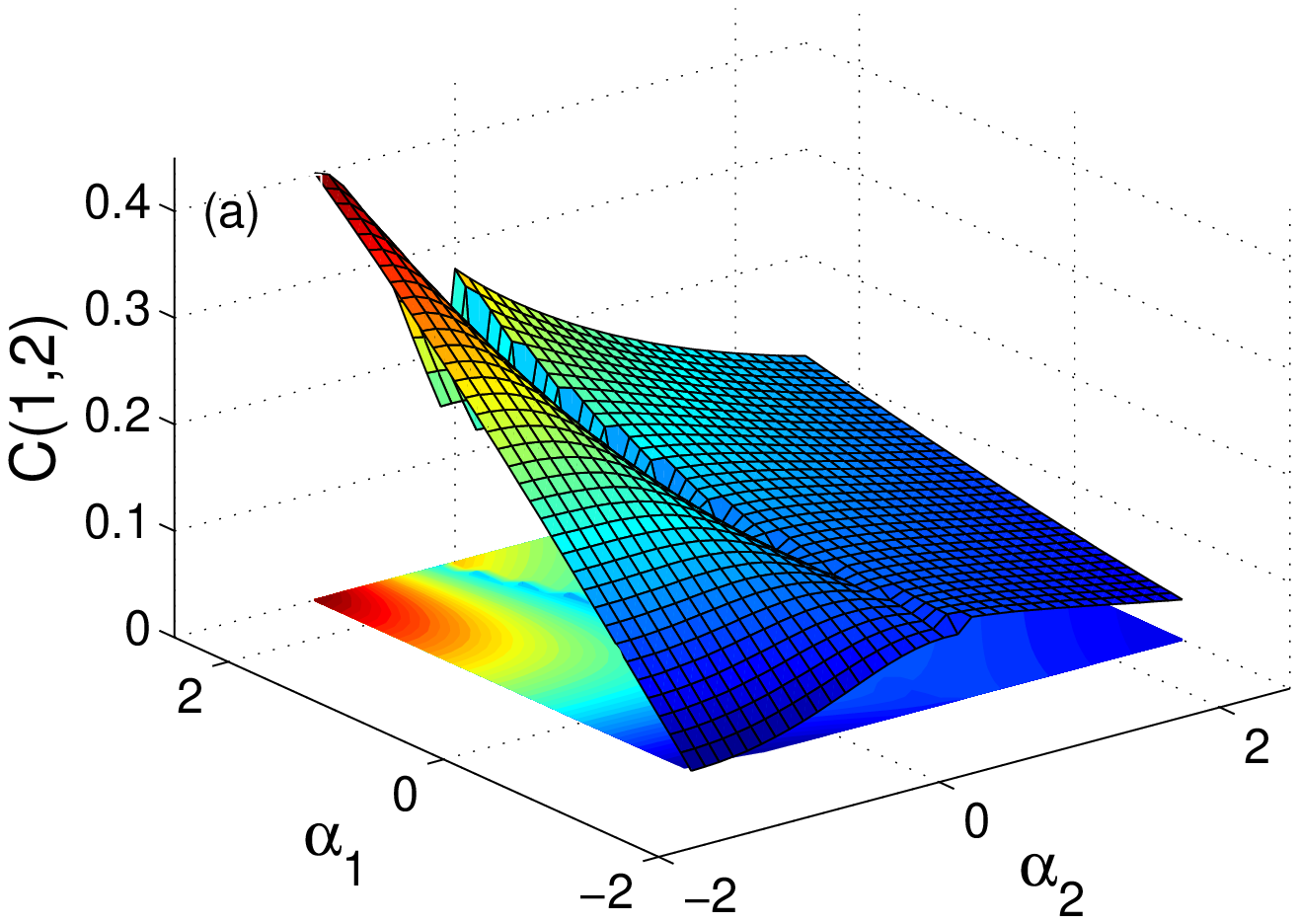}}\quad
   \subfigure{\includegraphics[width=8 cm]{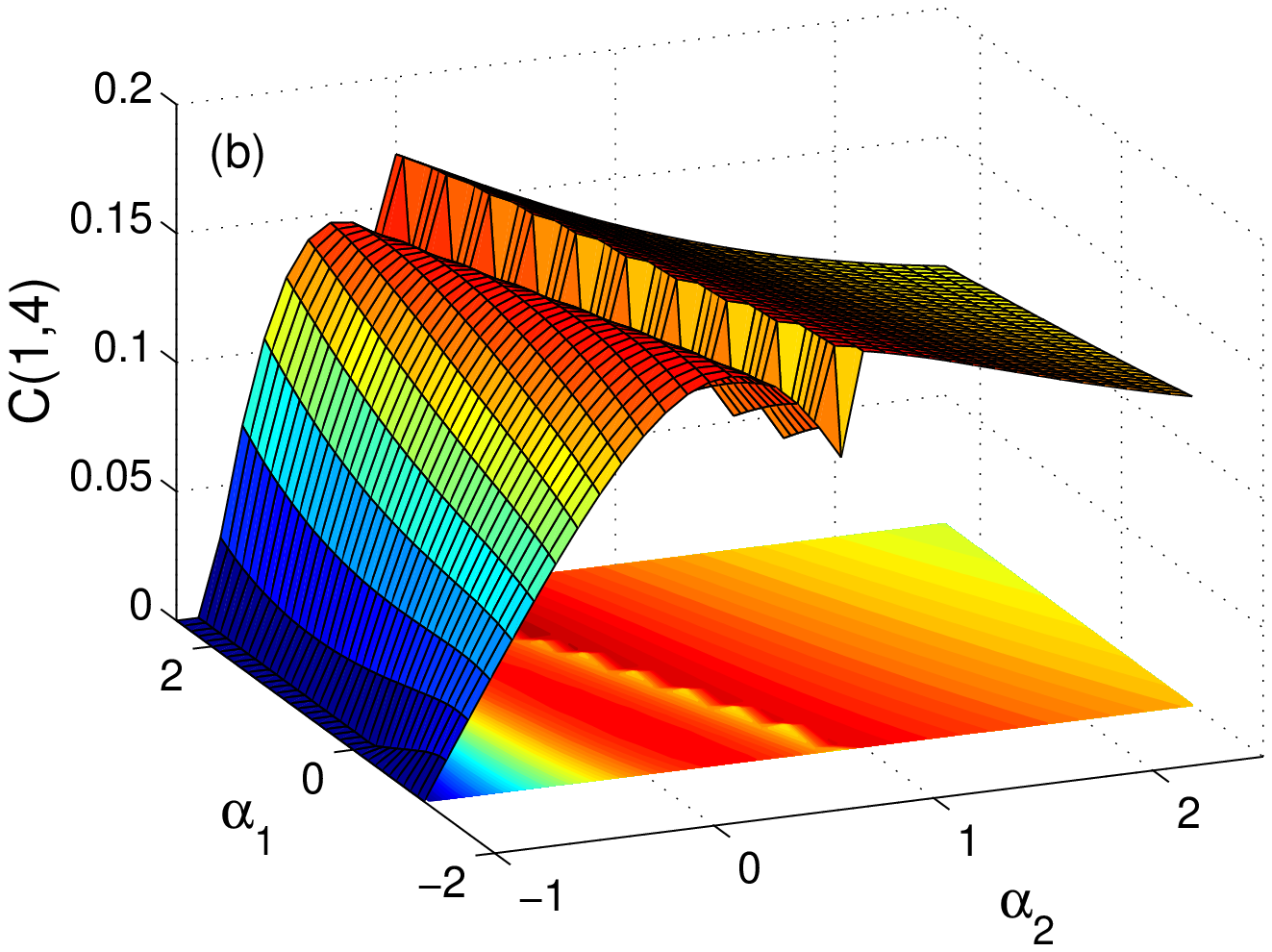}}\\
   \subfigure{\includegraphics[width=8 cm]{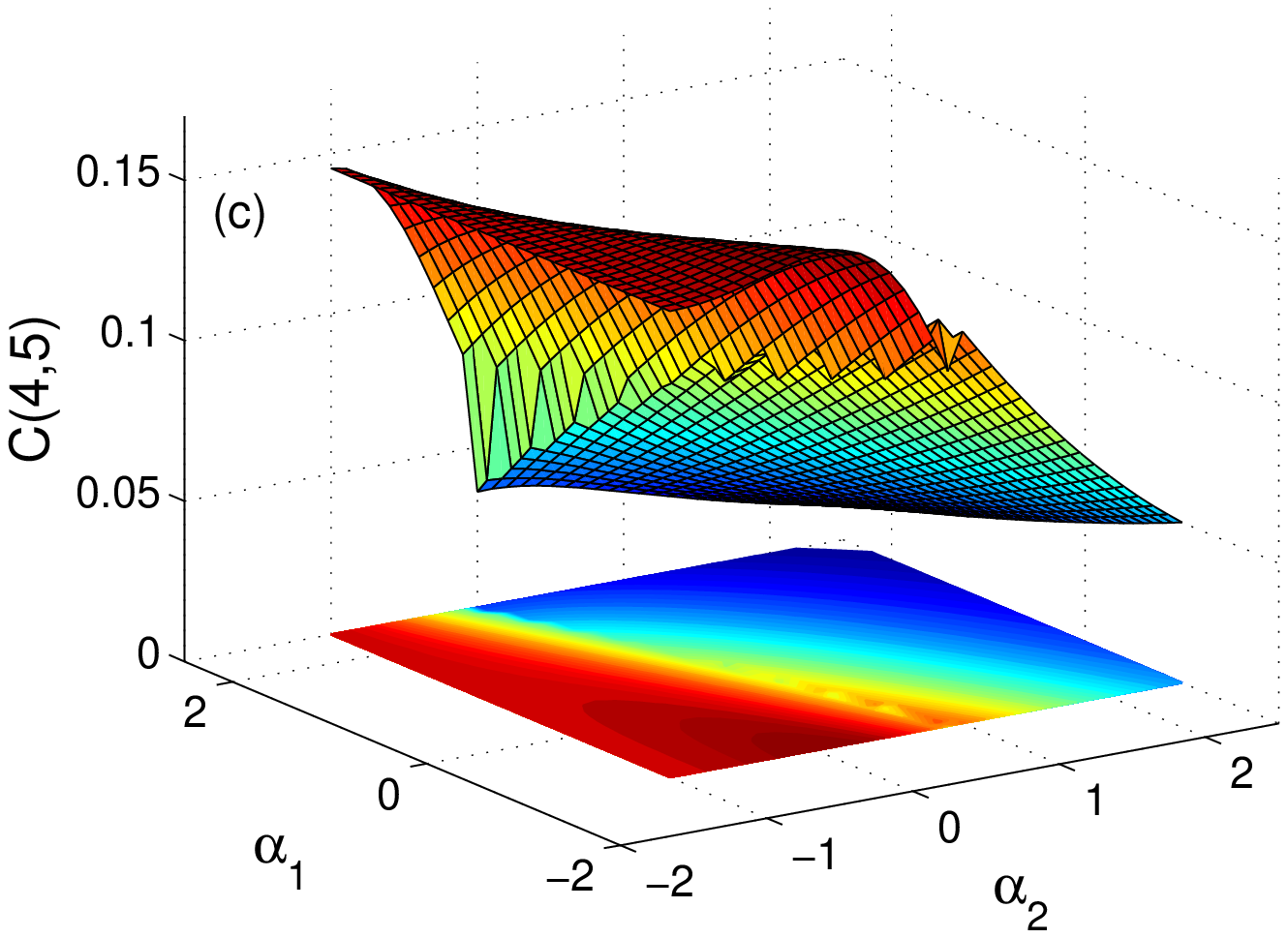}}\quad
   \subfigure{\includegraphics[width=8 cm]{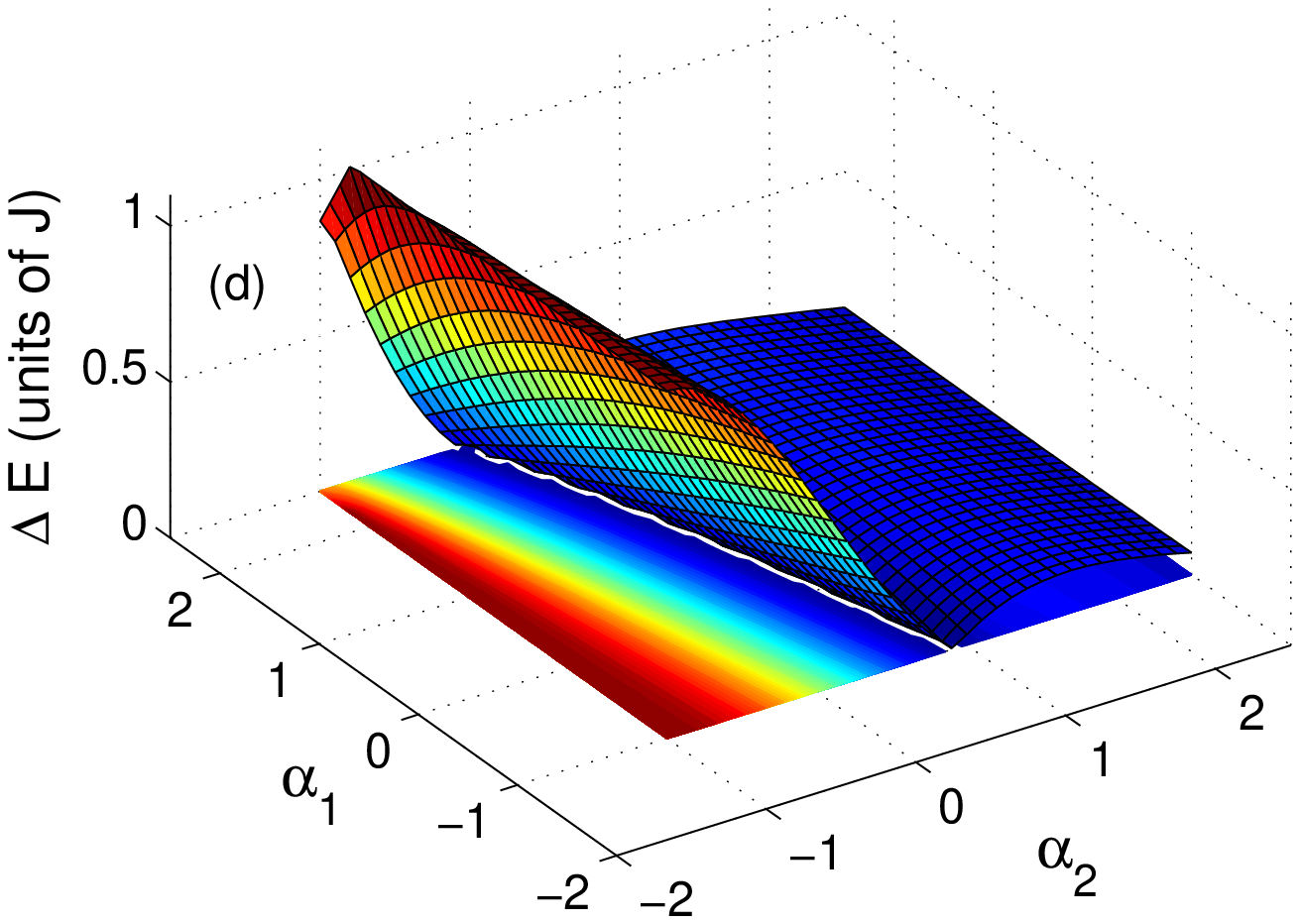}}
   \caption{{\protect\footnotesize (Color online) The concurrence $C(1,2)$, $C(1,4)$, $C(4,5)$ versus the impurity coupling strengths $\alpha_1$ and $\alpha_2$ with double impurities at sites 1 and 2 for the two dimensional partially anisotropic lattice ($\gamma = 0.5$) in an external magnetic field h=2.}}
 \label{Imp12_G05_h2}
 \end{minipage}
\end{figure}
\begin{figure}[htbp]
\begin{minipage}[c]{\textwidth}
\centering
   \subfigure{\includegraphics[width=8 cm]{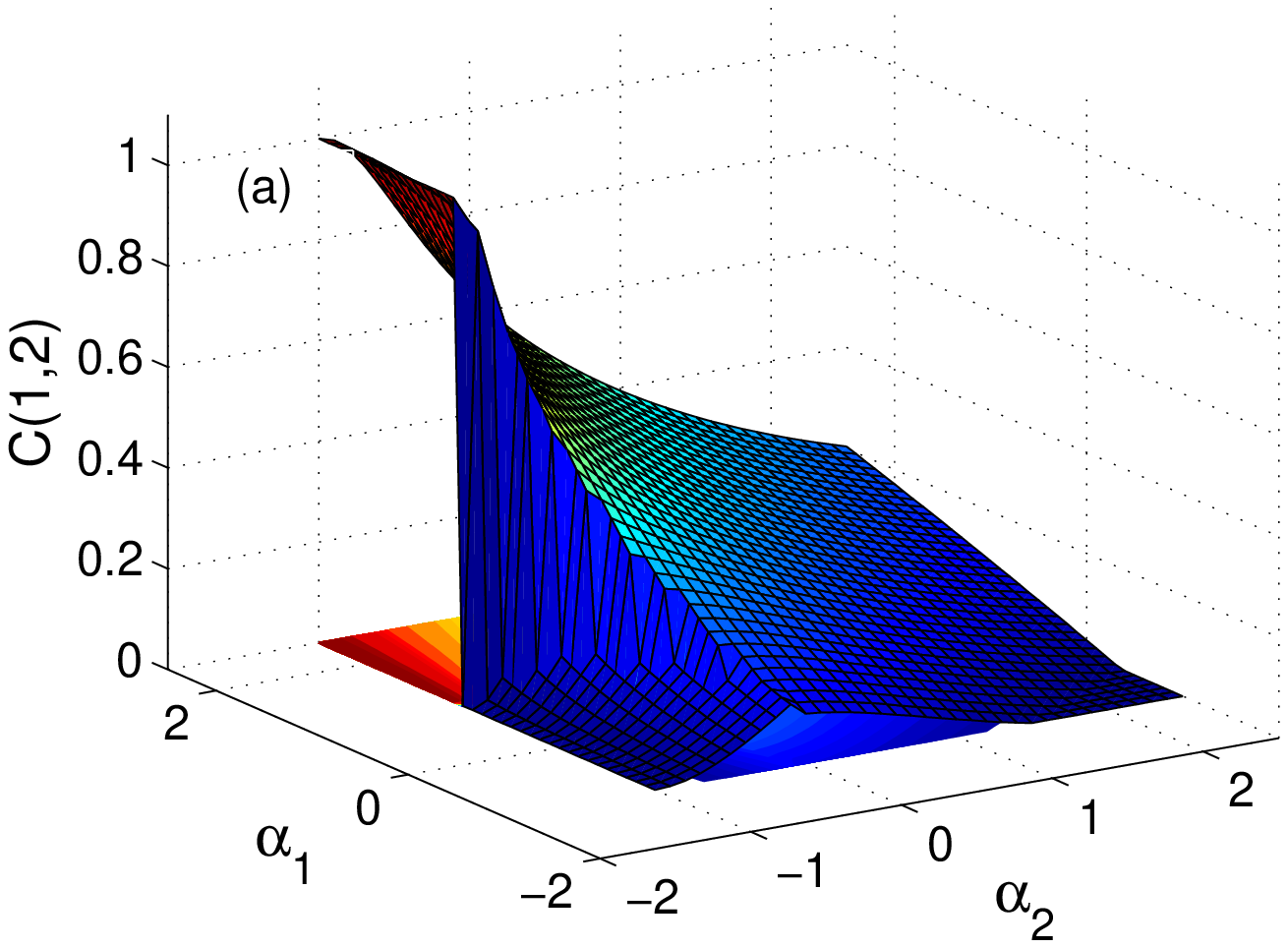}}\quad
   \subfigure{\includegraphics[width=8 cm]{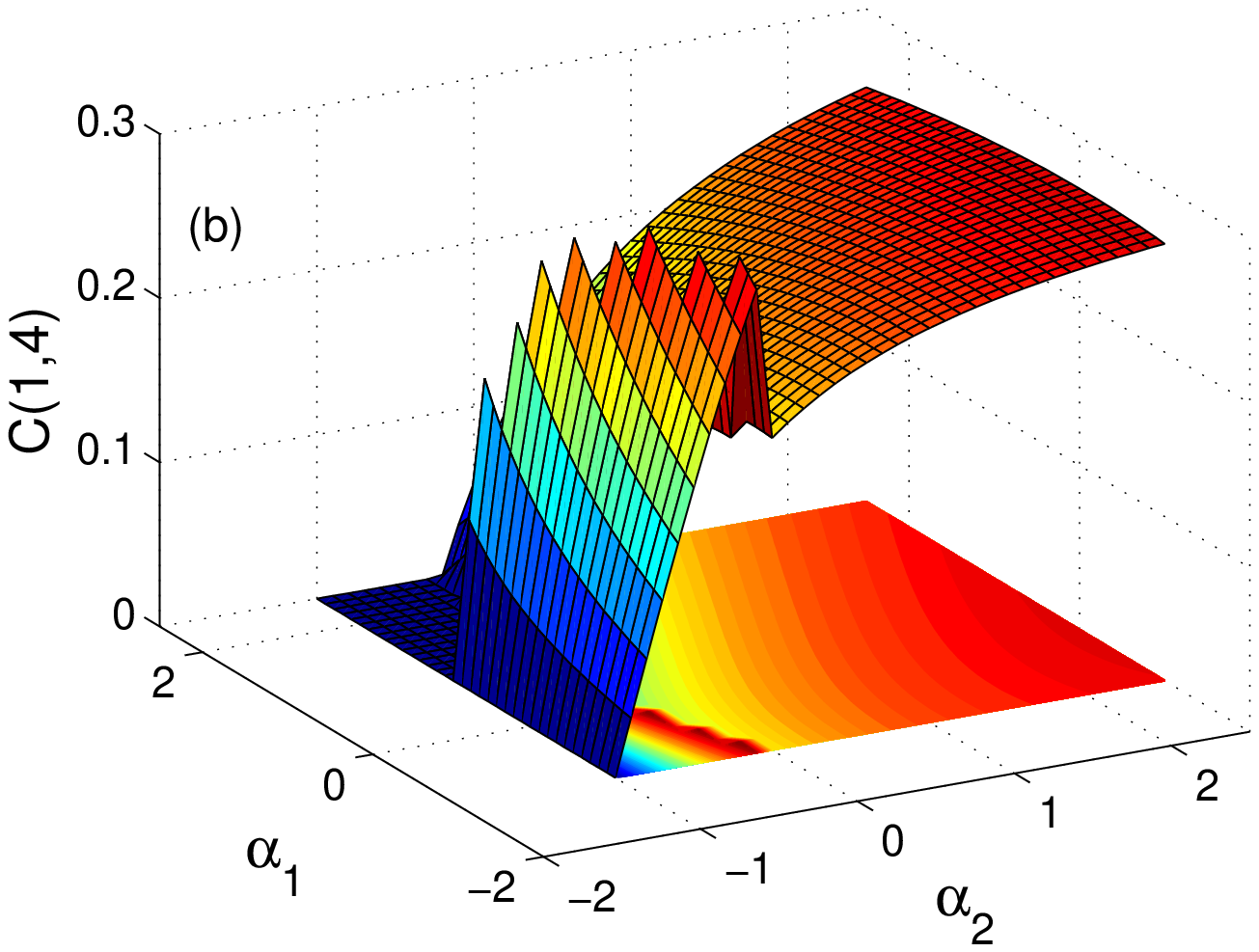}}\\
   \subfigure{\includegraphics[width=8 cm]{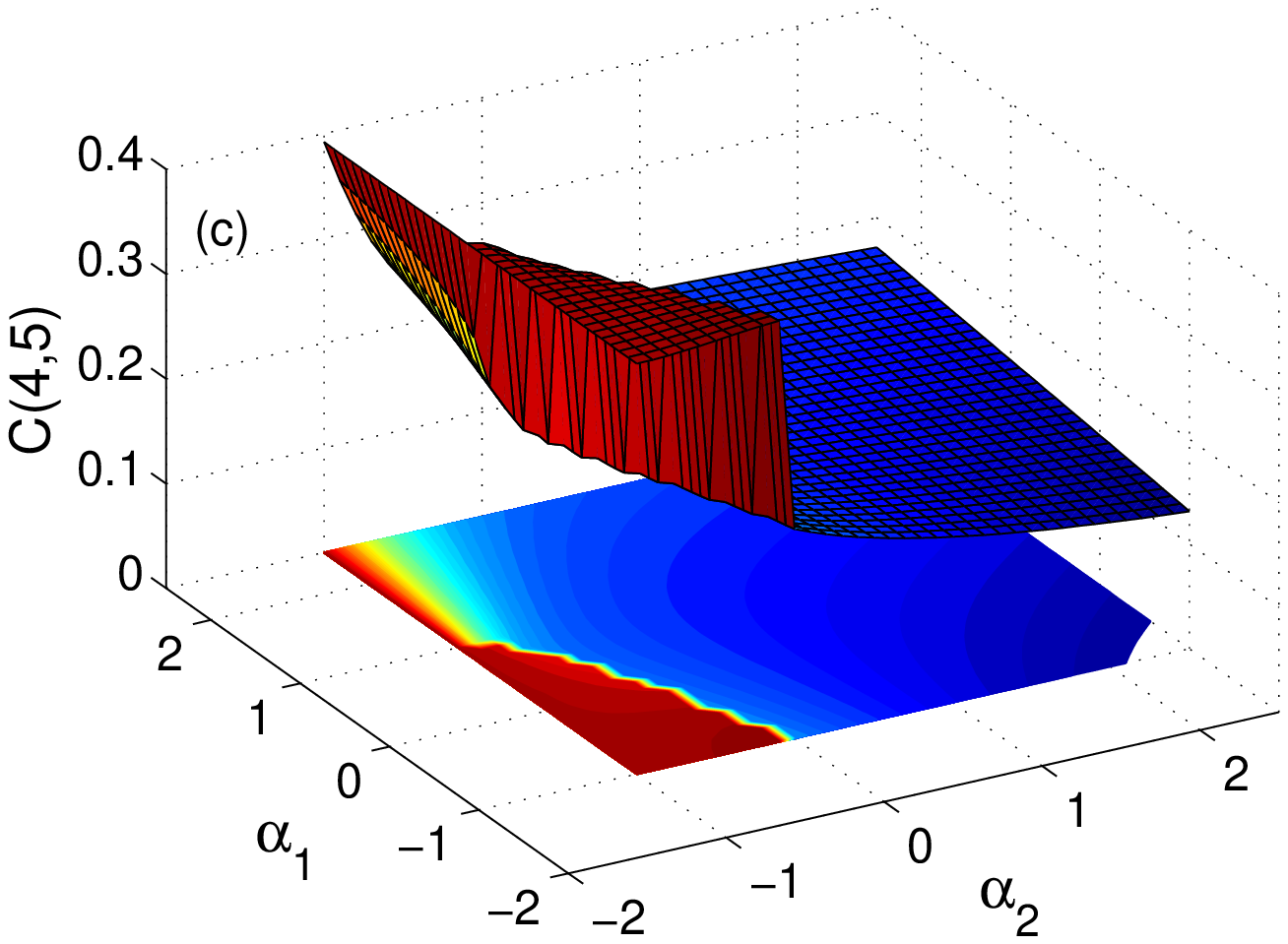}}\quad
   \subfigure{\includegraphics[width=8 cm]{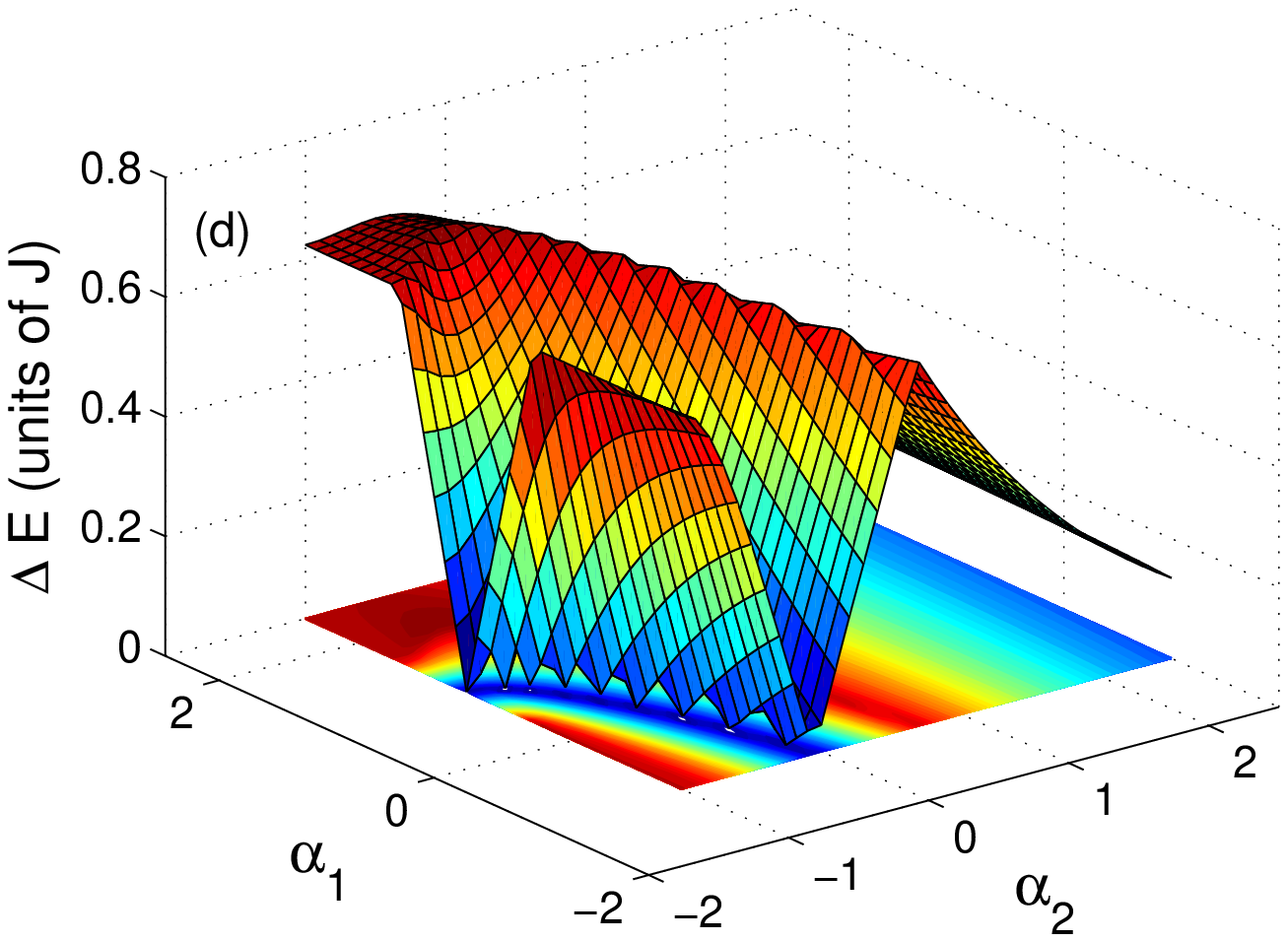}}
   \caption{{\protect\footnotesize (Color online) The concurrence $C(1,2)$, $C(1,4)$, $C(4,5)$ versus the impurity coupling strengths $\alpha_1$ and $\alpha_2$ with double impurities at sites 1 and 2 for the two dimensional XY lattice ($\gamma = 0$) in an external magnetic field h=1.}}
 \label{Imp12_G0_h1}
 \end{minipage}
\end{figure}
The concurrence C(1,4) is mainly controlled by the impurity strength $\alpha_2$ where it starts with a very small value when the impurity is very weak and increases monotonically until it reaches a maximum value at $\alpha_2=0$, i.e. with no impurity, and decays again as the impurity strength increases. The effect of $\alpha_1$ in that case is less significant and makes the concurrence slowly decreases as $\alpha_1$ increases which is expected since as the coupling between the two border sites 1 and 2 increases the entanglement between 1 and 4 decreases. It is important to note that in general $C(1,2)$ is much larger than $C(1,4)$ since the border entanglement is always higher than the central one as the entanglement is shared by many sites. The entanglement between two regular sites is shown in fig.~\ref{Imp12_G1_h2}(c) where the concurrence C(4,5) is depicted against $\alpha_1$ and $\alpha_2$, the entanglement decays gradually as $\alpha_2$ increases while $\alpha_1$ has a very small effect on the entanglement, which slightly decreases as $\alpha_1$ increases as shown. Interestingly, the behavior of the energy gap between the ground state and the first excited state of the Ising system $\Delta E$ versus the impurity strengths $\alpha_1$ and $\alpha_2$, which is explored in fig.~\ref{Imp12_G1_h2}(d) has a strong resemblance to that of the concurrence $C(4,5)$ except that the decay of $\Delta E$ against $\alpha_2$ is more rapid. The effect of changing the location of the impurities is considered in fig.~\ref{Imp14_G1_h2} where the two impurities exist at sites 1 and 4 in the Ising system. The behavior of the concurrences are very much the same as in the previous case except that the profiles of $C(1,2)$ and $C(1,4)$ have been exchanged as C(1,4) now represents the concurrence between the two impurity sites.
\begin{figure}[htbp]
\begin{minipage}[c]{\textwidth}
\centering
   \subfigure{\includegraphics[width=8 cm]{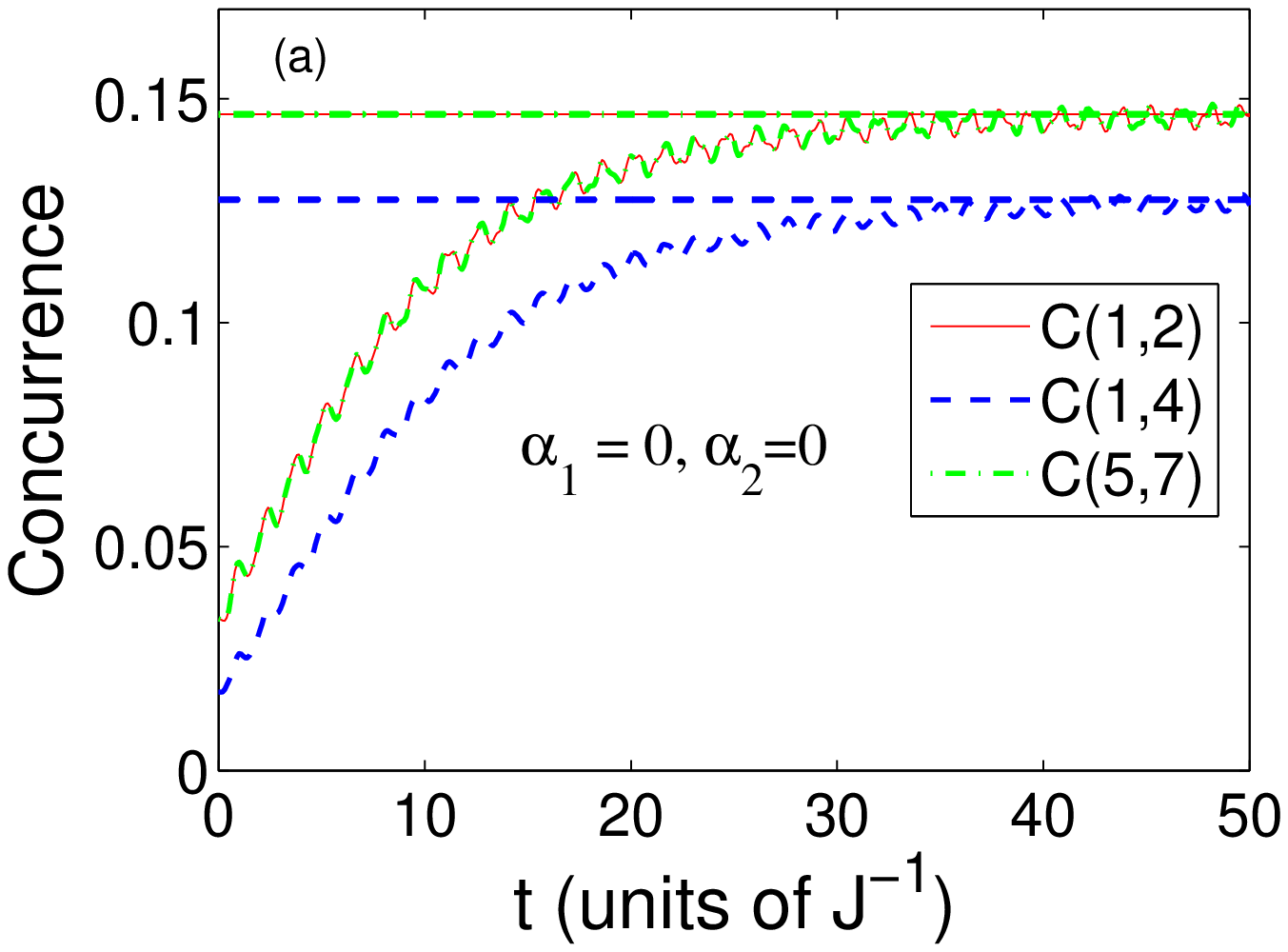}}\quad
   \subfigure{\includegraphics[width=8 cm]{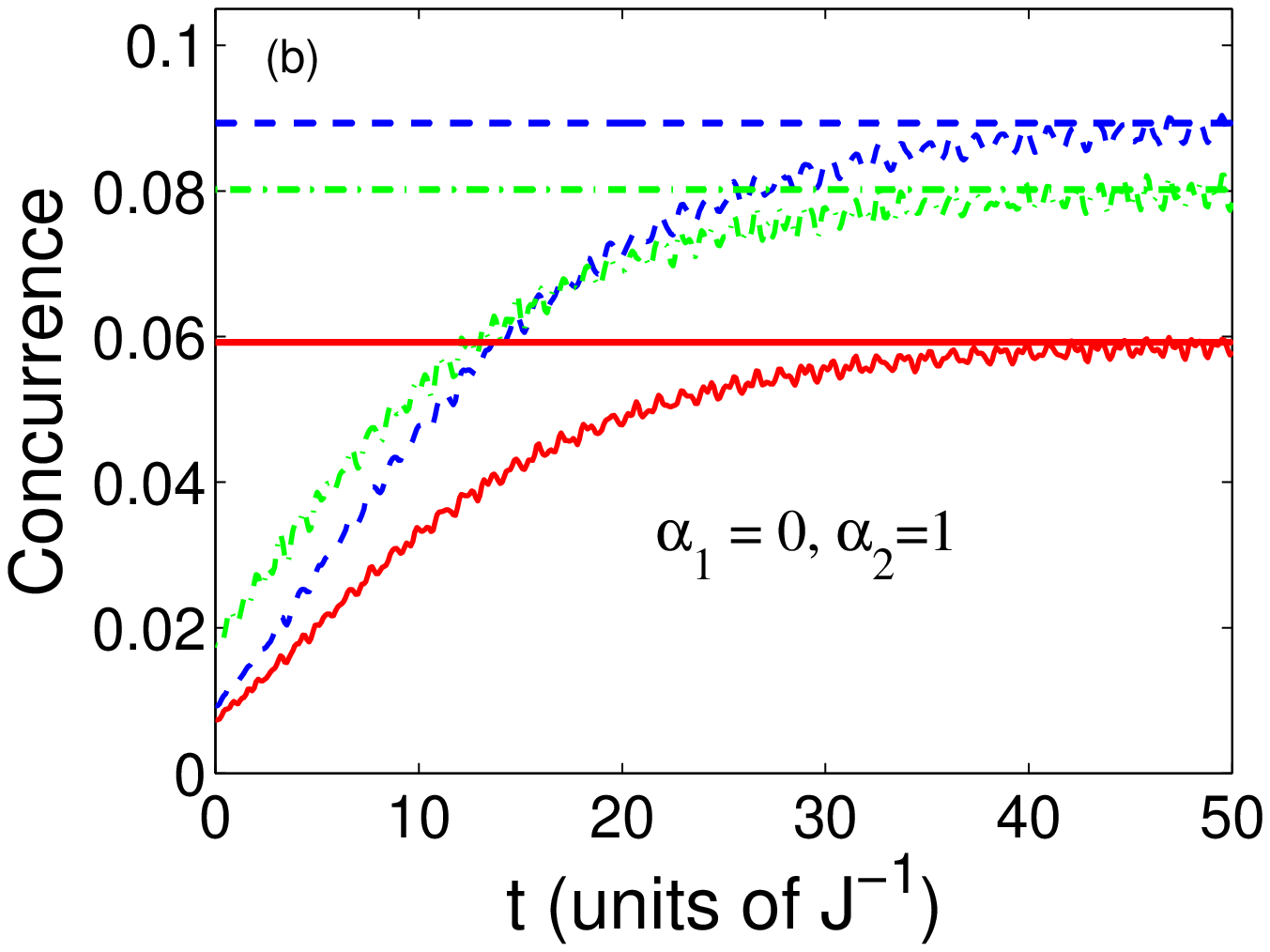}}\\
   \subfigure{\includegraphics[width=8 cm]{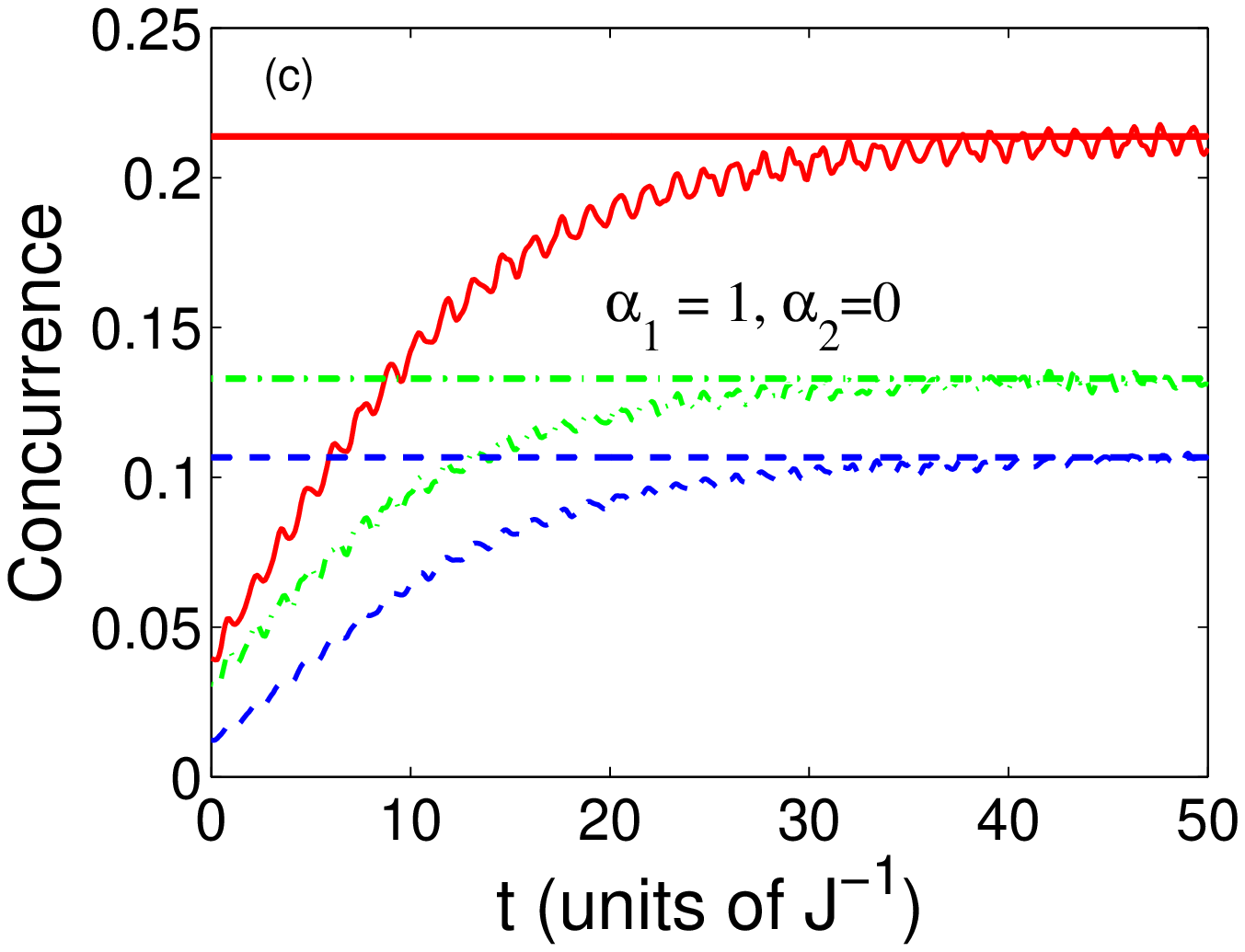}}\quad
   \subfigure{\includegraphics[width=8 cm]{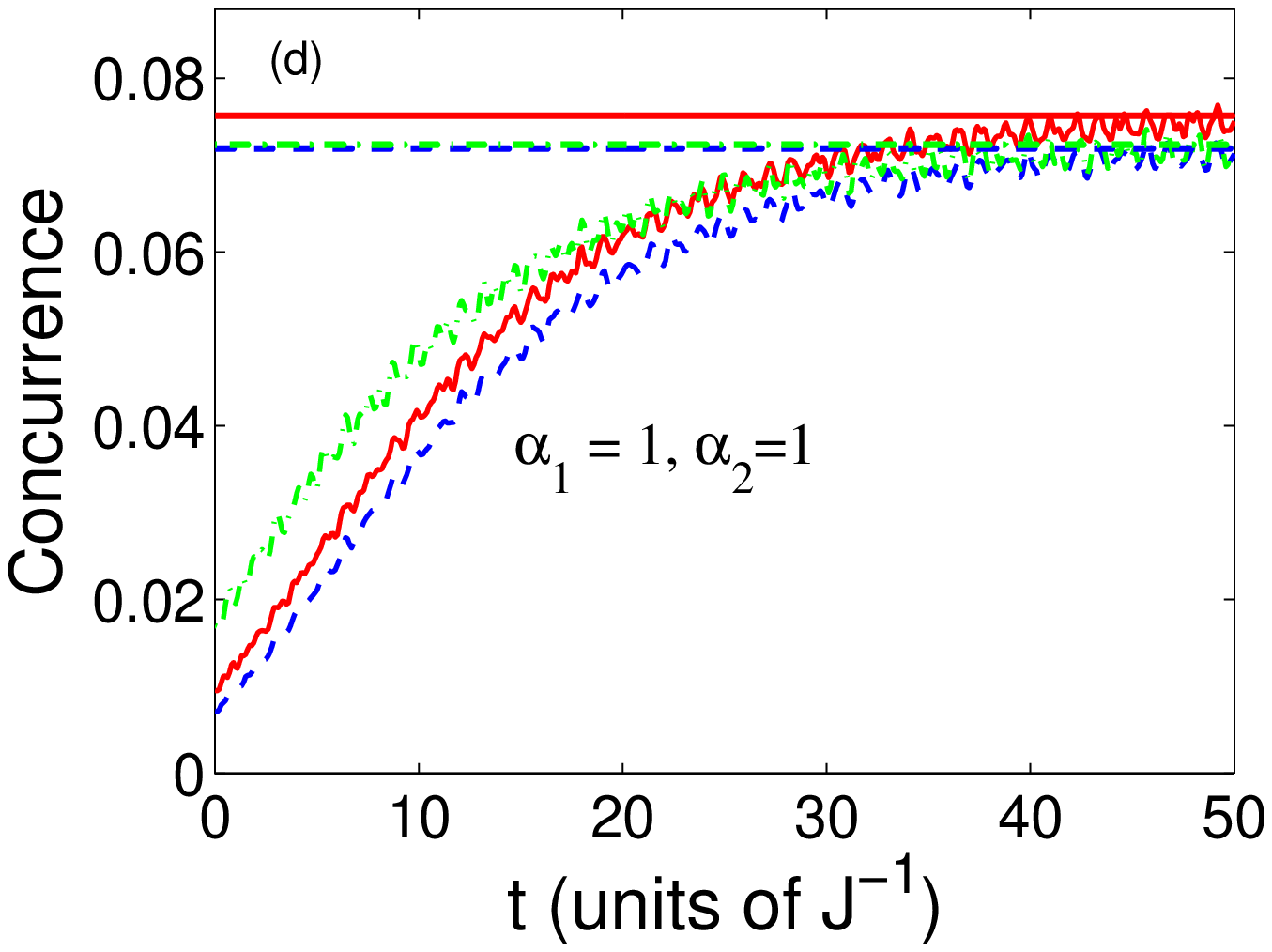}}
     \caption{{\protect\footnotesize (Color online) Dynamics of the concurrence $C(1,2)$, $C(1,4)$, $C(5,7)$ with double impurities at sites 1 and 2 for the two dimensional Ising lattice ($\gamma = 1$) in an exponential magnetic field where a=1, b=2, and w=0.1. The straight lines represent the equilibrium concurrences corresponding to constant magnetic field $h=2$. The legend for all subfigures is as shown in subfigure (a).}}
 \label{Imp12_Dyn_G1}
 \end{minipage}
\end{figure}
\begin{figure}[htbp]
\begin{minipage}[c]{\textwidth}
\centering
   \subfigure{\includegraphics[width=8 cm]{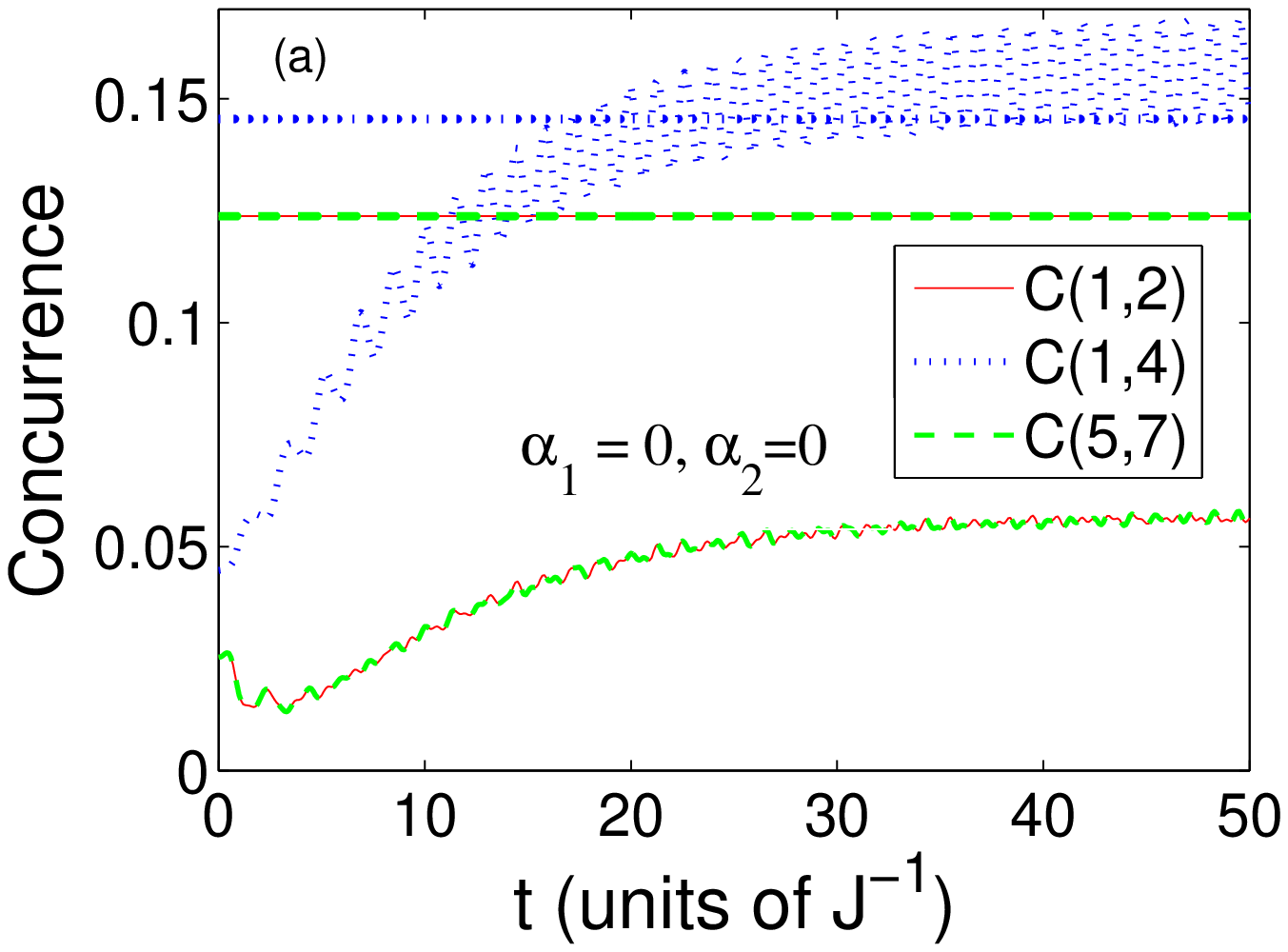}}\quad
   \subfigure{\includegraphics[width=8 cm]{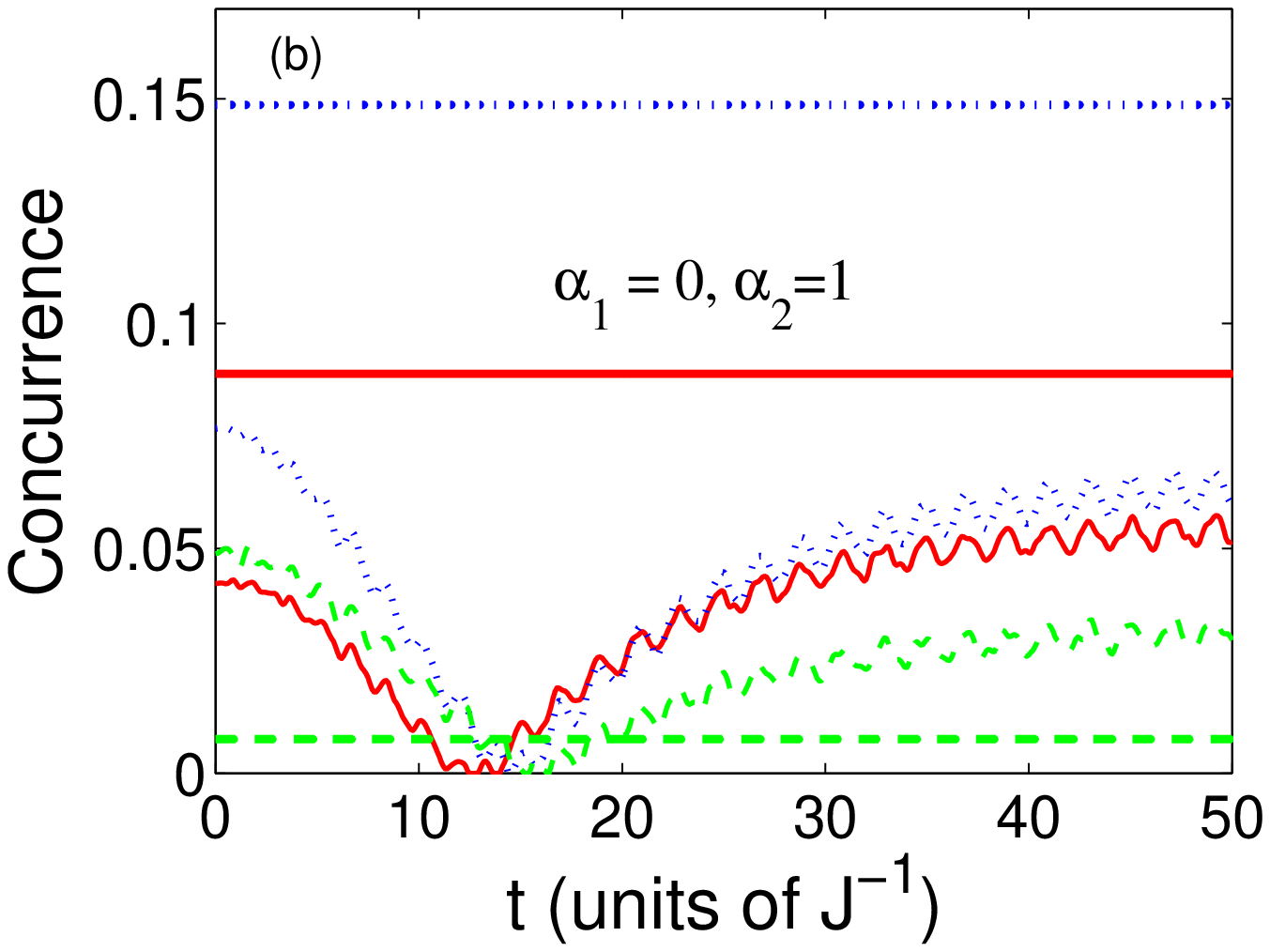}}\\
   \subfigure{\includegraphics[width=8 cm]{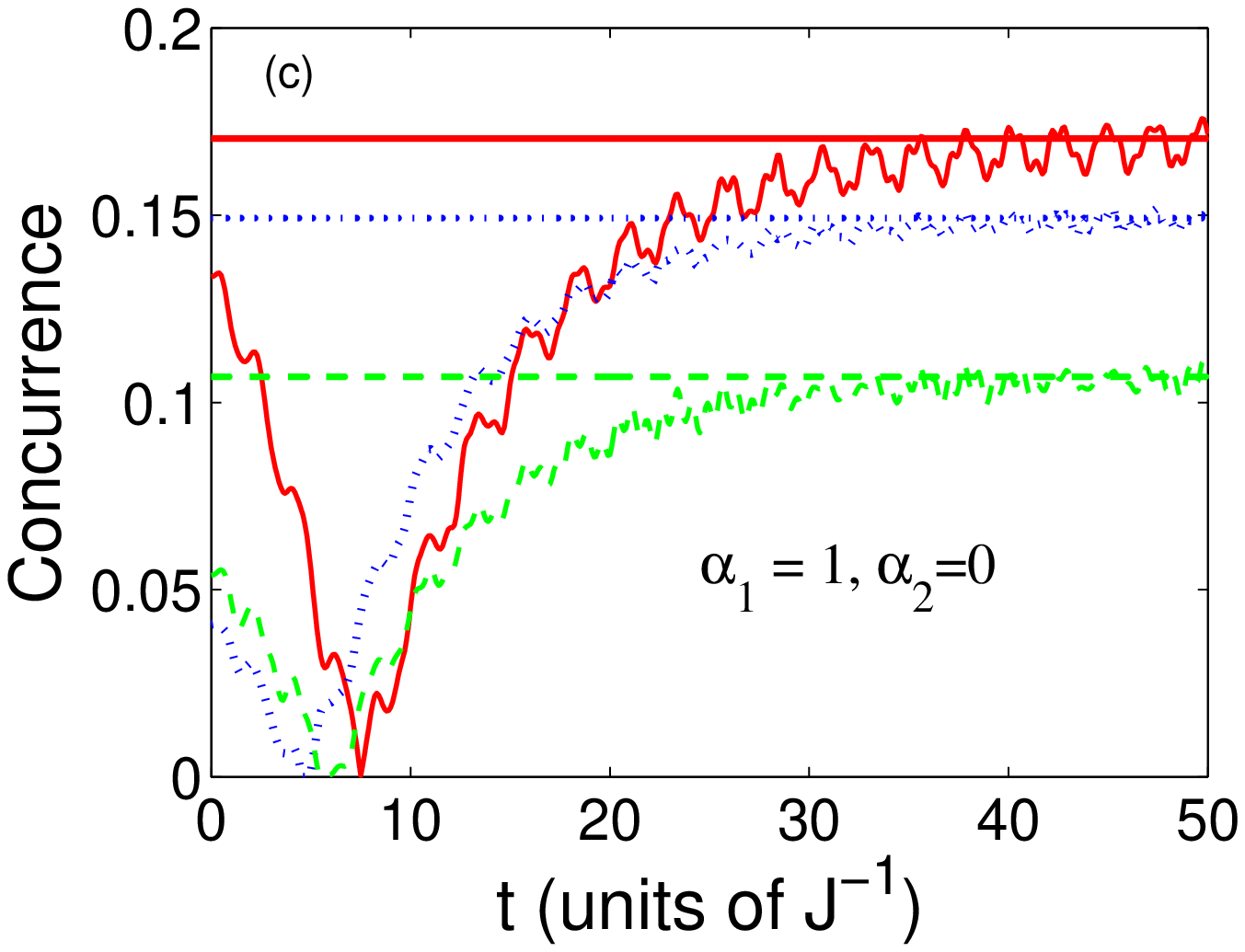}}\quad
   \subfigure{\includegraphics[width=8 cm]{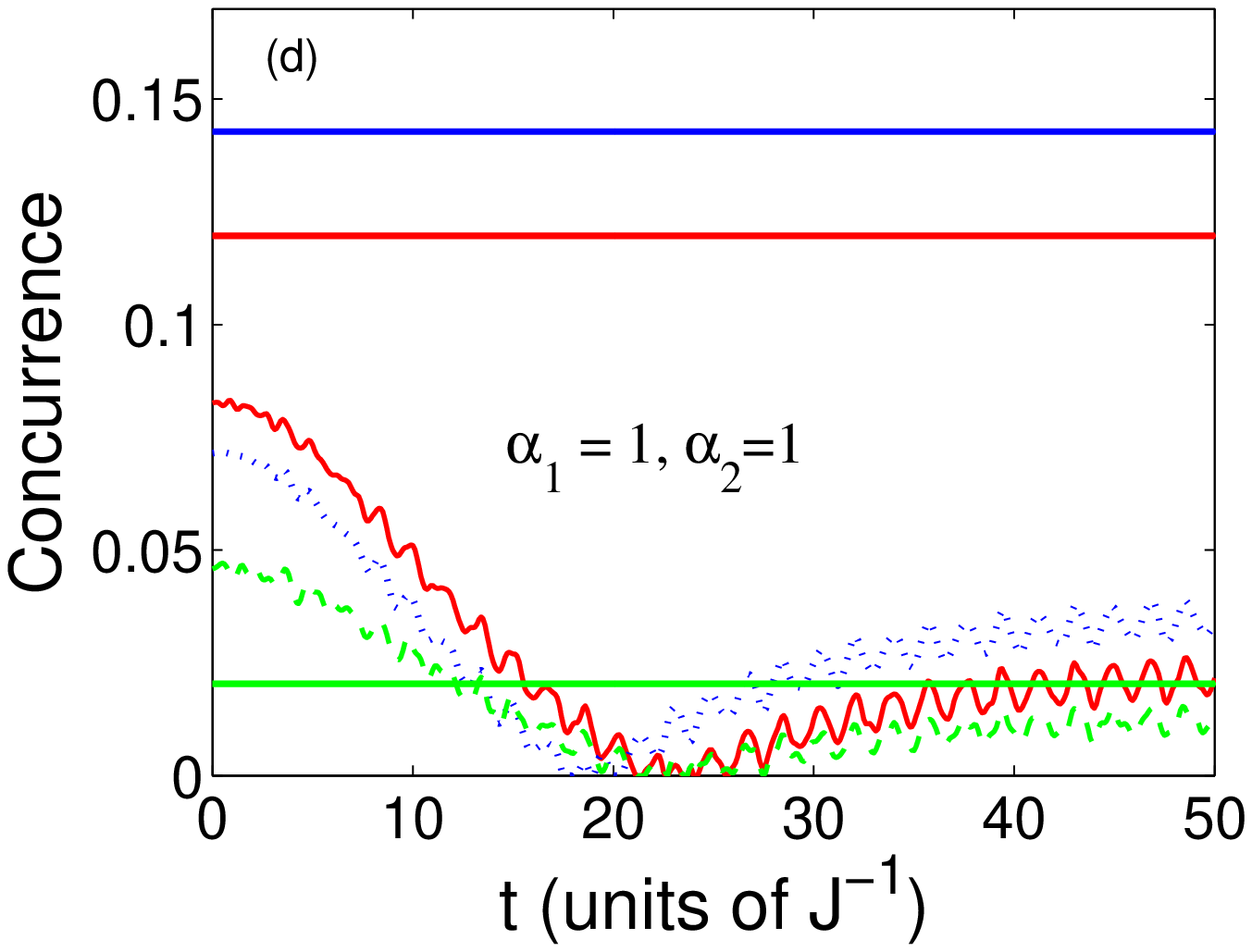}}
   \caption{{\protect\footnotesize (Color online) Dynamics of the concurrence $C(1,2)$, $C(1,4)$, $C(5,7)$ with double impurities at sites 1 and 2 for the two dimensional partially anisotropic lattice ($\gamma = 0.5$) in an exponential magnetic field where a=1, b=2, and w=0.1. The straight lines represent the equilibrium concurrences corresponding to constant magnetic field $h=2$. The legend for all subfigures is as shown in subfigure (a).}}
 \label{Imp12_Dyn_G05}
 \end{minipage}
\end{figure}

The partially anisotropic system, $\gamma=0.5$, with double impurity at sites 1 and 2 and under the effect of the external magnetic field $h=2$ is explored in fig.~\ref{Imp12_G05_h2}. As one can see, the overall behavior specially at the border values of the impurity strengths is the same as observed in the Ising case except that the concurrences suffer a local minimum within a small range of the impurity strength $\alpha_2$ between 0 and 1 while corresponding to the whole $\alpha_1$ range. The change of the entanglement around this local minimum takes a step-like profile which is very clear in the case of the concurrence $C(1,4)$ shown in fig.~\ref{Imp12_G05_h2}(b). Remarkably, the local minima in the plotted concurrences coincide with the line of vanishing energy gap as shown in fig.~\ref{Imp12_G05_h2}(d), which means that these minima correspond to a transition between a ground state and another one which takes place as the system parameters change. The anisotropic $XY$ model with two impurities at sites 1 and 2 in an external magnetic field $h=1.8$ is explored in fig.~\ref{Imp12_G0_h1}. The entanglement for this system shows much sharper changes as a function of the impurity strengths and the sharp step changes take place in a narrow region of both $\alpha_1$ and $\alpha_2$ specifically for $-1 \leq \alpha_1 \leq 1$ and $-1 \leq \alpha_2 \leq 0$. It is interesting to note that again the sharp step changes in the entanglement are corresponding to the line of minimum energy gap as shown in fig.~\ref{Imp12_G0_h1}(d) where this line varies continuously between a very small value and zero which explains the many steps appearing in the different concurrences and particularly $C(1,4)$ depicted in fig.~\ref{Imp12_G0_h1}(b).
\subsection{System dynamics with impurities}
\begin{figure}[htbp]
\begin{minipage}[c]{\textwidth}
\centering
   \subfigure{\includegraphics[width=8 cm]{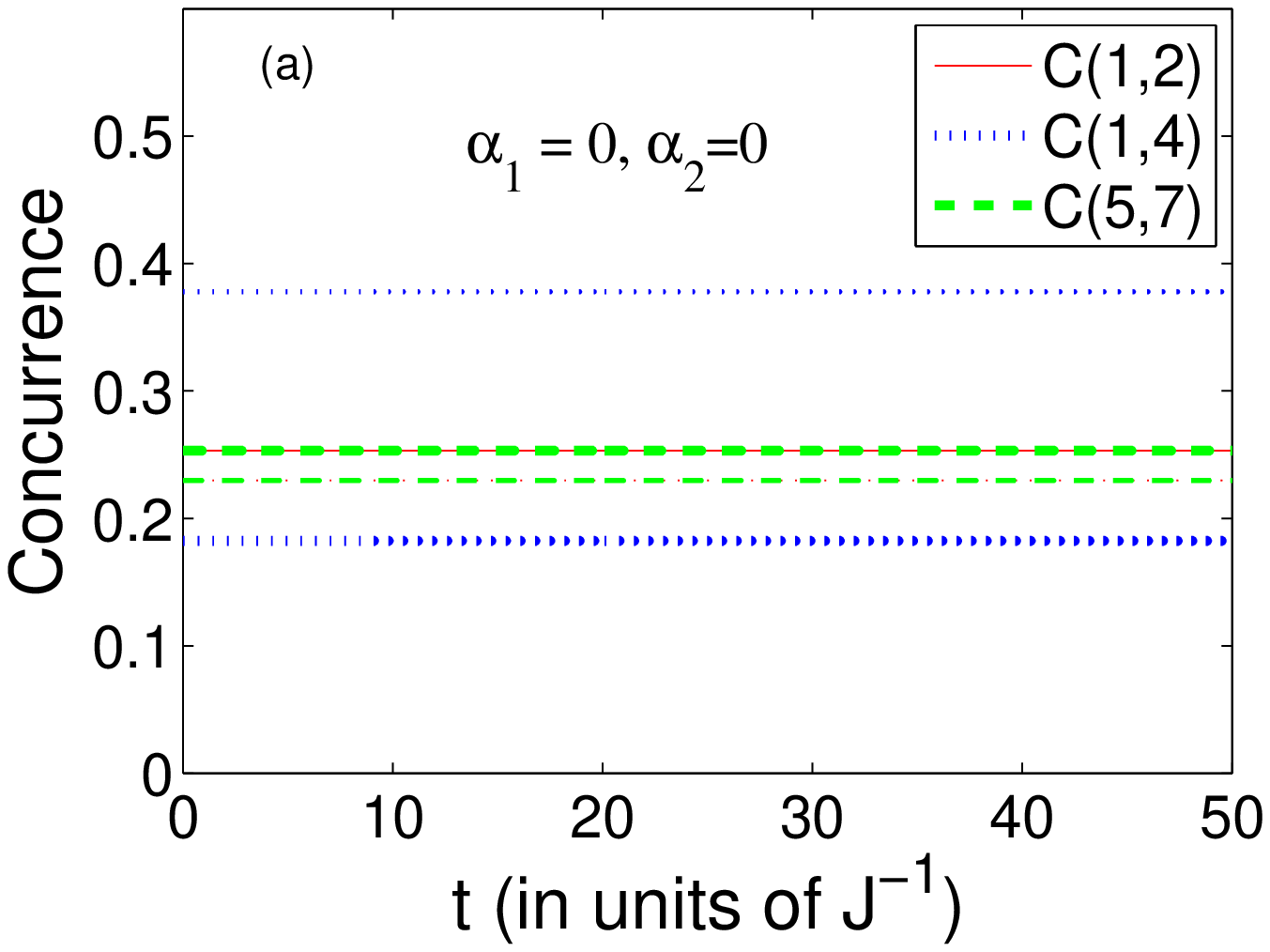}}\quad
   \subfigure{\includegraphics[width=8 cm]{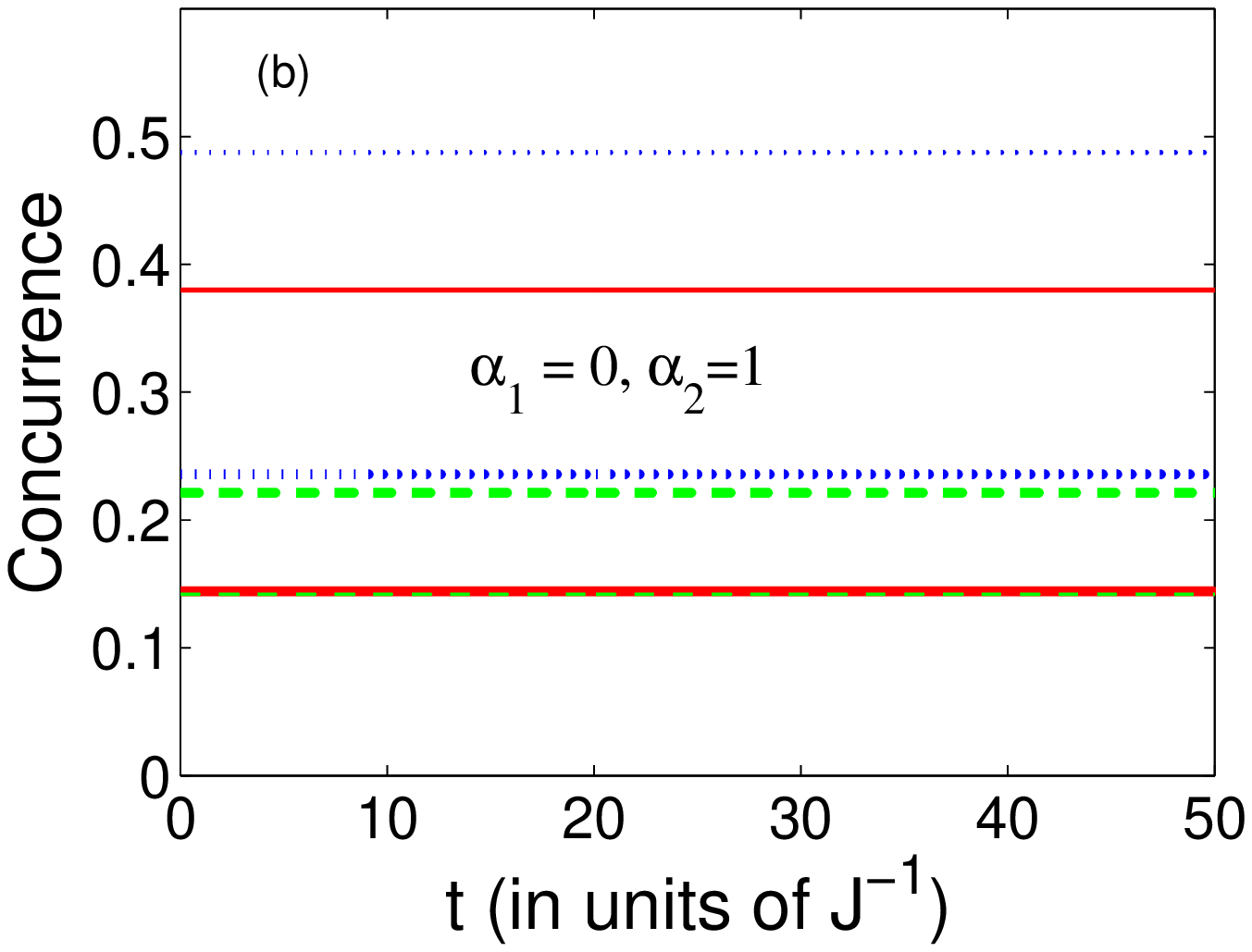}}\\
   \subfigure{\includegraphics[width=8 cm]{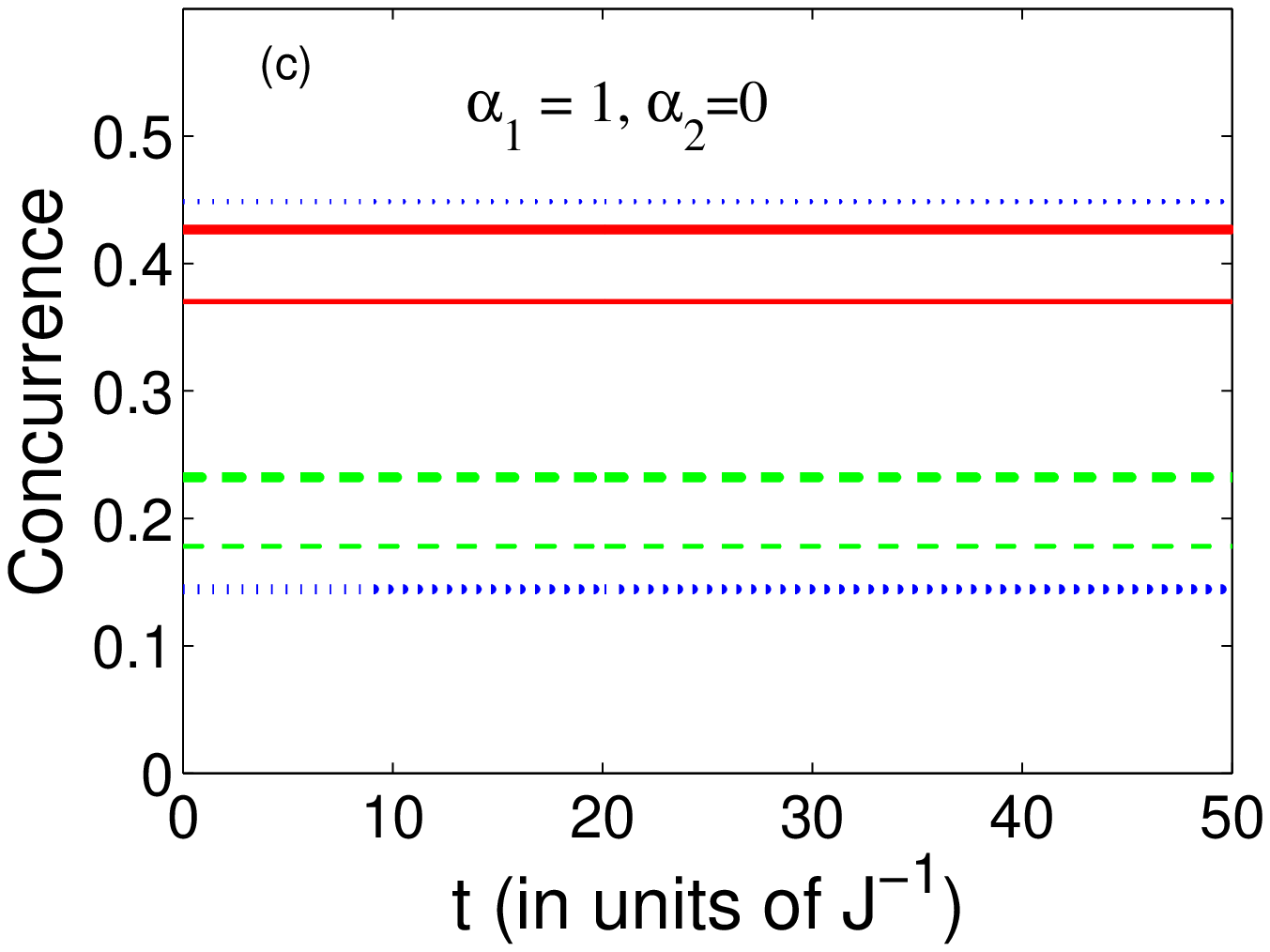}}\quad
   \subfigure{\includegraphics[width=8 cm]{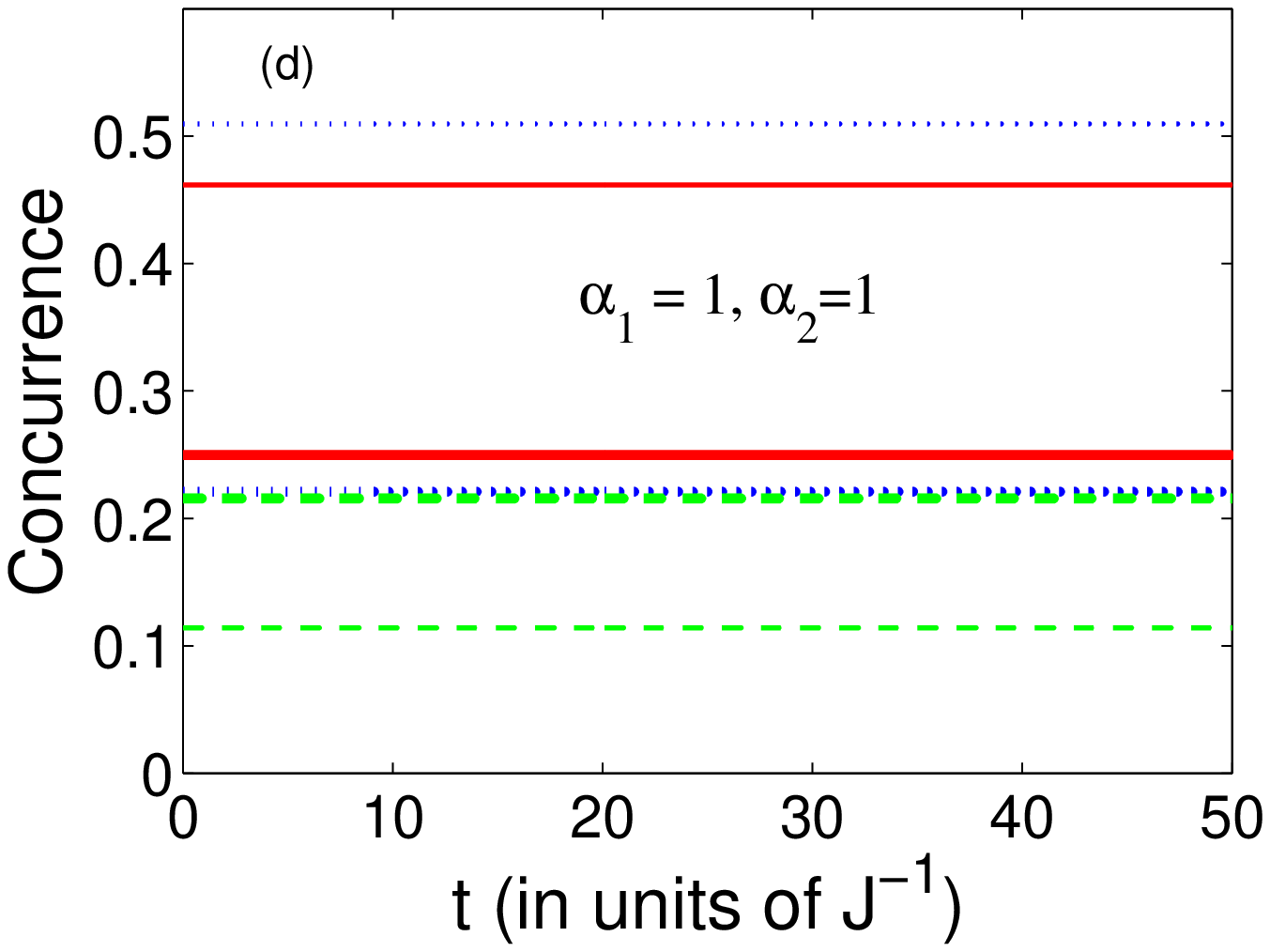}}
    \caption{{\protect\footnotesize (Color online) Dynamics of the concurrence $C(1,2)$, $C(1,4)$, $C(5,7)$ with double impurities at sites 1 and 2 for the two dimensional isotropic XY lattice ($\gamma = 0$) in an exponential magnetic field where a=1, b=1.8, and w=0.1. The straight (thicker) lines represent the equilibrium concurrences corresponding to constant magnetic field $h=1.8$. The legend for all subfigures is as shown in subfigure (a).}}
 \label{Imp12_Dyn_G0}
 \end{minipage}
\end{figure}
\begin{figure}[htbp]
\begin{minipage}[c]{\textwidth}
\centering
   \subfigure{\includegraphics[width=8 cm]{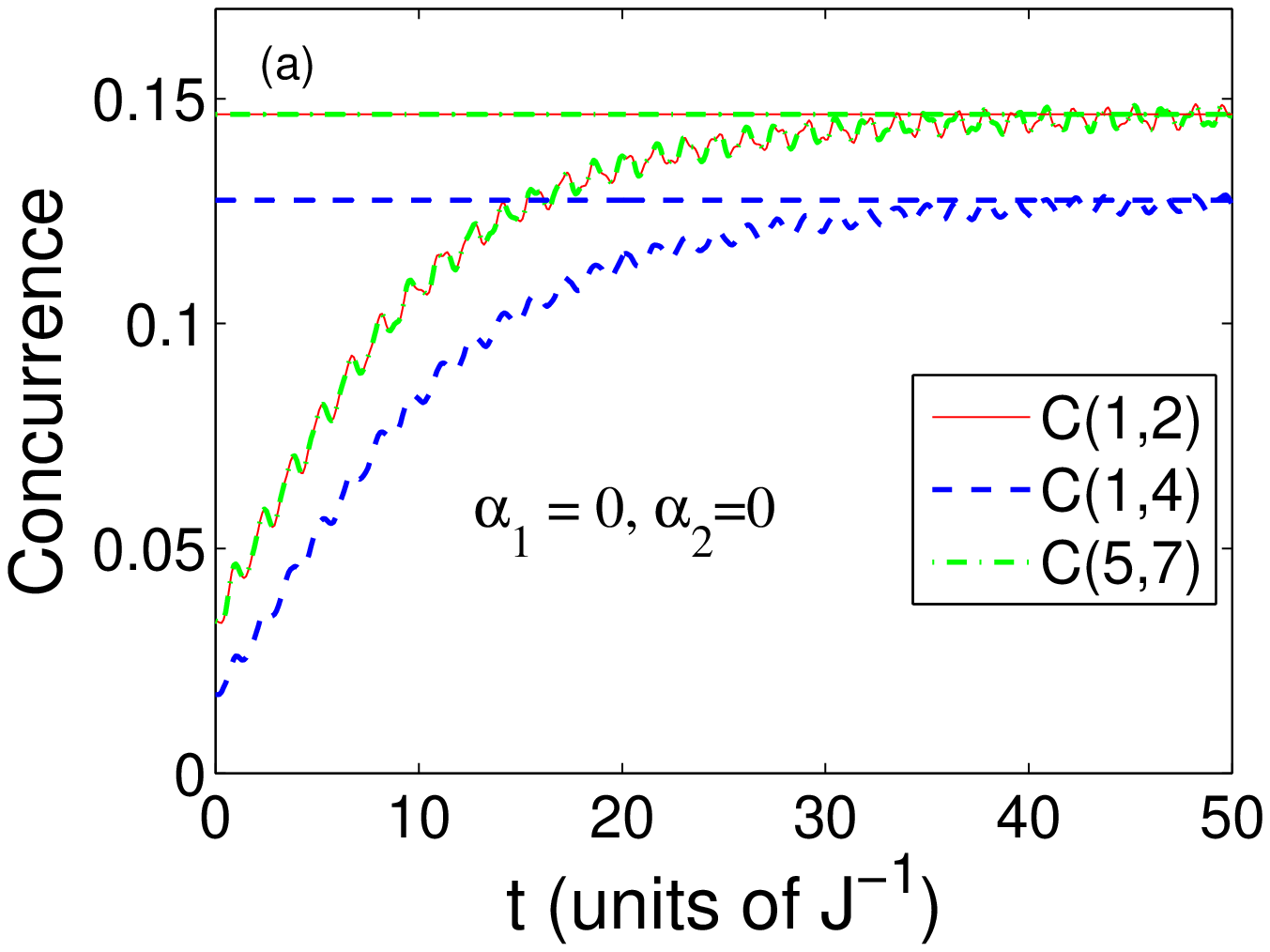}}\quad
   \subfigure{\includegraphics[width=8 cm]{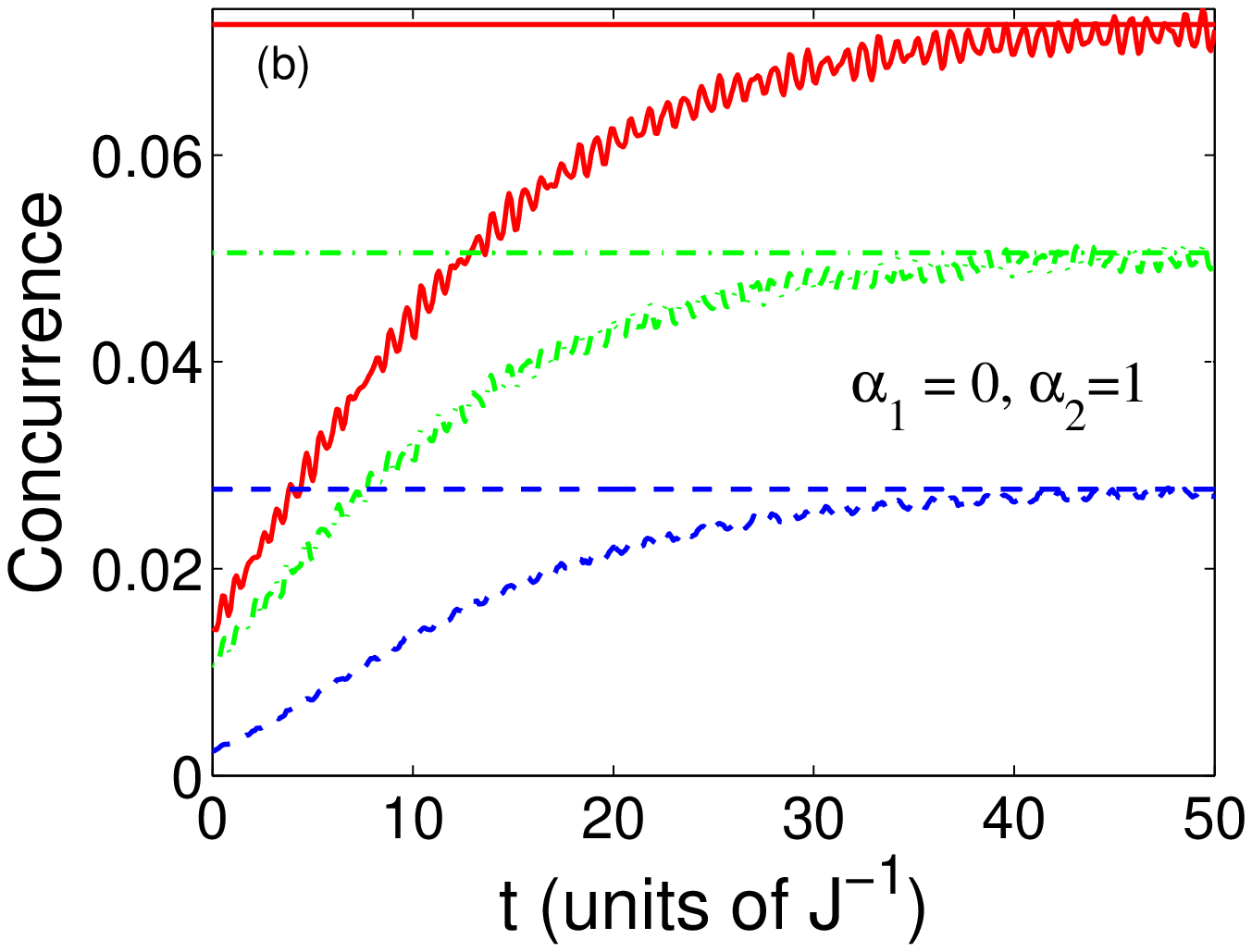}}\\
   \subfigure{\includegraphics[width=8 cm]{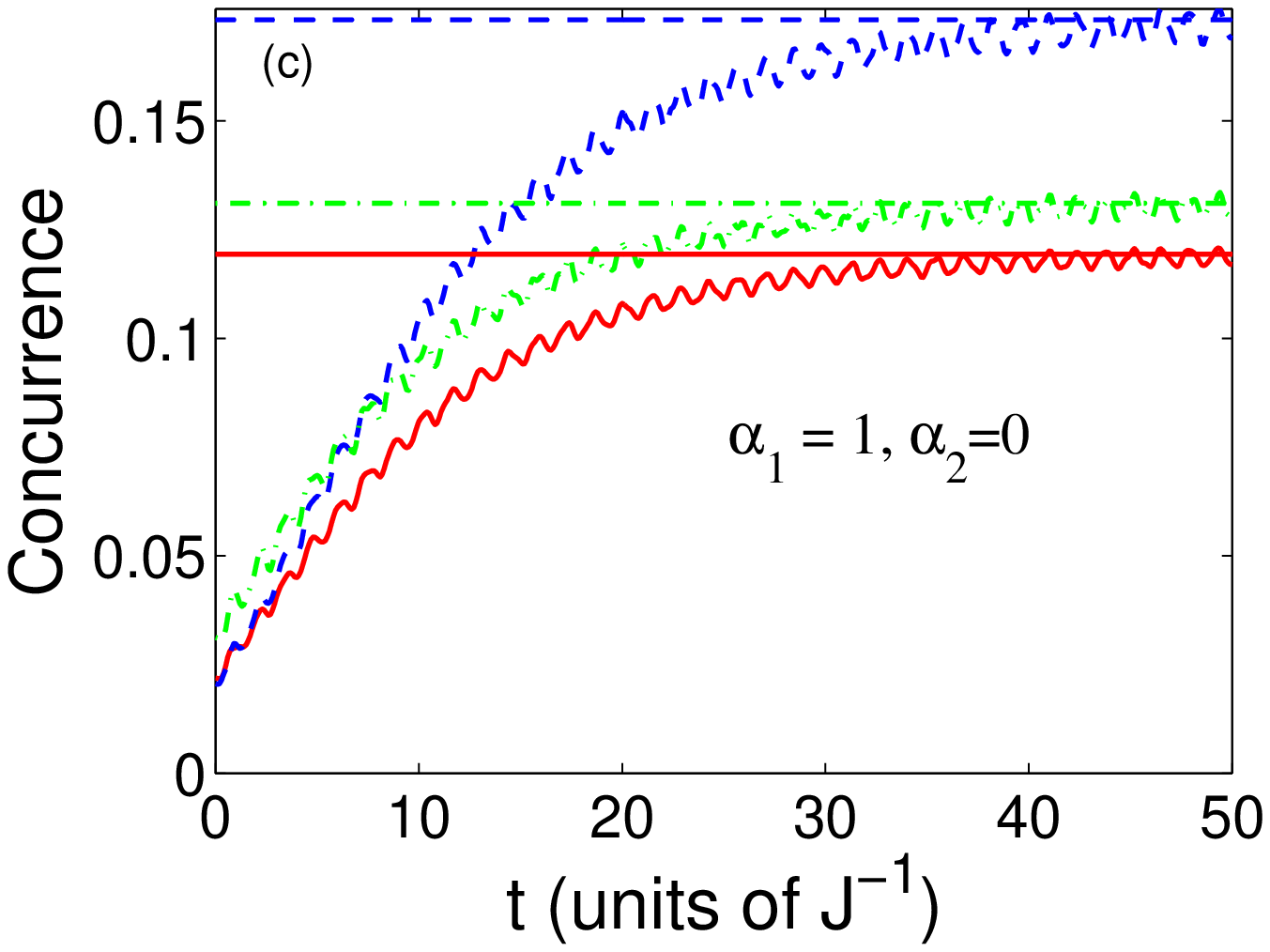}}\quad
   \subfigure{\includegraphics[width=8 cm]{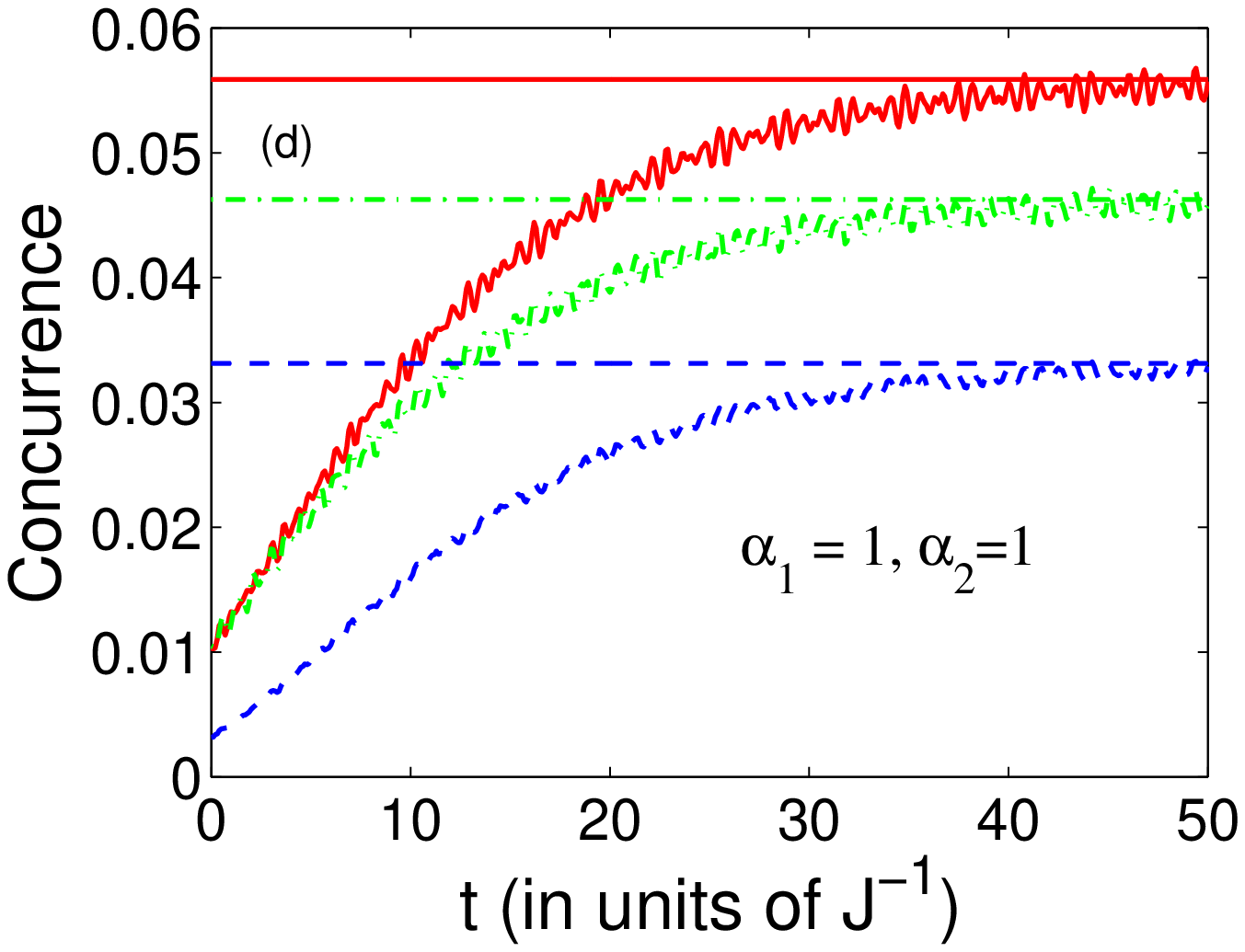}}
   \caption{{\protect\footnotesize (Color online) Dynamics of the concurrence $C(1,2)$, $C(1,4)$, $C(5,7)$ with double impurities at sites 1 and 4 for the two dimensional Ising lattice ($\gamma = 1$) in an exponential magnetic field where a=1, b=2, and w=0.1. The straight lines represent the equilibrium concurrences corresponding to constant magnetic field $h=2$. The legend for all subfigures is as shown in subfigure (a).}}
 \label{Imp14_Dyn_G1}
 \end{minipage}
\end{figure}
Now we turn to the dynamics of the two dimensional spin system with double impurity under the effect of an external exponential magnetic field to test the ergodicity of the system as we vary the degree of anisotropy or the location of the impurities. In fig.~\ref{Imp12_Dyn_G1} we consider the dynamics of the Ising system with two impurities at the sites 1 and 2 under the effect of an exponential magnetic field with parameters $a=1$, $b=2$ and $\omega=0.1$. As one can see the system shows an ergodic behavior for the different values of the impurities strengths $\alpha_1, \alpha_2= (0,0), (0,1), (1,0)$ and $(1,1)$. The Ising system sustains its ergodicity for all shown values of impurities strengths and other tested values. 
\begin{figure}[htbp]
\begin{minipage}[c]{\textwidth}
\centering
   \subfigure{\includegraphics[width=8 cm]{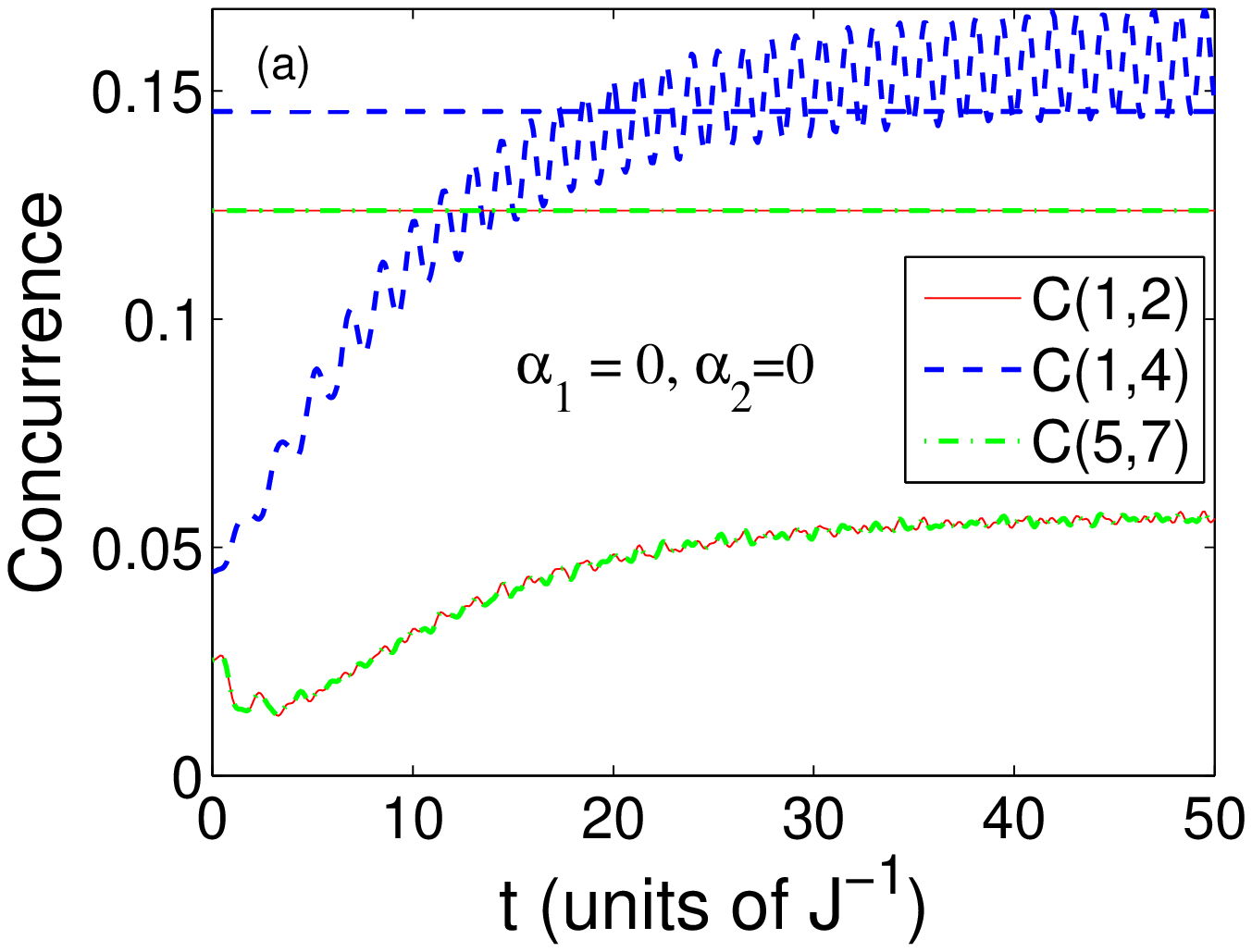}}\quad
   \subfigure{\includegraphics[width=8 cm]{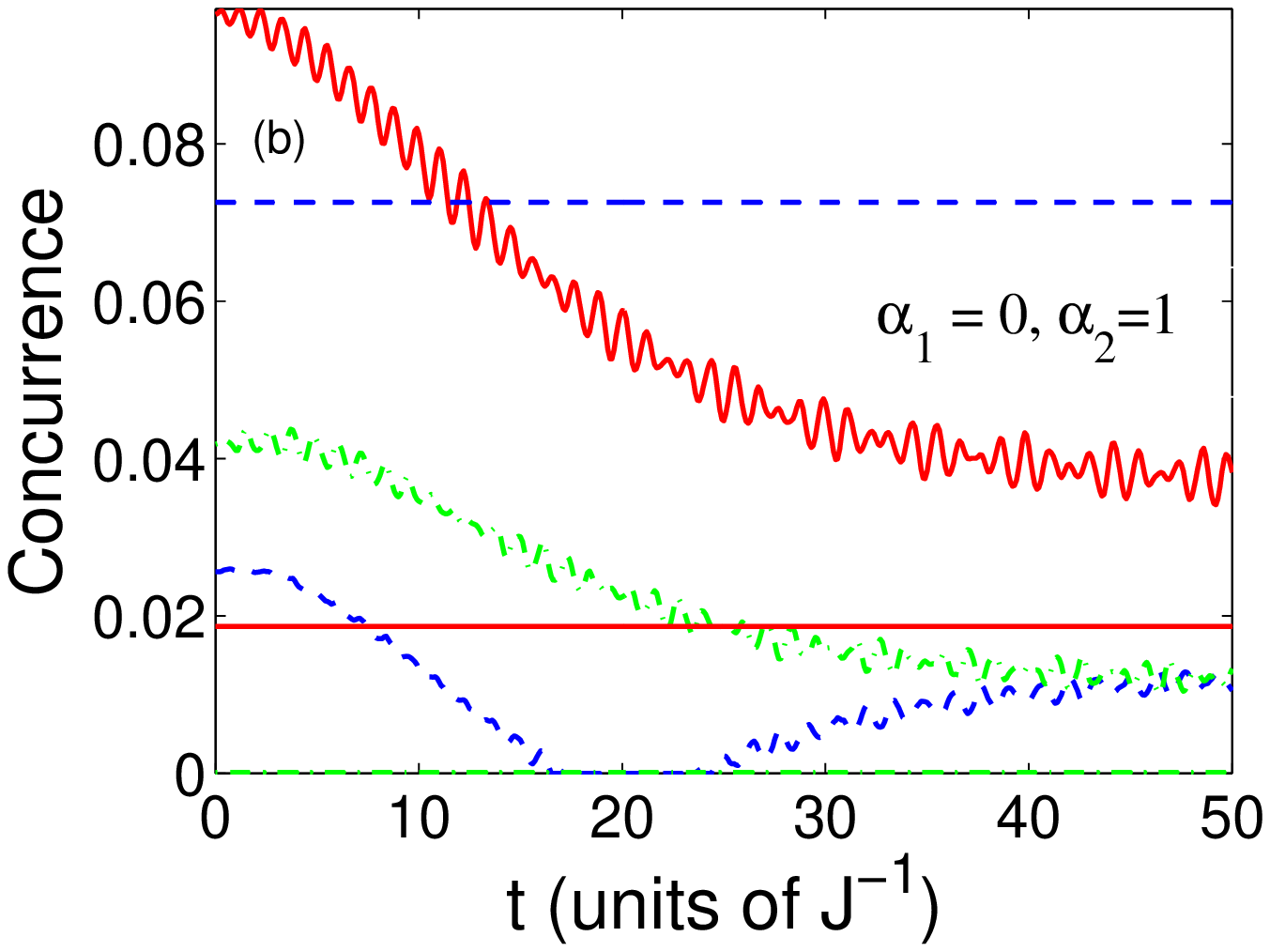}}\\
   \subfigure{\includegraphics[width=8 cm]{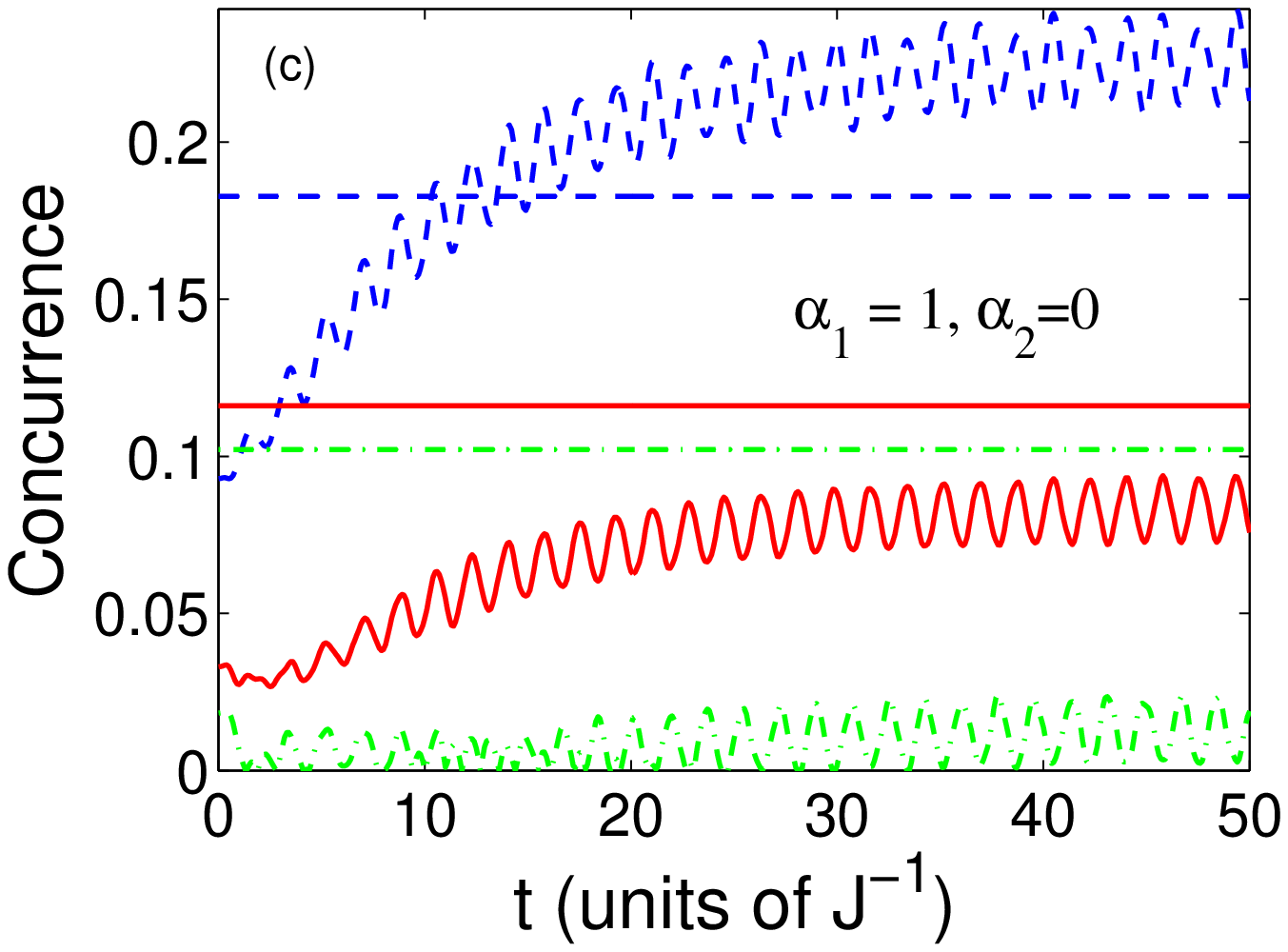}}\quad
   \subfigure{\includegraphics[width=8 cm]{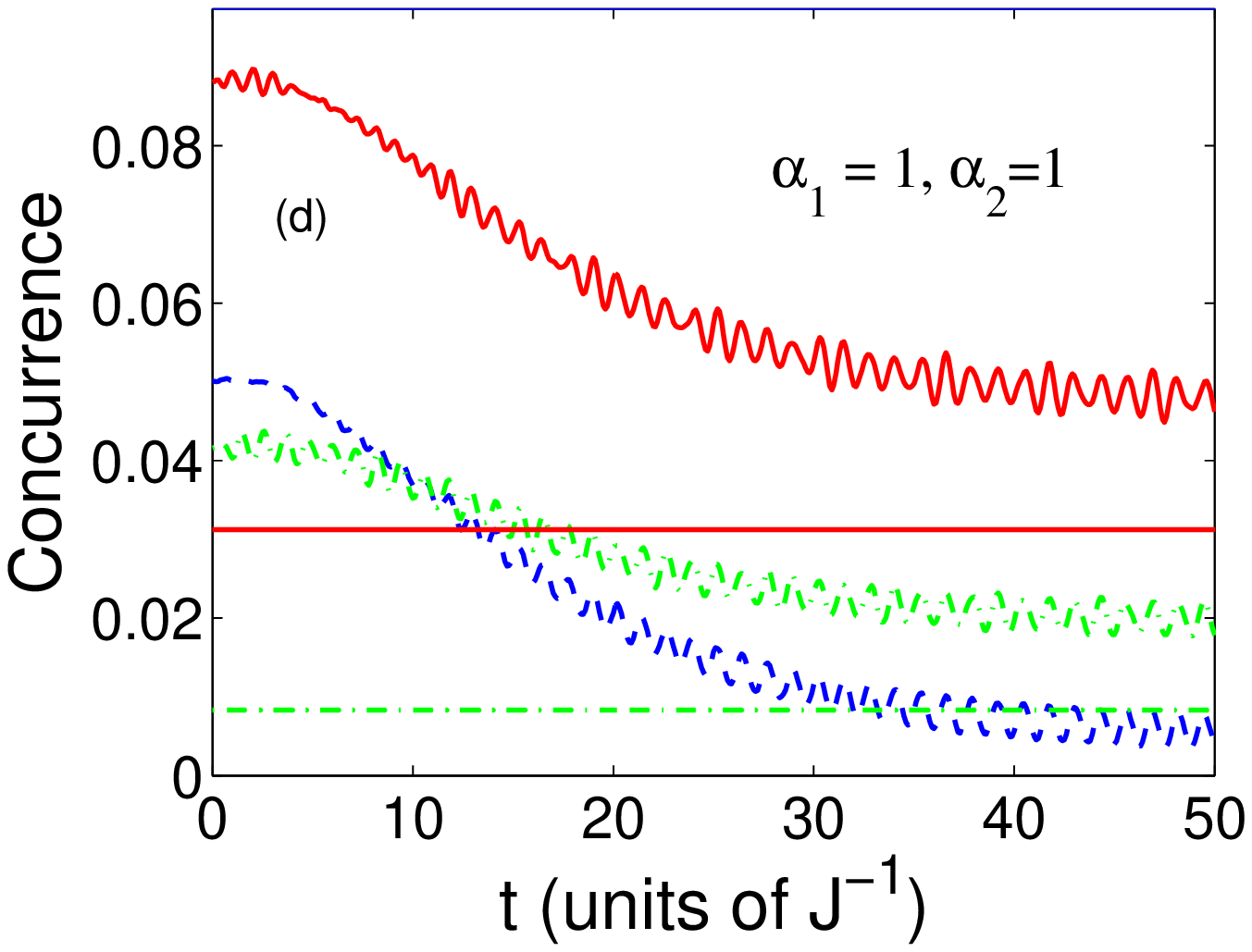}}
   \caption{{\protect\footnotesize (Color online) Dynamics of the concurrence $C(1,2)$, $C(1,4)$, $C(5,7)$ with double impurities at sites 1 and 4 for the two dimensional partially anisotropic lattice ($\gamma = 0.5$) in an exponential magnetic field where a=1, b=2, and w=0.1. The straight lines represent the equilibrium concurrences corresponding to constant magnetic field $h=2$. The legend for all subfigures is as shown in subfigure (a).}}
 \label{Imp14_Dyn_G05}
 \end{minipage}
\end{figure}
\begin{figure}[htbp]
\begin{minipage}[c]{\textwidth}
\centering
   \subfigure{\includegraphics[width=8 cm]{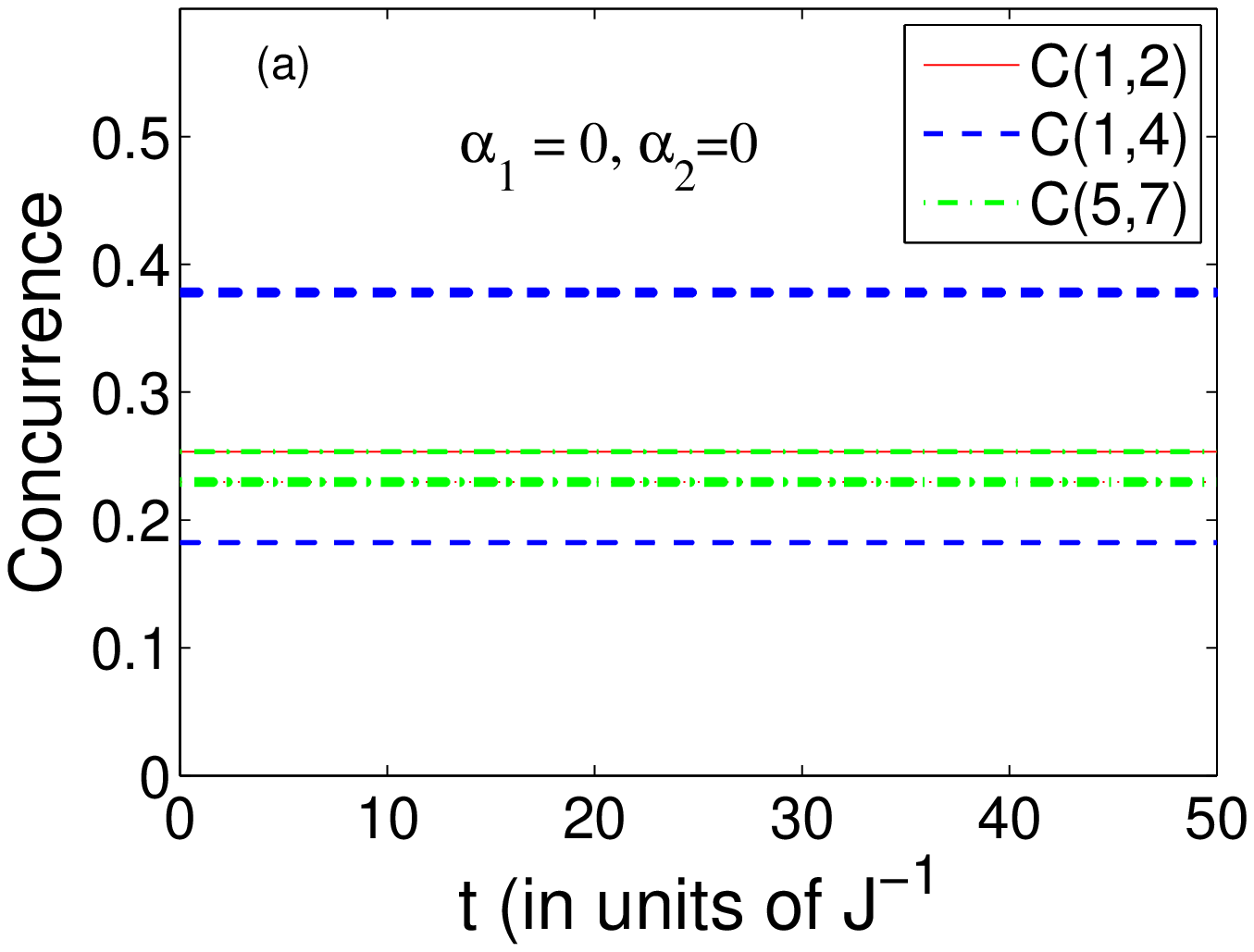}}\quad
   \subfigure{\includegraphics[width=8 cm]{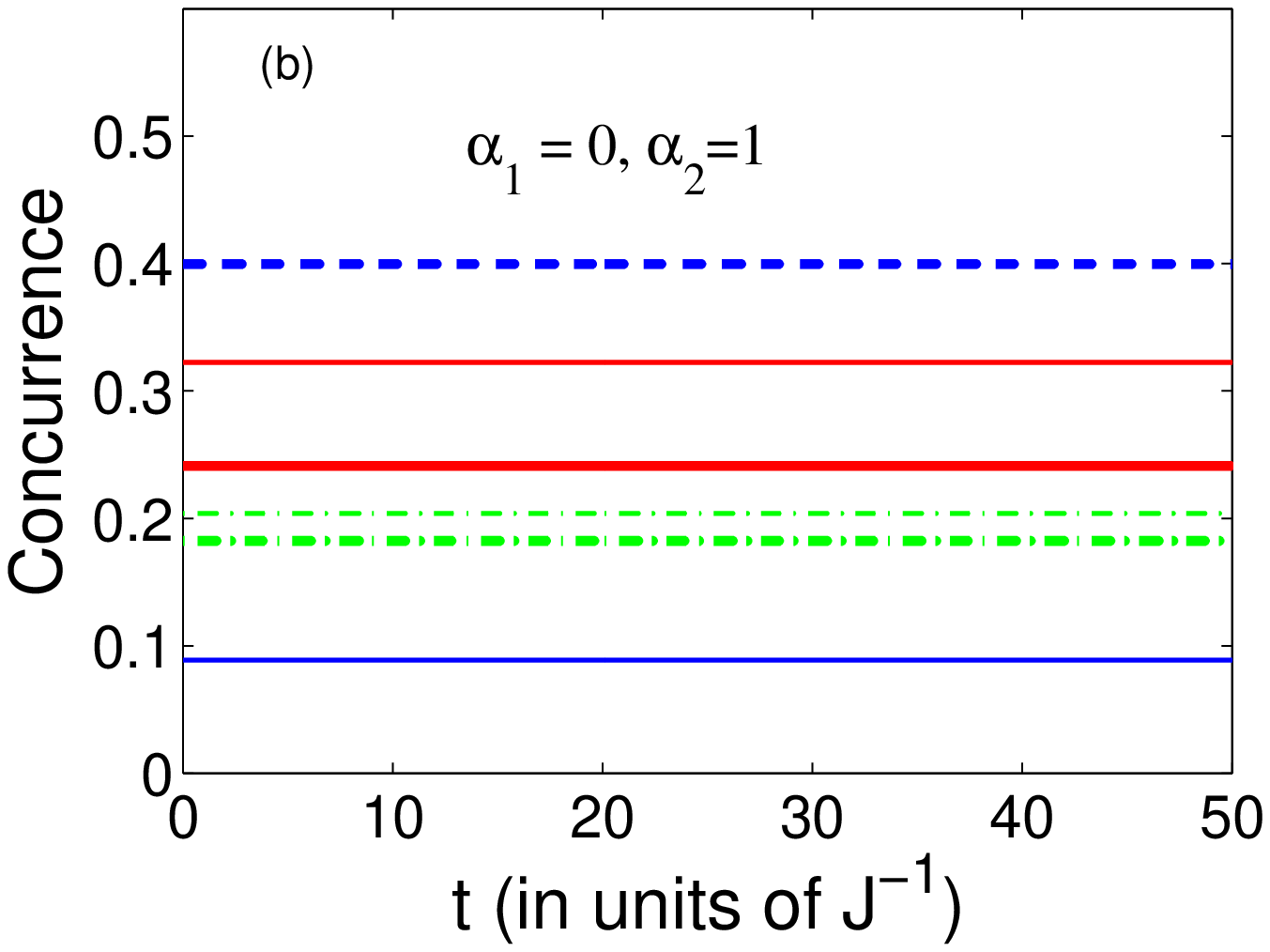}}\\
   \subfigure{\includegraphics[width=8 cm]{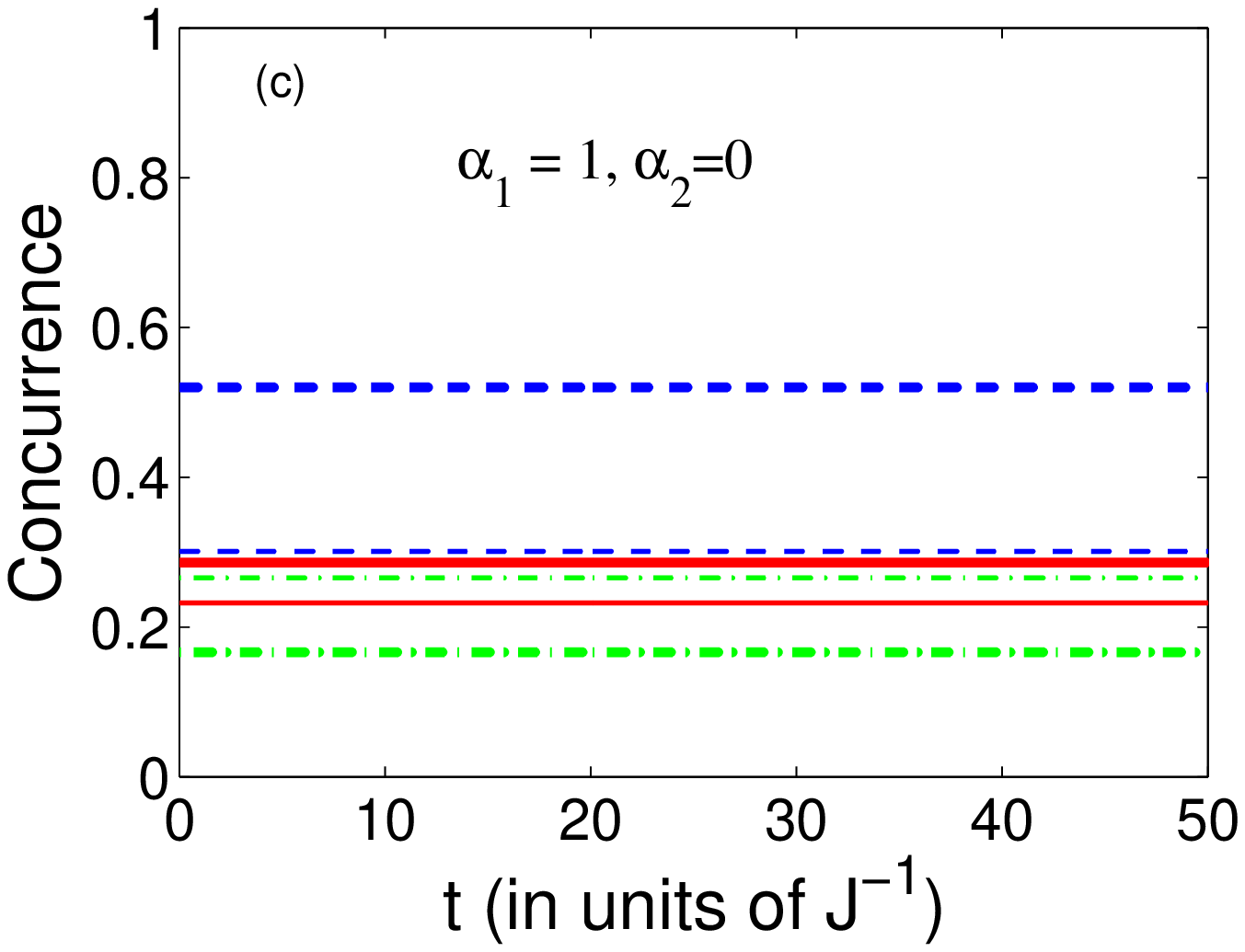}}\quad
   \subfigure{\includegraphics[width=8 cm]{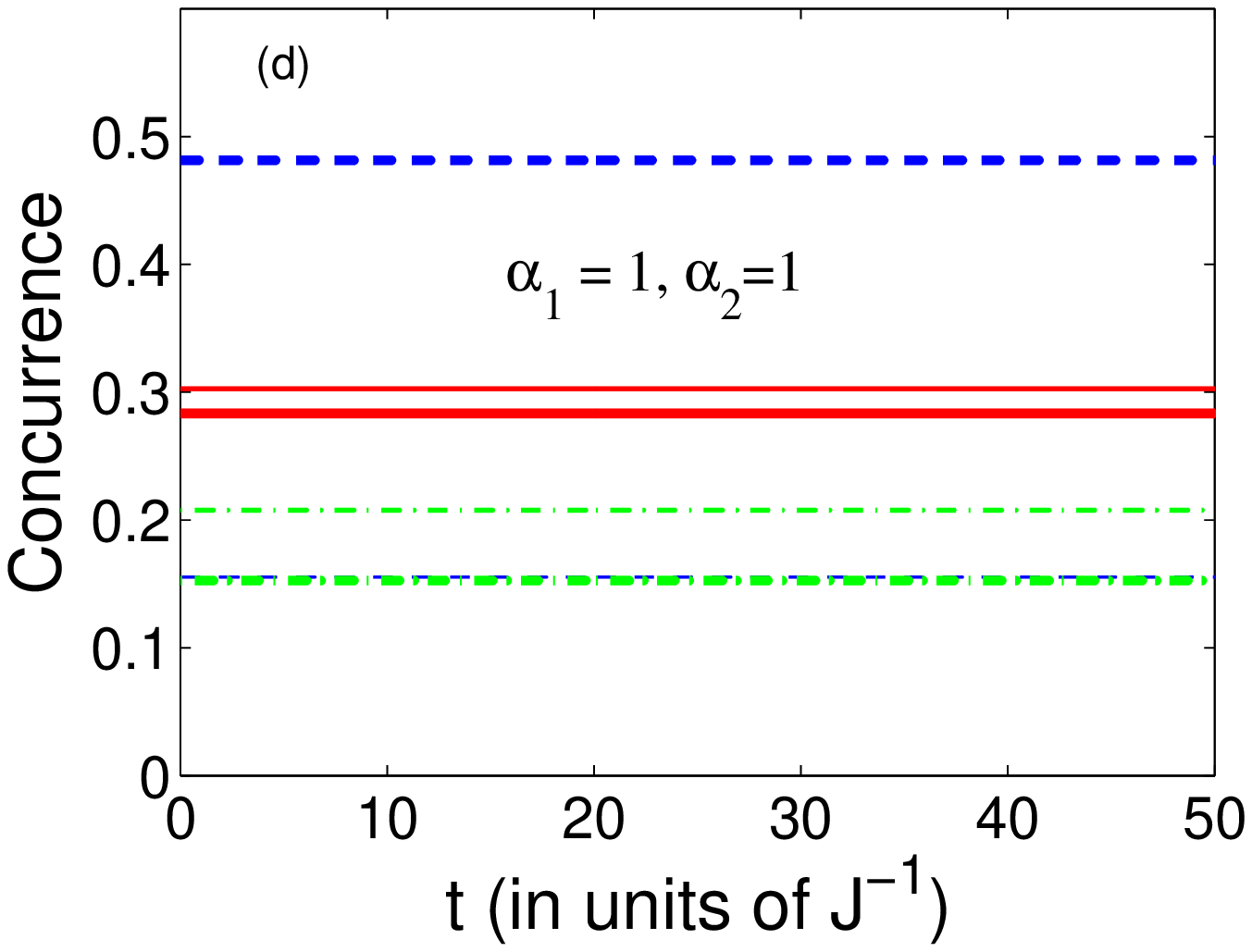}}
   \caption{{\protect\footnotesize (Color online) Dynamics of the concurrence $C(1,2)$, $C(1,4)$, $C(5,7)$ with double impurities at sites 1 and 4 for the two dimensional isotropic XY lattice ($\gamma = 0$) in an exponential magnetic field where a=1, b=1.8, and w=0.1. The straight (thicker) lines represent the equilibrium concurrences corresponding to constant magnetic field $h=1.8$. The legend for all subfigures is as shown in subfigure (a).}}
 \label{Imp14_Dyn_G0}
 \end{minipage}
\end{figure}
The partially anisotropic system under the same condition behaves differently. Its pure case, $\alpha_1,\,\alpha_2 = (0,1)$ case and $\alpha_1,\,\alpha_2 = (1,1)$ case are all nonergodic as dipicted in figs.~\ref{Imp12_Dyn_G05}(a) and \ref{Imp12_Dyn_G05}(b). Nevertheless, the system with impurity strengths $\alpha_1=1$ and $\alpha_2 = 0$ show ergodic behavior which means that the nonergodicity of the partially anisotropic system is sensitive for the strength and location of impurities. The isotropic system is explored in fig.~\ref{Imp12_Dyn_G0} which behaves nonergodically for all impurity strengths. Testing the effect of the impurity location we consider the same system with impurities at the sites 1 and 4. While the Ising system shows ergodicty at all impurity strengths as shown in fig.~\ref{Imp12_Dyn_G1}, the partially and isotropic $XY$ systems are nonergodic at the different impurity strengths as plotted in figs.~\ref{Imp14_Dyn_G05} and ~\ref{Imp14_Dyn_G0} respectively.  
\section{Conclusion and future directions}
We have investigated the nearest neighbor entanglement and ergodicity of a two-dimensional $XY$ spin lattice in an external magnetic field $h$. The spins are coupled to each other through nearest neighbor exchange interaction $J$. The number of spins in the lattice are 7 where we may consider one or two of them as impurities. We have found that the  completely anisotropic (the Ising), the partially anisotropic and isotropic systems behave in a very similar fashion to that of the one dimensional spin systems at the extreme, small and large, values of the parameter $\lambda=h/J$ but may deviate at the intermediate values. The first two systems show phase transition in the vicinity of the parameter critical value $\lambda=2$ and their entanglement vanishes as $\lambda$ increases. The entanglement of the isotropic system changes in a sharp step profile before suddenly vanishing in the vicinity of $\lambda=2$. 
The entanglement dynamics of the system with impurities was investigated under the effect of an external time-dependent magnetic field of exponential form. It was found that the ergodicity of the system can be tuned using the strength and location of the impurities as well as the degree of anisotropy of the coupling between the spins. It is interesting in future to investigate the same systems coupled to a dissipative environment and to examine the effect of impurity to tune the decoherence in the spin system and to investigate the ergodicity status under coupling to the environment. Furthermore we would like to investigate the same system with larger number of sites to test the system size effect and to clarify the critical value of the parameter $\lambda$ using finite size scaling \cite{Kais2003,Kais2007}. Previously, the 19 sites triangular static Ising lattice was treated exactly using the the trace minimization algorithm \cite{XuQ2011}. The dynamics of entanglement in the 19-site XY system is currently under consideration, and by taking advantage of parallel computing we can reach 34 spins by far.

\section*{Acknowledgments}
We are grateful to the Saudi NPST for support (project no. 11-MAT1492-02) and the deanship of scientific research, King Saud University. We are also grateful to the USA Army research office for partial support of this work at Purdue.

\end{document}